\documentclass[twocolumn,times]{aastex631}

\usepackage{amsmath}
\usepackage{appendix}

\shorttitle{Model for $r$-Process Production Patterns}
\shortauthors{Li et al.}

\begin{document}

\title{A Data-Driven Model for $r$-Process Production Patterns}
\author[0009-0006-5753-665X]{Chen-Qi Li}
\affiliation{School of Physics and Astronomy,
      University of Minnesota, Minneapolis, MN 55455, USA}
\author[0000-0002-3146-2668]{Yong-Zhong Qian}
\affiliation{School of Physics and Astronomy,
      University of Minnesota, Minneapolis, MN 55455, USA}
\author[0000-0002-7893-4183]{Axel Gross}
\affiliation{Theoretical Division and Center for Theoretical Astrophysics,
      Los Alamos National Laboratory, NM 87545, USA}
\author[0000-0002-2385-6771]{Zewei Xiong}
\affiliation{GSI Helmholtzzentrum f{\"u}r Schwerionenforschung, Planckstraße 1, D-64291 Darmstadt, Germany}  

\begin{abstract}
Using the elemental abundances in 68 metal-poor (MP) stars, we present a data-driven model for $r$-process production patterns covering Sr to U. We show that essentially all the $r$-process patterns in those and other test stars can be adequately explained as mixtures of Patterns 1 and 2, which provides theoretical insights into the empirical categories of limited-$r$, $r$-I, and $r$-II stars. We carry out an extensive survey of the yield templates produced by parametric $r$-process calculations. We propose that Pattern 2 may be dominated by a single template and points to regularity in $r$-process production by a subset of neutron star mergers (NSMs), while Pattern 1 is the average superposition of multiple templates and reflects production by other NSMs and perhaps also some magneto-rotational supernovae. We raise possible systematic issues with abundance ratios for elements measured in different ionization states, and highlight the need for examining the Os, Ir, and Pt measurements for HD~122563, which is dominated by Pattern 1 with a prominent Pt peak in our model. If this result is confirmed, the meaning of the limited-$r$ category requires drastic revision. Further measurements of a wider range of $r$-process elements in a larger sample of MP stars are critical to test and improve our model.
\end{abstract}

%%%%%%%%%%%%%%%%%%%%%%%%%%%%%%%%%%%%%%%%%%%%%%%%%%%%%%%%%%%%%%%%%%%%%%%%%%%%%%%%
\section{Introduction}
\label{sec:intro}

It is well known that Type Ia (SNe Ia) and core-collapse supernovae (CCSNe) are major sources for Fe, 
that elements heavier than the Fe group are mainly produced by the rapid ($r$) and slow ($s$) neutron-capture processes,
and that asymptotic giant branch (AGB) stars of low to intermediate masses are the site of the main $s$-process producing Sr 
and heavier elements (see e.g., \citealt{arcones2023} for a review). 
The spectacular multi-messenger observations of GW170817 \citep{abbott2017} provided strong support of binary neutron star mergers 
being a site of the $r$-process (e.g., \citealt{kasen2017}). Many theoretical studies have been devoted to this topic both before 
(see e.g., \citealt{thielemann2017} for a review) and after this event (e.g., \citealt{curtis2023,just2023,kiuchi2023}).
Similarly, black-hole--neutron-star mergers have also been studied as an $r$-process site (e.g., \citealt{2024PhRvL.133x1201W}).
For simplicity, we refer to the above two $r$-process sites as neutron star mergers (NSMs).
Theoretical studies also suggest that CCSNe may produce some elements heavier than the Fe group 
(e.g., \citealt{woosley1992,hoffman1997,wanajo2018,wang2023}) and 
that a subset of them may even be an $r$-process site (e.g., \citealt{nishimura2015,siegel2019,fischer2020,2023MNRAS.518.1557R}). 

Despite the above advances, we are still far from being able to make 
precise predictions for the nucleosynthesis of astrophysical sources. In particular, the extreme conditions 
in the dynamic environments of CCSNe and NSMs are inherently difficult to simulate, and there are large uncertainties in the nuclear 
input for simulating these sources and the associated $r$-process. On the other hand, because these two types of sources are associated with rapidly-evolving massive stars,
they are expected to have dominated the chemical evolution of the universe during the first $\sim 1$ Gyr, before Fe contributions from SNe Ia and $s$-process contributions from AGB
stars became significant. Consequently, metal-poor (MP) stars formed during this early epoch provide an excellent fossil record for deciphering the nucleosynthesis of CCSNe 
and NSMs. In particular, dedicated observational studies of $r$-process elements in MP stars have greatly improved our understanding of the $r$-process (see e.g., \citealt{2007PhR...442..237Q,2021RvMP...93a5002C} for reviews). Among the key findings, \cite{2000ApJ...533L.139S} showed that while Ba and heavier elements in CS~22892-052 closely follow the solar $r$-process pattern, the lighter elements Sr to Cd exhibit significant deviations. This result supports the earlier suggestion of \cite{1996ApJ...466L.109W} based on meteoritic data that multiple sources with distinct production patterns contributed to the solar $r$-process abundances. Further support is provided by observations of HD~122563 \citep{2006ApJ...643.1180H} and HD~88609 \citep{2007ApJ...666.1189H}. The abundances in these two MP stars are very similar, and those of Sr to Ag are greatly elevated relative to Ba and nearby heavier elements when compared to the solar $r$-process pattern.

While there is a general consensus on multiple $r$-process sources with distinct production patterns, robust identification of these patterns and the underlying sources from observations of MP stars is not straightforward. \cite{qian2001,2007PhR...442..237Q,qian2008} simply took the observed elemental abundance patterns in CS~22892-052 and HD~122563 as the production patterns of two sources without and with Fe production, respectively. They showed that the data on $r$-process elements in other MP stars could be largely explained as mixtures of those two patterns, which they attributed to low-mass and regular CCSNe with little and significant production of Fe, respectively. In hindsight, the patterns without and with Fe co-production are more likely associated with NSMs and CCSNe, respectively. The approach of \cite{qian2001,2007PhR...442..237Q,qian2008} was refined by \cite{2014ApJ...797..123H}, who obtained similar results by directly fitting the $r$-process abundance patterns in MP stars as mixtures of two production patterns based on CS~22892-052 and HD~122563. The latter authors also explored conditions in the neutrino-driven winds in CCSNe for producing the pattern dominated by the light $r$-process elements Sr to Ag. Further explorations of such conditions for explaining the patterns of these elements in MP stars were carried out by \cite{2022ApJ...935...27P,2024ApJ...966...11P}. A somewhat different view of the MP star data was taken by \cite{2022ApJ...936...84R}, who argued that just as Ba and heavier elements appear to follow the solar $r$-process pattern, the lighter elements Se, Sr, Y, Zr, Nb, Mo, and Te follow a different fixed pattern. From data in the literature, \cite{2023Sci...382.1177R} derived two baseline patterns for Se to Te and for Ba to Pt, respectively. They further argued that deviations from these two patterns, specifically in the regions of Ru to Ag and Eu to Pt, can be attributed to fission.

Using numerous MP stars in the literature [collected in the SAGA \citep{2008PASJ...60.1159S} and JINA \citep{2018ApJS..238...36A} databases], \cite{2022A&A...663A..70F} conducted a statistical study to determine if the abundance of Fe is correlated with those of other elements, especially the $r$-process elements. Based on the presence or absence of this correlation, they proposed up to five categories of astrophysical sources for the $r$-process, which include NSMs and a variety of CCSNe.
\cite{2025EPJA...61..207F} extended the correlation study to cover a wider range of elements and reached similar conclusions to \cite{2022A&A...663A..70F}.

Using the data on [Sr/Fe], [Ba/Fe], and [Eu/Fe] \citep{holmbeck2020} provided by the $R$-Process Alliance (RPA) search for $r$-process-enhanced stars in the Galactic halo, \cite{2023arXiv230909385G} demonstrated as a proof of concept that these data can be explained by mixtures of two production patterns: one with dominant production of Fe and Sr, which can be attributed to CCSNe, and the other with dominant production of Sr, Ba, and Eu, which can be attributed to NSMs. The above RPA data cover $\sim 200$ stars but only three $r$-process elements, so they provide very limited information on the $r$-process production patterns. 

In this paper, we propose a new data-driven approach to estimate the production patterns of $r$-process sources. In contrast to \cite{qian2001,2007PhR...442..237Q,qian2008}, \cite{2022A&A...663A..70F,2025EPJA...61..207F}, and \cite{2023arXiv230909385G}, we focus on the $r$-process elements exclusively. We assume that sources with similar astrophysical conditions produce an approximately fixed $r$-process pattern, at least in the average sense. This assumption is less likely to hold for production patterns including Fe because it is dominantly produced with very different nuclear systematics and astrophysical conditions from the $r$-process elements. While we lose some information on the underlying sources by ignoring Fe, the $r$-process production patterns inferred from data can still provide valuable insights into the astrophysical conditions in the ejecta undergoing the $r$-process, thereby shedding light on the sources.  

Our approach is greatly facilitated by the recent Chemical Evolution of $R$-process Elements in Stars (CERES) survey, which provided the data on Sr, Y, Zr \citep{2022A&A...665A..10L}, Ba to Eu \citep{2025A&A...693A.293L}, Hf, Os, Ir, and Pt \citep{2025A&A...693A.294A} for a homogeneously-analyzed sample of 52 MP stars. For statistical inference of $r$-process production patterns, such a sample has a clear advantage over a mixed ensemble of stars analyzed with non-uniform methods in the literature. However, elements such as Nb to Te and Gd to Re were not measured by CERES, so we have to use the data on these elements in the literature. We select 20 MP stars with a wide range of $r$-process elements measured, and check how these additional data affect the inferred $r$-process production patterns. We also test our inferred patterns using data outside the CERES and literature sample. In particular, we show that the recently published RPA data on a wide range of $r$-process elements in 10 MP stars \citep{2025A&A...704A.282R} can be accounted for by mixtures of our inferred patterns, just like those stars in the combined CERES and literature sample used to derive these patterns. Mixtures of these patterns also provide theoretical insights into the empirical categories of limited-$r$, $r$-I, and $r$-II stars \citep{2005ARA&A..43..531B,2018ARNPS..68..237F}.

Our main results are as follows. We derive Patterns 1 and 2 for $r$-process production from the data on the combined CERES and literature sample. Both these patterns cover the elements Sr to U, including those in the second (Te) and third (Pt) peak of the solar $r$-process pattern. Compared to Pattern 2, Patten 1 has much higher production of the light $r$-process elements relative to the Eu region. Pattern 2 is rather stable whether Eu or Y is used as the reference element. While Pattern 1 is also stable for the light $r$-process and the second and third peak elements, it is quite uncertain between the second and third peak. We carry out an extensive survey of the yield templates produced by parametric $r$-process calculations.  We propose that Pattern 2 may be dominated by a single template and points to regularity in $r$-process production by a subset of NSMs, while Pattern 1 is the average superposition of multiple templates and reflects production by other NSMs and perhaps also some magneto-rotational SNe.

We describe our approach to derive Patterns 1 and 2 from the CERES and literature data, and present the detailed results including tests by external data in \S\ref{sec:approach}. We present our parametric $r$-process calculations and explore the astrophysical conditions that might give rise to these patterns in \S\ref{sec:conditions}. We discuss our results and give conclusions in \S\ref{sec:discussion}. To avoid information overload in the main text, we present technical details of our approach including validation and other supporting results in Appendices~\ref{append:optimization}--\ref{append:cond}.

\section{Data-Driven Approach}
\label{sec:approach}

In this section, we derive two distinct $r$-process production patterns from data on MP stars, and present various results including tests by external data not used in the derivation. To simplify the presentation, we mention the pertinent information in the main text and put detailed description and discussion in the appendices. In particular, the choice of two patterns is mainly due to the relatively small number of MP stars used, and is justified by comparison with the results for three patterns in Appendix~\ref{append:3pattern} and by cross validation in Appendix~\ref{append:validation} (see also \S\S\ref{sec:uncertainty}--\ref{sec:2vs3}).

We define the patterns by number ratios of other elements relative to a reference element. We present results for two reference elements, Y and Eu, which have the most measurements in the CERES sample. Because these two elements correspond to the light end and the middle of the nuclear mass range of interest, comparison of the derived patterns with respect to them would provide a consistency test of our approach. In addition, because the solar abundances of Y and Eu are dominated by $s$-process and $r$-process contributions, respectively, consistency between the derived patterns would also confirm the $r$-process origin of the elements in those MP stars used in our approach.

Using Eu as the example reference element, we assume that the ratio of an $r$-process element E to Eu in an MP star is given by
\begin{align}
    ({\rm E/Eu})=x_{\rm Eu}({\rm E/Eu})_1+(1-x_{\rm Eu})({\rm E/Eu})_2,
    \label{eq:mix}
\end{align}
where (E/Eu)$_1$ and (E/Eu)$_2$ characterize the two production patterns, and $x_{\rm Eu}$ is the fraction of Eu contributed by Pattern 1. Observations give $\log\epsilon ({\rm E})=\log({\rm E/H})+12$, so it is more convenient to work with $\log({\rm E/Eu})=\log\epsilon ({\rm E})-\log\epsilon ({\rm Eu})$.

For a sample of $n$ stars each with $k$ elements (including Eu) measured, Eq.~(\ref{eq:mix}) provides $n(k-1)$ constraints, from which $n+2(k-1)$ parameters [$x_{\rm Eu}$ for each star plus (E/Eu)$_1$ and (E/Eu)$_2$ for each element other than Eu] can be optimized. The advantage of this approach to infer the $r$-process production patterns is that it addresses the complex problem of mixed contributions from different sources by deriving the optimal mixing parameter $x_{\rm Eu}$ for each star directly from the data. Clearly, the power of this approach depends on the excess of the number of constraints $n(k-1)$ over the number of parameters $n+2(k-1)$. While we focus on the case of two distinct production patterns here due to the limited data available now, it is straightforward to generalize the approach to more than two patterns (Appendix~\ref{append:3pattern}).

We optimize the parameters by minimizing
\begin{equation}
 Q=\sum_{j,{\rm E}} H\left(\frac{\delta\log({\rm E/Eu})_{*,j}}{\sigma_{\log({\rm E/Eu})_{*,j}}}\right),
 \label{eq:q}
\end{equation}
where $\delta\log({\rm E/Eu})_{*,j}=\log({\rm E/Eu})_{*,j}-\log({\rm E/Eu})^{\rm obs}_{*,j}$, $\log({\rm E/Eu})_{*,j}$ and $\log({\rm E/Eu})^{\rm obs}_{*,j}$ refer to the predicted value from Eq.~(\ref{eq:mix}) and the observed value for the $j$th star, respectively, $\sigma_{\log({\rm E/Eu})_{*,j}}$ is the measurement uncertainty in $\log({\rm E/Eu})^{\rm obs}_{*,j}$, and $H(y)$ is the Huber loss function defined by $H(y)=y^2/2$ for $|y|\leq 1$ and $H(y)=|y|-1/2$ for $|y|>1$. The form of $H(y)$ for $|y|>1$ reduces sensitivity to the outliers. In general, the number of measured elements varies from star to star. This nonuniformity is accommodated by summing over only the measured elements in each star to obtain $Q$ in Eq.~(\ref{eq:q}). Then the total number of constraints provided by Eq.~(\ref{eq:mix}) is $N=\sum_{\rm E}n_{\rm E}$, where $n_{\rm E}$ is the number of stars with E and Eu measurements. So long as $N$ is sufficiently large, the optimization procedure provides meaningful parameters $\log({\rm E/Eu})_1$ and $\log({\rm E/Eu})_2$ for Patterns 1 and 2, respectively, and the corresponding mixing parameters $x_{\rm Eu}$ for individual stars. More details on the procedure are given in Appendix~\ref{append:optimization}.

\subsection{Results for the CERES Data}
\begin{figure*}[htbp]
    \centering
    \includegraphics[width=1\textwidth]{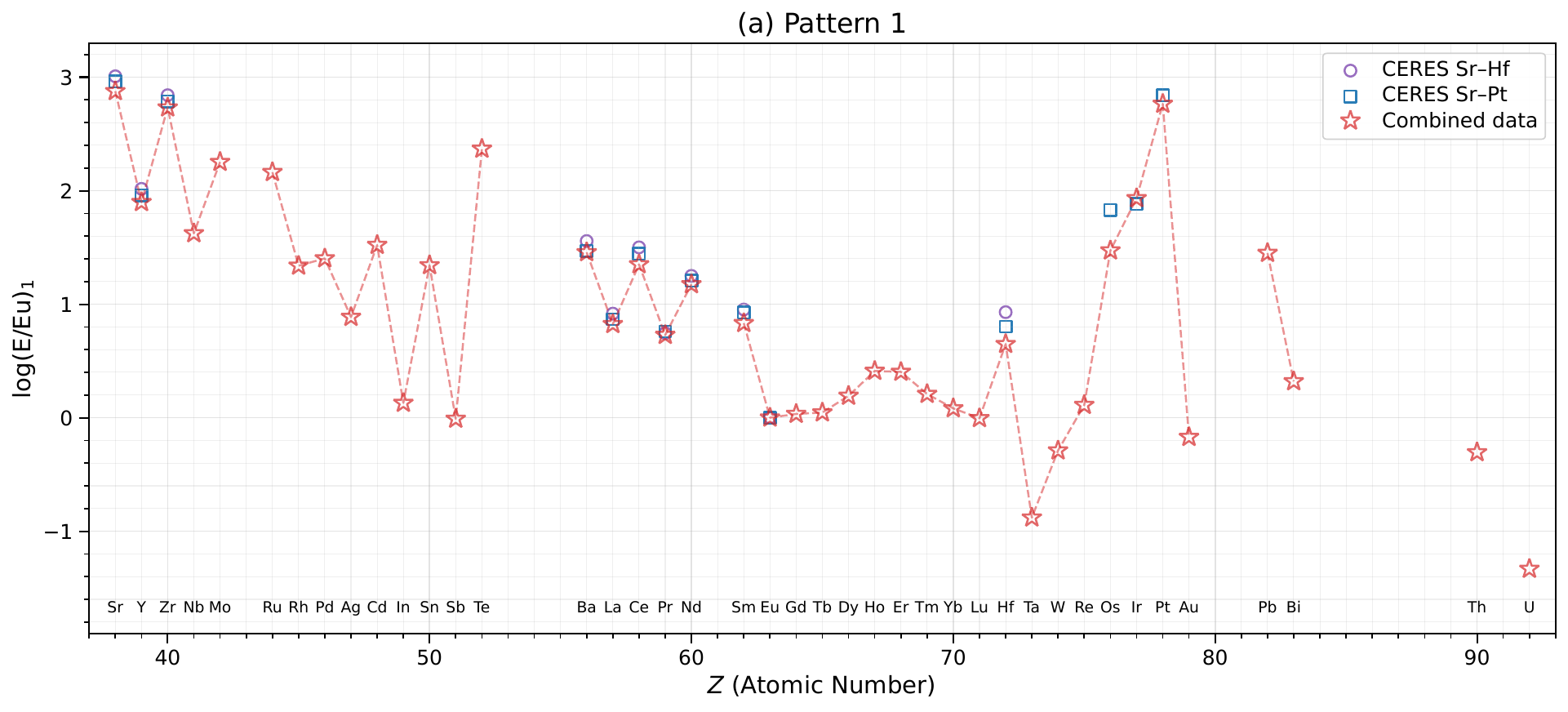}\\[1em]  
    \includegraphics[width=1\textwidth]{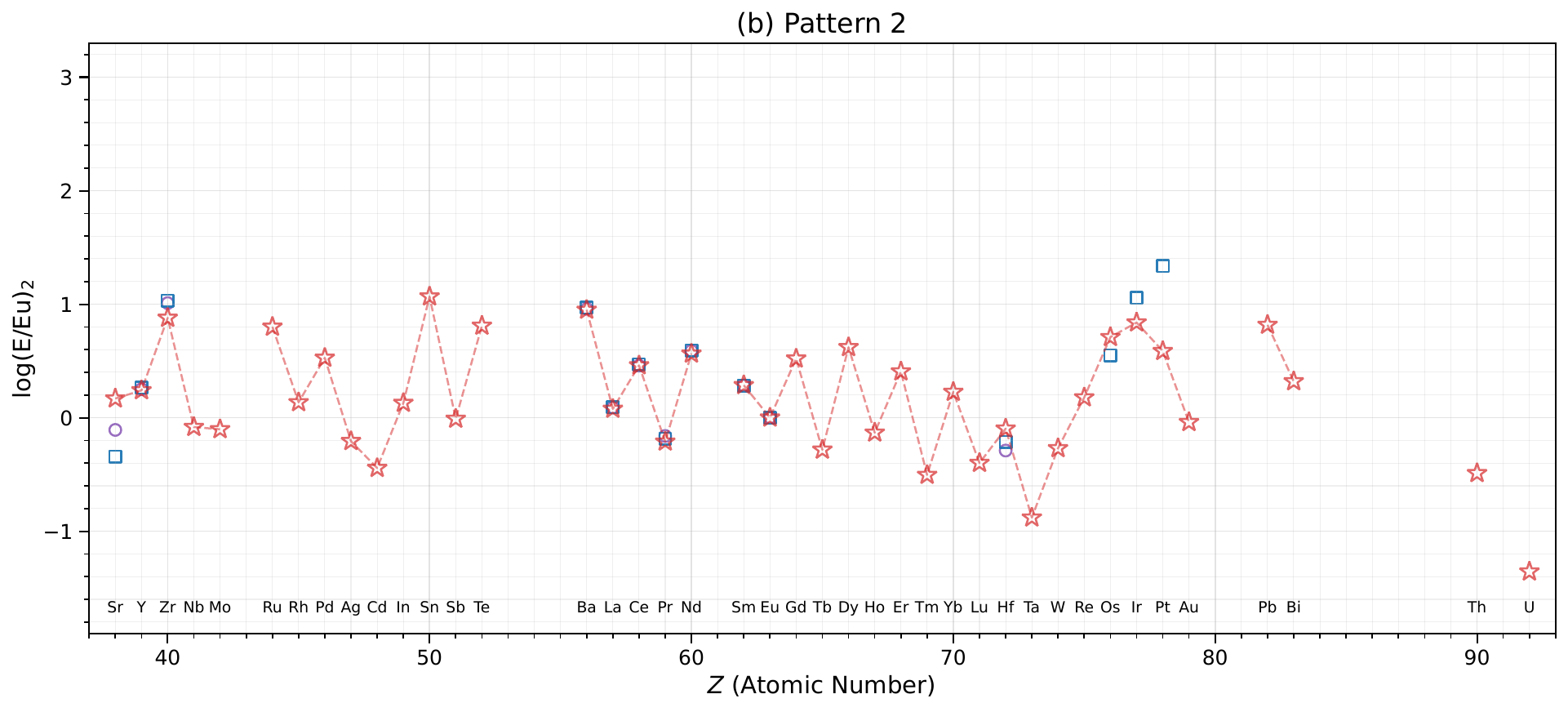}         
    \caption{Comparison of Patterns 1 and 2 for $r$-process production derived from the CERES data on Sr, Y, Zr, Ba to Eu, and Hf (violet circles), the CERES data including Os, Ir, and Pt (light blue squares), and the combined CERES and literature data (red stars).}
    \label{fig:pat}
\end{figure*}

We first apply our approach to the CERES data on Sr, Y, Zr \citep{2022A&A...665A..10L}, Ba to Eu \citep{2025A&A...693A.293L}, and Hf \citep{2025A&A...693A.294A}. These elements were all measured with lines of singly-ionized species and the corresponding ratios $\log(\rm E/Eu)$ are less affected by systematic uncertainties in the atmospheric model. We take the uncertainty in the measured $\log(\rm E/Eu)$ to be $\sigma({\rm E/Eu})=\sqrt{\sigma_{\rm E}^2+\sigma_{\rm Eu}^2}$\,, where $\sigma_{\rm E}$ is the line-to-line scatter in $\log\epsilon(\rm E)$. Because all the stars with E and Eu measurements have similar $\sigma({\rm E/Eu})$, we take the average and use it as $\sigma_{\log({\rm E/Eu})_{*,j}}$ in Eq.~(\ref{eq:q}). These average uncertainties are given in Table~\ref{tab:ceres}. The optimal parameters $\log({\rm E/Eu})_1$ and $\log({\rm E/Eu})_2$ are shown as violet circles in Fig.~\ref{fig:pat}.

In contrast to Eu, which was measured with lines of singly-ionized species, the elements Os, Ir, and Pt were measured with lines of neutral species \citep{2025A&A...693A.294A}. Consequently, the corresponding ratios $\log({\rm E/Eu})$ are expected to be more affected by the systematic uncertainties in the atmospheric model. We calculate the uncertainty in the measured $\log(\rm E/Eu)$ similarly to the other elements, but using the full uncertainty in $\log\epsilon({\rm E})$ for Os, Ir, and Pt. The resulting average uncertainties in the measured $\log(\rm E/Eu)$ are also given in Table~\ref{tab:ceres}. The optimal parameters $\log({\rm E/Eu})_1$ and $\log({\rm E/Eu})_2$ with inclusion of Os, Ir, and Pt are shown as light blue squares in Fig.~\ref{fig:pat}. It can be seen that this inclusion has little effect on the other elements in Patterns 1 and 2.

Using Patterns 1 and 2 including Os, Ir, and Pt, along with the corresponding mixing parameters $x_{\rm Eu}$ for stars in the CERES sample, we calculate the difference $\delta\log({\rm E/Eu})$ between the predicted and measured value for each element as a function of the measured [Fe/H]~$=\log({\rm Fe/H})-\log({\rm Fe/H})_\odot$ and show the results as light blue squares in Figs.~\ref{fig:residual1}--\ref{fig:residual4}. It can be seen that the mixtures of Patterns 1 and 2 provide good fits to the CERES data. For all the elements, the $\delta\log({\rm E/Eu})$ values are within the average uncertainties $\sigma_{\log(\rm E/Eu)}$ for the majority of the stars. In addition, there is no clear evolution of $\delta\log({\rm E/Eu})$ with [Fe/H].

\subsection{Results with Additional Literature Data}
\label{sec:lit}
Many $r$-process elements are not covered by the CERES data. To obtain more complete $r$-process production patterns, we select 20 MP stars with additional measured elements from the literature. These stars are CS~22892-052 \citep{2003ApJ...591..936S}, HD~88609 \citep{2007ApJ...666.1189H}, HD~122563, HD~126238 \citep{2012ApJS..203...27R}, CS~31082-001 \citep{2013A&A...550A.122S}, 2MASS~J15213995-3538094 \citep{2020ApJ...898...40C}, BD~17 3248, HD~108317, HD~160617, HD~84937, HD~19445, HD~128279, HD~140283 \citep{2022ApJ...936...84R}, HD~222925 \citep{2022ApJS..260...27R}, 2MASS~J00512646-1053170 \citep{2024MNRAS.529.1917S}, 2MASS~J20313531-3127319, 2MASS~J21402305-1227035, 2MASS~J00385967+2725516 \citep{2024A&A...688A.123X}, 2MASS~J22132050-5137385 \citep{2024ApJ...971..158R}, and 2MASS~J05383296-5904280 \citep{2025A&A...697A.127H}.

Among the above stars, HD~122563, HD~108317, and HD~128279 were also included in the CERES survey as CES1402+0941, CES1226+0518, and CES1436-2906, respectively. We have checked that when the $\log({\rm E/Eu})$ values are available from both the CERES survey and literature, they are in fair to good agreement except for Pt in HD~108317. For this star, the CERES survey gave $\log\epsilon({\rm Eu})=-1.19$ \citep{2025A&A...693A.293L} and $\log\epsilon({\rm Pt})=0.66$ \citep{2025A&A...693A.294A}, while \cite{2022ApJ...936...84R} gave $\log\epsilon({\rm Eu})=-1.37$ and $\log\epsilon({\rm Pt})=-0.40$. The discrepancy of 0.88 dex in $\log({\rm Pt/Eu})$ is mainly due to the large discrepancy in $\log\epsilon({\rm Pt})$, which cannot be easily understood. We exclude the Pt data for HD~108317 from the combined CERES and literature sample. In addition, we only use the literature data for the above three stars for consistency in treatment of all their measured elements. 

For uncertainties in $\log({\rm E/Eu})$ from the literature, we follow the suggestion of \cite{2022ApJS..260...27R} and use the uncertainties in [E/Fe] when available, and otherwise adopt the same treatment as for the CERES sample. The average uncertainties $\sigma_{\log({\rm E/Eu})}$ for the literature data are given in Table~\ref{tab:lit}. The combined CERES and literature sample contains 68 MP stars with $\log({\rm E/Eu})$ data for a varying range of elements (Tables~\ref{tab:ceres} and \ref{tab:lit}). The optimal parameters $\log({\rm E/Eu})_1$ and $\log({\rm E/Eu})_2$ derived from this sample are given in Table~\ref{tab:pat} and shown as red stars in Fig.~\ref{fig:pat}. 

It can be seen from Fig.~\ref{fig:pat} that for the elements common to both the CERES (light blue squares) and combined sample, the results are essentially the same with several exceptions. The most prominent exception is Pt in Pattern 2, with $\log({\rm Pt/Eu})_2$ from the CERES sample being $\approx 0.75$ dex higher than that from the combined sample. This shift may be another indication that there are large systematic uncertainties in Pt measurements, as exhibited by the large discrepancy in $\log\epsilon({\rm Pt})$ for HD~108317 between the CERES and literature data. Note that we have excluded the Pt data for this star from the combined sample. The above shift in $\log({\rm Pt/Eu})_2$ suggests that there might be large systematic uncertainties in Pt measurements across the combined sample. We urge that these measurements be reexamined to provide reliable Pt data for further studies of the $r$-process production patterns. There is also a large shift of $\approx 0.51$ dex in $\log({\rm Sr/Eu})_2$. However, we note that when $({\rm E/Eu})_1$ greatly exceeds $({\rm E/Eu})_2$ as in the case of Sr, $({\rm E/Eu})_2$ is likely obtained as the small difference between two large numbers [Eq.~(\ref{eq:mix})], and therefore, may be susceptible to large uncertainties. In contrast to Pattern 2, the only significant shift in Pattern 1 is for Os, with $\log({\rm Os/Eu})_1$ from the CERES sample being $\approx 0.36$ dex higher than that from the combined sample.

\begin{figure*}[htbp]
    \centering
    \includegraphics[width=1\textwidth]{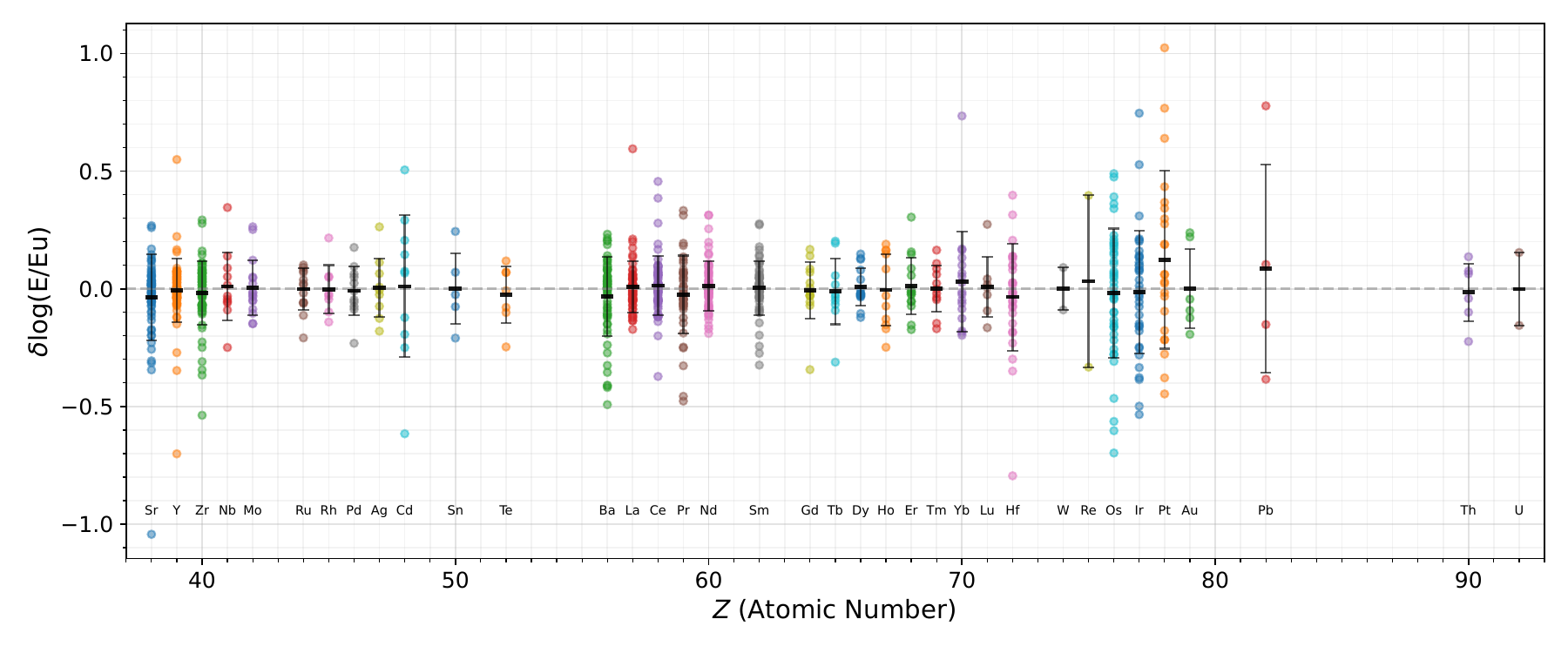}\\[1em]          
    \caption{The differences $\delta\log({\rm E/Eu})$ between the predicted and measured values for the combined sample. For each element, the thick horizontal bar indicates the mean $\delta\log({\rm E/Eu})$ across the sample and the error bar represents the $1\sigma$ scatter around the mean.}
    \label{fig:residual_all}
\end{figure*}

Using Patterns 1 and 2 (red stars in Fig.~\ref{fig:pat}) along with the corresponding mixing parameters $x_{\rm Eu}$ for stars in the combined sample, we calculate the difference $\delta\log({\rm E/Eu})$ between the predicted and measured value for each element as a function of the measured [Fe/H] and show the results as red stars in Figs.~\ref{fig:residual1}--\ref{fig:residual4}. It can be seen that for the elements common to both the CERES and combined sample, the same overall good agreement between predictions and measurements is obtained with the patterns derived from either sample. In addition, the mixtures of the patterns derived from the combined sample provide good fits to a wide range of other elements. For these mixtures, the differences $\delta\log({\rm E/Eu})$ are also summarized for all the stars in Fig.~\ref{fig:residual_all}, where the thick horizontal bar indicates the mean $\delta\log({\rm E/Eu})$ for an element across the combined sample and the error bar represents the $1\sigma$ scatter around the mean. It can be seen that the mean $\delta\log({\rm E/Eu})$ is very close to zero for all the elements and the typical scatter is a few tenths of a dex, comparable to the average measurement uncertainties. Large scatter of $\delta\log({\rm E/Eu})$ mainly occurs for elements measured in a small number of stars. In addition, as noted above, the comparison between the predicted and measured values of Pt is subject to unresolved measurement issues and should be viewed with caution. 

Based on the above results, we conclude that with few exceptions [e.g., $\delta\log({\rm Sr/Eu})\approx-1$ for 2MASS~J21402305-1227035 (J2140-1227, \citealt{2024A&A...688A.123X})], the $r$-process patterns in MP stars of the combined sample can be adequately described by mixtures of Patterns 1 and 2 derived from this sample. Note that although Th and U are radioactive, their $\log({\rm E/Eu})_1$ and $\log({\rm E/Eu})_2$ values derived from the present-day data can be used to predict the present-day Th/Eu and U/Eu ratios in MP stars. This approximation is valid because the age differences among all MP stars with [Fe/H]~$\lesssim-1.5$ are small compared to the long lifetimes of $^{232}$Th and $^{238}$U while the relatively short-lived $^{235}$U in these stars had all decayed.

\subsection{Patterns with Different Reference Elements}
\label{sec:diff-ref}
It can be seen from Table~\ref{tab:pat} that $\log({\rm E/Eu})_1$ exceeds $\log({\rm E/Eu})_2$ by $\approx0.9$--2.7 dex for the light $r$-process elements Sr to Cd. In the region of Te and heavier elements, $\log({\rm E/Eu})_1-\log({\rm E/Eu})_2$ is $\approx 0$--1.1 dex for most of the elements, and is 1.6 and 2.2 dex for Te and Pt, respectively, but is $-0.5$ and $-0.4$ dex for Gd and Dy, respectively. Therefore, compared to Pattern 2, Pattern 1 is especially prominent in production of the light $r$-process elements Sr to Cd, as well as the second and third peak elements, Te and Pt, respectively. Note that In, Sb, Ta, and Bi each have a single measurement (Table~\ref{tab:lit}), and the measured value is assigned to both $\log({\rm E/Eu})_1$ and $\log({\rm E/Eu})_2$ in each case by the optimization procedure. In addition, Rh, Ag, Sn, Te, Lu, W, Re, Au, Pb, Th, and U each have fewer than 10 measurements (Table~\ref{tab:lit}). So more data on these elements are especially needed to confirm their parameters for Patterns 1 and 2.

\begin{figure*}[htbp]
    \centering
    \includegraphics[width=1\textwidth]{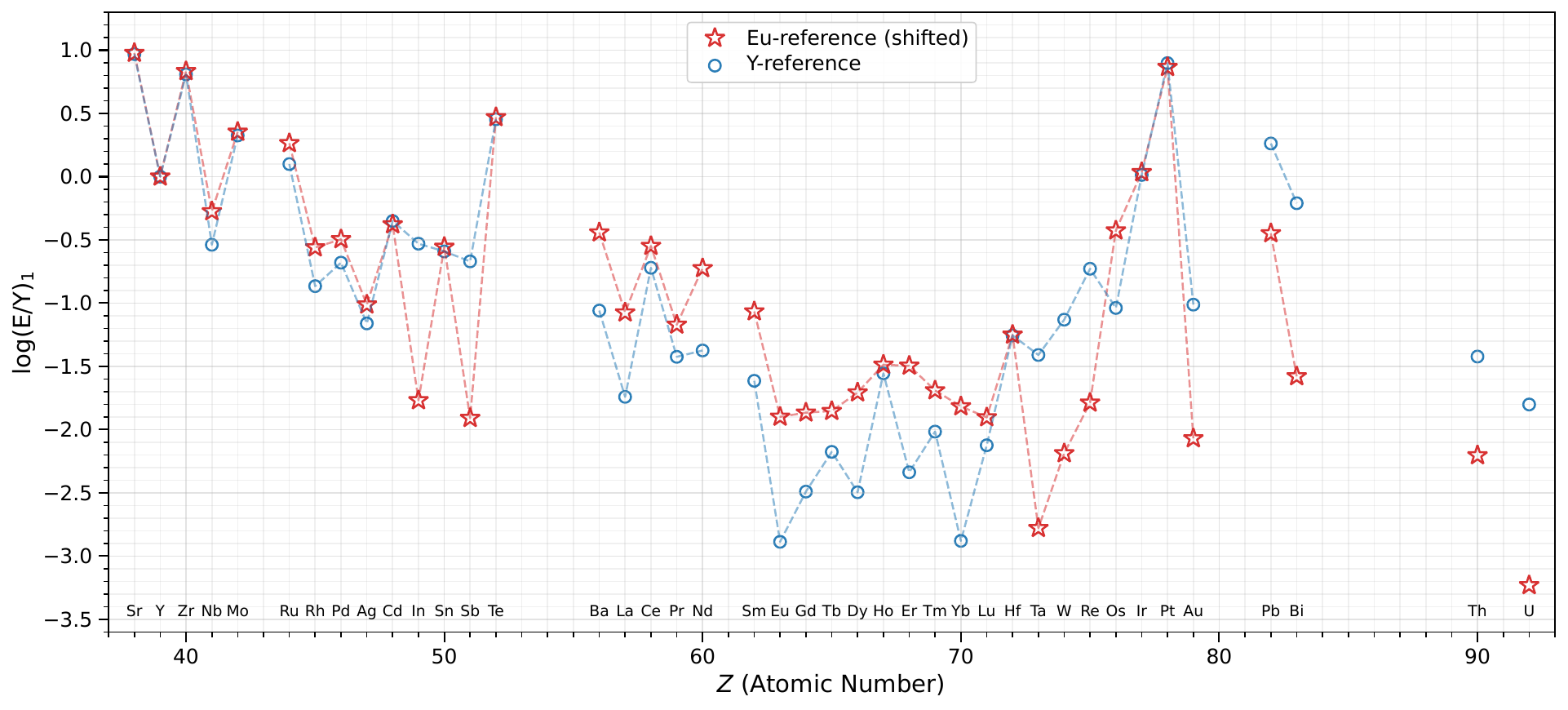}\\[1em]  
    \includegraphics[width=1\textwidth]{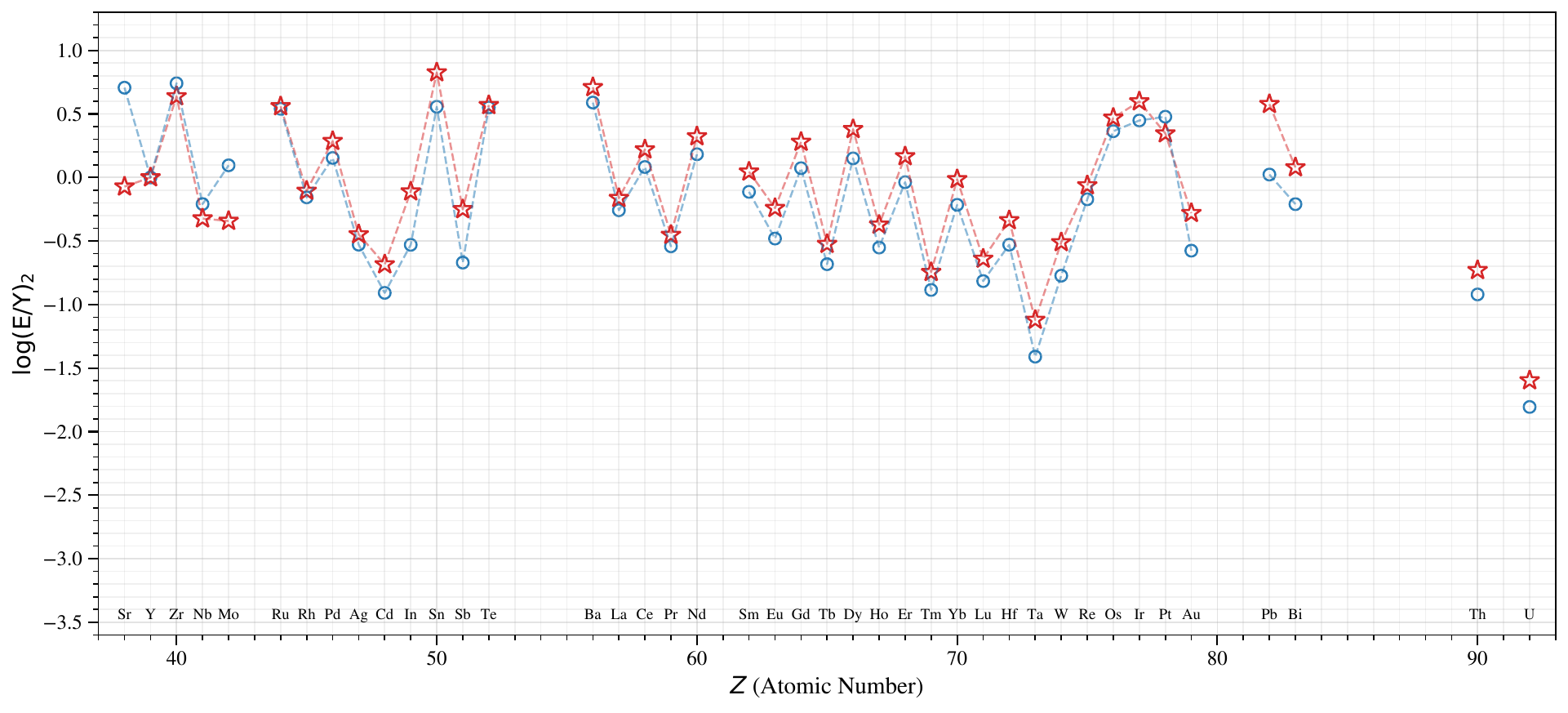}         
    \caption{Patterns~1 and~2 derived from the combined sample in terms of $\log({\rm E/Y})_1$ and $\log({\rm E/Y})_2$ (light blue circles). 
    The red stars show the patterns derived with Eu as the reference element but shifted by $-\log({\rm Y/Eu})_1$ and $-\log({\rm Y/Eu})_2$, respectively, for comparison.}
    \label{fig:pat_Y_Eu}
\end{figure*}

To check the stability of Patterns 1 and 2, we rederive them in terms of $\log({\rm E/Y})_1$ and $\log({\rm E/Y})_2$ from the combined sample. The relevant average measurement uncertainties $\sigma_{\log({\rm E/Y})}$ are given in Tables~\ref{tab:ceres} and \ref{tab:lit}. The optimal parameters $\log({\rm E/Y})_1$ and $\log({\rm E/Y})_2$ are given in Table~\ref{tab:pat} and shown as light blue circles in Fig.~\ref{fig:pat_Y_Eu}. The corresponding difference $\delta\log({\rm E/Y})$ between the predicted and measured value for each element is summarized for all the stars in the combined sample in Fig.~\ref{fig:residual_all_Y}, where the mean $\delta\log({\rm E/Y})$ for an element across the entire sample and the $1\sigma$ scatter around the mean are also indicated. Comparison of Figs.~\ref{fig:residual_all} and \ref{fig:residual_all_Y} shows that the same level of agreement between the predictions and data is achieved whether Eu or Y is used as the reference element.

The red stars in Fig.~\ref{fig:pat_Y_Eu} show Patterns 1 and 2 derived with Eu as the reference element but shifted by $-\log({\rm Y/Eu})_1$ and $-\log({\rm Y/Eu})_2$, respectively, for comparison with the corresponding patterns derived with Y as the reference element. It can be seen that Pattern 2 is rather stable when the reference element is changed from Eu to Y, and so are the parts of Pattern 1 covering the light $r$-process elements Sr to Cd and the second and third peak elements. The significant changes fall into three categories. The first category covers those elements for which either $({\rm E/Eu})_1\gg ({\rm E/Eu})_2$ or $({\rm E/Y})_2\gg ({\rm E/Y})_1$. As mentioned above, when for example, $({\rm E/Eu})_1$ greatly exceeds $({\rm E/Eu})_2$, $({\rm E/Eu})_2$ is likely obtained as the small difference between two large numbers, and therefore, may be susceptible to large uncertainties. Such uncertainties occur for $\log({\rm Sr/Eu})_2$ and $\log({\rm Mo/Eu})_2$ with $({\rm E/Eu})_1\gg ({\rm E/Eu})_2$ and for $\log({\rm E/Y})_1$ from Ba to Yb with $({\rm E/Y})_2\gg ({\rm E/Y})_1$ (Table~\ref{tab:pat}). The second category consists of In, Sb, Ta, and Bi, each with only a single measurement that is used to obtain $\log({\rm E/Eu})_1=\log({\rm E/Eu})_2=\log\epsilon(\rm E)-\log\epsilon(\rm Eu)$ and $\log({\rm E/Y})_1=\log({\rm E/Y})_2=\log\epsilon(\rm E)-\log\epsilon(\rm Y)$. For these elements, discrepancies occur when the reference element is changed from Eu to Y because $\log\epsilon(\rm Y)-\log\epsilon(\rm Eu)$ for the star with the relevant measurements neither coincides with $\log(\rm Y/Eu)_1$ nor with $\log(\rm Y/Eu)_2$. The third category consists of W, Re, Au, Pb, Th, and U, for which large discrepancies occur in Pattern 1 when the reference element is changed from Eu to Y (a significant discrepancy also occurs for Pb in Pattern 2). All the elements in this category each have a small number of measurements (Table~\ref{tab:lit}), which could cause large uncertainties in their derived parameters.

Based on the uncertainties discussed above  (see also \S\ref{sec:uncertainty} and Appendix~\ref{append:bayesian}), the choice of the reference element matters for those elements with large differences between the parameters for Patterns 1 and 2. Examination of Table~\ref{tab:pat} suggests that as a reference element, Eu is better for Pattern 1 as $({\rm E/Y})_2\gg({\rm E/Y})_1$ for the Eu region, while Y is better for Pattern 2 as $({\rm E/Eu})_1\gg({\rm E/Eu})_2$ for the light $r$-process elements Sr to Cd. However, this approximate rule is most useful when dealing with stars dominated by Pattern 1 or 2. For stars with more balanced contributions from both patterns, the same reference element must be used for both patterns.

An exception to the above approximate rule is the star J2140-1227 \citep{2024A&A...688A.123X}, which is dominated by Pattern 1. The best fit (violet circles) using Patterns 1 and 2 with Y as the reference element is in fair agreement with the data (Fig.~\ref{fig:J2140-1227}). This fit coincides with Pattern 1 (light blue circles) except for some differences at Eu, Gd, Dy, Er, and Yb. In contrast, when Eu is used as the reference element, although the best fit is still dominated by Pattern 1, it severely underestimates Sr, Y, and Zr (Fig.~\ref{fig:residual1}). 

J2140-1227 clearly illustrates the uncertainties in Pattern 1, and observing many of its unmeasured elements can help quantify these uncertainties better. On the other hand, the uncertainties in Pattern 1 may not be entirely due to the causes discussed above. For example, Pattern 1 may well represent the average of production by multiple sets of astrophysical conditions with different yield templates, so it is reasonable to expect fluctuations in the production between the second and third peak. In this case, J2140-1227 simply reflects deviations from the average.

In general, with either Eu or Y as the reference element, the overall stability of Pattern 2 and the partial stability of Pattern 1 covering Sr to Cd, Te, and Pt supports our approach to derive the $r$-process production patterns from data on MP stars. Of course, this approach can be greatly improved by a much larger sample of stars for which the full range of elements are homogeneously analyzed. Patterns 1 and 2 derived from such comprehensive data should be much more reliable no matter which reference element is used.

\subsection{Fits to Example Stars}
We now show fits to the data on some example stars using Patterns 1 and 2 derived above. For convenience of presentation, we focus on Patterns 1 and 2 with Eu as the reference element. The results with Y as the reference element are presented in Appendix~\ref{append:Yref}.

\subsubsection{2MASS~J22132050-5137385 (J2213-5137)}
\begin{figure*}[htbp]
    \centering
\includegraphics[width=1\textwidth]{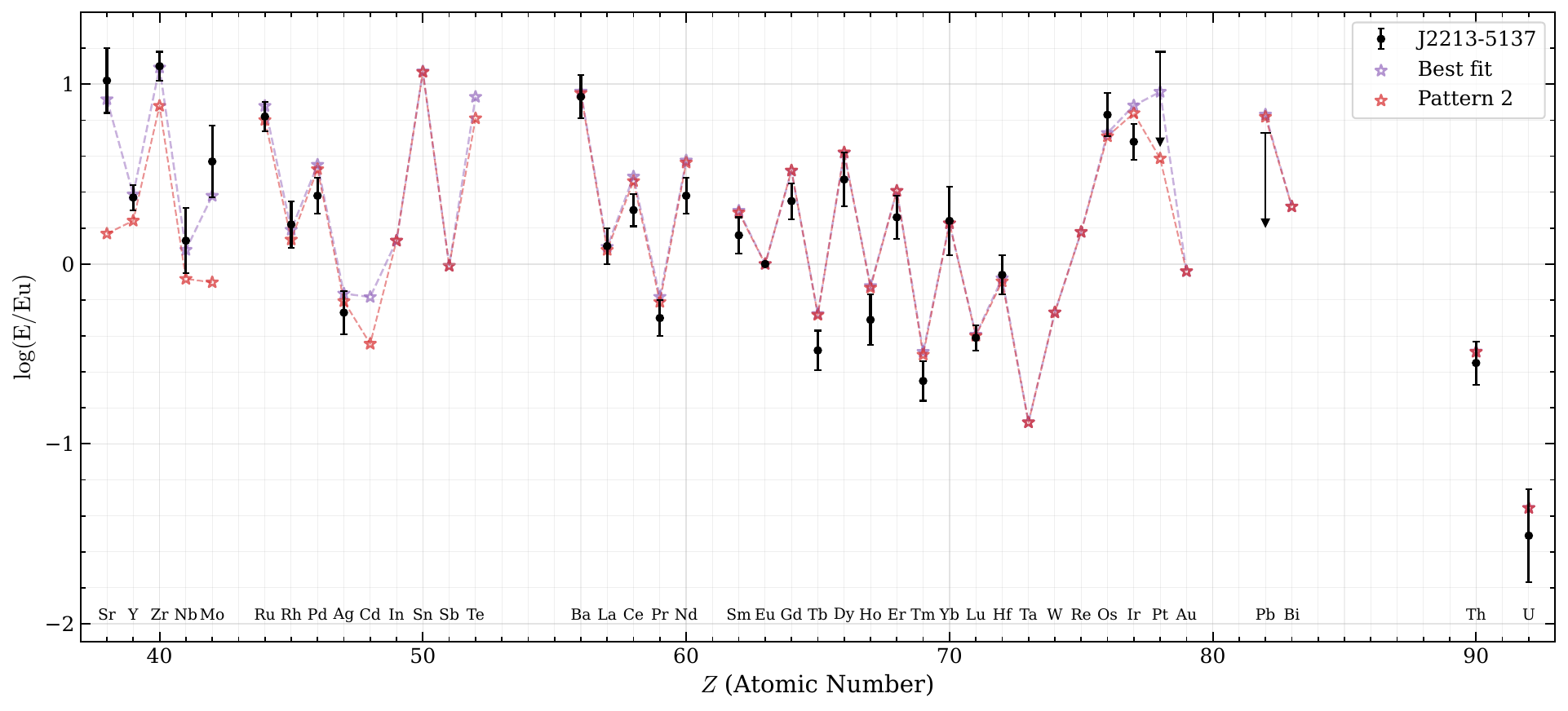}
\hfill
\caption{Comparison of the best fit (violet stars) and Pattern 2 (red stars) with data (black symbols, \citealt{2024ApJ...971..158R}) on $\log({\rm E/Eu})$ for J2213-5137. The best fit with a mixing parameter $x_{\rm Eu} = 9\times10^{-3}$ is dominated by contributions from Pattern 2 except for the light $r$-process elements and Pt. See text for details.}
\label{fig:pat2}
\end{figure*}

In Fig.~\ref{fig:pat2}, the black filled circles with error bars show the data on $\log({\rm E/Eu})$ for J2213-5137 with [Fe/H]~$=-2.2$ and the highest-known [Eu/Fe]~$=2.45$ \citep{2024ApJ...971..158R}. The violet stars correspond to the best fit with a mixing parameter $x_{\rm Eu}=9\times 10^{-3}$ and the red stars show Pattern 2. It can be seen that the best fit is in very good agreement with the data and is dominated by contributions from Pattern 2 except for the light $r$-process elements (especially Sr and Mo) and Pt.

As noted above, the $\log({\rm E/Eu})_2$ values for Sr and Mo may have large uncertainties. Because Y is the better reference element for the light $r$-process elements in Pattern 2, we compare the corresponding best fit (violet circles) and Pattern 2 (light blue circles) with the data on $\log({\rm E/Y})$ for J2213-5137 in Fig.~\ref{fig:s16_Y}. It can be seen that although the best fit is obtained with a mixing parameter $x_{\rm Y}=0.16$, Pattern 2 alone provides an almost equally good fit to the data. The same is also true of CS~22892-052 (Fig.~\ref{fig:s19_Y}) with [Fe/H]~$=-3.1$ and [Eu/Fe]~$=1.64$ \citep{2003ApJ...591..936S}, CS~31082-001 (Fig.~\ref{fig:s20_Y}) with [Fe/H]~$=-2.9$ and [Eu/Fe]~$=1.69$ \citep{2013A&A...550A.122S}, and 2MASS~J15213995-3538094 (J1521-3538, Fig.~\ref{fig:s18_Y}) with [Fe/H]~$=-2.8$ and [Eu/Fe]~$=2.2$ \citep{2020ApJ...898...40C}. Because Pattern 2 covers some elements that have not been detected in these stars yet, future measurements of these elements will provide further tests of this pattern for $r$-process production.

\subsubsection{HD~122563 (CES1402+094)}
\begin{figure*}[htbp]
    \centering
\includegraphics[width=1\textwidth]{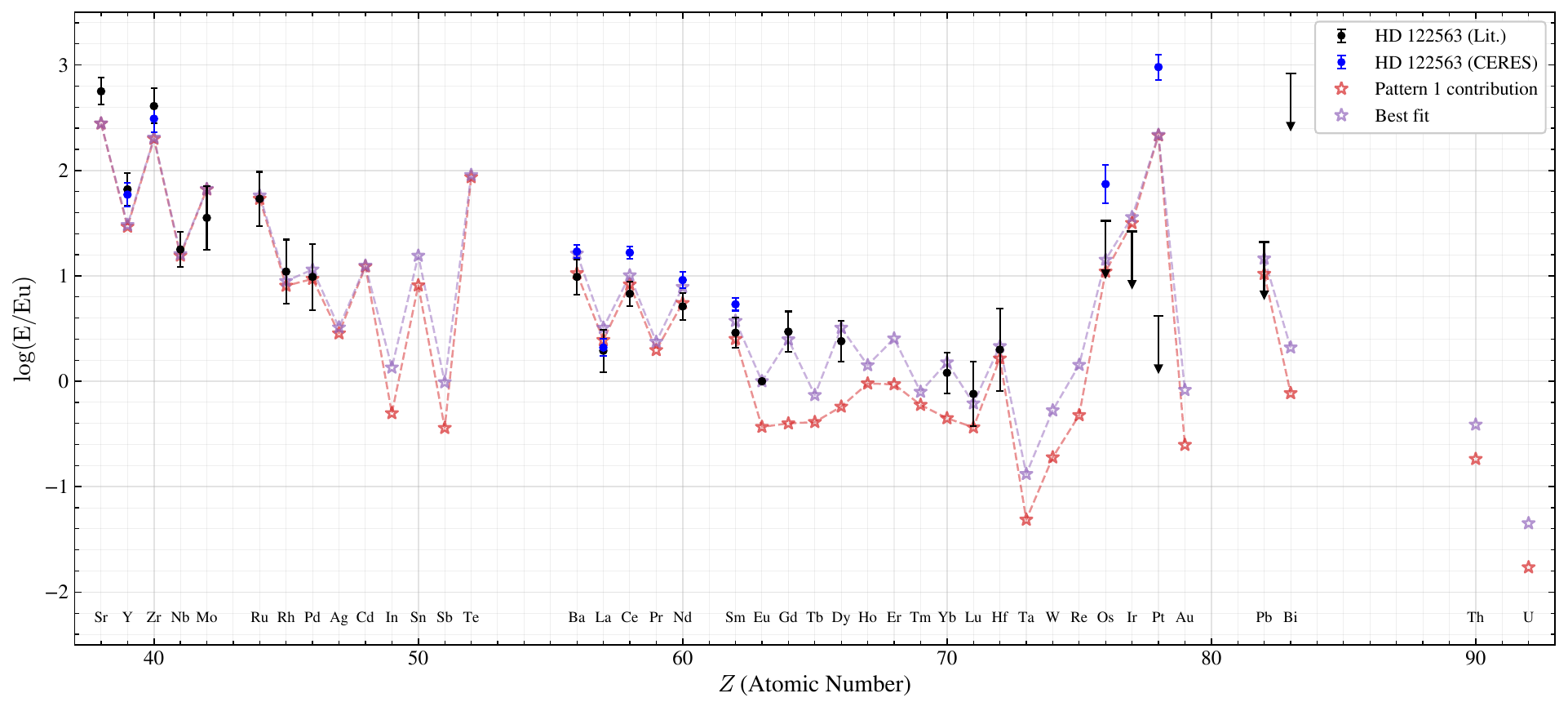}
\hfill
\caption{Comparison of the best fit (violet stars) and contributions from Pattern 1 (red stars) with data on $\log({\rm E/Eu})$ for HD~122563. The best fit with a mixing parameter $x_{\rm Eu} = 0.37$ is dominated by contributions from Pattern 1 for the light $r$-process and the second and third peak elements. Contributions from Pattern 2 are mostly limited to the Eu region. The best fit predicts a prominent third peak, with Os and Ir close to the upper limits from the literature data (black symbols, \citealt{2012ApJS..203...27R}) but with Pt way above the upper limit. While the CERES data (blue symbols, \citealt{2022A&A...665A..10L,2025A&A...693A.293L,2025A&A...693A.294A}) are consistent with the literature data on Y, Zr, and Ba to Sm, the former data on Os and Pt are significantly above the best fit. The large discrepancy in Pt between the CERES and literature data requires careful examination.}
\label{fig:HD122563}
\end{figure*}

In Fig.~\ref{fig:HD122563}, the black filled circles with error bars show the data on $\log({\rm E/Eu})$ for HD~122563 with [Fe/H]~$=-2.6$ and [Eu/Fe]~$=-0.68$ \citep{2012ApJS..203...27R}. This star has long been regarded as deficient in heavy $r$-process elements because its [Eu/Fe] is well below zero and the abundances in the Eu region are well below those of the light $r$-process elements Sr, Y, and Zr \citep{2006ApJ...643.1180H,2007ApJ...666.1189H}. The blue filled circles with error bars in Fig.~\ref{fig:HD122563} represent the new CERES data \citep{2022A&A...665A..10L,2025A&A...693A.293L,2025A&A...693A.294A}, and are consistent with the known high enrichment of the light $r$-process elements relative to the Eu region. However, they also show that the third peak elements Os and Pt are more enriched than Y and Zr, respectively. As mentioned above, Os and Pt are measured with neutral species, in contrast to Eu and other elements measured with singly-ionized species. Consequently, the ratios Os/Eu and Pt/Eu are subject to systematic uncertainties in stellar atmospheric models. We note that $\log({\rm Os/Eu})=1.87$ and $\log({\rm Pt/Eu})=2.98$ from CERES are incompatible with the upper limits $\log({\rm Os/Eu})<1.52$ and $\log({\rm Pt/Eu})<0.62$ from \cite{2012ApJS..203...27R}. While the moderate discrepancy in $\log({\rm Os/Eu})$ might be expected from differences in the atmospheric models, the discrepancy of $>2.36$ dex in $\log({\rm Pt/Eu})$, which exceeds even that of 0.88 dex for HD~108317 noted above, suggests that other factors (e.g., different adopted atomic lines) in spectroscopic analysis might be involved. Clearly, data on Pt in particular, and other elements measured with neutral species in general, should be carefully checked.

As mentioned above, we only use the literature data (excluding upper limits) from \cite{2012ApJS..203...27R} on HD~122563 in the combined sample to derive the $r$-process production patterns. In this sense, the fit to the data also makes predictions for unmeasured or undetected elements. The violet stars in Fig.~\ref{fig:HD122563} show the best fit with a mixing parameter $x_{\rm Eu}=0.37$. Compared to the literature data, this fit is in good agreement for Sr to Hf and predicts a prominent third peak with Os and Ir close to the upper limits but with Pt way above the upper limit. In contrast, while the fit is also consistent with the CERES data \citep{2022A&A...665A..10L,2025A&A...693A.293L,2025A&A...693A.294A} on Y, Zr, and Ba to Sm, the predictions for Os and Pt are significantly below the CERES measurements. Clearly, resolving the observational discrepancies in Os and Pt and measuring Ir in HD~122563 will provide a strong test of Pattern 1. It would be extremely striking if this MP star has prominent third-peak elements despite deficiencies in the Eu region relative to the light $r$-process elements. 

The red stars in Fig.~\ref{fig:HD122563} show the contributions from Pattern 1 to HD~122563. Compared to Pattern 2, the light $r$-process as well as second and third peak elements in Pattern 1 have much higher production relative to Eu (Table~\ref{tab:pat}). Therefore, with $x_{\rm Eu}=0.37$ for the best fit to HD~122563, contributions from Pattern 2 are limited mostly to the Eu region, while those from Pattern 1 dominate for the other elements. This result is also obtained when Y is used as the reference element (Fig.~\ref{fig:HD122563_Y}).

\subsubsection{HD~222925}
\begin{figure*}[htbp]
    \centering
\includegraphics[width=1\textwidth]{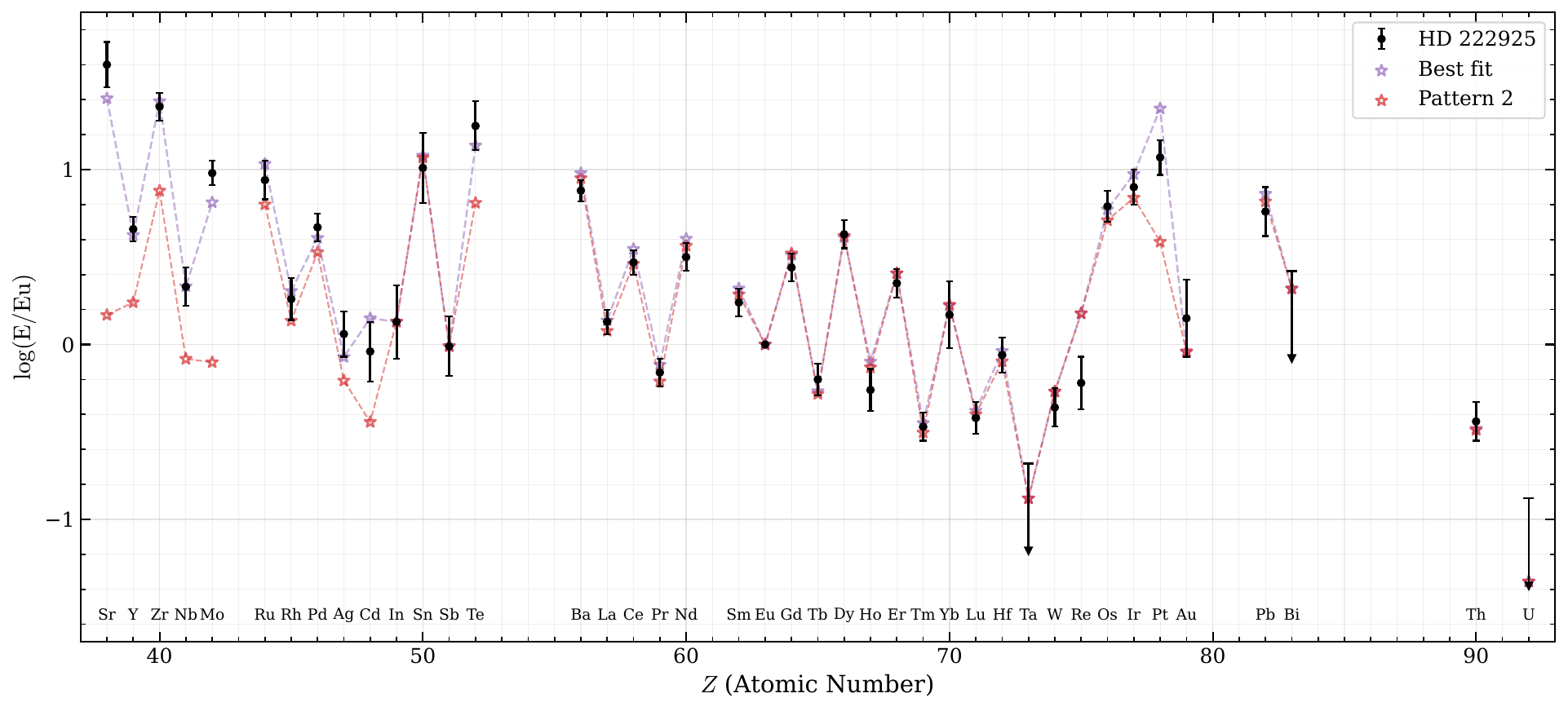}
\hfill
\caption{Comparison of the best fit (violet stars) and Pattern 2 (red stars) with data (black symbols, \citealt{2022ApJS..260...27R}) on $\log({\rm E/Eu})$ for HD~222925. The best fit corresponds to a mixing parameter $x_{\rm Eu} = 3.2\times10^{-2}$, and its difference from Pattern 2 shows that contributions from Pattern 1 are mostly limited to the light $r$-process elements, Te, and Pt.}
\label{fig:s1}
\end{figure*}

The violet stars in Fig.~\ref{fig:s1} show the best fit to the data on $\log({\rm E/Eu})$ for HD~222925 \citep{2022ApJS..260...27R}. With [Fe/H]~$=-1.46$ and [Eu/Fe]~$=1.32$, this star has the most comprehensive measurements of $r$-process elements among MP stars. The best fit with a mixing parameter $x_{\rm Eu}=3.2\times10^{-2}$ is in good agreement with the data. As can be seen from the difference between the best fit and Pattern 2 (red stars), contributions from Pattern 1 are mostly limited to the light $r$-process elements, Te, and Pt. The same result is also obtained when Y is used as the reference element (Fig.~\ref{fig:s1_Y}).

\subsection{Mixing parameters}
\label{sec:mix}
\begin{figure*}[htbp]

    \centering
    
    % -------- Row 1 --------
    \begin{minipage}[t]{0.49\textwidth}
        \centering
        \includegraphics[width=\textwidth]{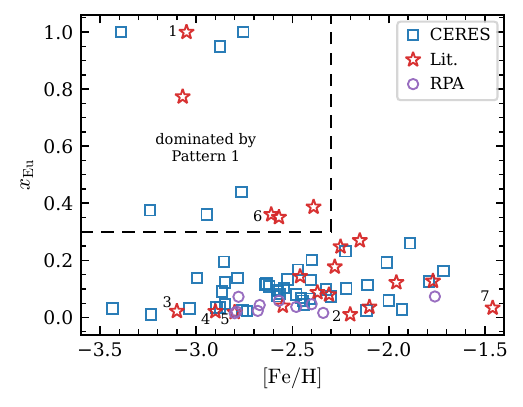}
    \end{minipage}\hfill
    \begin{minipage}[t]{0.49\textwidth}
        \centering
        \includegraphics[width=\textwidth]{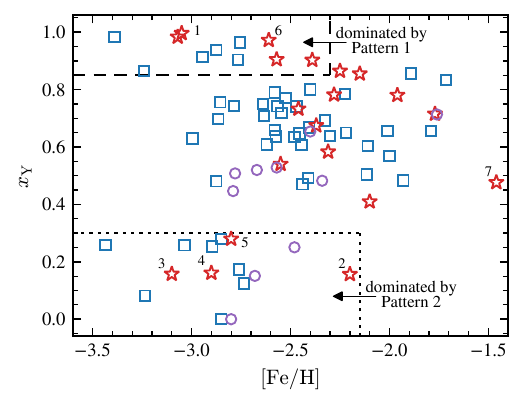}
    \end{minipage}
    
    \vspace{0.1em} 
    
    % -------- Row 2 --------
    \begin{minipage}[t]{0.49\textwidth}
        \centering
        \includegraphics[width=\textwidth]{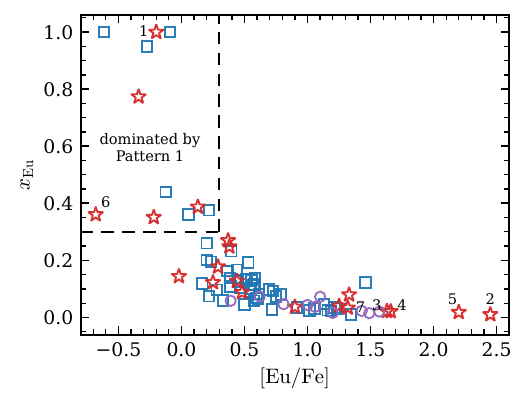}
    \end{minipage}\hfill
    \begin{minipage}[t]{0.49\textwidth}
        \centering
        \includegraphics[width=\textwidth]{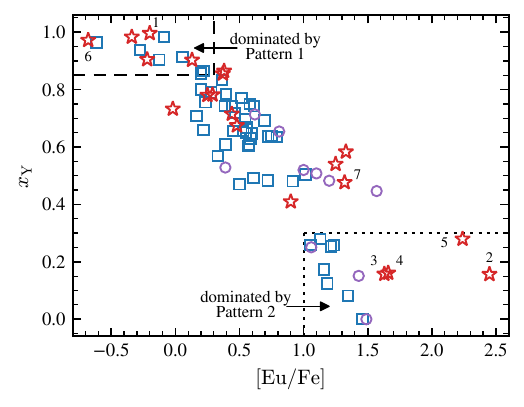}
    \end{minipage}
    
    \caption{
    Mixing parameters $x_{\rm Eu}$ (left panels) and $x_{\rm Y}$ (right panels) as functions of [Fe/H] (upper panels) and [Eu/Fe] (lower panels). 
    Light blue squares and red stars correspond to the CERES and literature sample, respectively. Violet circles represent the 10 stars in a recently published RPA sample \citep{2025A&A...704A.282R}, which is not used to derive Patterns 1 and 2, and therefore, provides an external test of our model.
    The fraction $x_{\rm Eu}$ ($x_{\rm Y}$) of Eu (Y) contributed by Pattern 1 is calculated with Eu (Y) as the reference element.
    The upper left corner of each panel corresponds to stars dominated by Pattern 1.
    The lower left corner of the upper right panel and the lower right corner of the lower right panel correspond to stars dominated by Pattern 2.
    Labels 1--7 next to red star symbols correspond to J2140-1227, J2213-5137, CS~22892-052, CS~31082-001, J1521-3538, HD~122563, and HD~222925, respectively.
    }
\label{fig:mix}
\end{figure*}

The mixing parameter $x_{\rm Eu}$ ($x_{\rm Y}$) is the fraction of Eu (Y) contributed by Pattern 1. Compared to Pattern 2, Pattern 1 has much higher production of the light $r$-process elements relative to the Eu region (Table~\ref{tab:pat} and Fig.~\ref{fig:pat_Y_Eu}). Therefore, the same $r$-process pattern may correspond to very different values of $x_{\rm Eu}$ and $x_{\rm Y}$. The upper panels of Fig.~\ref{fig:mix} show $x_{\rm Eu}$ and $x_{\rm Y}$ as functions of [Fe/H] for all the stars in the combined sample, while the lower panels show them as functions of [Eu/Fe]. The stars discussed above, J2140-1227, J2213-5137, CS~22892-052, CS~31082-001, and J1521-3538, HD~122563, and HD~222925 are labeled 1--7 in Fig.~\ref{fig:mix}. 

By examining the distributions of stars in the four panels of Fig.~\ref{fig:mix}, we distinguish three groups of stars: those with $r$-process patterns dominated by Pattern 1 or 2 and those with more balanced contributions from both patterns. Stars dominated by Pattern 1, of which J2140-1227 (star 1) and HD~122563 (star 6) are prominent examples, are in the upper left corners of the four panels of Fig.~\ref{fig:mix}. They can be distinguished from the rest of the stars by any one of the following criteria: ($x_{\rm Eu}>0.3$, [Fe/H]~$<-2.3$), ($x_{\rm Eu}>0.3$, [Eu/Fe]~$<0.3$), ($x_{\rm Y}>0.85$, [Fe/H]~$<-2.3$), or ($x_{\rm Y}>0.85$, [Eu/Fe]~$<0.3$). In contrast, stars dominated by Pattern 2, of which J2213-5137, CS~22892-052, CS~31082-001, and J1521-3538 (stars 2--5) are prominent examples, are best distinguished with ($x_{\rm Y}<0.3$, [Fe/H]~$<-2.15$) or ($x_{\rm Y}<0.3$, [Eu/Fe]~$>1$), which corresponds to the lower left corner of the upper right panel or the lower right corner of the lower right panel of Fig.~\ref{fig:mix}. While they also correspond to $x_{\rm Eu}\leq 0.12$ (all except one have $x_{\rm Eu}\leq0.05$), this criterion along with [Fe/H]~$<-2.15$ or [Eu/Fe]~$>1$ cannot separate them from some stars with more balanced contributions to Y from Patterns 1 and 2 (cf. left and right panels of Fig.~\ref{fig:mix}). For example, although HD~222925 (star 7) with $x_{\rm Y}=0.48$ can be distinguished by its higher [Fe/H]~$=-1.46$, it also has $x_{\rm Eu}=3.2\times10^{-2}$ and [Eu/Fe]~$=1.32$, which puts it in the same region of ($x_{\rm Eu}\leq 0.12$, [Eu/Fe]~$>1$) as those stars dominated by Pattern 2 (lower left panel of Fig.~\ref{fig:mix}).

Stars dominated by Pattern 1 or 2 are listed in Table~\ref{tab:star_pattern_1_2}, which also gives their [Fe/H], [Eu/Fe], [Y/Fe], [Y/Eu], $x_{\rm Eu}$, and $x_{\rm Y}$. Further observations of many unmeasured elements in these stars can provide clear tests of Patterns 1 and 2 derived here. 

Table~\ref{tab:star_pattern_1_2} shows that stars dominated by Pattern 1 has [Y/Eu]~$=0.05\pm0.35$. Using $\log\epsilon_\odot({\rm Y})=2.21$ and $\log\epsilon_\odot({\rm Eu})=0.52$ \citep{2009ARA&A..47..481A}, we obtain [Y/Eu]$_1=0.21$ (1.20) corresponding to $\log({\rm Y/Eu})_1=1.90$ [$\log({\rm Eu/Y})_1=-2.89$] for Pattern 1 with Eu (Y) as the reference element. The above comparison confirms the approximate rule that Eu is the better reference element for Pattern 1. However, [Y/Eu]~$=0.89$ for J2140-1227 \citep{2024A&A...688A.123X} is closer to [Y/Eu]$_1=1.20$ for Pattern 1 with Y as the reference element (Fig.~\ref{fig:J2140-1227}). As discussed above, this outlier clearly illustrates the uncertainties in the Eu region for Pattern 1, which may be regarded more appropriately as deviations from some average pattern. If Pattern 1 is indeed the average of production by multiple sets of astrophysical conditions with different yield templates, then the above scatter of 0.35 dex in [Y/Eu] for stars dominated by Pattern 1, which is much larger than the average measurement uncertainty of 0.11 dex, may also indicate intrinsic variations in the production ratio Y/Eu among sources for Pattern 1.

Table~\ref{tab:star_pattern_1_2} shows that stars dominated by Pattern 2 has [Y/Eu]~$=-1.18\pm0.09$, which is in very good agreement with [Y/Eu]$_2=-1.21$ corresponding to $\log({\rm Eu/Y})_2=-0.48$ for Pattern 2 with Y as the reference element. While [Y/Eu]$_2=-1.45$ corresponding to $\log({\rm Y/Eu})_2=0.24$ for Pattern 2 with Eu as the reference element is not too far off, Y is clearly the better reference element for Pattern 2. More importantly, the above scatter of 0.09 dex in [Y/Eu] for stars dominated by Pattern 2 is consistent with the average measurement uncertainty of 0.11 dex, and therefore, supports that Pattern 2 is very stable and likely dominated by a single yield template.

Stars dominated by Pattern 2 also have [Eu/Fe]~$=1.48\pm0.44$ and [Y/Fe]~$=0.30\pm0.41$. Being much larger than the average measurement uncertainties of 0.11 dex, the scatters for [Eu/Fe] and [Y/Fe] indicate strong intrinsic variations. It is possible that the source for Pattern 2 has widely variable production of $r$-process elements relative to Fe while maintaining a stable $r$-process pattern. However, given the observational evidence from GW170817 for NSMs being an $r$-process source (e.g., \citealt{kasen2017}), it is more natural for at least a subset of NSMs to be the source for Pattern 2. In this scenario, as NSMs produce no Fe, enrichment of Fe is completely decoupled from that of the $r$-process elements for stars dominated by Pattern 2. This scenario becomes especially compelling with the discovery of J2213-5137 (J1521-3538), which has extremely high values of [Eu/Fe]~$=2.45$ (2.20) and [Y/Fe]~$=1.13$ (1.03).

It is interesting to compare the criteria for MP stars dominated by Pattern 1 or 2 discussed above with those for categorizing MP stars based on observations alone. MP stars are often referred to as limited-$r$, $r$-I, and $r$-II if they have [Eu/Fe]~$<0.3$, $0.3\leq$~[Eu/Fe]~$\leq 1$, and [Eu/Fe]~$>1$, respectively. Additional criteria are [Sr/Ba]~$>0.5$ and [Sr/Eu]~$>0$ for being limited-$r$ \citep{2018ARNPS..68..237F}, and [Ba/Eu]~$<0$ for being $r$-I or $r$-II \citep{2005ARA&A..43..531B}. Based on [Eu/Fe] alone, MP stars dominated by Pattern 1 are potentially limited-$r$, and those dominated by Pattern 2 are potentially $r$-II. 

Using the parameters for Patterns 1 and 2 in Table~\ref{tab:pat} along with $\log\epsilon_\odot({\rm Sr})=2.87$, $\log\epsilon_\odot({\rm Ba})=2.18$, and $\log\epsilon_\odot({\rm Eu})=0.52$ \citep{2009ARA&A..47..481A}, we obtain [Sr/Ba]~$>0.5$ and [Sr/Eu]~$>0.01$ for $x_{\rm Eu}>0.3$ or [Sr/Ba]~$>0.43$ and [Sr/Eu]~$>-0.12$ for $x_{\rm Y}>0.85$. Considering the uncertainties in measurements and in our derived production patterns, we conclude that our criteria 
($x_{\rm Eu}>0.3$, [Eu/Fe]~$<0.3$) or ($x_{\rm Y}>0.85$, [Eu/Fe]~$<0.3$) for MP stars dominated by Pattern 1 are equivalent to the empirical criteria for limited-$r$ stars.

For MP stars dominated by Pattern 2, we obtain [Ba/Eu]~$=-0.71$ to $-0.61$ for $x_{\rm Eu}<0.12$ or [Ba/Eu]~$\approx -0.59$ for $x_{\rm Y}<0.3$. Therefore, these stars have well-defined [Ba/Eu] reflecting the stability of Pattern 2. While they definitely belong to the $r$-II category with [Ba/Eu]~$<0$ and [Eu/Fe]~$>1$, their most prominent characteristics are the essentially fixed $r$-process patterns. In this regard, J2213-5137 and J1521-3538 with extremely high values of [Eu/Fe]~$=2.45$ and 2.20, respectively, are just like other MP stars dominated by Pattern 2. While their extremely high [Eu/Fe] values provide strong support for NSMs producing Pattern 2, they do not call for another category of $r$-process-enriched MP stars.

For MP stars with more balanced contributions from Patterns 1 and 2, we obtain $-0.61\leq$~[Ba/Eu]~$\leq-0.49$ for $0.12\leq x_{\rm Eu}\leq0.3$ or $-0.59\leq$~[Ba/Eu]~$\leq-0.55$ for $0.3\leq x_{\rm Y}\leq0.85$. While these stars easily satisfy [Ba/Eu]~$<0$, their range of [Ba/Eu] is rather narrow. Most of these stars also have $0.3\leq$~[Eu/Fe]~$\leq 1$ (Fig.~\ref{fig:mix}) and therefore, belong to the $r$-I category. However, some stars with $0.12\leq x_{\rm Eu}\leq0.3$ or $0.3\leq x_{\rm Y}\leq0.85$ lie outside the range $0.3\leq$~[Eu/Fe]~$\leq 1$ (Fig.~\ref{fig:mix}). They still represent mixtures of Patterns 1 and 2, but would not be classified as $r$-I. 

Based on the above discussion, our model of $r$-process production patterns provides theoretical insights into the empirical categories of limited-$r$, $r$-I, and $r$-II stars. While the empirical categories are convenient, our model addresses $r$-process patterns in MP stars in a unified framework and can provide more detailed constraints on the underlying astrophysical sources. The mixing parameters $x_{\rm Eu}$ and especially $x_{\rm Y}$ are clearly correlated with [Eu/Fe], at least for the MP stars studied in this paper (lower panels of Fig.~\ref{fig:mix}). Stars dominated by Pattern 1 or 2 tend to have lower [Fe/H] than those with more balanced contributions from both patterns (upper right panel of Fig.~\ref{fig:mix}), which is consistent with the general expectation that the former stars were enriched with $r$-process elements by few events at earlier times and the latter stars by multiple events at later times. We plan to further explore the above correlations and trends in a separate work.

\subsection{Tests by external data}
Our data-driven approach can be further tested by fitting mixtures of Patterns 1 and 2 to the $r$-process patterns in stars not included in the combined sample. Specifically, we discuss the fits to the 10 MP stars in a recently published RPA sample \citep{2025A&A...704A.282R}. We also show the fits to an MP star discovered by the SkyMapper telescope in the southern sky \citep{2021Natur.595..223Y} and to the solar $r$-process pattern.

\subsubsection{RPA sample}
We find that the $r$-process patterns in the RPA sample reported by \cite{2025A&A...704A.282R} can be readily accounted for by mixtures of Patterns 1 and 2. For illustration, we show the best fits to the data on $\log({\rm E/Y})$ for J1430-2371 and J1432-4125 in Figs.~\ref{fig:J1430-2371} and \ref{fig:J1432-4125}, respectively. These two stars correspond to the largest and smallest mixing parameter $x_{\rm Y}$ for the RPA sample, with $x_{\rm Y}=0.71$ and 0 for J1430-2371 and J1432-4125, respectively. The mixing parameters $x_{\rm Eu}$ and $x_{\rm Y}$ for the 10 stars (violet circles) in this sample are shown in Fig.~\ref{fig:mix}. It can be seen that none of these stars are dominated by Pattern 1, three stars are dominated by Pattern 2, and the rest received significant contributions from both patterns. We also show the difference $\delta\log({\rm E/Y})$ between the predicted and measured value for each element for all the stars in Fig.~\ref{fig:residual_new_stars}, where the mean $\delta\log({\rm E/Y})$ for an element across the entire sample and the $1\sigma$ scatter around the mean are also indicated. Comparison of Figs.~\ref{fig:residual_new_stars} and \ref{fig:residual_all_Y} shows that the agreement between the predictions and data for the external RPA sample is at the same level as for the combined sample used to derive Patterns 1 and 2. The majority of the elements have mean differences of $-0.1\leq\langle\delta\log({\rm E/Y})\rangle\leq 0.1$ dex for the RPA sample. The mean differences with the largest magnitudes are $\langle\delta\log({\rm Rh/Y})\rangle\approx-0.35$, $\langle\delta\log({\rm Pd/Y})\rangle\approx-0.3$, and $\langle\delta\log({\rm Pb/Y})\rangle\approx0.25$. We note that [Rh/Fe], [Pd/Fe], [Pb/Fe] in the RPA sample \citep{2025A&A...704A.282R} have relatively large measurement uncertainties of 0.31--0.37, 0.24--0.64, and 0.29--0.38 dex, respectively, which suggests that [Rh/Y], [Pd/Y], and [Pb/Y] also have significant measurement uncertainties \citep{2022ApJS..260...27R}. On the other hand, the combined sample used to derive Patterns 1 and 2 contains only 8, 11, and 4 stars with data on Rh, Pd, and Pb (Table~\ref{tab:lit}), respectively, and therefore, the pattern parameters for these three elements might also have large uncertainties. While a larger sample of MP stars with a wider range of measured elements is needed to improve our data-driven approach to derive the $r$-process production patterns, we regard that it passes the test by the above RPA sample.

\subsubsection{SMASSJ200322.54-114203.3 (SMASS~2003-1142)}
\label{sec:test-p2}
\begin{figure*}[htbp]
    \centering
\includegraphics[width=1\textwidth]{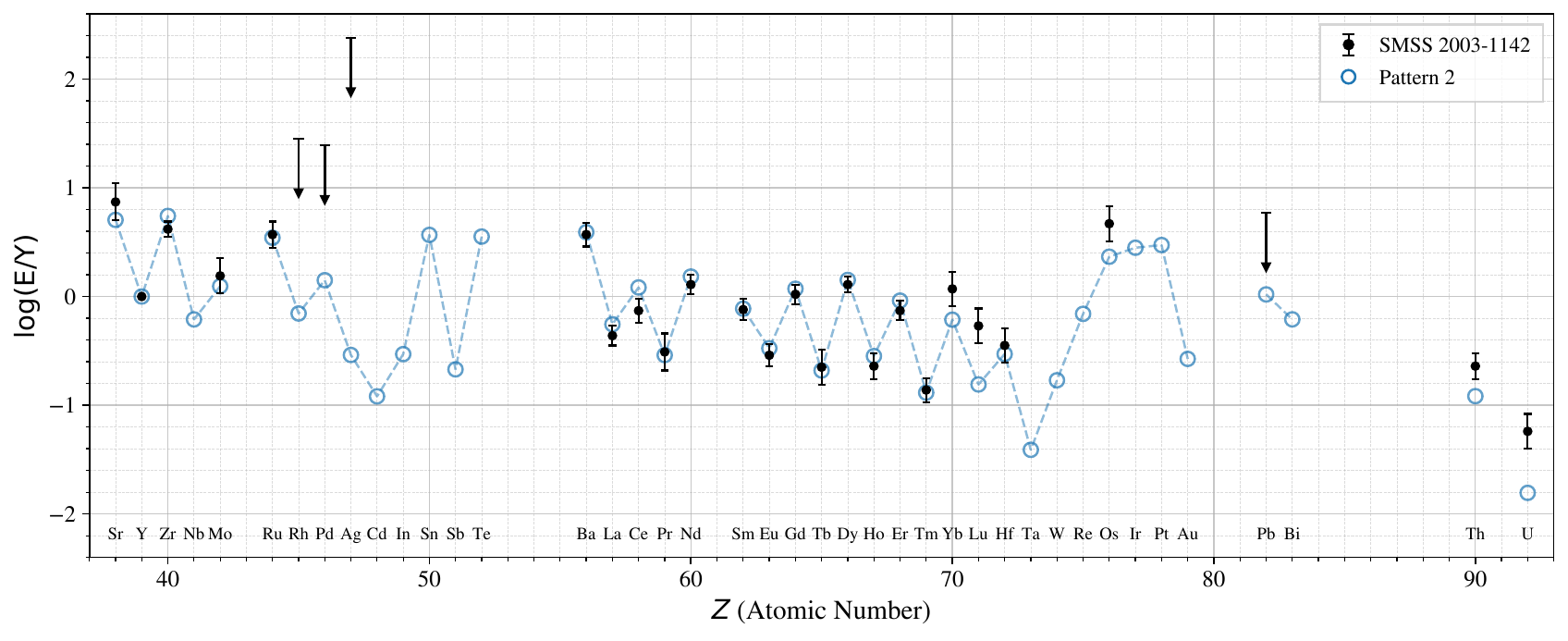}
\hfill
\caption{Comparison of Pattern 2 (light blue circles) with the data (black symbols, \citealt{2021Natur.595..223Y}) on $\log({\rm E/Y})$ for SMSS~2003-1142.}
\label{fig:SMSS_2003_Y}
\end{figure*}

With [Fe/H]~$=-3.5$, [Eu/Fe]~$=1.7$, and [Y/Eu]~$=-1.15$ \citep{2021Natur.595..223Y}, SMASS~2003-1142 should be dominated by Pattern 2 based on the discussion in \S\ref{sec:mix}. Figure~\ref{fig:SMSS_2003_Y} compares Pattern 2 (light blue circles) with the data (black symbols) on $\log({\rm E/Y})$. It can be seen that Pattern 2 fits the data very well, with significant deviations for Lu and U only. As discussed above, Y is the better reference element for Pattern 2 due to uncertainties in $\log({\rm E/Eu})_2$ for some light $r$-process elements. Indeed, as shown in Fig.~\ref{fig:SMSS_2003_Eu}, the best fit with Eu as the reference element is slightly worse and requires a fraction $x_{\rm Eu}=2.5\times10^{-2}$ of Eu from Pattern 1 to account for the observed light $r$-process elements Sr to Ru.

Based on the results shown in Figs.~\ref{fig:SMSS_2003_Y}, \ref{fig:s16_Y}--\ref{fig:s18_Y}, and \ref{fig:J1432-4125}, we conclude that the $r$-process pattern in SMASS 2003-1142 is the same as those in J2213-5137, CS~22892-052, CS~31082-001, J1521-3538, and J1432-4125. All of the above patterns can essentially be accounted for by Pattern 2, which is better defined with Y as the reference element. 

\subsubsection{solar \texorpdfstring{$r$}{r}-process pattern}
\begin{figure*}[htbp]
    \centering
\includegraphics[width=1\textwidth]{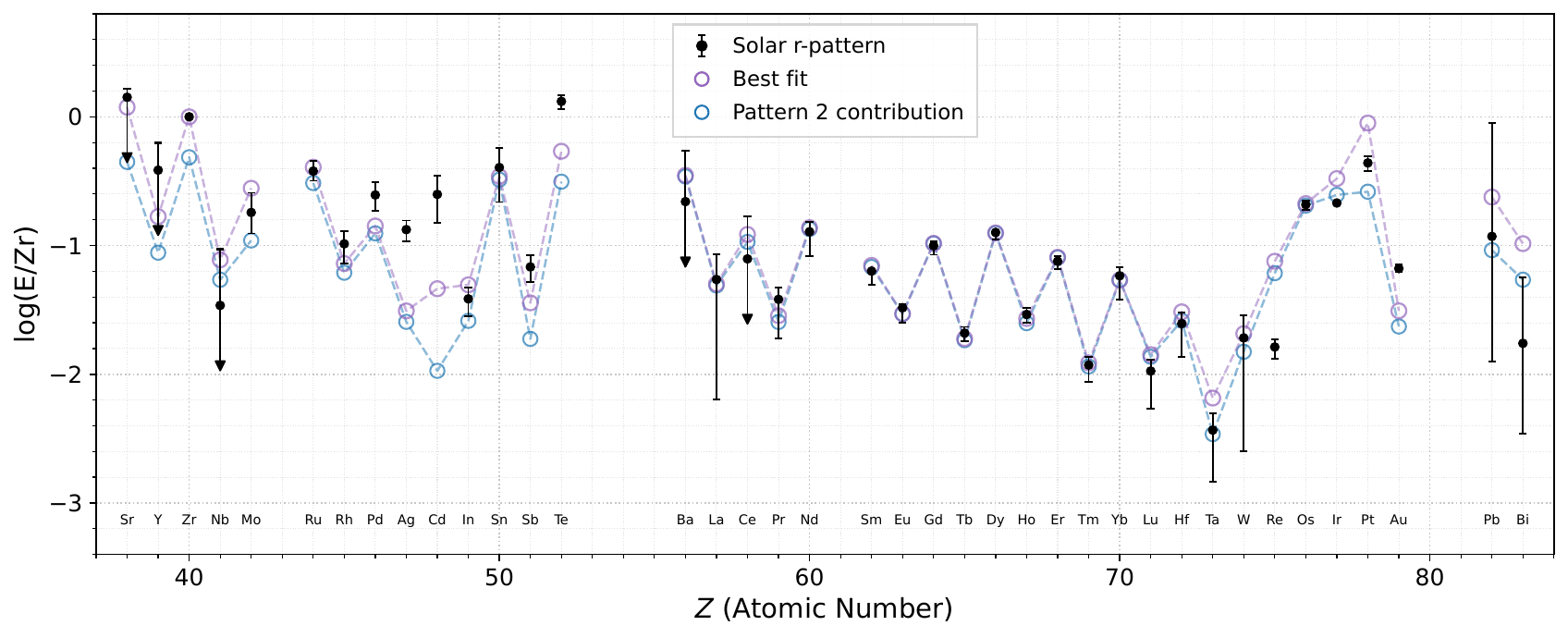}
\hfill
\caption{Comparison of the best fit (violet circles) and contributions from Pattern 2 (light blue circles) with the solar $r$-process pattern (black symbols, \citealt{goriely1999}) in terms of $\log({\rm E/Zr})$. The best fit corresponds to a mixing parameter $x_{\rm Zr}=0.51$, and is dominated by contributions from Pattern 2 with those from Pattern 1 mostly limited to the light $r$-process elements, Te, and Pt. The large discrepancies between the best fit and the solar $r$-process pattern at Ag, Cd, Te, Re, and Au suggest that the parameters for these elements for Patterns 1 and 2 have large uncertainties.}
\label{fig:solar_Y}
\end{figure*}

Because the solar $r$-process abundance of Y is very uncertain, we shift Patterns 1 and 2 with Y as the reference element by $-\log({\rm Zr/Y})_1$ and $-\log({\rm Zr/Y})_2$, respectively, and obtain Patterns 1 and 2 with Zr as the reference element. We ignore Th and U because their solar abundances received contributions over the Galactic history prior to solar system formation and cannot be simply compared with their relative abundances in Patterns 1 and 2 derived for MP stars. Figure~\ref{fig:solar_Y} compares the best fit (violet circles) and Pattern 2 contributions (light blue circles) with the solar $r$-process pattern (black symbols, \citealt{goriely1999}) in terms of $\log({\rm E/Zr})$. It can be seen that relative to the solar $r$-process pattern, the best fit with a mixing parameter $x_{\rm Zr} = 0.51$ is significantly lower at Ag, Cd, Te, and Au but higher at Re. The same discrepancies also occur for the best fit using Patterns 1 and 2 with Eu as the reference element (Fig.~\ref{fig:solar}). We note that the discrepant elements have only 10 or fewer measurements in the combined sample (Table~\ref{tab:lit}), and therefore, their parameters for Patterns 1 and 2 may be subject to large uncertainties. Further, in contrast to the reference elements Y and Eu (as well as Zr and many other elements) that are measured with singly-ionized species, Ag, Cd, Te, and Au are usually measured with neutral species. So there might be systematic errors in the data used to derive the parameters for Patterns 1 and 2 for these latter elements. Clearly, the resolution of the above discrepancies rely on a large uniformly-analyzed sample with many careful measurements of the pertinent elements.

\subsection{Uncertainties in the Two-Pattern Model}
\label{sec:uncertainty}
In Appendix~\ref{append:bayesian}, we use Bayesian inference to estimate the uncertainties in Patterns 1 and 2 and the corresponding mixing parameters. We assume uniform distributions within $\pm 2$ dex of the optimized values as the priors for Patterns 1 and 2. We also include unknown scatters for the observed elemental ratios in addition to the reported measurement uncertainties. Although broad uniform priors are used, the median pattern parameters are in very good agreement with the optimized patterns (Table~\ref{tab:pat}). In addition, the Bayesian analysis confirms that Pattern 1 for Sr to Cd, Te, and Pt and Pattern 2 for the majority of the elements from Sr to U are well defined whether Eu or Y is used as the reference element (Fig.~\ref{fig:bayesian_patterns}). For both patterns, large uncertainties in In, Sb, Ta, Bi, and U are clearly due to the few measurements used (Table~\ref{tab:lit}). For Sn, Lu, W, Re, Au, and Th in Pattern 1 and Te and Pb in Pattern 2, large uncertainties can be similarly attributed to the fewer than 10 measurements used. Because the pattern parameters are obtained effectively from a subtraction procedure, large uncertainties also occur when the relative production of an element in one pattern greatly exceeds that in the other (see discussion in \S\ref{sec:diff-ref}), and these uncertainties are exacerbated by the lack of measurements. Examples are Gd, Dy, Ho, Er, and Tm in Pattern 1 and Cd in Pattern 2. While large uncertainties obscure the changes in Pattern 1 for Gd, Dy, Ho, Er, Tm, Lu, and Ta when the reference element is changed from Eu to Y, the changes for Ba to Eu and Yb remain clear.

The Bayesian analysis shows that the large additional scatters in measurements across the combined sample (Fig.~\ref{fig:additional_scatter}) cause large uncertainties in Pt for Pattern 2 (see also discussion in \S\ref{sec:lit}). For the majority of the other elements, the additional scatters are less than 0.1 dex, which suggests that although not ideal, the combined sample is reasonable to use. 

Finally, the Bayesian analysis shows that the mixing parameters $x_{\rm Eu}$ and $x_{\rm Y}$ are rather well determined (Fig.~\ref{fig:mixing_para_baye}) and are in good agreement with those obtained using the optimized Patterns 1 and 2 (Fig.~\ref{fig:mix}).

\subsection{Two vs. Three Patterns}
\label{sec:2vs3}
Because Pattern 1 likely represents some average superposition, we present in Appendix~\ref{append:3pattern} a model involving Patterns $1'$, $2'$, and $3'$ with Y as the reference element (Fig.~\ref{fig:three_patterns}). It can be seen from Fig.~\ref{fig:model_pattern_comparison} that Pattern $1'$ is close to Pattern 1 up to Hf, Pattern $2'$ is close to Pattern 2 except for Sr, and Pattern $3'$ is close to Pattern 1 up to Cd. Note that the stability of Pattern 1 for Sr to Cd and that of Pattern 2 in the 2-pattern model are reproduced in the 3-pattern model. If we accept the three-pattern model, then Pattern 1 in the two-pattern model is effectively some average superposition of Patterns $1'$ and $3'$. However, the production of Ba and heavier elements (especially those beyond Hf) for Patterns $1'$ and $3'$ (e.g., the extremely high production of Au and Pb without co-production of Th for Pattern $1'$), and very little production of Sr for Pattern $2'$, are difficult to understand, which suggests that the three-pattern model might be only a mathematical solution instead of a physical one. 

The residuals for the three-pattern fits to the combined sample are shown in Fig.~\ref{fig:three_pattern_residuals}. As an external check, Fig.~\ref{fig:three_pattern_rpa_residuals} shows the residuals for the 10-star RPA sample \citep{2025A&A...704A.282R}. For both samples, the agreement between the model and data is at the same level as achieved by the 2-pattern model for nearly all the elements (cf. Figs.~\ref{fig:residual_all_Y} and \ref{fig:residual_new_stars}). Although the three-pattern model gives much better fits to Pt, Au, and Pb, this improvement is achieved at the cost of an additional pattern.

In Appendix~\ref{append:validation}, we perform 10-fold cross validation to further compare models involving two and three patterns. We calculate $\langle\chi^2\rangle_{\rm test}$ and $\langle\chi^2\rangle_{\rm train}$, which are the mean squares of the standardized residuals for all the entries in each test fold and the corresponding training folds, respectively. The predictive performance of the model is measured by $\langle\chi^2\rangle_{\rm test}$. As can be seen from Fig.~\ref{fig:cv_compare}, the three-pattern model has overall worse predictive performance (higher $\langle\chi^2\rangle_{\rm test}$) than the two-pattern model despite that the former model has lower $\langle\chi^2\rangle_{\rm train}$.

When different training folds are used, the labels of the inferred patterns become ambiguous in general. However, as shown in Fig.~\ref{fig:cv_2patterns}, the two-pattern model remains stable across the training folds, and therefore, its patterns can be labeled consistently. Note that the average inferred Patterns 1  and 2 are in excellent agreement with the corresponding patterns inferred from the full sample, which in turn are very close to Patterns 1 and 2 optimized using the original combined sample (Fig.~\ref{fig:pat_Y_Eu}).

In contrast, the three-pattern model varies widely across different training folds. In view of this unstable behavior and its worse predictive performance (Fig.~\ref{fig:cv_compare}), we conclude that the three-pattern model is not supported by our combined sample. However, it is still possible that such a model is preferred when a much larger sample of stars with high-quality measurements is available.

\section{Astrophysical Conditions for Producing Patterns 1 and 2}
\label{sec:conditions}
We now focus on the optimized Patterns 1 and 2 (Fig.~\ref{fig:pat_Y_Eu}), and explore the astrophysical conditions that may produce them using a parametric model for $r$-process nucleosynthesis. We consider the expansion of some ejecta from an initial state specified by the temperature $T_0=9$~GK, the density $\rho_0$, and the electron fraction $Y_{e,0}$ at time $t=0$. We adopt the time evolution of density suggested by \cite{lippuner2015},
\begin{equation}
    \rho(t) = \begin{cases}
        \rho_0 \exp(-t/\tau_{\rm exp}), & t\leq 3\tau_{\rm exp}, \\
        \rho_0 [3\tau_{\rm exp}/(et)]^3, & t>3\tau_{\rm exp},
    \end{cases}
\end{equation}
where the expansion timescale $\tau_{\rm exp}$ is the third adjustable parameter in addition to $\rho_0$ and $Y_{e,0}$ for the ejecta. We note that both $\rho(t)$ and $d\rho/dt$ are continuous at $t=3\tau_{\rm exp}$. For simplicity, we assume that the temperature evolves with time as
\begin{equation}
T(t)=T_0[\rho(t)/\rho_0]^{1/3}.
\end{equation}

For each set of parameters $\rho_0$, $Y_{e,0}$, and $\tau_{\rm exp}$ for the ejecta, we start the nucleosynthesis run from $T_0=9$~GK at $t=0$, and calculate the initial nuclear composition from $T_0$, $\rho_0$, and $Y_{e,0}$ assuming nuclear statistical equilibrium and using the SFHo equation of state \citep{2013ApJ...774...17S}. We follow the subsequent nucleosynthesis during the expansion of the ejecta using version 1.6.0 of the Portable Routines for Integrated nucleoSynthesis Modeling (PRISM) reaction network \citep{2021PhRvC.104a5803S}. The nuclear input to PRISM is based on the 2012 version of the Finite Range Droplet Model \citep{2012PhRvL.108e2501M,2016ADNDT.109....1M}. The radiative capture and fission rates are calculated with the CoH$_3$ statistical Hauser-Feshbach code \citep{2019arXiv190105641K,2021EPJA...57...16K}. The $\beta$-decay rates, along with probabilities of delayed neutron emission, are calculated assuming statistical de-excitation from excited states \citep{2016PhRvC..94f4317M,2018ApJ...869...14M}. The rates for the remaining reactions (e.g., $\alpha$-decay) are obtained from the REACLIB database \citep{2010ApJS..189..240C}. Nuclear fission is handled as in \cite{Vassh2019}. 

We logarithmically sample $\rho_0\in[10^6,10^{9.5}]$~g~cm$^{-3}$ and $\tau_{\rm exp}\in[10^{-3.6},10^{-1}]$~s, and uniformly sample $Y_{e,0}\in[0.005,0.5]$. Each parameter takes 25 values over the sampled interval, which results in a total of 15,625 runs. Each run includes decay of synthesized nuclei out to $t=1$~Gyr and the nuclear abundances at that time are used below, except for $^{232}$Th, $^{235}$U, and $^{238}$U. For these long-lived nuclei, we add an extra period of 12~Gyr for their decay assuming that the $r$-process elements in the MP stars of interest were produced 13~Gyr ago. The result from each run is referred to as a yield template.

\subsection{Pattern 2}
\begin{figure*}[htbp]
    \centering \includegraphics[width=1\textwidth]{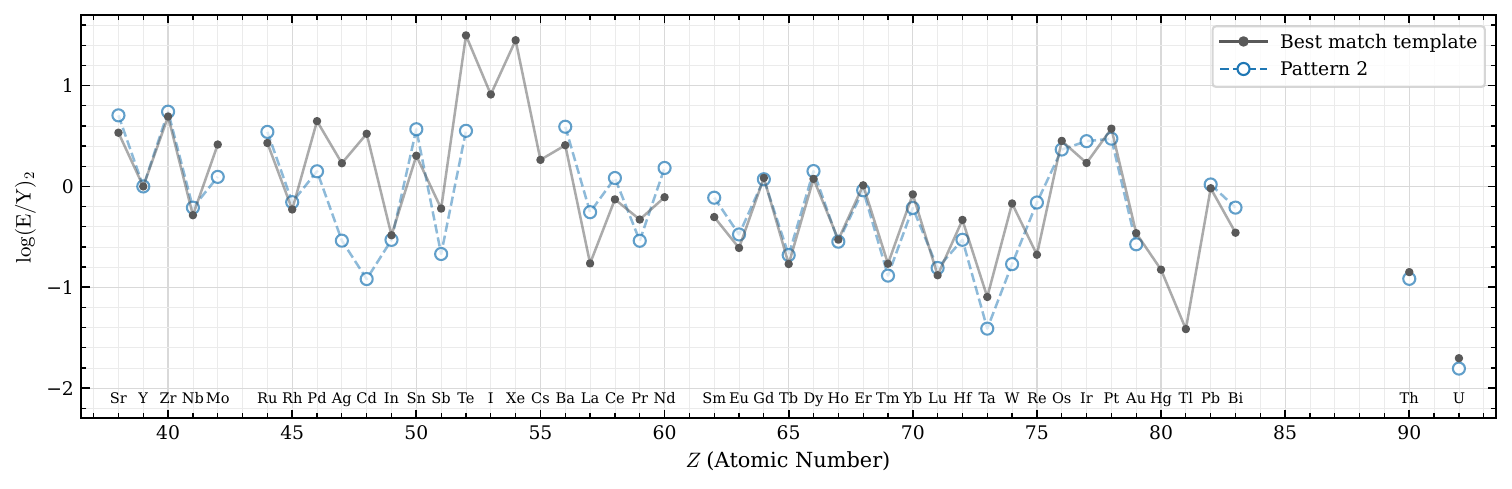}\\[1em]    
    \caption{The yield template (black filled circles) from a parametric $r$-process calculation that best matches Pattern 2 (light blue circles) in terms of $\log({\rm E/Y})_2$. The calculation assumes $\rho_0 = 5.89 \times 10^8$~g~cm$^{-3}$, $Y_{e,0} = 0.19$, and $\tau_{\rm exp} = 0.25$~ms.}
\label{fig:cond-p2}
\end{figure*}

\begin{figure*}[htbp]
    \centering
\includegraphics[width=1\textwidth]{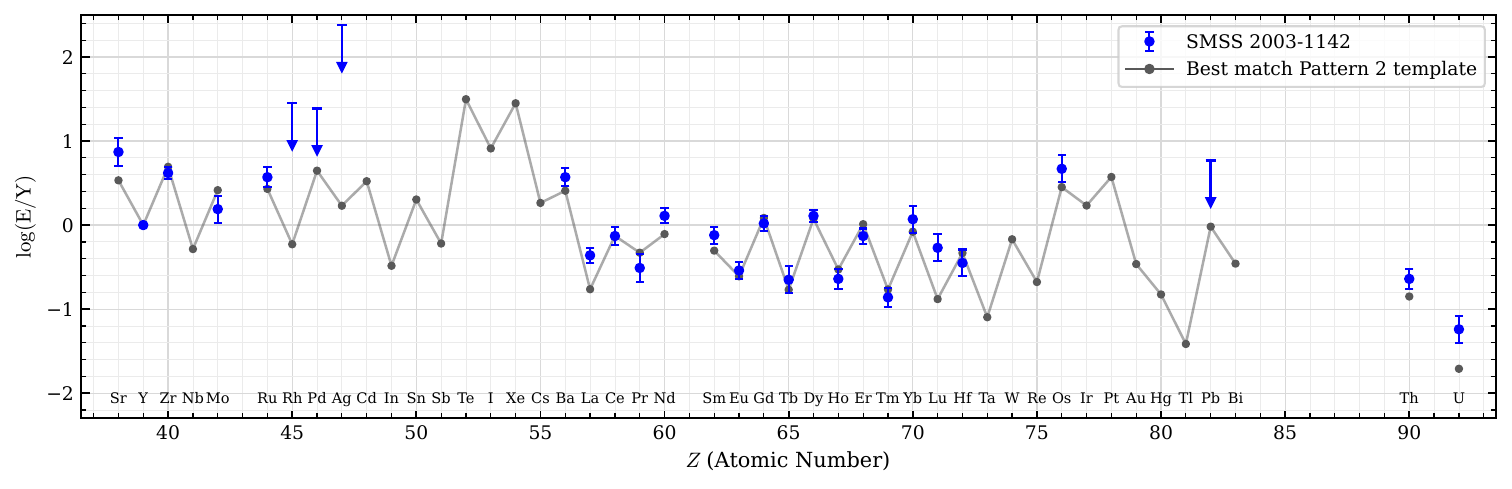}
\caption{Comparison of the template in Fig.~\ref{fig:cond-p2} (black filled circles) with the data (blue symbols, \citealt{2021Natur.595..223Y}) on $\log({\rm E/Y})$ for SMSS 2003-1142.}
\label{fig:condition_SMSS_2003_Y}
\end{figure*}

As Pattern 2 is rather stable (Fig.~\ref{fig:pat_Y_Eu}  and Appendix~\ref{append:validation}) and can essentially account for the $r$-process patterns in a number of MP stars in terms of $\log(\rm E/Y)$ (\S\ref{sec:test-p2}), we try to match it with a single yield template. The best match is found for $\rho_0 = 5.89 \times 10^8$~g~cm$^{-3}$, $Y_{e,0} = 0.19$, and $\tau_{\rm exp} = 0.25$~ms. As can be seen from Fig.~\ref{fig:cond-p2}, the corresponding template provides a good overall match to Pattern 2 in terms of $\log(\rm E/Y)_2$ for the elements from Sr to U. Clear exceptions are Pd, Ag, Cd, and Te, for which the $\log(\rm E/Y)_2$ values may have large uncertainties as noted above. The star SMASS~2003-1142 \citep{2021Natur.595..223Y} is not used to derive Pattern 2. Comparison of its data with the above template shows a very good match (Fig.~\ref{fig:condition_SMSS_2003_Y}). Future observations of Pd, Ag, Cd, and Te in this and other stars dominated by Pattern 2 can test whether these elements are better represented by the above template, thereby shedding more light on the astrophysical conditions for producing Pattern 2.

We note that the template in Fig.~\ref{fig:cond-p2} contains additional elements lighter than Sr, including a peak at Se (not shown). The total mass fraction of Sr and heavier elements is 0.8, and that of Se, Br, Kr, and Rb is $\approx 0.2$. Measurements of e.g., Se and Rb in MP stars dominated by Pattern 2 can further test the above template. 

We also note that there is a high degree of degeneracy in the astrophysical conditions for producing the above template. Between this template and Pattern 2, the root mean square deviation (RMSD) across the elements is 0.39 dex. Requiring that the RMSD be less than 0.45 dex and that the total mass fraction of Sr and heavier elements in the template exceed 0.1, we find 37 sets of parameters and show them in terms of $Y_{e,0}$, $\rho_0$, and $\tau_{\rm exp}$ in the left panel of Fig.~\ref{fig:para_pattern2}. The right panel shows the same sets of conditions with $\rho_0$ replaced by the initial entropy $S_0$. All the 37 sets of conditions give similar fits to Pattern 2. These conditions correspond to $Y_{e,0}=0.19$--0.21, $\rho_0\approx 8\times10^7$ to $10^9$~g~cm$^{-3}$, and $\tau_{\rm exp}\approx 0.25$--0.87~ms. Note that $S_0$ is dominated by non-relativistic nucleons and nuclei, so it scales as $\ln(T_0^{3/2}/\rho_0)$ and the above range of $\rho_0$ corresponds to a relatively narrow range of $S_0\approx 9$--15 in units of Boltzmann constant $k_{\rm B}$ per nucleon.

In principle, Pattern 2 could also represent a superposition of multiple yield templates. In this case, the relative amounts of ejecta producing distinct yield templates must be rather restricted to account for the coincidence of Pattern 2 with the $r$-process patterns in a number of MP stars. A possible way to provide such regularity of $r$-process production is for the yield template in Fig.~\ref{fig:cond-p2} to be a dominant component of any superposition. In addition, as widely noted in the literature, fission cycling can explain at least the regular production of the heavy $r$-process elements. Our parametric studies show that fission cycling is able to produce yield templates close to Pattern 2 for multiple sets of conditions (Fig.~\ref{fig:para_pattern2}).

\subsection{Pattern 1}
\label{sec:cond-p1}
\begin{figure*}[htbp]
    \centering \includegraphics[width=1\textwidth]{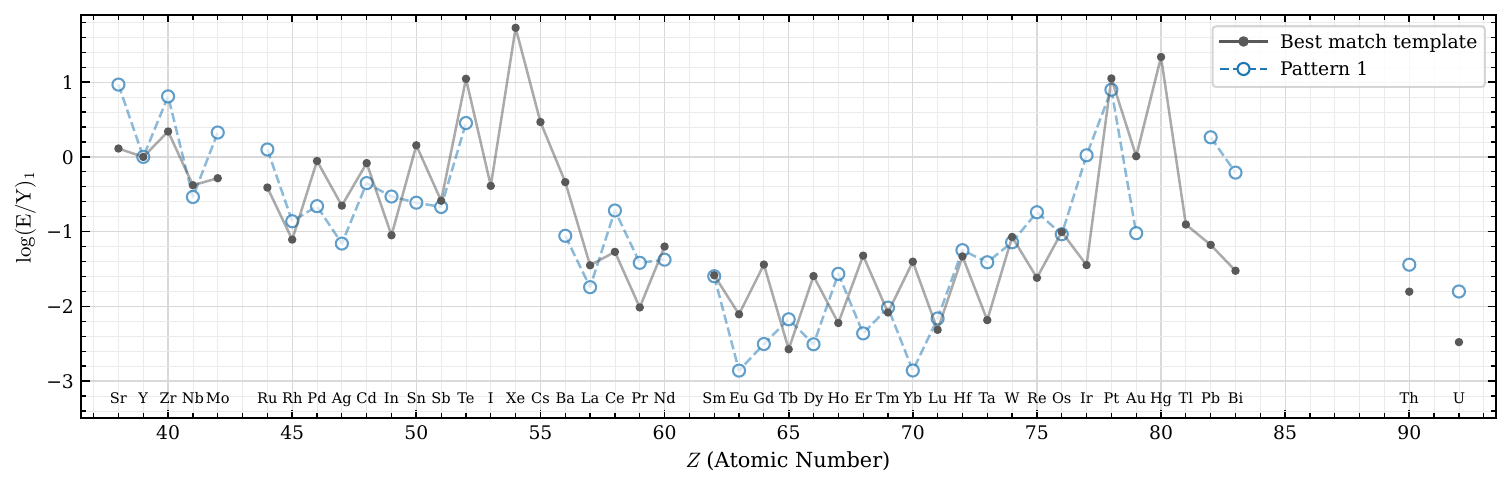}\\[1em]    
    \caption{The yield template (black filled circles) from a parametric $r$-process calculation that best matches Pattern 1 (light blue circles) in terms of $\log({\rm E/Y})_1$. The calculation assumes $\rho_0 = 2.74 \times 10^6$~g~cm$^{-3}$, $Y_{e,0} = 0.48$, and $\tau_{\rm exp} = 0.25$~ms.}
    \label{fig:para_1_Y_14210}
\end{figure*}

Based on the discussion in \S\ref{sec:mix}, Pattern 1 likely represents the average of production by multiple sets of astrophysical conditions with different yield templates. We now show that it cannot be matched with any single template from our parametric $r$-process calculations. The best match in terms of $\log({\rm E/Y})_1$ is found for $\rho_0 = 2.74 \times 10^6$~g~cm$^{-3}$ ($S_0=102$), $Y_{e,0} = 0.48$, and $\tau_{\rm exp} = 0.25$~ms. As can be seen from Fig.~\ref{fig:para_1_Y_14210}, while the corresponding template qualitatively matches Pattern 1, there is a general lack of quantitative agreement. The same is also true of the template from a slightly different set of conditions $\rho_0 = 1.95 \times 10^6$~g~cm$^{-3}$ ($S_0=137$), $Y_{e,0} = 0.46$, and $\tau_{\rm exp} = 0.25$~ms (Fig.~\ref{fig:para_1_Eu_Y}), which is the best match to Pattern 1 in terms of $\log({\rm E/Eu})_1$ and close to the best in terms of $\log({\rm E/Y})_1$. In contrast to the astrophysical conditions for producing Pattern 2, the above conditions predominantly produce $\alpha$ particles, with the total mass fraction of Sr and heavier elements being only $\approx 0.01$. In this sense, these conditions, if they exist, are extremely inefficient at producing the $r$-process elements.

\begin{figure*}[htbp]
    \centering \includegraphics[width=1\textwidth]{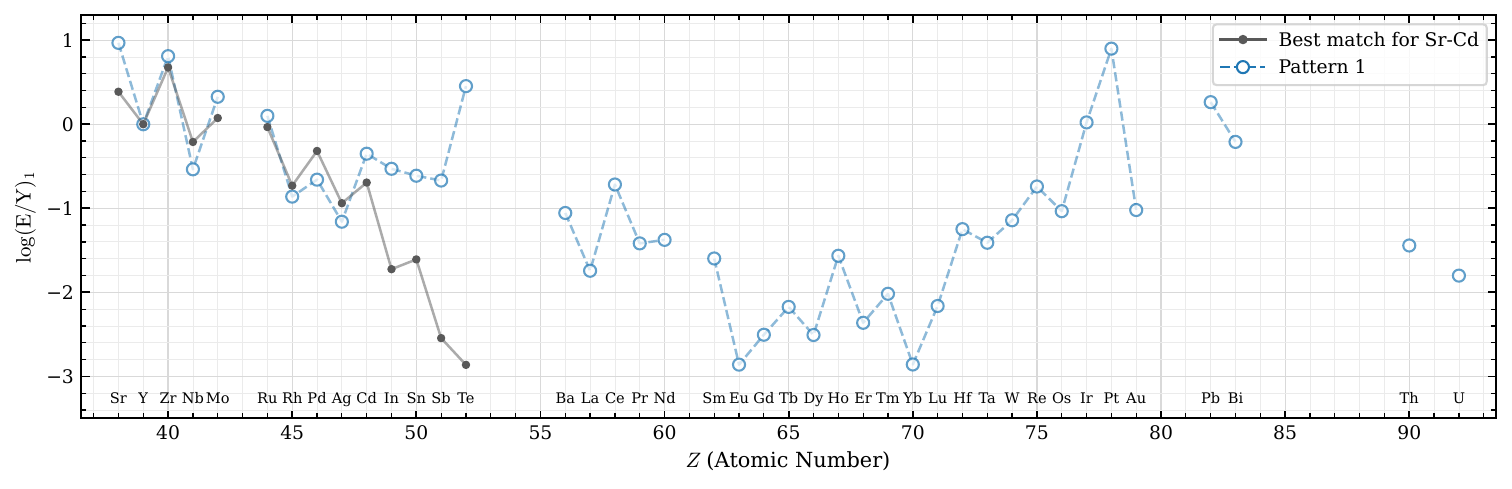}\\[1em]    
    \caption{
    The yield template (black filled circles) from a parametric $r$-process calculation that best matches Pattern 1 (light blue circles) for Sr to Cd in terms of $\log({\rm E/Y})_1$. The calculation assumes $\rho_0 = 2.26 \times 10^9$~g~cm$^{-3}$, $Y_{e,0} = 0.31$, and $\tau_{\rm exp} = 0.87$~ms.
    }
    \label{fig:para_1}
\end{figure*}

We next consider Pattern 1 as the superposition of multiple yield templates. Because Pattern 1 is dominated by the light $r$-process and the second and third peak elements (Fig.~\ref{fig:pat_Y_Eu}), it is reasonable to assume that one template covers the light $r$-process elements and the other templates cover the heavier elements. As Pattern 1 is uncertain between the second and third peak (Fig.~\ref{fig:pat_Y_Eu}), we focus on the former template covering Sr to Cd. Pattern 1 is rather stable for these elements whether Y or Eu is used as the reference element (Fig.~\ref{fig:pat_Y_Eu}). The best match in terms of $\log({\rm E/Y})_1$ is shown in Fig.~\ref{fig:para_1} and corresponds to $\rho_0 = 2.26 \times 10^9$~g~cm$^{-3}$, $Y_{e,0} = 0.31$, and $\tau_{\rm exp} = 0.87$~ms. The template in Fig.~\ref{fig:para_1} also contains elements lighter than Sr, including a peak at Se (not shown). The total mass fraction of Sr to Cd is 0.12, and that of Ti to Rb is 0.88. Measurements of e.g., Se and Rb in MP stars dominated by Pattern 1 can further test the above template.

As in the case of Pattern 2, there are also numerous sets of conditions with yield templates matching Pattern 1 for Sr to Cd. The RMSD across these elements for the above best match is 0.31 dex. Requiring that the RMSD be less than 0.4 dex and that the total mass fraction of Sr to Cd in the template exceed 0.1, we find 41 sets of parameters as shown in Fig.~\ref{fig:para_pattern1}. All the 41 sets of conditions give similar fits. These conditions correspond to $Y_{e,0}=0.29$--0.34, $S_0\approx 7$--21, and $\tau_{\rm exp}\approx 0.68$--17~ms. Compared to the conditions for producing Pattern 2, those for producing Pattern 1 for Sr to Cd not only are less neutron rich, but also have much slower expansion, and almost a half of them have higher entropy. We note that the conditions shown in Figs.~\ref{fig:para_pattern2} and \ref{fig:para_pattern1} are similar to those obtained by \cite{2026arXiv260711323K} through directly fitting the $r$-process patterns in three MP stars and the Sun.

\begin{figure*}[htbp]
    \centering \includegraphics[width=1\textwidth]{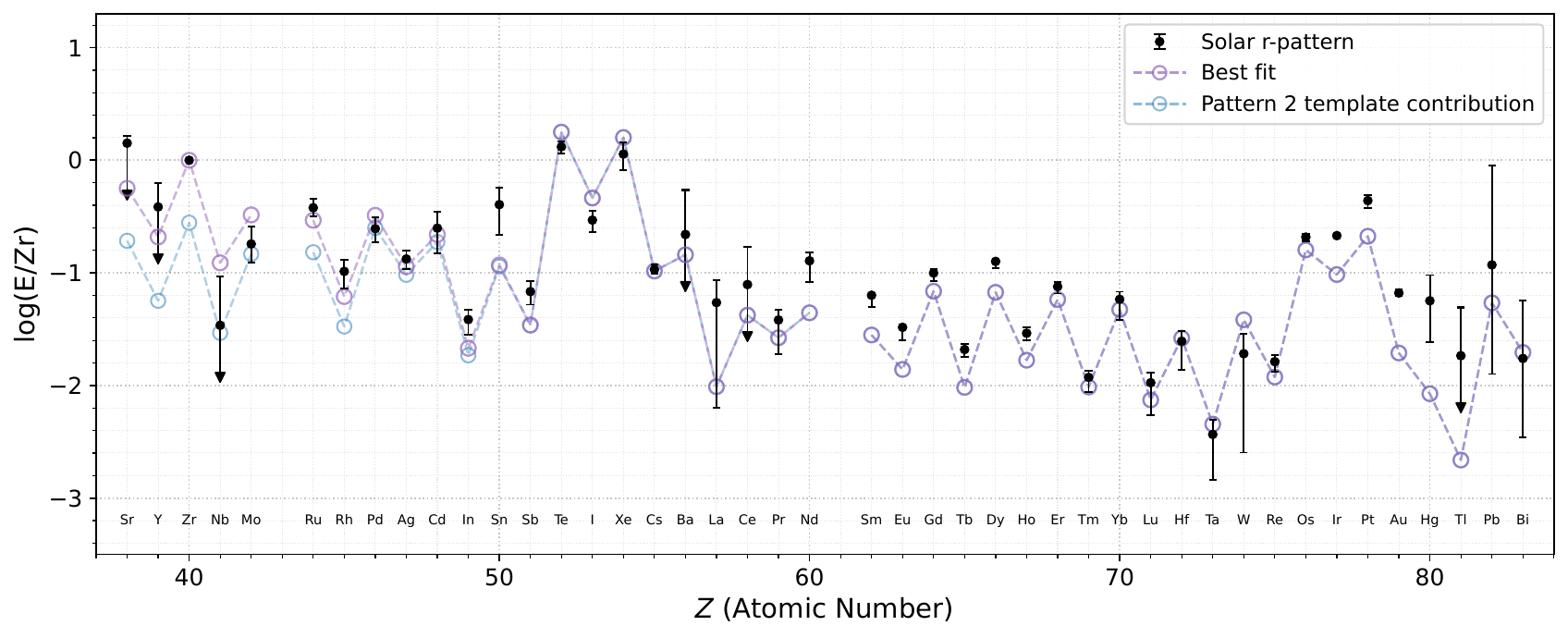}\\[1em]    
    \caption{Comparison of the best fit (violet circles) and contributions from the template in Fig.~\ref{fig:cond-p2} (light blue circles) with the solar $r$-process pattern (black symbols, \citealt{goriely1999}) in terms of $\log({\rm E/Zr})$. The best fit is a mixture of the templates in Figs.~\ref{fig:cond-p2} and \ref{fig:para_1}, with a fraction $x_{\rm Zr} = 0.72$ of Zr contributed by the latter. The violet and blue circles coincide beyond Sn. The deficiencies beyond Cd require additional production with different templates from the one in Fig.~\ref{fig:cond-p2}.}
    \label{fig:solar-12}
\end{figure*}

As a simple test of the yield template in Fig.~\ref{fig:para_1}, we use it along with the template in Fig.~\ref{fig:cond-p2} to fit the solar $r$-process pattern in terms of $\log({\rm E/Zr})$. The best fit is shown in Fig.~\ref{fig:solar-12} and corresponds to a mixing parameter $x_{\rm Zr}=0.72$ (the fraction of Zr contributed by the template in Fig.~\ref{fig:para_1}). It can be seen that the fit is quite good for the light $r$-process and second peak elements. With only the contributions from the template in Fig.~\ref{fig:cond-p2}, the fit is not as good beyond the second peak. Clearly, additional templates for the elements beyond the second peak, or more generally beyond Cd, are needed to fit the overall solar $r$-process pattern. Better determination of Pattern 1 from more comprehensive data on MP stars will help find these templates and the associated astrophysical conditions. Of course, when these templates are included, the true best fit to the solar $r$-process pattern would differ from the above illustrative fit.

Compared with the best fit to the solar $r$-process pattern using Patterns 1 and 2 derived from the data on MP stars (Fig.~\ref{fig:solar_Y}), the above illustrative fit using the calculated templates removes the deficiencies at Ag, Cd, and Te (Fig.~\ref{fig:solar-12}). This removal is due to the much higher production of these three elements by the template in Fig.~\ref{fig:cond-p2} relative to Pattern 2. We emphasize again that future measurements of these elements in MP stars dominated by Pattern 2 will help determine whether their $r$-process patterns are better represented by our calculated template and shed important light on the astrophysical conditions for producing Pattern 2.

\section{Discussion and Conclusions}
\label{sec:discussion}
Using the elemental abundances in 68 MP stars from the combined CERES and literature sample, we have derived Patterns 1 and 2 for $r$-process production (Fig.~\ref{fig:pat_Y_Eu} and Table~\ref{tab:pat})  with estimated uncertainties (Appendix~\ref{append:bayesian}) and cross validation (Appendix~\ref{append:validation}), and have shown that essentially all the $r$-process patterns in the combined sample can be adequately explained as mixtures of these patterns (Figs.~\ref{fig:residual_all}, \ref{fig:residual1}--\ref{fig:residual4}, and \ref{fig:residual_all_Y}). We also have confirmed that the same explanation applies to other MP stars not used for deriving Patterns 1 and 2 (Figs.~\ref{fig:SMSS_2003_Y}, \ref{fig:J1430-2371}, \ref{fig:J1432-4125}, \ref{fig:residual_new_stars}, and \ref{fig:SMSS_2003_Eu}), as well as to the solar $r$-process pattern (Figs.~\ref{fig:solar_Y} and \ref{fig:solar}). Compared to Pattern 2, Patten 1 has much higher production of the light $r$-process elements relative to the Eu region. Pattern 2 is rather stable whether Eu or Y is used as the reference element. While Pattern 1 for the light $r$-process and the second and third peak elements is also stable, Pattern 1 between the second and third peak is quite uncertain (Fig.~\ref{fig:pat_Y_Eu}). Data on [Y/Eu] indicate that as the reference element, Eu is generally better for Pattern 1 while Y is better for Pattern 2 (\S\ref{sec:mix}). Stars dominated by Pattern 1 or 2 (Table~\ref{tab:star_pattern_1_2}) occupy distinct regions bounded by their mixing parameters ($x_{\rm Eu}$ or $x_{\rm Y}$), [Fe/H], and [Eu/Fe] (Fig.~\ref{fig:mix}). These results and our unified framework to account for the $r$-process patterns in MP stars provide theoretical insights into the empirical categories of limited-$r$, $r$-I, and $r$-II stars (\S\ref{sec:mix}). While it is straightforward to extend our approach to more than two $r$-process production patterns, a three-pattern model is not supported by the limited data for our combined sample (Appendices~\ref{append:3pattern} and \ref{append:validation}).

Comparisons with the data on $\log(\rm E/Y)$ for J2213-5137, CS~22892-052, CS~31082-001, and J1521-3538 (Figs.~\ref{fig:s16_Y}--\ref{fig:s18_Y}) in the combined sample, as well as external stars J1432-4125 (Fig.~\ref{fig:J1432-4125}) and SMASS 2003-1142 (Fig.~\ref{fig:SMSS_2003_Y}), show that the $r$-process patterns in all these stars can be essentially accounted for by Pattern 2, which points to regularity in $r$-process production by the corresponding sources. This regularity tightly constrains the relative amounts of ejecta producing distinct yield templates if Pattern 2 is the superposition of such templates. On the other hand, an extensive survey of parametric $r$-process calculations shows that Pattern 2  may be dominated by a single yield template (Figs.~\ref{fig:cond-p2} and \ref{fig:condition_SMSS_2003_Y}), which can be produced by a range of astrophysical conditions (Fig.~\ref{fig:para_pattern2}). The corresponding ejecta is very neutron-rich with well-defined $Y_{e,0}=0.19$--0.21, has entropy $S_0\approx 9$--15, and rapidly expands with a timescale $\tau_{\rm exp} = 0.25$--0.87~ms. While our $r$-process calculations differ in some aspects of parametrization and nuclear input, these results are at least in qualitative agreement with other parametric studies (e.g., yield templates of Group 5 in \citealt{2025ApJ...990...37K}).

Although Pattern 1 can be qualitatively matched by a single yield template (Figs.~\ref{fig:para_1_Y_14210} and \ref{fig:para_1_Eu_Y}, see also templates of Group 3 in \citealt{2025ApJ...990...37K}), there is a general lack of quantitative agreement. We conclude that Pattern 1 is the superposition of multiple templates, and therefore, most likely represents an average. Because Pattern 1 for the light $r$-process elements Sr to Cd is rather stable whether Eu or Y is used as the reference element (Fig.~\ref{fig:pat_Y_Eu}), this part likely corresponds to a single template. Superposition of this template and others for heavier elements would lead to variations of the overall pattern between the second and third peak. Consequently, even as an average, Pattern 1 for the Eu region would depend on whether Eu or Y is the reference element (Fig.~\ref{fig:pat_Y_Eu}), especially when the sample of stars used to derive Patterns 1 and 2 is not very large. Deviations from the average are expected and can explain the high value of [Y/Eu]~$=0.89$ for the outlier J2140-1227 in comparison with [Y/Eu]~$=0.05\pm0.35$ for all the MP stars dominated by Pattern 1 in Table~\ref{tab:star_pattern_1_2}. 

As in the case of Pattern 2, the template (Fig.~\ref{fig:para_1}, see also patterns of Group 1 in \citealt{2025ApJ...990...37K}) matching Pattern 1 for the light $r$-process elements Sr to Cd can be produced by a range of astrophysical conditions (Fig.~\ref{fig:para_pattern1}). The corresponding ejecta is moderately neutron rich with well-defined $Y_{e,0}=0.29$--0.34, has entropy $S_0\approx7$--21, and expands with a timescale $\tau_{\rm exp}\approx 0.68$--17~ms. This expansion is generally much slower than that for producing the likely dominant template for Pattern 2. Large uncertainties in Pattern 1 between the second and third peak (Fig.~\ref{fig:pat_Y_Eu}) prevent us from finding the corresponding templates and the associated astrophysical conditions. Nevertheless, additional production beyond Cd with different templates from Pattern 2 is required to explain the data on MP stars (e.g., Figs.~\ref{fig:HD122563}, \ref{fig:J2140-1227}, and \ref{fig:HD122563_Y}) and the solar $r$-process pattern (e.g., Fig.~\ref{fig:solar-12}). 

The astrophysical conditions from a wide range of CCSN and NSM models are summarized in Fig.~9 and Table~3 of \cite{2025ApJ...990...37K}. Compared to the conditions for producing the template matching Pattern 2 or Pattern 1 for the light $r$-process elements Sr to Cd, standard neutrino-driven winds in regular CCSNe have much higher $Y_{e,0}\gtrsim 0.4$ (e.g., \citealt{1996ApJ...471..331Q,wang2023}), and therefore, cannot be the source for either template. Further, as these winds can significantly produce the elements up to Cd at most (e.g., \citealt{hoffman1997}), they cannot contribute to Pattern 1 beyond Cd. So standard neutrino-driven winds do not play any significant role in our explanation of the $r$-process patterns in MP stars. Ejecta from magneto-rotational SNe cannot be the source for our templates, either, as the expansion timescale is too long ($\tau_{\rm exp}\gg 10$~ms, \citealt{2017MNRAS.469L..43O,2021MNRAS.501.5733R}) or there are other mismatches ($Y_{e,0}=0.18$ and $\tau_{\rm exp}\sim 7$~ms, \citealt{2012ApJ...750L..22W}). However, at least some of this ejecta can contribute to Pattern 1 beyond Cd.

Conditions for producing the template matching Pattern 2 appear to exist in shock-heated ejecta in some NSM simulations \citep{2017PhRvD..96l4005B}, while those for producing the template matching Pattern 1 for the light $r$-process elements Sr to Cd are found in shock-heated and disk ejecta in other NSM simulations \citep{2024MNRAS.527.8812J,2024MNRAS.533.2096R}, as well as in neutrino-driven winds from the hypermassive NS formed in NSMs \citep{2014MNRAS.443.3134P,2015ApJ...813....2M}. Some disk ejecta from NSMs has very long expansion timescales ($\tau_{\rm exp}\gg10$~ms, \citealt{2013MNRAS.435..502F,2016MNRAS.463.2323W}), and can only contribute to Pattern 1 beyond Cd. The tidal ejecta from NSMs has very low $Y_{e,0}<0.1$ and entropy $S_0\ll 1$ \citep{2012MNRAS.426.1940K,2013MNRAS.430.2121P,2013MNRAS.430.2585R}, and can also contribute to Pattern 1 beyond Cd. In terms of mass (e.g., \citealt{2024MNRAS.533.2096R}), the tidal ejecta [${\cal{O}}(10^{-4})$ $M_\odot$] is much less than the shock-heated ejecta [${\cal{O}}(10^{-3})$ $M_\odot$], which in turn is much less than the disk ejecta [${\cal{O}}(10^{-2})$ $M_\odot$]. Neutrino-driven winds from a sufficiently long-lived hypermassive NS can eject a total mass up to $\sim 10^{-2}\,M_\odot$ \citep{2015ApJ...813....2M}.

Based on the above discussion, Pattern 2 is likely produced by some NSMs whose $r$-process material is dominated by shock-heated ejecta with the conditions shown in Fig.~\ref{fig:para_pattern2}. As discussed in \S\ref{sec:mix}, all the MP stars dominated by Pattern 2 in Table~\ref{tab:star_pattern_1_2} have well-defined [Y/Eu]~$=-1.18\pm0.09$, but wide ranges of [Eu/Fe]~$=1.48\pm0.44$ and [Y/Fe]~$=0.30\pm0.41$, with extremely high values of [Eu/Fe]~$=2.45$ (2.20) and [Y/Fe]~$=1.13$ (1.03) for the outlier J2213-5137 (J1521-3538). These results strongly support that Pattern 2 is decoupled from Fe and therefore, is produced by the subset of NSMs proposed above.

In contrast, Pattern 1 represents the average superposition of multiple templates produced by a variety of NSM ejecta and perhaps also by some magneto-rotational SN ejecta. This superposition is consistent with the significant scatter in [Y/Eu]~$=0.05\pm0.35$ for all the MP stars dominated by Pattern 1 in Table~\ref{tab:star_pattern_1_2}. As discussed above, for another subset of NSMs, shock-heated and disk ejecta as well as neutrino-driven winds can produce the template matching Pattern 1 for the light $r$-process elements Sr to Cd, while tidal and some disk ejecta can contribute to Pattern 1 beyond Cd. Based on the relative amounts of different NSM ejecta, it is possible that Pattern 1 should have much higher production of the light $r$-process elements relative to the Eu region compared to Pattern 2, especially if magneto-rotational SNe contribute very little to that region.

 We are well aware of the uncertainties in CCSN and NSM simulations, and expect that improved simulations will lead us closer to the sources for Patterns 1 and 2. As discussed in \S\ref{sec:mix}, MP stars dominated by Pattern 1 or 2 belong to the limited-$r$ or $r$-II category, respectively, and most of those with more balanced contributions from both patterns are $r$-I stars. The extensive RPA survey \citep{2018ApJ...858...92H,2018ApJ...868..110S,2020ApJ...898..150E} has discovered 56 limited-$r$, 163 $r$-I, and 28 $r$-II stars. We estimate that NSMs with a wide range of ejecta contributing to Pattern 1 are $\sim 2$ times more common than those with $r$-process material dominated by shock-heated ejecta producing Pattern 2. 

The above results and conclusions demonstrate what has been achieved by our data-driven model for $r$-process production patterns and also illustrate how it can be improved. As emphasized throughout this paper, the foundation for our model is the data on a wide range of $r$-process elements in a large sample of MP stars. We recognize that the combined sample used here is not ideal: it is not uniformly analyzed and lacks coverage for many $r$-process elements.  The latter factor leads to significant uncertainties especially in Pattern 1 (Appendix~\ref{append:bayesian}). There are also potential systematic errors in abundance ratios for elements measured with neutral species (e.g., Pd, Ag, Cd, Te, Os, Ir, Pt, Au) relative to those measured with singly-ionized species (e.g., Sr, Y, Zr, Ba to Hf). A particular important issue is the discrepancies between the CERES measurements and literature data on Os and Pt for HD~122563 (Fig.~\ref{fig:HD122563}). If the CERES measurements are confirmed, HD~122563 should no longer be considered as poor in heavy $r$-process elements. Instead, it should be an outstanding MP star dominated by Pattern 1 with a prominent third peak. In that case, the meaning of the limited-$r$ category should be drastically revised. In general, Os, Ir, and Pt measurements for MP stars should be examined for possible systematic errors so that Patterns 1 and 2 for these elements can be determined more precisely (see Figs.~\ref{fig:residual_all}, \ref{fig:residual1}--\ref{fig:residual4}, \ref{fig:residual_all_Y}, and \ref{fig:additional_scatter} for comparison with other elements). It takes tremendous efforts to observe and analyze a large number of $r$-process elements in a large sample of MP stars. We hope that this paper further motivates such efforts, which in turn will greatly advance the understanding of the $r$-process.

\section*{Acknowledgments}
We thank the anonymous referee for helpful suggestions that have improved the paper. Y.Z.Q. is grateful for the hospitality of the ExtreMe Matter Institute (EMMI) and the Nuclear Astrophysics and Structure Department at the GSI Helmholtzzentrum f{\"u}r Schwerionenforschung GmbH, Darmstadt, Germany, where this work was initiated. This work was supported in part by the US Department of Energy under grant DE-FG02-87ER40328 (C.L. and Y.Z.Q.) and by the Alexander von Humboldt Foundation via a Humboldt Research Award to Y.Z.Q. A.G. acknowledges support from the Laboratory Directed Research and Development program of Los Alamos National Laboratory (LANL) under Project Nos. 20230052ER and 20240004DR. LANL is operated by Triad National Security, LLC, for the National Nuclear Security Administration of U.S. Department of Energy (Contract No. 89233218CNA000001). Z.X. acknowledges support of the European Research Council (ERC) under the ERC Advanced Grant KILONOVA (No. 885281) through the European Union’s Horizon 2020 research and innovation program and under the ERC Grant NeuTrAE (No. 101165138). The parametric nucleosynthesis calculations were performed at LANL. The results from these calculations and the data on MP stars were analyzed with resources of the Minnesota Supercomputing Institute.

This work is partially funded by the European Union. Views and opinions expressed are however those of the authors only and do not necessarily reflect those of the European Union or the European Research Council Executive Agency. Neither the European Union nor the granting authority can be held responsible for them.

\software{NumPy \citep{harris2020numpy}, pandas \citep{mckinney2010pandas}, Matplotlib \citep{hunter2007matplotlib}, SciPy \citep{virtanen2020scipy}, PyMC and PyTensor \citep{abrilpla2023pymc}, and ArviZ \citep{kumar2019arviz}.}

\appendix

\section{Optimization}
\label{append:optimization}
\restartappendixnumbering
\renewcommand{\theHfigure}{A\arabic{figure}}
\renewcommand{\theHtable}{A\arabic{table}}

Taking Eu as the example reference element, we optimize the parameters $\log({\rm E/Eu})_1$ and $\log({\rm E/Eu})_2$ for Patterns 1 and 2, respectively, along with the mixing parameter $x_{\rm Eu}$ for each star, by minimizing the quantity $Q$ in Eq.~(\ref{eq:q}). We perform the optimization with \texttt{scipy.optimize.minimize} using the L-BFGS-B algorithm ($\texttt{maxiter}=20000$, $\texttt{maxfun}=20000$, and $\texttt{ftol}=10^{-9}$). L-BFGS-B is a bounded quasi-Newton algorithm that uses a limited-memory approximation to the inverse Hessian, making it computationally efficient for the large parameter space considered here. Clearly, $0\leq x_{\rm Eu}\leq 1$. The range of $\log({\rm E/Eu})_1$ and $\log({\rm E/Eu})_2$ is set to be $[-6,6]$.

For initialization, we take $x_{\rm Eu}=0.5$, $\log({\rm E/Eu})_1= m({\rm E/Eu})+0.25p({\rm E/Eu})$, and $\log({\rm E/Eu})_2= m({\rm E/Eu})-0.25p({\rm E/Eu})$, where $m({\rm E/Eu})$ is the median of the observed $\log({\rm E/Eu})^{\rm obs}_*$ values for the relevant sample, and $p({\rm E/Eu})$ corresponds to the first principal-component direction for the values of $\log({\rm E/Eu})^{\rm obs}_*-m({\rm E/Eu)}$. We exclude upper limits for elemental abundances in our procedure because they lack constraining power (and criteria for setting upper limits may differ for different observational studies). Even with definite measurements of element E, significant uncertainties in $\log({\rm E/Eu})_1$ and $\log({\rm E/Eu})_2$ are expected when the number of measurements is insufficient (Appendix~\ref{append:bayesian}).

The results of the above optimization procedure are confirmed by Bayesian inference in Appendix~\ref{append:bayesian} and cross validated in Appendix~\ref{append:validation}.

%\clearpage
\section{Data Samples, Derived Patterns, and Representative Stars}
\label{append:data}
\restartappendixnumbering
\renewcommand{\theHfigure}{B\arabic{figure}}
\renewcommand{\theHtable}{B\arabic{table}}

We give the average measurement uncertainties for individual elements in the CERES and literature samples in Tables~\ref{tab:ceres} and \ref{tab:lit}, respectively. Table~\ref{tab:pat} gives Patterns 1 and 2 derived from the combined sample, and Table~\ref{tab:star_pattern_1_2} lists the stars dominated by either pattern.

\begin{table}[htbp]
\centering
\caption{Average uncertainties for the CERES sample}
\begin{tabular}{ccccc}
\hline\hline
$Z$ & E & $n_{\rm E}$ & $\sigma_{\log({\rm E/Eu})}$ & $\sigma_{\log({\rm E/Y})}$ \\
\hline
38 & Sr & 41 & 0.06 & 0.11 \\
39 & Y  & 51 & 0.11 & --   \\
40 & Zr & 50 & 0.13 & 0.16 \\
56 & Ba & 42 & 0.06 & 0.11 \\
57 & La & 45 & 0.08 & 0.13 \\
58 & Ce & 46 & 0.06 & 0.11 \\
59 & Pr & 30 & 0.06 & 0.11 \\
60 & Nd & 43 & 0.08 & 0.13 \\
62 & Sm & 38 & 0.06 & 0.11 \\
63 & Eu & 51 & --   & 0.11   \\
72 & Hf & 19 & 0.14 & 0.17 \\
76 & Os & 33 & 0.18 & 0.20 \\
77 & Ir & 31 & 0.17 & 0.20 \\
78 & Pt & 18 & 0.12 & 0.15 \\
\hline
\end{tabular}
\tablecomments{
$n_{\rm E}$ is the number of stars with measurements of E and the relevant reference element.
}
\label{tab:ceres}
\end{table}

\begin{table}[htbp]
\centering
\caption{Average uncertainties for the literature sample}
\begin{tabular}{ccccc}
\hline\hline
$Z$ & E & $n_{\rm E}$ & $\sigma_{\log({\rm E/Eu})}$ & $\sigma_{\log({\rm E/Y})}$ \\
\hline
38 & Sr & 20 & 0.14 & 0.15 \\
39 & Y  & 20 & 0.10 & -- \\
40 & Zr & 20 & 0.11 & 0.12 \\
41 & Nb & 11 & 0.16 & 0.16 \\
42 & Mo & 16 & 0.18 & 0.19 \\
44 & Ru & 14 & 0.15 & 0.16 \\
45 & Rh & 8  & 0.20 & 0.21 \\
46 & Pd & 11 & 0.15 & 0.15 \\
47 & Ag & 9  & 0.19 & 0.19 \\
48 & Cd & 10 & 0.21 & 0.22 \\
49 & In & 1  & 0.21 & 0.21 \\
50 & Sn & 5  & 0.20 & 0.20 \\
51 & Sb & 1  & 0.17 & 0.17 \\
52 & Te & 7  & 0.14 & 0.14 \\
56 & Ba & 20 & 0.12 & 0.13 \\
57 & La & 17 & 0.12 & 0.13 \\
58 & Ce & 14 & 0.10 & 0.12 \\
59 & Pr & 11 & 0.13 & 0.13 \\
60 & Nd & 17 & 0.12 & 0.13 \\
62 & Sm & 14 & 0.13 & 0.13 \\
63 & Eu & 20 & -- & 0.11 \\
64 & Gd & 13 & 0.12 & 0.13 \\
65 & Tb & 10 & 0.15 & 0.16 \\
66 & Dy & 15 & 0.12 & 0.13 \\
67 & Ho & 11 & 0.14 & 0.15 \\
68 & Er & 15 & 0.14 & 0.15 \\
69 & Tm & 11 & 0.14 & 0.15 \\
70 & Yb & 17 & 0.18 & 0.18 \\
71 & Lu & 7  & 0.20 & 0.20 \\
72 & Hf & 9  & 0.19 & 0.19 \\
73 & Ta & 1  & 0.24 & 0.24 \\
74 & W  & 2  & 0.18 & 0.18 \\
75 & Re & 2  & 0.18 & 0.19 \\
76 & Os & 10 & 0.20 & 0.21 \\
77 & Ir & 6  & 0.12 & 0.13 \\
78 & Pt & 8  & 0.22 & 0.22 \\
79 & Au & 6  & 0.28 & 0.29 \\
82 & Pb & 4  & 0.19 & 0.19 \\
83 & Bi & 1  & 0.33 & 0.34 \\
90 & Th & 6  & 0.14 & 0.14 \\
92 & U  & 2  & 0.22 & 0.22 \\
\hline
\end{tabular}
\tablecomments{
$n_{\rm E}$ is the number of stars with measurements of E and the relevant reference element.
}
\label{tab:lit}
\end{table}

\begin{table*}[htbp]
\centering
\scriptsize
\setlength{\tabcolsep}{2.0pt}
\renewcommand{\arraystretch}{1.05}
\caption{Optimized Patterns 1 and 2 derived from the combined sample along with the Bayesian posterior medians and 16th--84th percentile credible intervals.}
\noindent\makebox[\textwidth][c]{%
\hspace*{-0.25in}%
\begin{tabular}{ccrrrrrrrr}
\hline\hline
Z & E &
$\log({\rm E/Eu})_1$ & $\log({\rm E/Eu})_2$ &
$\log({\rm E/Eu})_{1,\rm Bayes}$ &
$\log({\rm E/Eu})_{2,\rm Bayes}$ &
$\log({\rm E/Y})_1$  & $\log({\rm E/Y})_2$ &
$\log({\rm E/Y})_{1,\rm Bayes}$ &
$\log({\rm E/Y})_{2,\rm Bayes}$ \\
\hline
38 & Sr & 2.88 & 0.17 & $2.91_{-0.03}^{+0.02}$ & $-0.50_{-0.88}^{+0.79}$ & 0.97 & 0.70 & $1.00_{-0.03}^{+0.03}$ & $0.64_{-0.08}^{+0.07}$ \\
39 & Y  & 1.90 & 0.24 & $1.93_{-0.03}^{+0.03}$ & $0.15_{-0.11}^{+0.09}$ & 0.00 & 0.00 & 0.00 & 0.00 \\
40 & Zr & 2.73 & 0.88 & $2.77_{-0.03}^{+0.03}$ & $0.83_{-0.15}^{+0.12}$ & 0.81 & 0.74 & $0.81_{-0.03}^{+0.03}$ & $0.75_{-0.06}^{+0.05}$ \\
41 & Nb & 1.62 & $-0.08$ & $1.64_{-0.10}^{+0.10}$ & $-0.14_{-0.47}^{+0.22}$ & $-0.54$ & $-0.21$ & $-0.51_{-0.13}^{+0.12}$ & $-0.19_{-0.11}^{+0.10}$ \\
42 & Mo & 2.25 & $-0.10$ & $2.26_{-0.07}^{+0.06}$ & $-0.35_{-1.11}^{+0.61}$ & 0.33 & 0.10 & $0.24_{-0.10}^{+0.09}$ & $0.16_{-0.17}^{+0.14}$ \\
44 & Ru & 2.16 & 0.80 & $2.20_{-0.08}^{+0.08}$ & $0.76_{-0.14}^{+0.12}$ & 0.10 & 0.54 & $0.11_{-0.11}^{+0.09}$ & $0.54_{-0.08}^{+0.08}$ \\
45 & Rh & 1.34 & 0.14 & $1.39_{-0.21}^{+0.19}$ & $0.09_{-0.24}^{+0.16}$ & $-0.87$ & $-0.16$ & $-0.85_{-0.27}^{+0.22}$ & $-0.16_{-0.11}^{+0.11}$ \\
46 & Pd & 1.40 & 0.53 & $1.50_{-0.11}^{+0.10}$ & $0.47_{-0.10}^{+0.09}$ & $-0.68$ & 0.15 & $-0.64_{-0.13}^{+0.12}$ & $0.14_{-0.07}^{+0.07}$ \\
47 & Ag & 0.89 & $-0.21$ & $0.89_{-0.17}^{+0.16}$ & $-0.21_{-0.16}^{+0.12}$ & $-1.16$ & $-0.53$ & $-1.16_{-0.20}^{+0.17}$ & $-0.55_{-0.10}^{+0.09}$ \\
48 & Cd & 1.52 & $-0.44$ & $1.56_{-0.09}^{+0.08}$ & $-1.35_{-0.75}^{+0.82}$ & $-0.35$ & $-0.91$ & $-0.29_{-0.10}^{+0.09}$ & $-1.68_{-0.83}^{+0.88}$ \\
49 & In & 0.13 & 0.13 & $0.72_{-1.75}^{+0.86}$ & $-0.05_{-0.94}^{+0.34}$ & $-0.53$ & $-0.53$ & $-0.60_{-1.22}^{+0.46}$ & $-0.62_{-1.18}^{+0.45}$ \\
50 & Sn & 1.34 & 1.07 & $1.13_{-1.08}^{+0.51}$ & $1.12_{-0.27}^{+0.15}$ & $-0.59$ & 0.56 & $-1.33_{-0.86}^{+0.77}$ & $0.62_{-0.15}^{+0.12}$ \\
51 & Sb & $-0.01$ & $-0.01$ & $0.61_{-1.74}^{+0.82}$ & $-0.18_{-0.95}^{+0.29}$ & $-0.67$ & $-0.67$ & $-0.74_{-1.23}^{+0.43}$ & $-0.72_{-1.20}^{+0.40}$ \\
52 & Te & 2.37 & 0.81 & $2.46_{-0.12}^{+0.08}$ & $0.45_{-1.00}^{+0.51}$ & 0.45 & 0.55 & $0.54_{-0.19}^{+0.13}$ & $0.41_{-0.77}^{+0.31}$ \\
56 & Ba & 1.46 & 0.95 & $1.49_{-0.08}^{+0.07}$ & $0.99_{-0.04}^{+0.04}$ & $-1.06$ & 0.59 & $-1.01_{-0.16}^{+0.14}$ & $0.61_{-0.03}^{+0.03}$ \\
57 & La & 0.82 & 0.08 & $0.76_{-0.05}^{+0.05}$ & $0.11_{-0.03}^{+0.02}$ & $-1.74$ & $-0.26$ & $-1.76_{-0.12}^{+0.11}$ & $-0.26_{-0.02}^{+0.02}$ \\
58 & Ce & 1.35 & 0.46 & $1.36_{-0.05}^{+0.05}$ & $0.44_{-0.04}^{+0.04}$ & $-0.72$ & 0.08 & $-0.85_{-0.09}^{+0.08}$ & $0.11_{-0.03}^{+0.03}$ \\
59 & Pr & 0.73 & $-0.21$ & $0.79_{-0.09}^{+0.08}$ & $-0.22_{-0.08}^{+0.07}$ & $-1.42$ & $-0.54$ & $-1.40_{-0.12}^{+0.11}$ & $-0.52_{-0.04}^{+0.03}$ \\
60 & Nd & 1.18 & 0.56 & $1.18_{-0.06}^{+0.06}$ & $0.55_{-0.03}^{+0.03}$ & $-1.37$ & 0.18 & $-1.38_{-0.12}^{+0.11}$ & $0.19_{-0.02}^{+0.02}$ \\
62 & Sm & 0.83 & 0.29 & $0.88_{-0.07}^{+0.07}$ & $0.27_{-0.03}^{+0.03}$ & $-1.61$ & $-0.11$ & $-1.62_{-0.21}^{+0.16}$ & $-0.11_{-0.02}^{+0.02}$ \\
63 & Eu & 0.00 & 0.00 & 0.00 & 0.00 & $-2.89$ & $-0.48$ & $-2.84_{-0.30}^{+0.18}$ & $-0.47_{-0.02}^{+0.02}$ \\
64 & Gd & 0.03 & 0.52 & $-0.82_{-0.76}^{+0.62}$ & $0.55_{-0.04}^{+0.04}$ & $-2.49$ & 0.07 & $-3.44_{-0.72}^{+0.82}$ & $0.08_{-0.04}^{+0.04}$ \\
65 & Tb & 0.05 & $-0.28$ & $0.13_{-0.21}^{+0.19}$ & $-0.29_{-0.07}^{+0.07}$ & $-2.18$ & $-0.68$ & $-2.07_{-0.34}^{+0.24}$ & $-0.68_{-0.06}^{+0.06}$ \\
66 & Dy & 0.19 & 0.62 & $-0.80_{-0.69}^{+0.76}$ & $0.63_{-0.03}^{+0.03}$ & $-2.50$ & 0.15 & $-3.33_{-0.80}^{+0.88}$ & $0.15_{-0.04}^{+0.04}$ \\
67 & Ho & 0.41 & $-0.13$ & $0.58_{-0.76}^{+0.27}$ & $-0.17_{-0.12}^{+0.10}$ & $-1.55$ & $-0.55$ & $-2.30_{-0.86}^{+0.79}$ & $-0.52_{-0.06}^{+0.06}$ \\
68 & Er & 0.40 & 0.41 & $-0.08_{-0.71}^{+0.34}$ & $0.44_{-0.05}^{+0.05}$ & $-2.34$ & $-0.04$ & $-3.21_{-0.78}^{+0.82}$ & $-0.03_{-0.04}^{+0.04}$ \\
69 & Tm & 0.21 & $-0.50$ & $0.17_{-0.71}^{+0.29}$ & $-0.50_{-0.12}^{+0.09}$ & $-2.02$ & $-0.89$ & $-2.60_{-0.96}^{+0.79}$ & $-0.87_{-0.06}^{+0.05}$ \\
70 & Yb & 0.08 & 0.23 & $0.08_{-0.15}^{+0.15}$ & $0.19_{-0.06}^{+0.06}$ & $-2.88$ & $-0.22$ & $-3.02_{-1.09}^{+0.53}$ & $-0.25_{-0.05}^{+0.05}$ \\
71 & Lu & 0.00 & $-0.40$ & $-0.02_{-0.74}^{+0.36}$ & $-0.41_{-0.11}^{+0.10}$ & $-2.12$ & $-0.82$ & $-2.39_{-0.89}^{+0.44}$ & $-0.80_{-0.09}^{+0.09}$ \\
72 & Hf & 0.65 & $-0.10$ & $0.74_{-0.12}^{+0.11}$ & $-0.10_{-0.10}^{+0.09}$ & $-1.26$ & $-0.53$ & $-1.24_{-0.18}^{+0.14}$ & $-0.52_{-0.07}^{+0.07}$ \\
73 & Ta & $-0.88$ & $-0.88$ & $-0.17_{-1.84}^{+0.91}$ & $-1.06_{-0.88}^{+0.35}$ & $-1.41$ & $-1.41$ & $-1.20_{-1.43}^{+0.63}$ & $-1.61_{-1.03}^{+0.41}$ \\
74 & W  & $-0.29$ & $-0.27$ & $-0.01_{-1.56}^{+1.17}$ & $-0.35_{-0.65}^{+0.19}$ & $-1.13$ & $-0.77$ & $-1.83_{-0.88}^{+0.91}$ & $-0.73_{-0.20}^{+0.16}$ \\
75 & Re & 0.11 & 0.18 & $-0.03_{-1.27}^{+1.33}$ & $0.10_{-0.27}^{+0.19}$ & $-0.73$ & $-0.17$ & $-1.72_{-0.70}^{+0.81}$ & $-0.28_{-0.17}^{+0.17}$ \\
76 & Os & 1.47 & 0.71 & $1.64_{-0.11}^{+0.10}$ & $0.63_{-0.09}^{+0.08}$ & $-1.04$ & 0.36 & $-0.48_{-0.27}^{+0.16}$ & $0.32_{-0.07}^{+0.07}$ \\
77 & Ir & 1.93 & 0.84 & $1.86_{-0.12}^{+0.11}$ & $0.92_{-0.10}^{+0.09}$ & 0.01 & 0.45 & $-0.25_{-0.31}^{+0.19}$ & $0.56_{-0.08}^{+0.07}$ \\
78 & Pt & 2.76 & 0.59 & $2.69_{-0.08}^{+0.07}$ & $-0.06_{-0.93}^{+0.81}$ & 0.90 & 0.48 & $0.81_{-0.14}^{+0.11}$ & $0.45_{-0.62}^{+0.30}$ \\
79 & Au & $-0.17$ & $-0.04$ & $-0.30_{-1.25}^{+1.30}$ & $-0.08_{-0.24}^{+0.14}$ & $-1.01$ & $-0.58$ & $-1.80_{-0.82}^{+0.82}$ & $-0.52_{-0.16}^{+0.14}$ \\
82 & Pb & 1.45 & 0.82 & $2.16_{-0.19}^{+0.14}$ & $-0.32_{-0.58}^{+0.65}$ & 0.26 & 0.02 & $0.37_{-0.32}^{+0.21}$ & $-0.25_{-0.88}^{+0.37}$ \\
83 & Bi & 0.32 & 0.32 & $0.91_{-1.76}^{+1.01}$ & $0.13_{-0.84}^{+0.44}$ & $-0.21$ & $-0.21$ & $-0.02_{-1.42}^{+0.72}$ & $-0.44_{-1.03}^{+0.53}$ \\
90 & Th & $-0.30$ & $-0.49$ & $-0.06_{-1.48}^{+0.93}$ & $-0.52_{-0.28}^{+0.09}$ & $-1.42$ & $-0.92$ & $-2.08_{-0.91}^{+0.77}$ & $-0.89_{-0.09}^{+0.07}$ \\
92 & U  & $-1.33$ & $-1.36$ & $-0.22_{-2.05}^{+0.64}$ & $-1.55_{-1.00}^{+0.28}$ & $-1.80$ & $-1.81$ & $-1.54_{-1.50}^{+0.53}$ & $-1.97_{-1.08}^{+0.30}$ \\
\hline
\end{tabular}%
\hspace*{0.25in}%
}
\label{tab:pat}
\end{table*}

\begin{table*}[htbp]
\centering
\caption{Stars dominated by Pattern 1 or 2}
\label{tab:star_pattern_1_2}
\begin{tabular}{lcrrrrcc}
\hline\hline
Star (label in Fig.~\ref{fig:mix}) & Dominant Pattern & [Fe/H] & [Eu/Fe] & [Y/Fe] & [Y/Eu] & $x_{{\rm Eu}}$ & $x_{{\rm Y}}$ \\
\hline
HD 88609                & 1 & $-3.07$ & $-0.34$ & $-0.12$ &  0.22 & 0.77 & 0.98 \\
HD 122563 (6)           & 1 & $-2.61$ & $-0.68$ & $-0.55$ &  0.13 & 0.36 & 0.97 \\
HD 140283               & 1 & $-2.57$ & $-0.22$ & $-0.41$ & $-0.19$ & 0.35 & 0.91 \\
J0038+2755              & 1 & $-2.39$ &  0.13 & $-0.08$ & $-0.21$ & 0.39 & 0.90 \\
J2140-1227 (1)          & 1 & $-3.05$ & $-0.20$ &  0.69 &  0.89 & 1.00 & 1.00 \\
CES0045-0932            & 1 & $-2.88$ & $-0.28$ & $-0.32$ & $-0.04$ & 0.95 & 0.94 \\
CES0338-2402            & 1 & $-2.77$ & $-0.13$ & $-0.33$ & $-0.20$ & 0.44 & 0.90 \\
CES0419-3651            & 1 & $-2.76$ & $-0.62$ & $-0.40$ &  0.22 & 1.00 & 0.96 \\
CES1942-6103            & 1 & $-3.24$ &  0.22 & $-0.11$ & $-0.33$ & 0.38 & 0.86 \\
CES2019-6130            & 1 & $-2.95$ &  0.06 & $-0.15$ & $-0.21$ & 0.36 & 0.91 \\
CES2103-6505            & 1 & $-3.39$ & $-0.09$ &  0.21 &  0.30 & 1.00 & 0.98 \\
J1521-3538 (5)         & 2 & $-2.80$ &  2.20 &  1.03 & $-1.17$ & 0.02 & 0.28 \\
J2213-5137 (2)          & 2 & $-2.20$ &  2.45 &  1.13 & $-1.32$ & 0.01 & 0.15 \\
CS 22892-052 (3)        & 2 & $-3.10$ &  1.63 &  0.44 & $-1.19$ & 0.02 & 0.16 \\
CS 31082-001 (4)        & 2 & $-2.90$ &  1.66 &  0.53 & $-1.13$ & 0.02 & 0.16 \\
CES0102-6143            & 2 & $-2.85$ &  1.13 &  0.11 & $-1.02$ & 0.05 & 0.28 \\
CES0109-0443            & 2 & $-3.24$ &  1.35 &  0.20 & $-1.15$ & 0.01 & 0.08 \\
CES0301+0616            & 2 & $-2.90$ &  1.22 &  0.14 & $-1.08$ & 0.03 & 0.25 \\
CES1221-0328            & 2 & $-2.85$ &  1.46 &  0.16 & $-1.30$ & 0.12 & 0.00 \\
CES2231-3238            & 2 & $-2.74$ &  1.19 & $-0.05$ & $-1.24$ & 0.02 & 0.12 \\
CES2254-4209            & 2 & $-2.76$ &  1.16 & $-0.13$ & $-1.29$ & 0.03 & 0.17 \\
CES2330-5626            & 2 & $-3.04$ &  1.24 &  0.04 & $-1.20$ & 0.03 & 0.26 \\
CES2334-2642            & 2 & $-3.44$ &  1.06 & $-0.04$ & $-1.10$ & 0.03 & 0.26 \\
\hline
\end{tabular}
\end{table*}

\clearpage
\section{Element-by-Element Comparisons between Model and Data}
\label{append:deviations}
\restartappendixnumbering
\renewcommand{\theHfigure}{C\arabic{figure}}
\renewcommand{\theHtable}{C\arabic{table}}

\begin{figure*}[htbp]
\centering
\includegraphics[width=\textwidth]{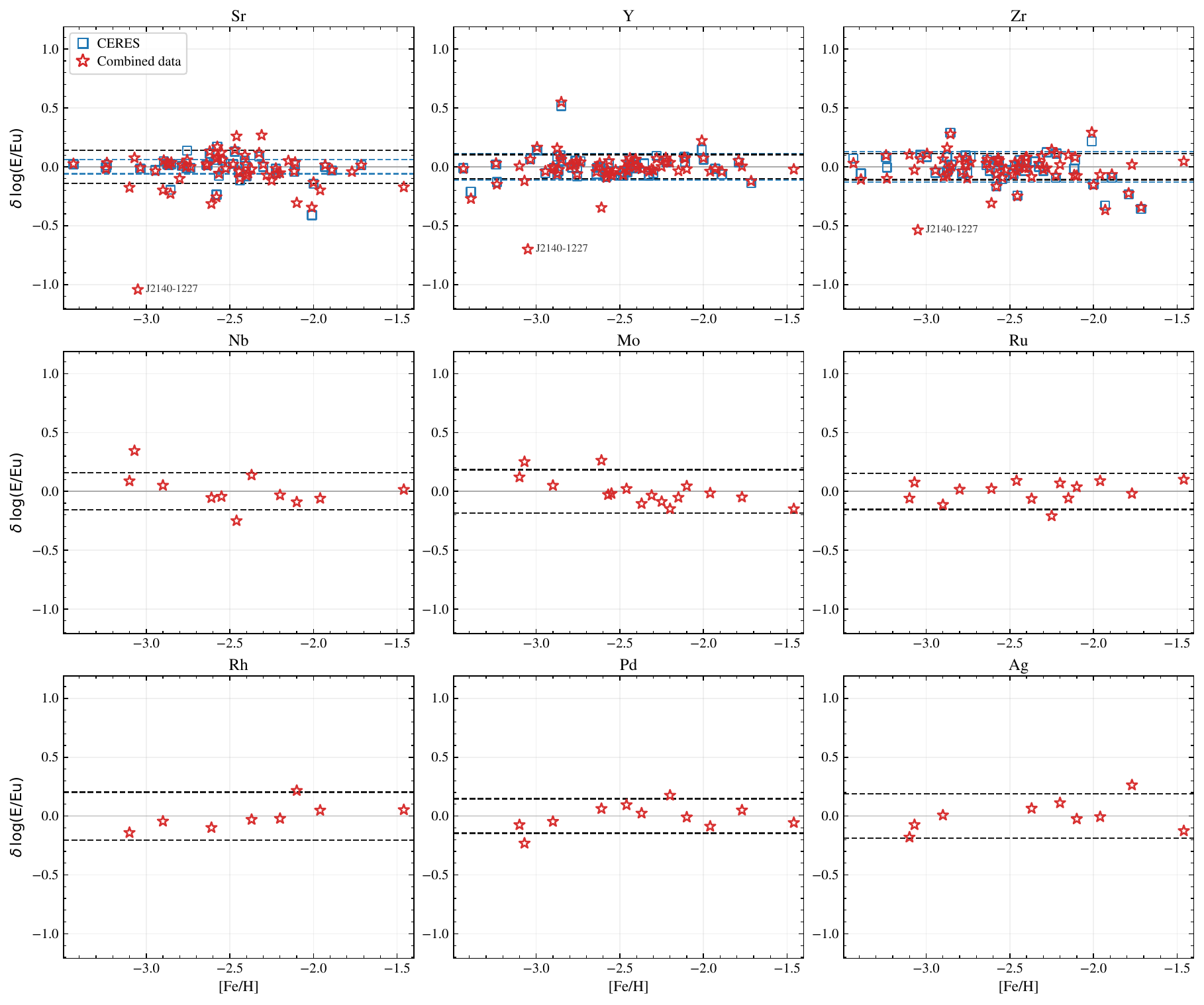}
\caption{Residuals $\delta\log({\rm E/Eu})$, defined as the predicted value minus the measured value, for Sr to Ag as a function of the measured [Fe/H]. Light blue squares show the model based on the CERES sample, and red stars show the model based on the combined CERES and literature sample. Light blue and black dashed lines mark the average measurement uncertainties $\sigma_{\log({\rm E/Eu})}$ for the CERES and literature samples, respectively.}
\label{fig:residual1}
\end{figure*}

\begin{figure*}[htbp]
\centering
\includegraphics[width=\textwidth]{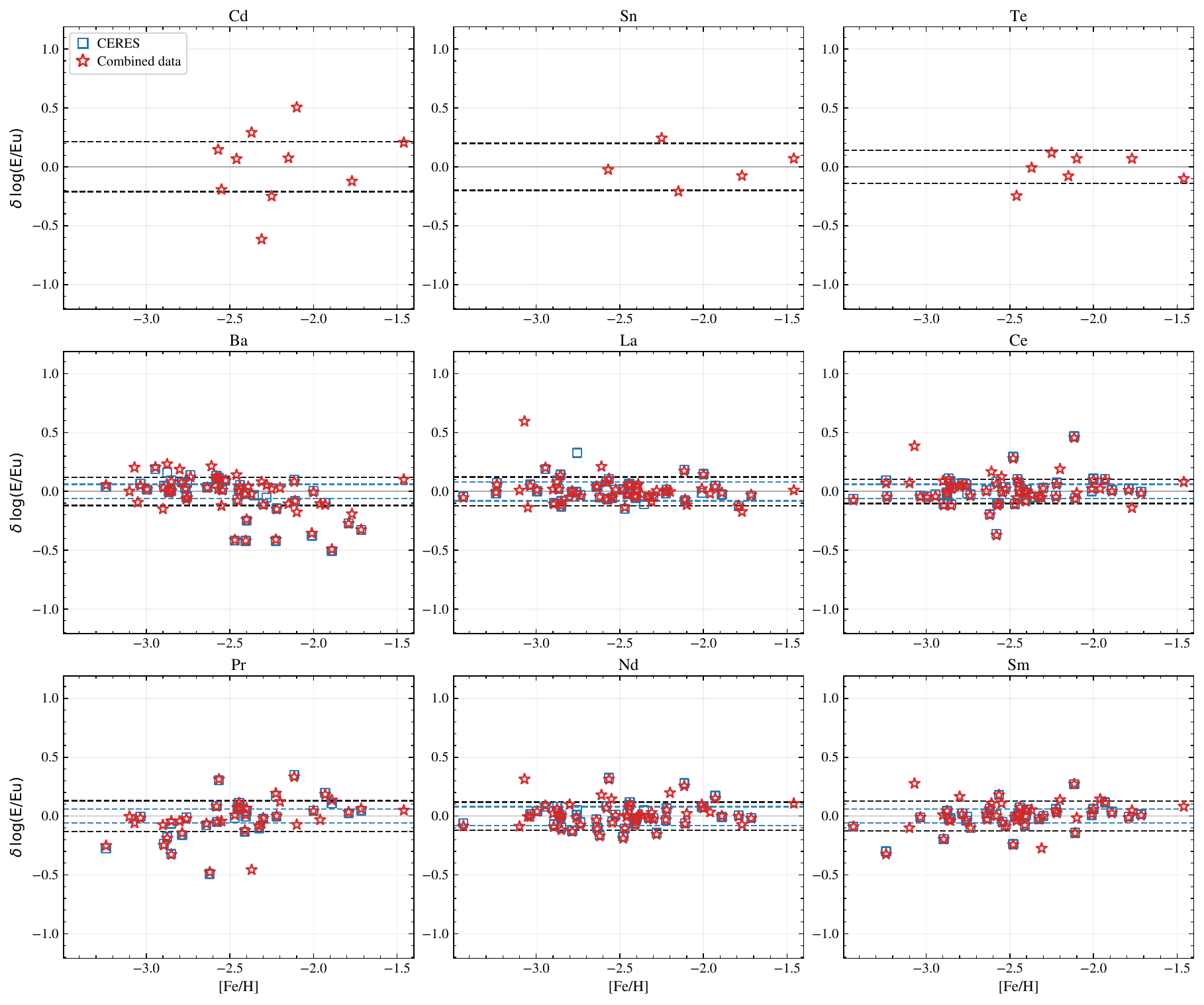}
\caption{Same as Fig.~\ref{fig:residual1}, but for Cd to Sm.}
\label{fig:residual2}
\end{figure*}

\begin{figure*}[htbp]
\centering
\includegraphics[width=\textwidth]{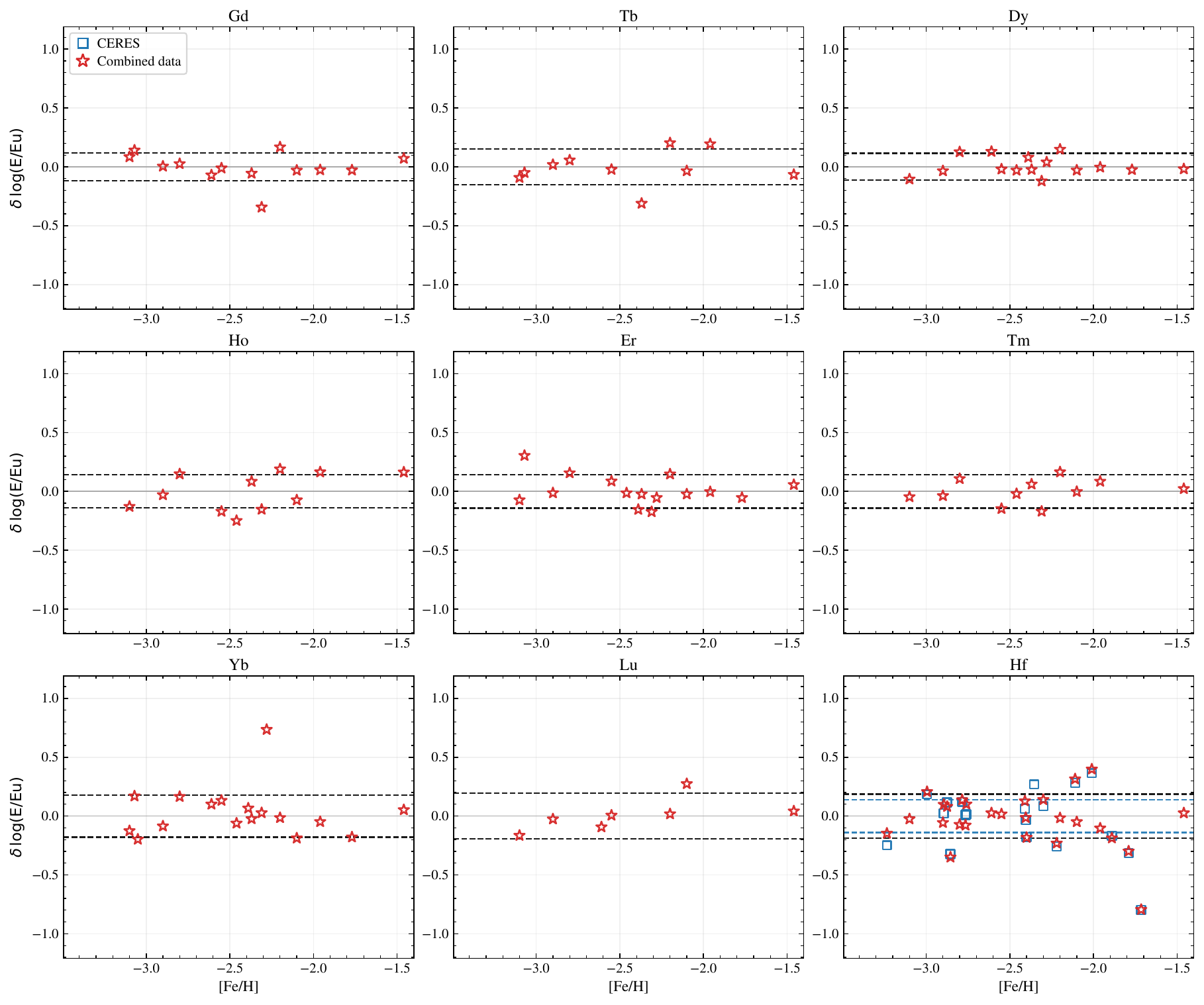}
\caption{Same as Fig.~\ref{fig:residual1}, but for Gd to Hf.}
\label{fig:residual3}
\end{figure*}

\begin{figure*}[htbp]
\centering
\includegraphics[width=\textwidth]{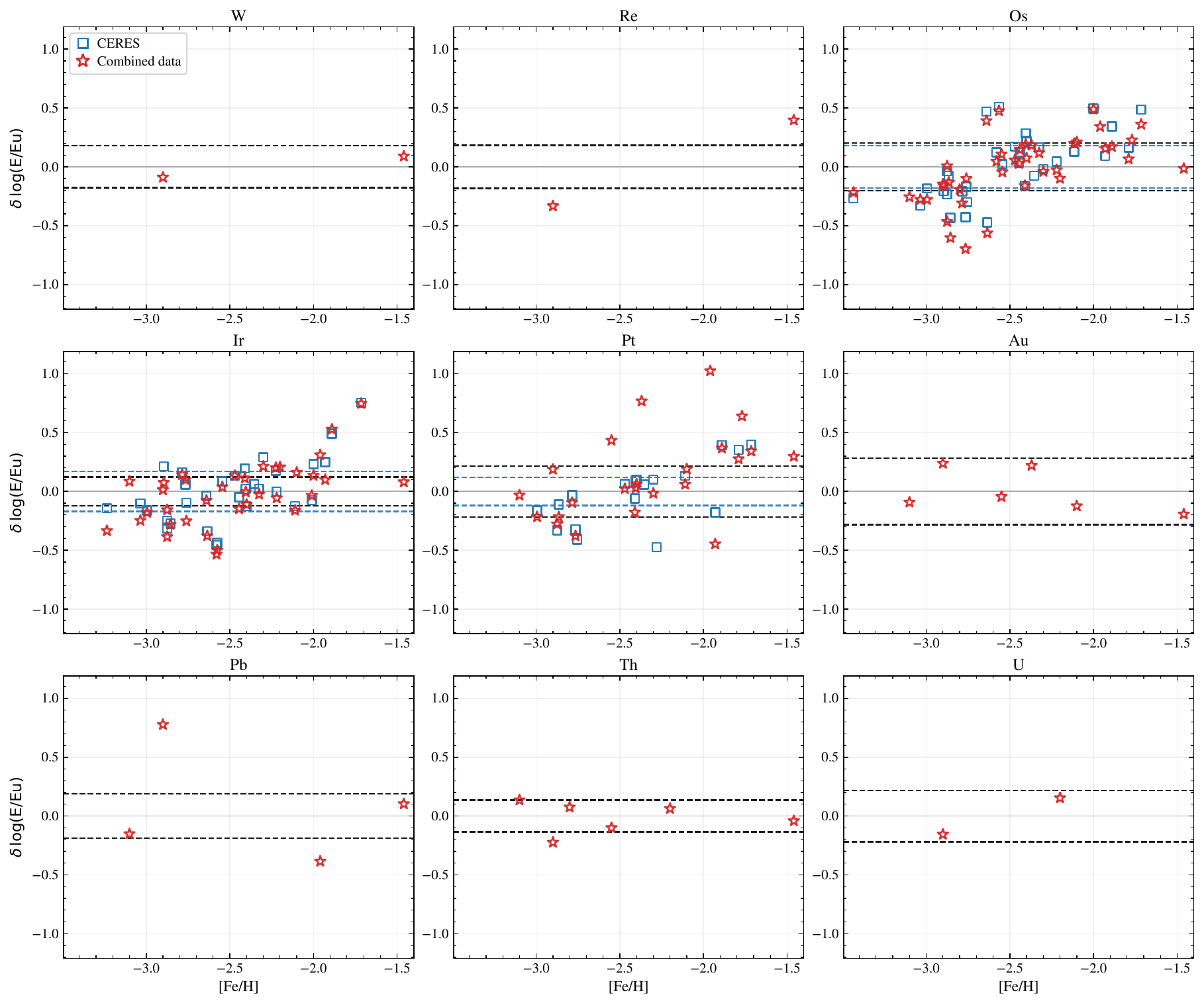}
\caption{Same as Fig.~\ref{fig:residual1}, but for W to U.}
\label{fig:residual4}
\end{figure*}

Figures~\ref{fig:residual1}--\ref{fig:residual4} compare the model predictions with the MP-star data element by element. For most stars, the predictions agree with the data within the average measurement uncertainties. Several exceptions are notable. J2140-1227 \citep{2024A&A...688A.123X} has extremely high Sr, Y, and Zr abundances relative to Eu that cannot be reproduced when Eu is the reference element (Fig.~\ref{fig:residual1}). In addition, Os, Ir, and Pt show relatively large residuals in several stars (Fig.~\ref{fig:residual4}). These three elements were measured with lines of neutral species, whereas Eu was measured with lines of singly-ionized species. Systematic uncertainties associated with the different ionization states may therefore affect the inferred $\log({\rm E/Eu})_1$ and $\log({\rm E/Eu})_2$ values.

\clearpage

\section{Results with Y as the Reference Element}
\label{append:Yref}
\restartappendixnumbering
\renewcommand{\theHfigure}{D\arabic{figure}}
\renewcommand{\theHtable}{D\arabic{table}}

\begin{figure*}[htbp]
\centering
\includegraphics[width=\textwidth]{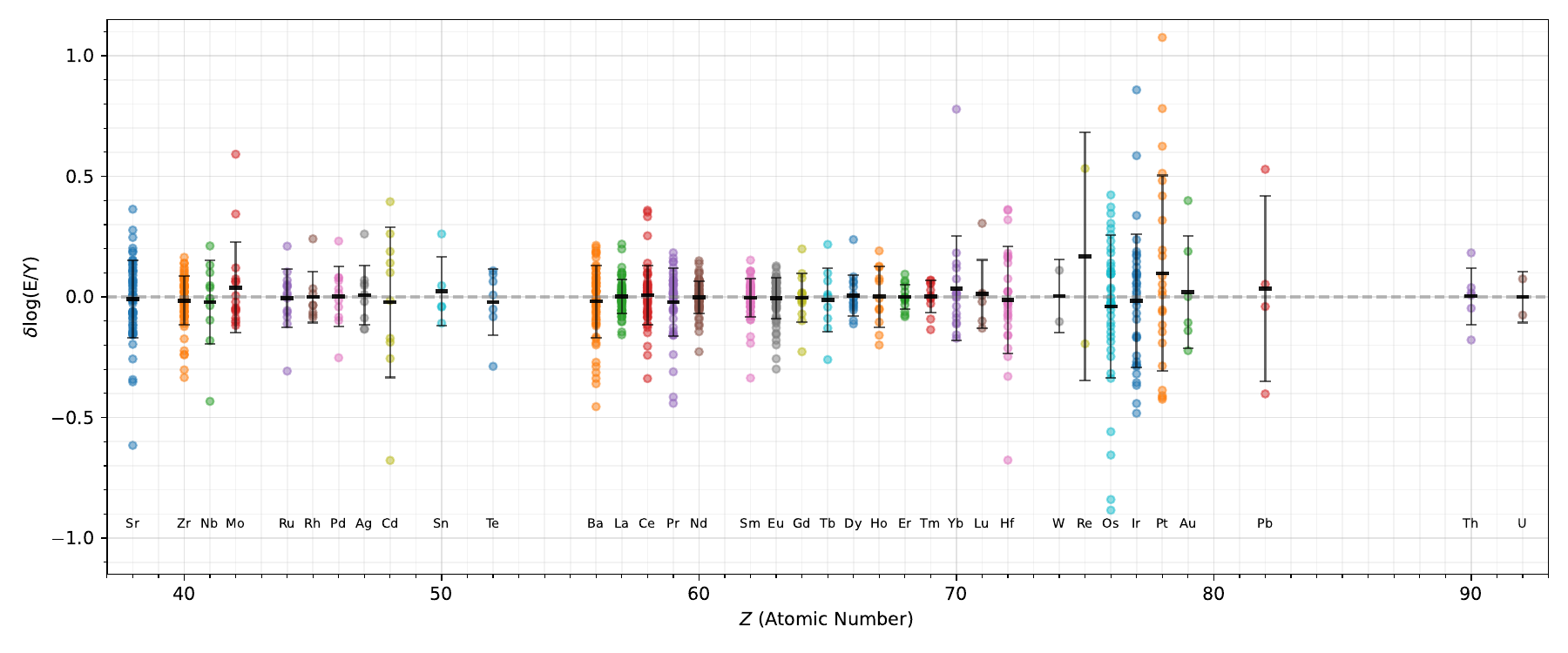}\\[1em]
\caption{Residuals $\delta\log({\rm E/Y})$ for the combined sample. For each element, the thick horizontal bar marks the mean residual and the error bar shows the $1\sigma$ scatter about the mean.}
\label{fig:residual_all_Y}
\end{figure*}

\begin{figure*}[tbp]
\centering
\includegraphics[width=\textwidth]{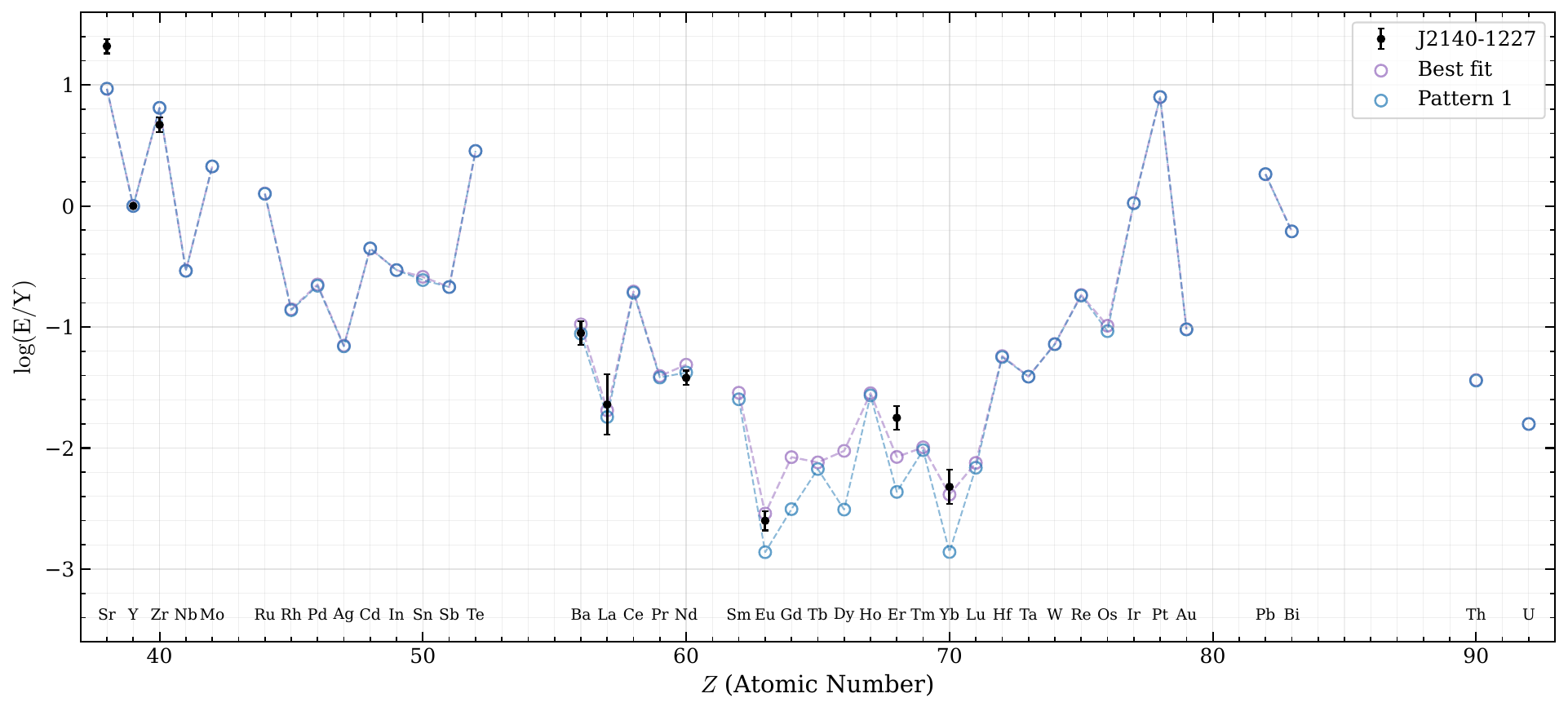}
\caption{Comparison of the best fit (violet circles) and Pattern 1 (light blue circles) with the data on $\log({\rm E/Y})$ for J2140-1227 \citep{2024A&A...688A.123X}. The best fit, with $x_{\rm Y}=0.996$, coincides with Pattern 1 except in the Eu-to-Yb region.}
\label{fig:J2140-1227}
\end{figure*}

\begin{figure*}[tbp]
\centering
\includegraphics[width=\textwidth]{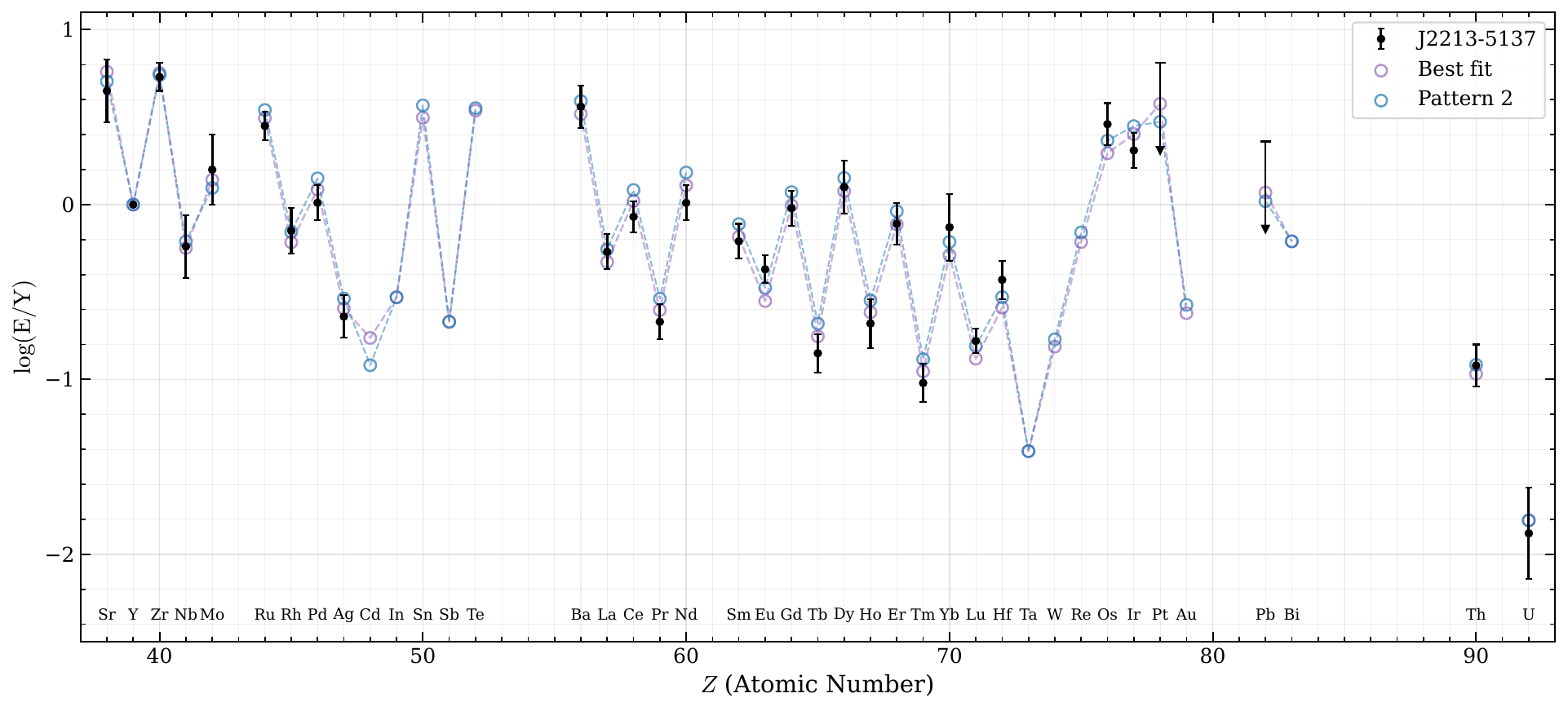}
\caption{Comparison of the best fit (violet circles) and Pattern 2 (light blue circles) with the data (black symbols, \citealt{2024ApJ...971..158R}) on $\log({\rm E/Y})$ for J2213-5137. The best fit, with $x_{\rm Y}=0.16$, is nearly identical to Pattern 2.}
\label{fig:s16_Y}
\end{figure*}

\begin{figure*}[htbp]
\centering
\includegraphics[width=\textwidth]{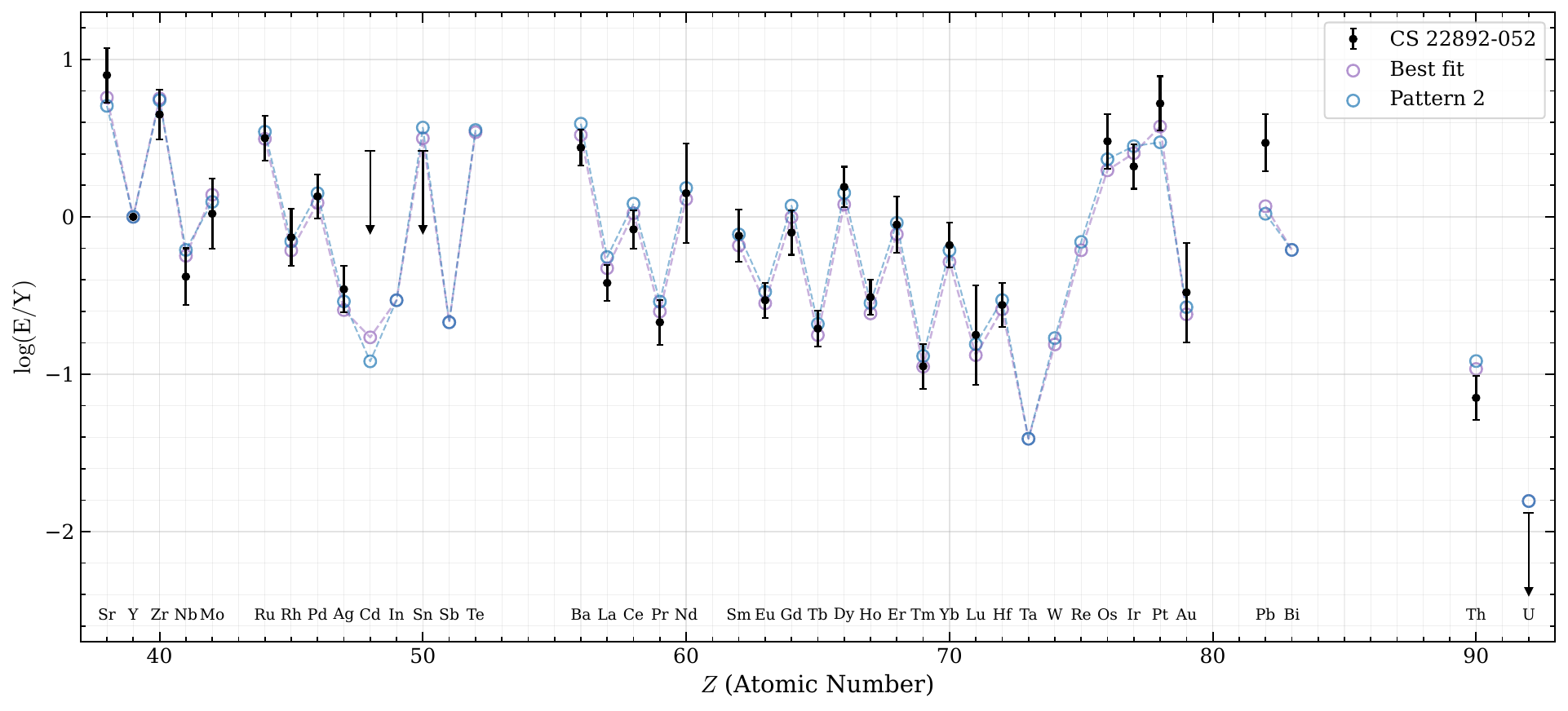}
\caption{Comparison of the best fit (violet circles) and Pattern 2 (light blue circles) with the data (black symbols, \citealt{2003ApJ...591..936S}) on $\log({\rm E/Y})$ for CS~22892-052. The best fit, with $x_{\rm Y}=0.16$, is nearly identical to Pattern 2.}
\label{fig:s19_Y}
\end{figure*}

\begin{figure*}[htbp]
\centering
\includegraphics[width=\textwidth]{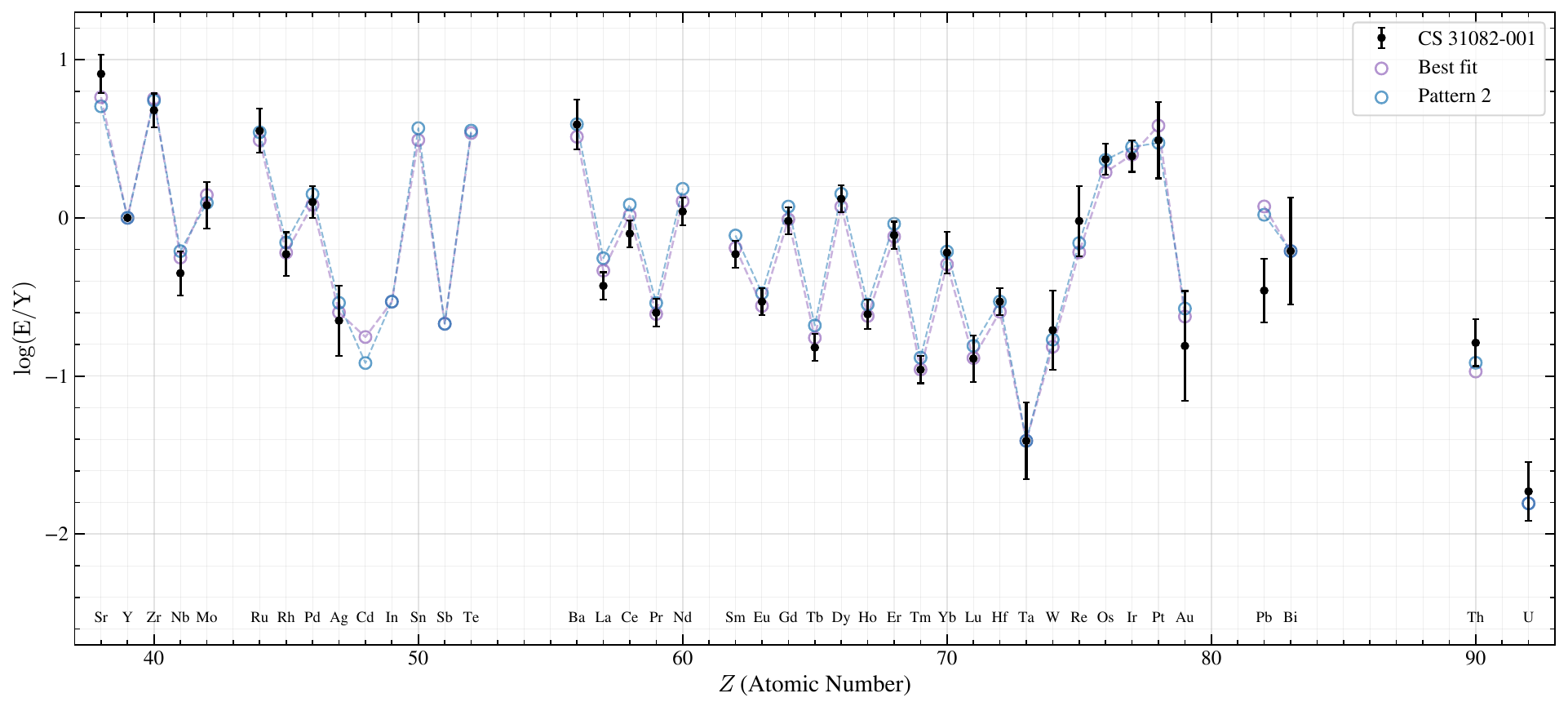}
\caption{Comparison of the best fit (violet circles) and Pattern 2 (light blue circles) with the data (black symbols, \citealt{2013A&A...550A.122S}) on $\log({\rm E/Y})$ for CS~31082-001. The best fit, with $x_{\rm Y}=0.17$, is nearly identical to Pattern 2.}
\label{fig:s20_Y}
\end{figure*}

\begin{figure*}[htbp]
\centering
\includegraphics[width=\textwidth]{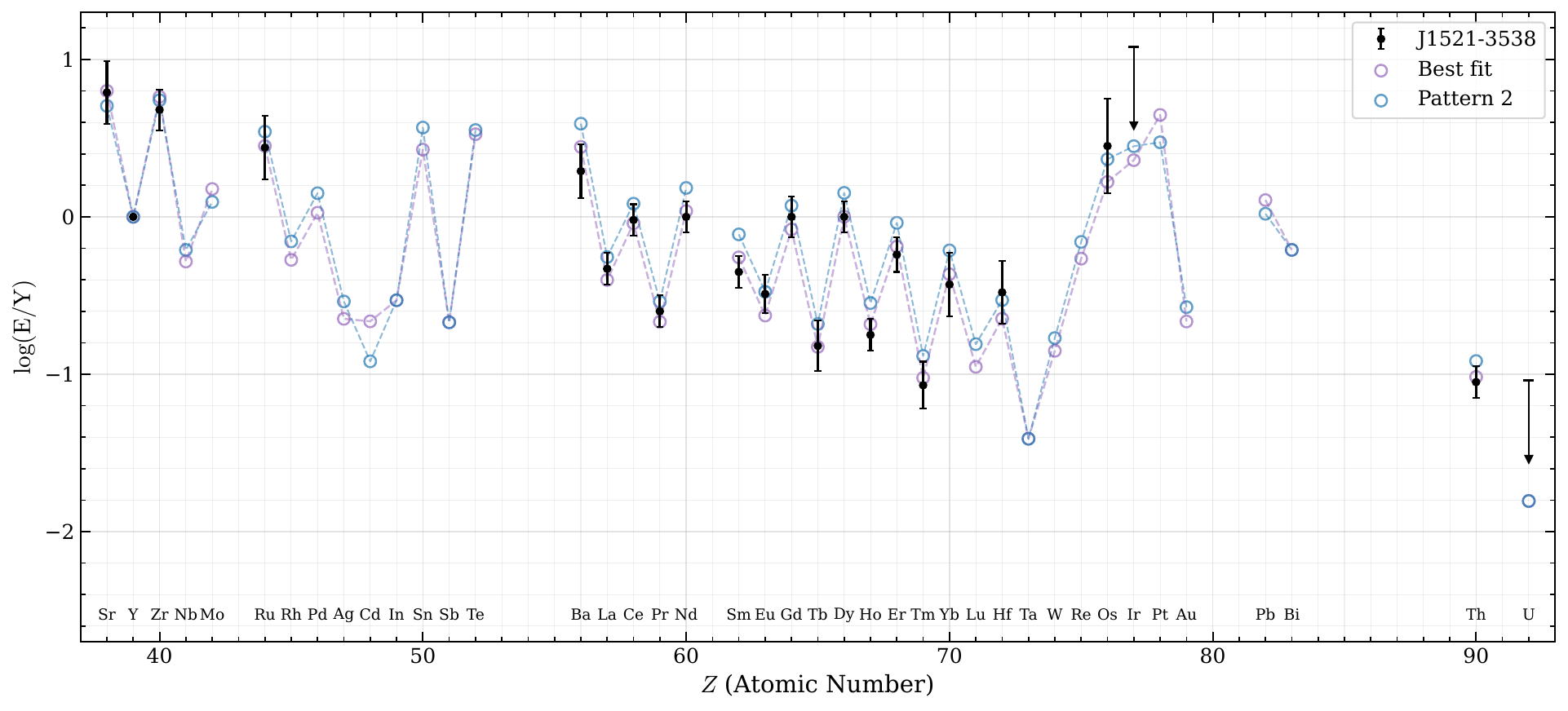}
\caption{Comparison of the best fit (violet circles) and Pattern 2 (light blue circles) with the data (black symbols, \citealt{2020ApJ...898...40C}) on $\log({\rm E/Y})$ for J1521-3538. The best fit, with $x_{\rm Y}=0.28$, is nearly identical to Pattern 2.}
\label{fig:s18_Y}
\end{figure*}

\begin{figure*}[htbp]
\centering
\includegraphics[width=\textwidth]{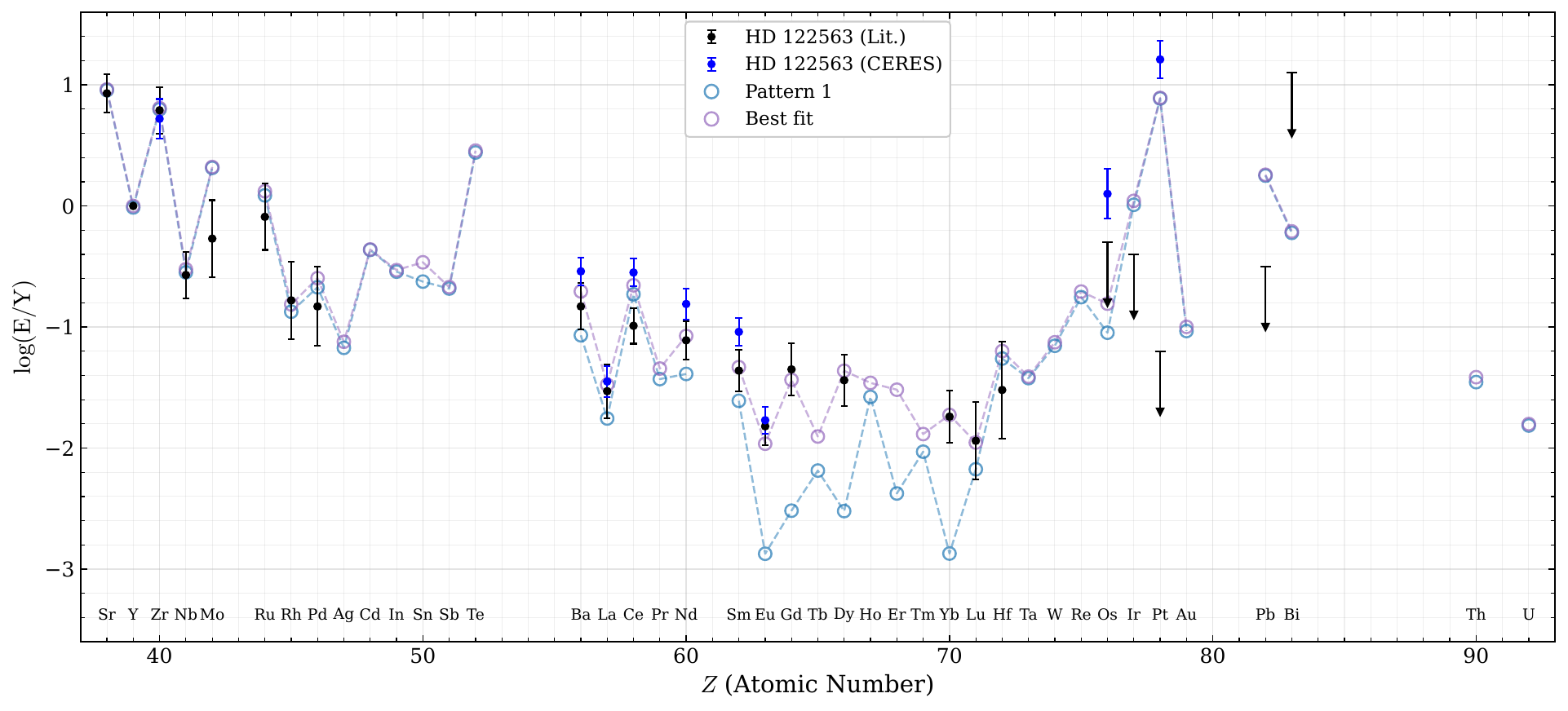}
\caption{Comparison of the best fit (violet circles) and Pattern 1 (light blue circles) with the data on $\log({\rm E/Y})$ for HD~122563. The best fit, with $x_{\rm Y}=0.97$, is dominated by Pattern 1 for the light $r$-process and second and third peak elements; contributions from Pattern 2 are concentrated near Eu. The CERES measurements (blue symbols, \citealt{2022A&A...665A..10L,2025A&A...693A.293L,2025A&A...693A.294A}) agree with the literature measurements of Zr and Ba to Eu. The best fit predicts a prominent third peak, with the Os prediction being well below the CERES measurement. The Pt prediction is close to the CERES measurement but lies well above the literature upper limit (black symbols, \citealt{2012ApJS..203...27R}). The discrepancy between the CERES and literature constraints on Pt requires further examination.}
\label{fig:HD122563_Y}
\end{figure*}

\begin{figure*}[htbp]
\centering
\includegraphics[width=\textwidth]{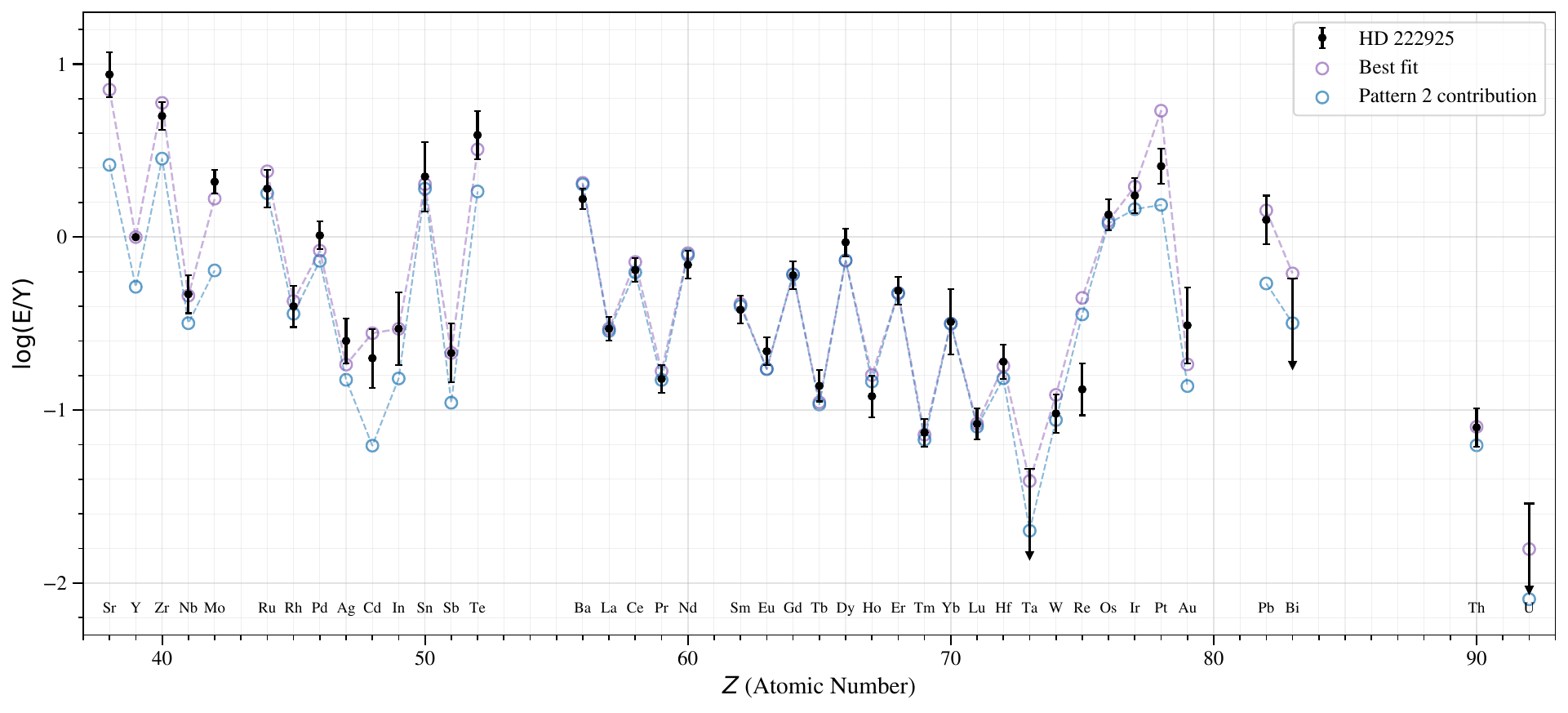}
\caption{Comparison of the best fit (violet circles) and Pattern 2 contributions (light blue circles) with the data (black symbols, \citealt{2022ApJS..260...27R}) on $\log({\rm E/Y})$ for HD~222925. The difference between the best fit, with $x_{\rm Y}=0.48$, and Pattern 2 contributions shows that Pattern 1 contributes primarily to the light $r$-process elements, Te, Pt, and Pb.}
\label{fig:s1_Y}
\end{figure*}

We present the element-by-element residuals and some representative stellar fits obtained with Y as the reference element in Figs.~\ref{fig:residual_all_Y}--\ref{fig:s1_Y}.

\clearpage

\section{Tests with External Data}
\label{append:test}
\restartappendixnumbering
\renewcommand{\theHfigure}{E\arabic{figure}}
\renewcommand{\theHtable}{E\arabic{table}}

\begin{figure*}[htbp]
\centering
\includegraphics[width=\textwidth]{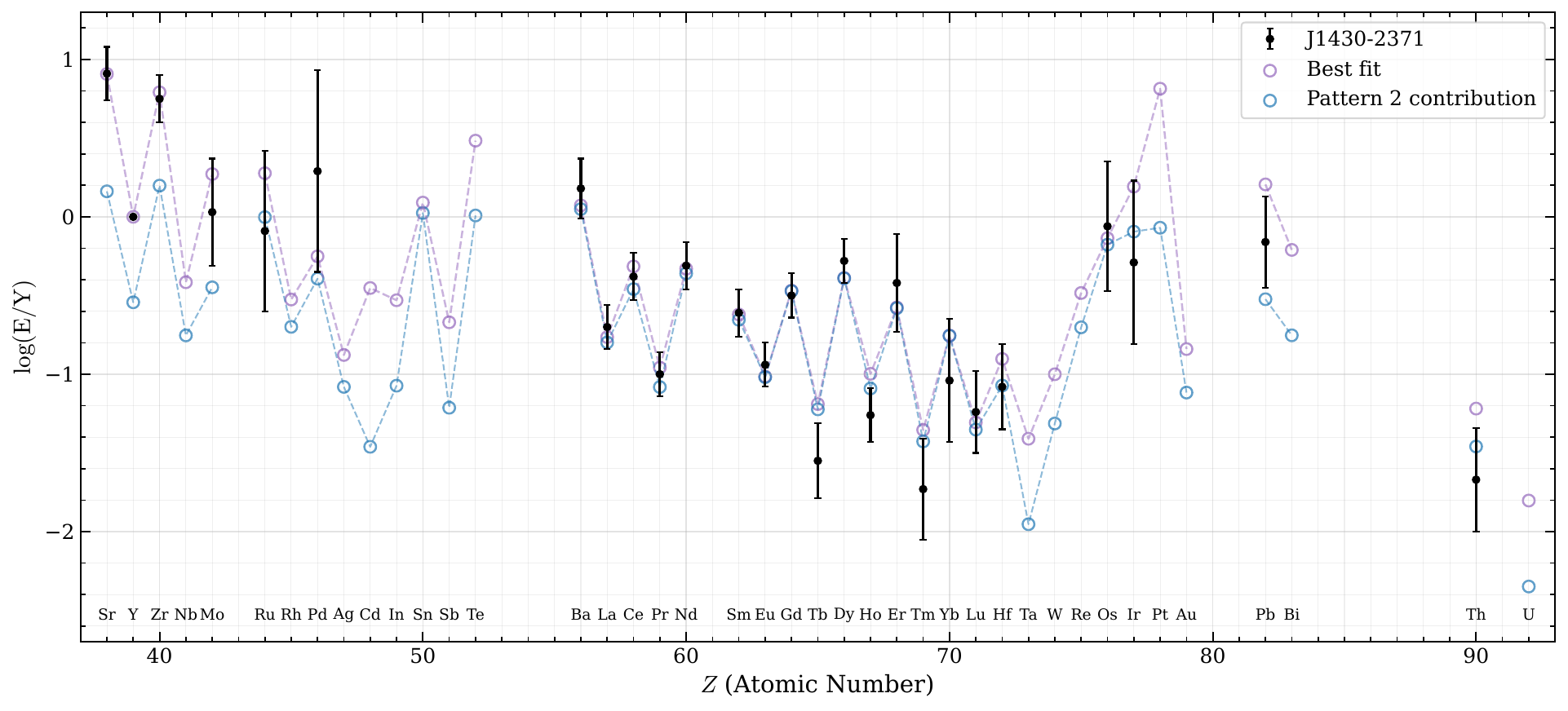}
\caption{Comparison of the best fit (violet circles) and contributions from Pattern 2 (light blue circles) with the data (black symbols, \citealt{2025A&A...704A.282R}) on $\log({\rm E/Y})$ for J1430-2371. The difference between the best fit, with $x_{\rm Y}=0.71$, and Pattern 2 contributions shows that Pattern 1 contributes primarily to the light $r$-process elements, Te, Pt, and Pb.}
\label{fig:J1430-2371}
\end{figure*}

\begin{figure*}[htbp]
\centering
\includegraphics[width=\textwidth]{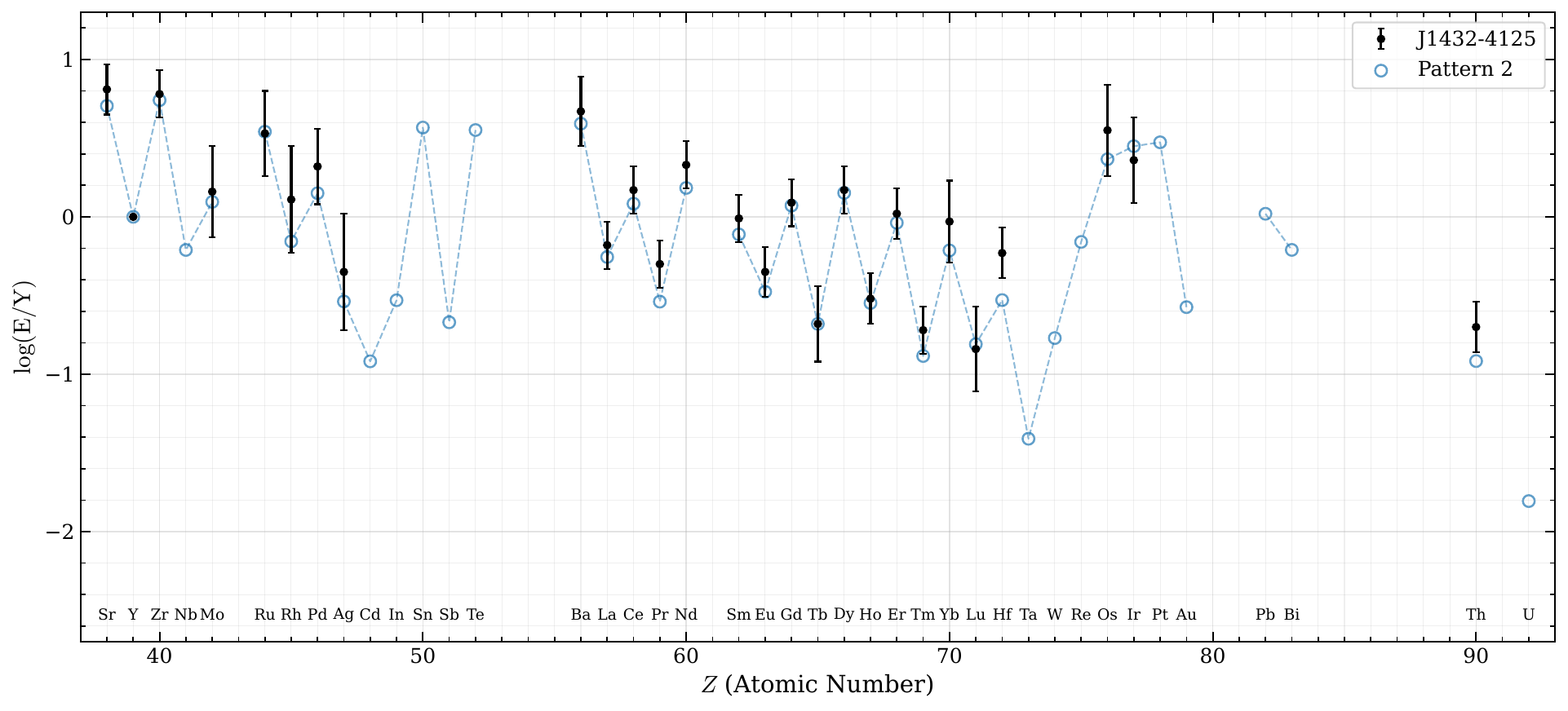}
\caption{Comparison of Pattern 2 (light blue circles) with the data (black symbols, \citealt{2025A&A...704A.282R}) on $\log({\rm E/Y})$ for J1432-4125. Pattern 2 is the best fit for this star, corresponding to $x_{\rm Y}=0$.}
\label{fig:J1432-4125}
\end{figure*}

\begin{figure*}[htbp]
\centering
\includegraphics[width=\textwidth]{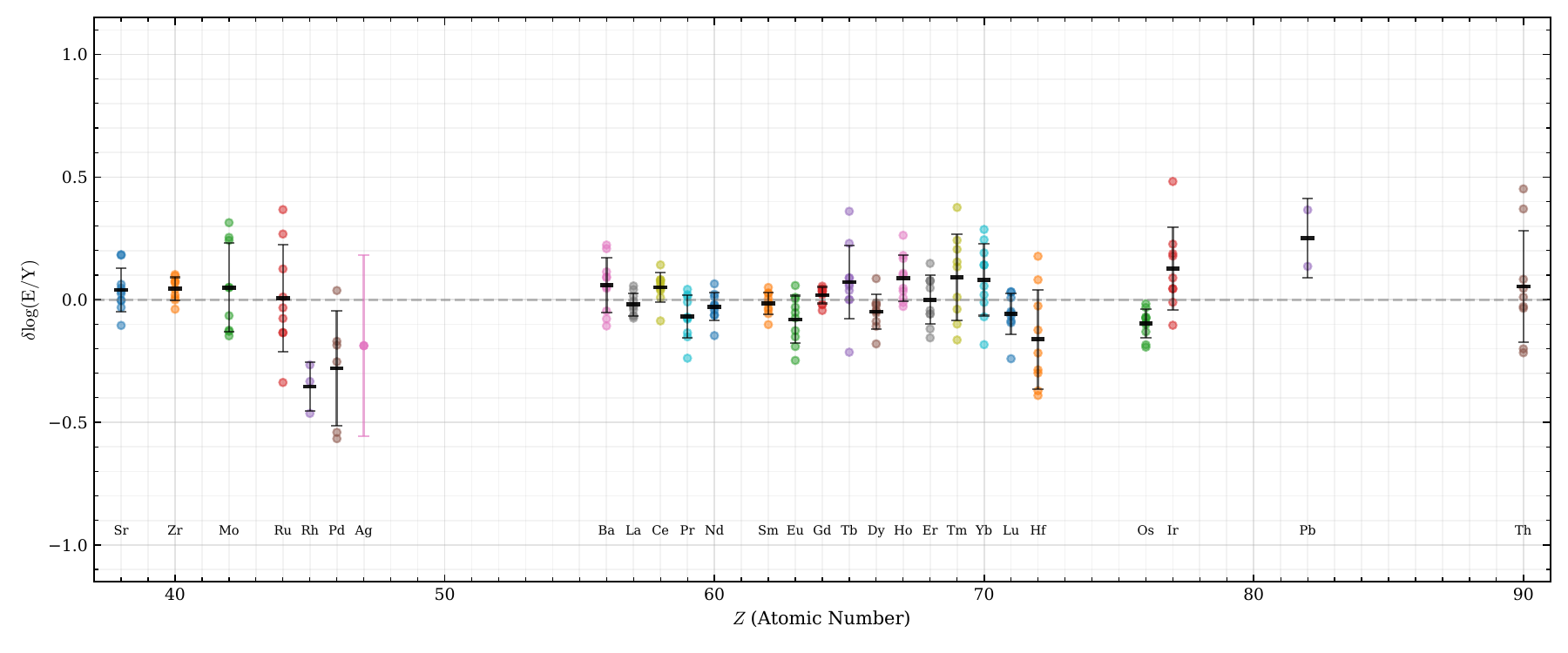}\\[1em]
\caption{Residuals $\delta\log({\rm E/Y})$ for the 10 stars in the RPA sample \citep{2025A&A...704A.282R}. For each element except Ag, the thick horizontal bar marks the mean residual and the error bar shows the $1\sigma$ scatter about the mean. Ag has only one measurement, so the pink error bar gives its measurement uncertainty.}
\label{fig:residual_new_stars}
\end{figure*}

\begin{figure*}[htbp]
\centering
\includegraphics[width=\textwidth]{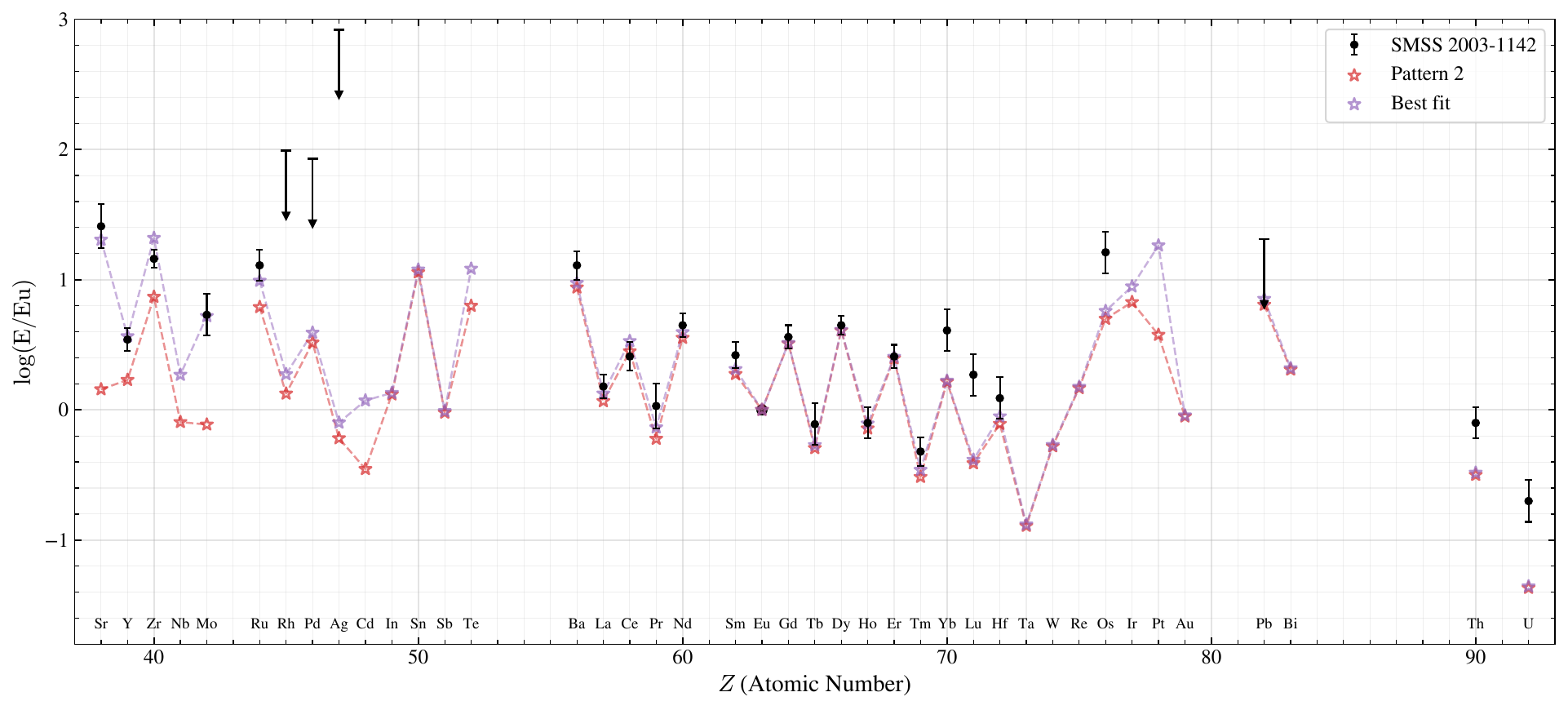}
\caption{Comparison of the best fit (violet stars) and Pattern 2 (red stars) with the data (black symbols, \citealt{2021Natur.595..223Y}) on $\log({\rm E/Eu})$ for SMSS~2003-1142. The difference between the best fit, with $x_{\rm Eu}=2.5\times10^{-2}$, and Pattern 2 shows that Pattern 1 contributes primarily to the light $r$-process elements, Te, and Pt.}
\label{fig:SMSS_2003_Eu}
\end{figure*}

\begin{figure*}[htbp]
\centering
\includegraphics[width=\textwidth]{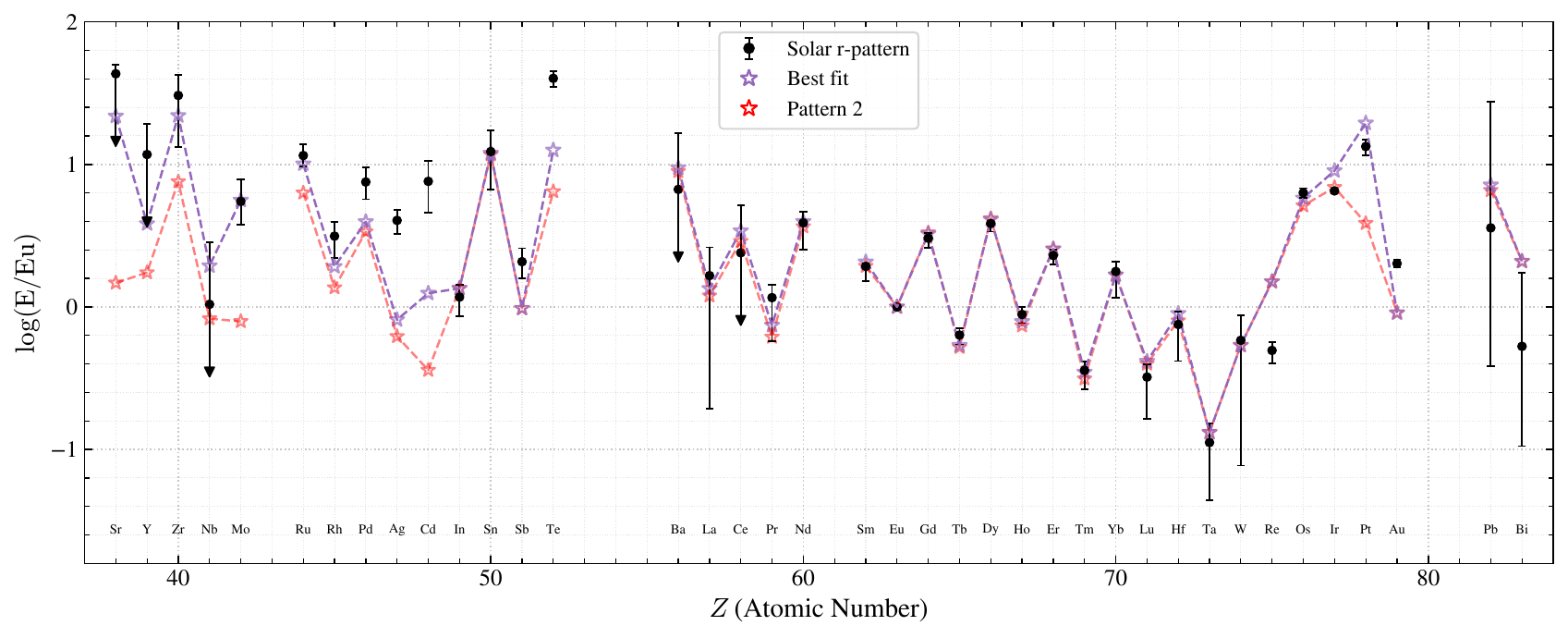}
\caption{Comparison of the best fit (violet stars) and Pattern 2 (red stars) with the solar $r$-process pattern (black symbols, \citealt{goriely1999}) on the Eu-reference scale. The difference between the best fit, with $x_{\rm Eu}=2.7\times10^{-2}$, and Pattern 2 shows that Pattern 1 contributes primarily to the light $r$-process elements, Te, and Pt. The large residuals at Ag, Cd, Te, Re, and Au indicate that the inferred pattern parameters for these elements are relatively uncertain.}
\label{fig:solar}
\end{figure*}

We test the model against data not used to derive Patterns 1 and 2. Figures~\ref{fig:J1430-2371} and \ref{fig:J1432-4125} show fits to the two stars with the largest and smallest $x_{\rm Y}$ values in the RPA sample, and Fig.~\ref{fig:residual_new_stars} summarizes the residuals for all 10 RPA stars. We also compare the model with SMSS~2003-1142 and the solar $r$-process pattern in Figs.~\ref{fig:SMSS_2003_Eu} and \ref{fig:solar}, respectively.

\clearpage

\section{Bayesian Inference of Uncertainties in Two-Pattern Model}
\label{append:bayesian}
\restartappendixnumbering
\renewcommand{\theHfigure}{F\arabic{figure}}
\renewcommand{\theHtable}{F\arabic{table}}

Patterns 1 and 2 presented in the main text are obtained by optimization. Here we use Bayesian inference to estimate the uncertainties in these patterns. Taking Eu as the example reference element, we need to specify the priors for the pattern parameters $\log({\rm E/Eu})_1$ and $\log({\rm E/Eu})_2$ and for the mixing parameters $x_{{\rm Eu},j}$, where $j$ runs over all the stars in the combined sample. We assume uniform distributions within $\pm 2$ dex of the optimized values as the priors for $\log({\rm E/Eu})_1$ and $\log({\rm E/Eu})_2$. We take a normal distribution of $\mathrm{logit}(x_{{\rm Eu},j})$, whose mean corresponds to the optimized value of $x_{{\rm Eu},j}$ and whose standard deviation is $\sigma_x=0.2$, to be the prior for $x_{{\rm Eu},j}$. Note that $\mathrm{logit}(x)=\ln[x/(1-x)]$ increases from $-\infty$ to $\infty$ as $x$ increases from 0 to 1.

Using $\log({\rm E/Eu})_1$, $\log({\rm E/Eu})_2$, and $x_{{\rm Eu},j}$, we calculate the predicted value $\log({\rm E/Eu})_{*,j}$ for the $j$th star from Eq.~(\ref{eq:mix}). The likelihood for measuring the observed value $\log({\rm E/Eu})^{\rm obs}_{*,j}$ is taken to be a normal distribution whose mean is $\log({\rm E/Eu})_{*,j}$ and whose standard deviation is $\sigma_{{\rm tot},j}({\rm E/Eu})$. Here $\sigma_{{\rm tot},j}({\rm E/Eu})$ is the total uncertainty given by
\begin{equation}
\sigma^{2}_{{\rm tot},j}({\rm E/Eu})=\sigma^{2}_j({\rm E/Eu})+\tau^{2}({\rm E/Eu}),
\label{eq:total_uncertainty}
\end{equation}
where $\tau({\rm E/Eu})$ is an unknown scatter in addition to the reported measurement uncertainty $\sigma_j({\rm E/Eu})=\sigma_{\log({\rm E/Eu})_*,j}$. The additional scatter $\tau({\rm E/Eu})$ depends on element E and is expected because our combined sample is not uniformly analyzed. For each element, we take $\tau({\rm E/Eu})$ to be the same for all the stars in the combined sample and assume a half-normal distribution with $\sigma_\tau=0.1$ dex as its prior. 

\begin{figure*}[htbp]
\centering
\includegraphics[width=\textwidth]{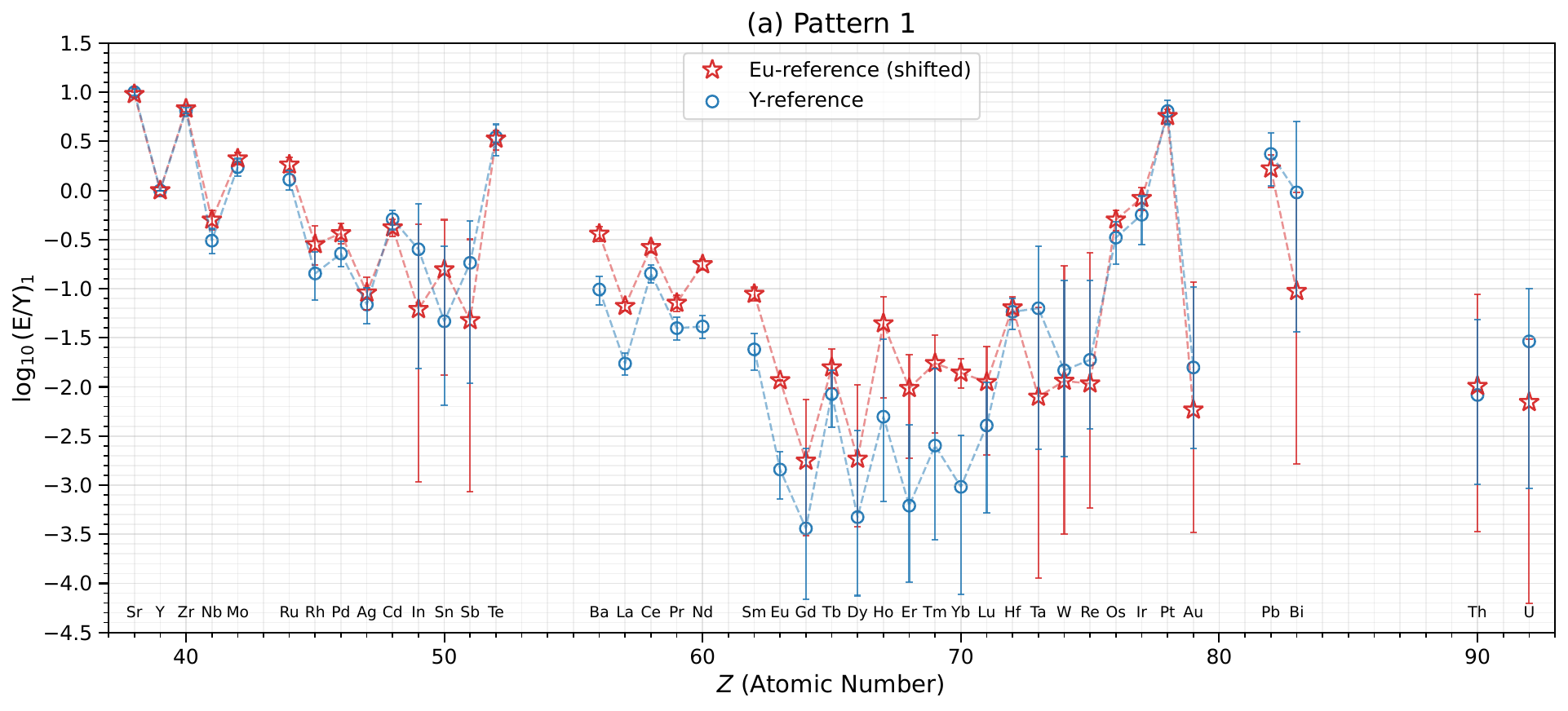}\\[1em]
\includegraphics[width=\textwidth]{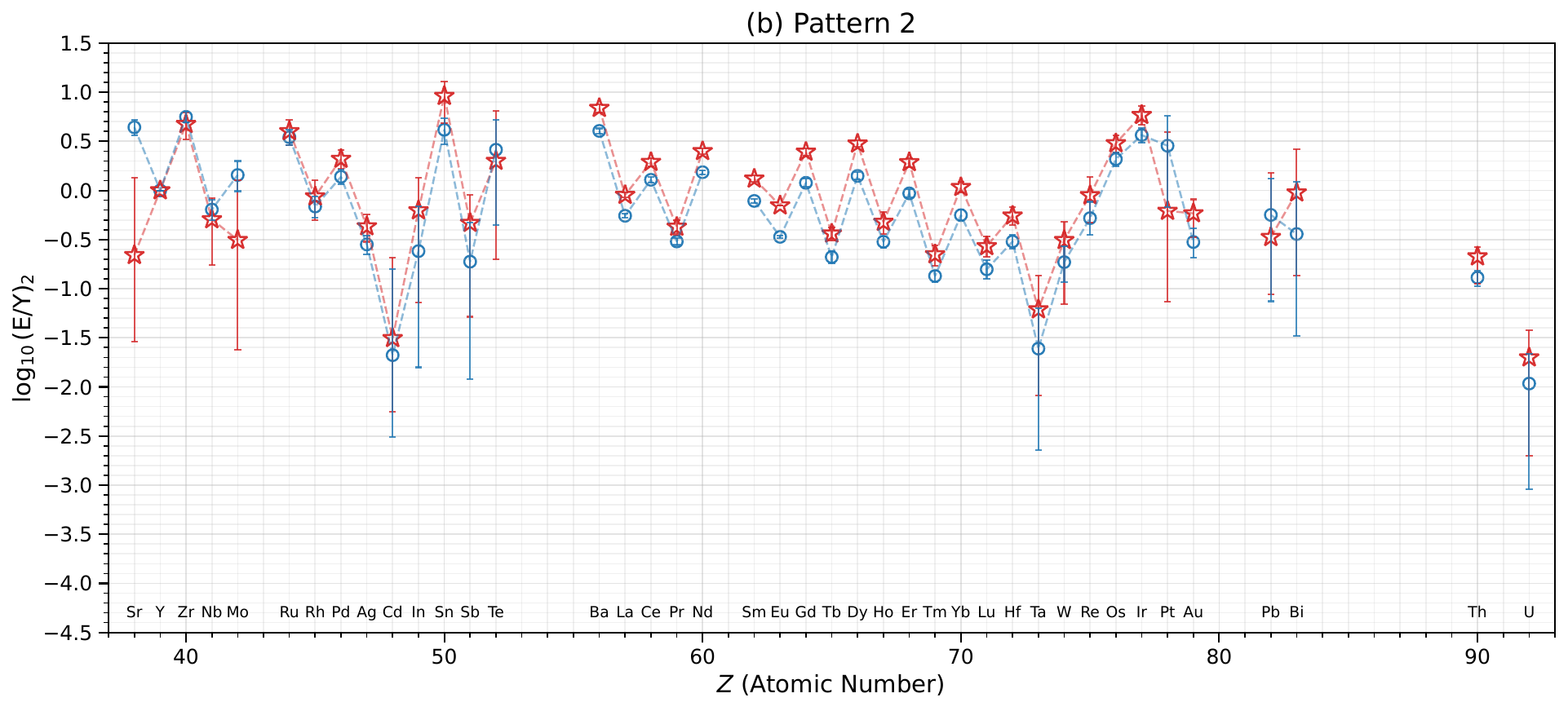}
\caption{Bayesian constraints on Patterns 1 (top panel) and 2 (bottom panel) inferred from the combined sample. Light blue circles show posterior medians for $\log({\rm E/Y})_1$ or $\log({\rm E/Y})_2$. Red stars show posterior medians for $\log({\rm E/Eu})_1$ or $\log({\rm E/Eu})_2$ that are shifted by subtracting the median $\log({\rm Y/Eu})_1$ or $\log({\rm Y/Eu})_2$, respectively. In all cases, error bars indicate 16th--84th percentile credible intervals.}
\label{fig:bayesian_patterns}
\end{figure*}

\begin{figure*}[htbp]
\centering
\includegraphics[width=0.48\textwidth]{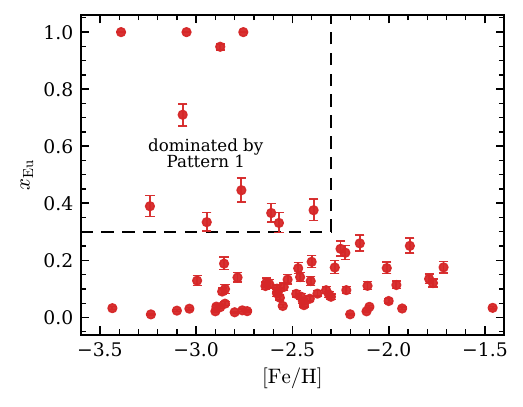}
\hfill
\includegraphics[width=0.48\textwidth]{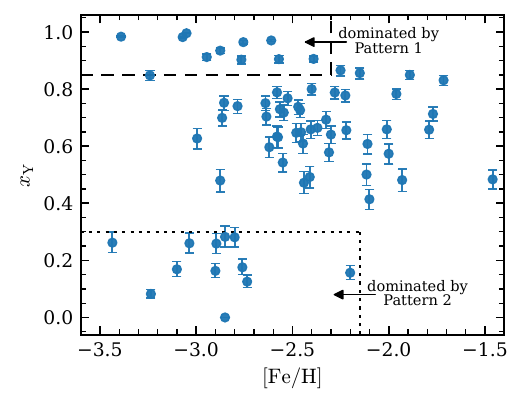}
\caption{Mixing parameters $x_{\rm Eu}$ (left panel) and $x_{\rm Y}$ (right panel) inferred from the combined sample shown as functions of the measured [Fe/H] in terms of posterior medians and 16th--84th percentile credible intervals. The upper left corner of each panel corresponds to stars dominated by Pattern 1. The lower left corner of the right panel corresponds to stars dominated by Pattern 2.}
\label{fig:mixing_para_baye}
\end{figure*}

\begin{figure*}[htbp]
\centering
\includegraphics[width=\textwidth]{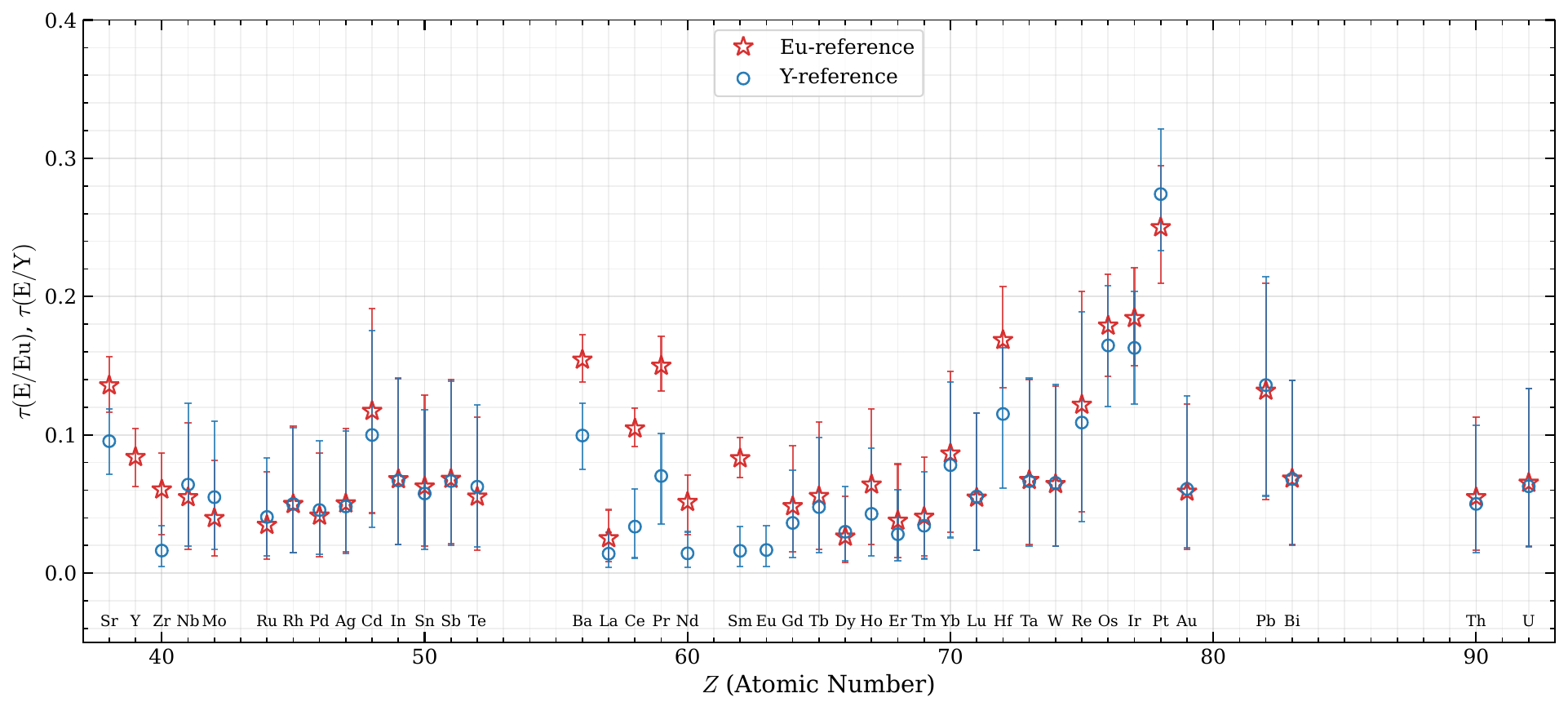}\\[1em]
\caption{Posterior medians and 16th--84th percentile credible intervals for additional scatters $\tau({\rm E/Eu})$ (red stars with error bars) or $\tau({\rm E/Y})$ (light blue circles with error bars) inferred from the combined sample.}
\label{fig:additional_scatter}
\end{figure*}

We sample the posterior using the No-U-Turn Sampler implemented in \texttt{PyMC}. We run five independent Markov chains, each with 3000 tuning (burn-in) steps followed by 3000 retained draws, yielding a total of 15,000 post-burn-in samples. The target acceptance probability is set to 0.9. The posterior medians and 16th--84th percentile credible intervals are shown for the pattern parameters $\log({\rm E/Eu})_1$ and $\log({\rm E/Eu})_2$ [$\log({\rm E/Y})_1$ and $\log({\rm E/Y})_2$], the mixing parameters $x_{{\rm Eu},j}$ ($x_{{\rm Y},j}$), and the additional scatters $\tau({\rm E/Eu})$ [$\tau({\rm E/Y})$] in Figs.~\ref{fig:bayesian_patterns}--\ref{fig:additional_scatter}. The results for the pattern parameters are also given in Table~\ref{tab:pat}. 

\clearpage
\section{Three-Pattern Model}
\label{append:3pattern}
\restartappendixnumbering
\renewcommand{\theHfigure}{G\arabic{figure}}
\renewcommand{\theHtable}{G\arabic{table}}

Here we present a three-pattern model for the number ratio (E/Y) in an MP star:
\begin{equation}
\left(\frac{\rm E}{{\rm Y}}\right)
=
x\left(\frac{\rm E}{{\rm Y}}\right)_{1'}
+
y\left(\frac{\rm E}{{\rm Y}}\right)_{2'}
+
(1-x-y)\left(\frac{\rm E}{{\rm Y}}\right)_{3'},
\label{eq:three_pattern_mixture}
\end{equation}
where $x$, $y$, and $1-x-y$ are the fractions of Y contributed by Patterns $1'$, $2'$, and $3'$, respectively. Clearly, $x\geq0$, $y\geq0$, and $x+y\leq1$. We exclude elements with fewer than three measurements (In, Sb, Ta, W, Re, Bi, and U, see Table~\ref{tab:lit}) from the analysis because they do not provide effective constraints. We use the sequential least-squares quadratic programming (SLSQP) optimizer to derive Patterns $1'$, $2'$, and $3'$. This method directly enforces the coupled constraints on $x$ and $y$. We note that the L-BFGS-B method (used in the two-pattern analysis, see Appendix~\ref{append:optimization}) frequently gives two-pattern-like solutions, for which two of the three patterns are nearly identical. Although the SLSQP method finds three distinct patterns, this sensitivity to the optimizer indicates that the combined sample of MP stars used does not robustly constrain three independent patterns.

\begin{figure*}[htbp]
\centering
\includegraphics[width=\textwidth]{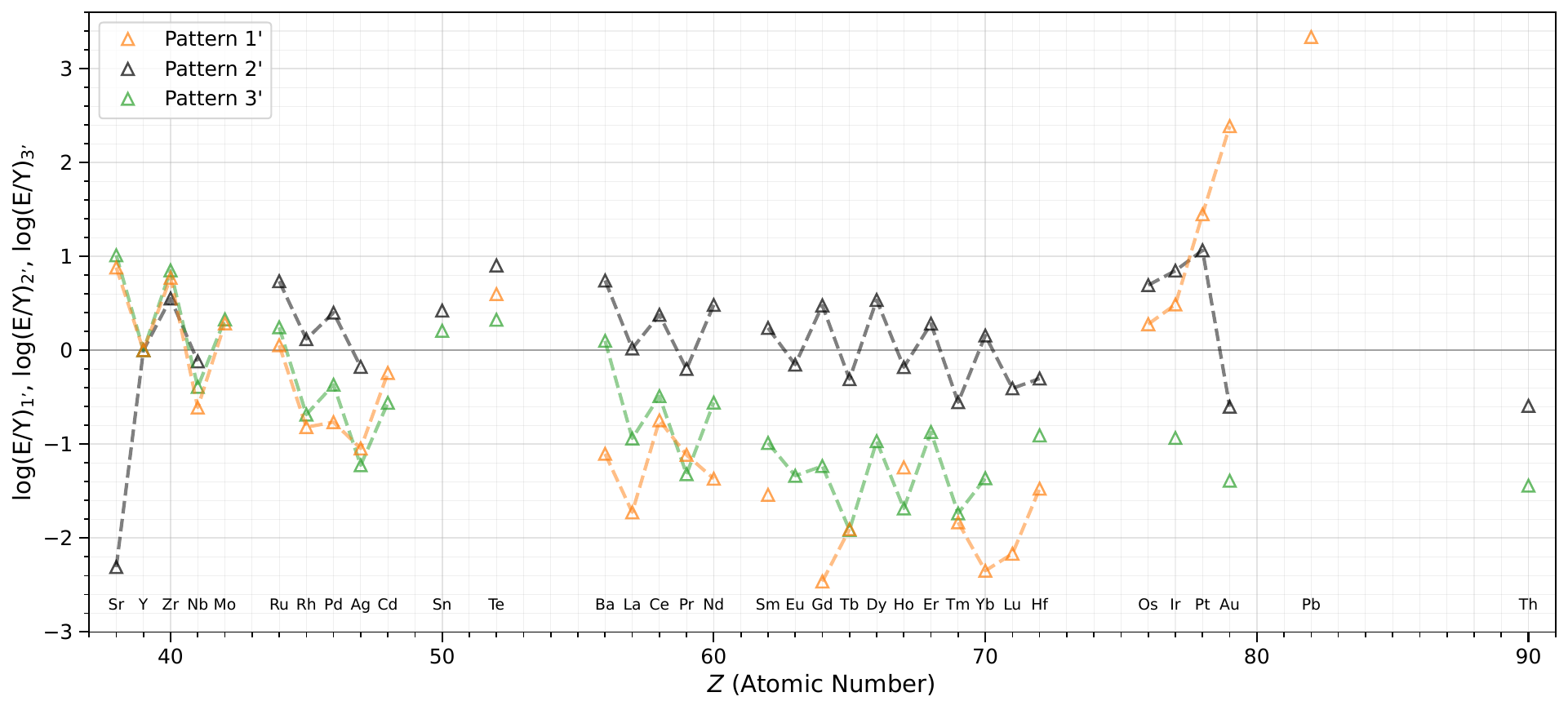}
\caption{Patterns $1'$, $2'$, and $3'$ inferred from the combined sample on the Y-reference scale. Elements with $\log({\rm E/Y})_{i'}<-2.5$ ($i=1$, 2, and 3) are omitted for clarity.}
\label{fig:three_patterns}
\end{figure*}

\begin{figure*}[htbp]
\centering
\includegraphics[width=0.97\textwidth]{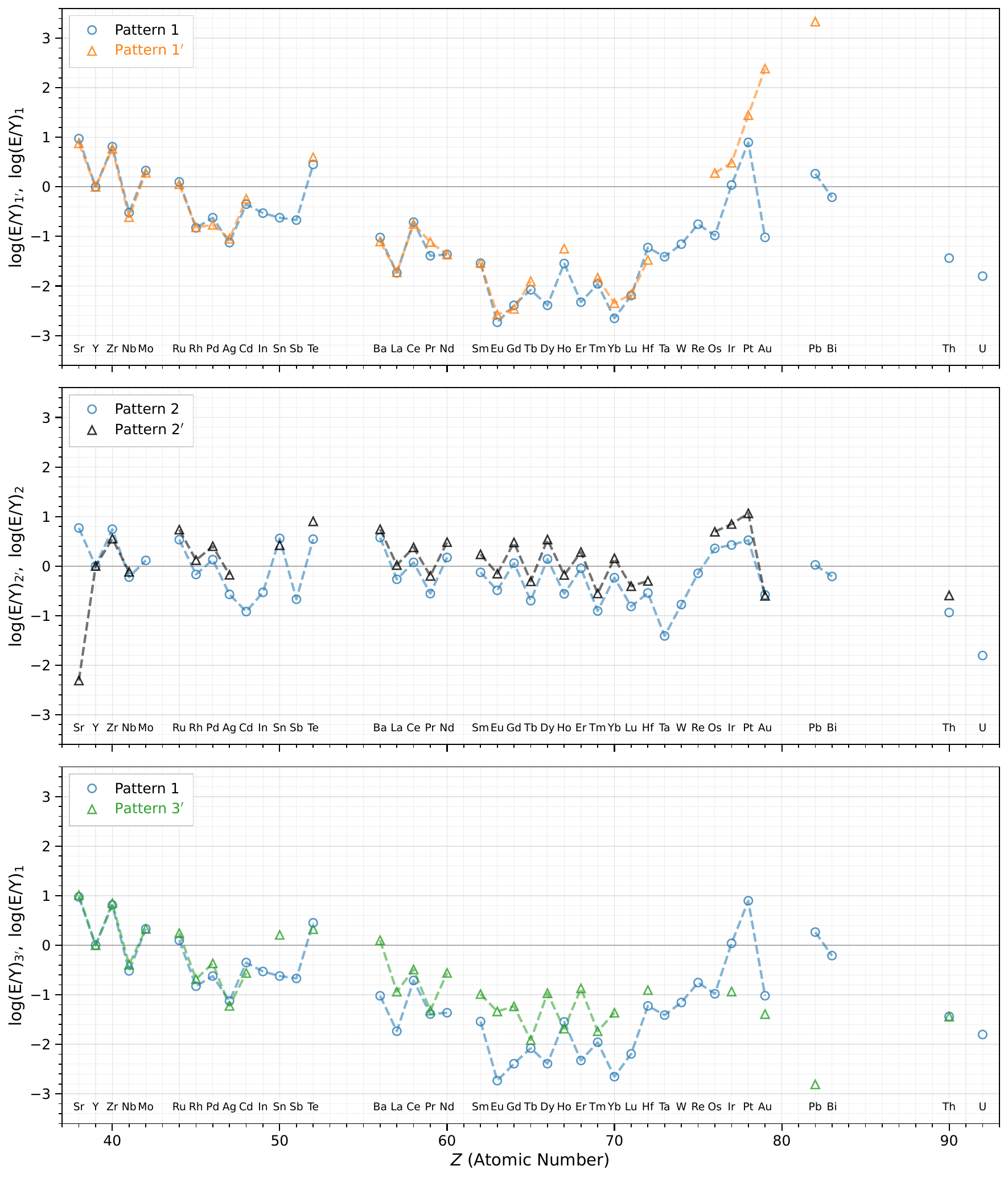}
\caption{Comparison of models involving two and three patterns on the Y-reference scale. Patterns 1 and 2 are for the two-pattern model, while Patterns $1'$, $2'$, and $3'$ are for the three-pattern model.}
\label{fig:model_pattern_comparison}
\end{figure*}

\begin{figure*}[htbp]
\centering
\includegraphics[width=\textwidth]{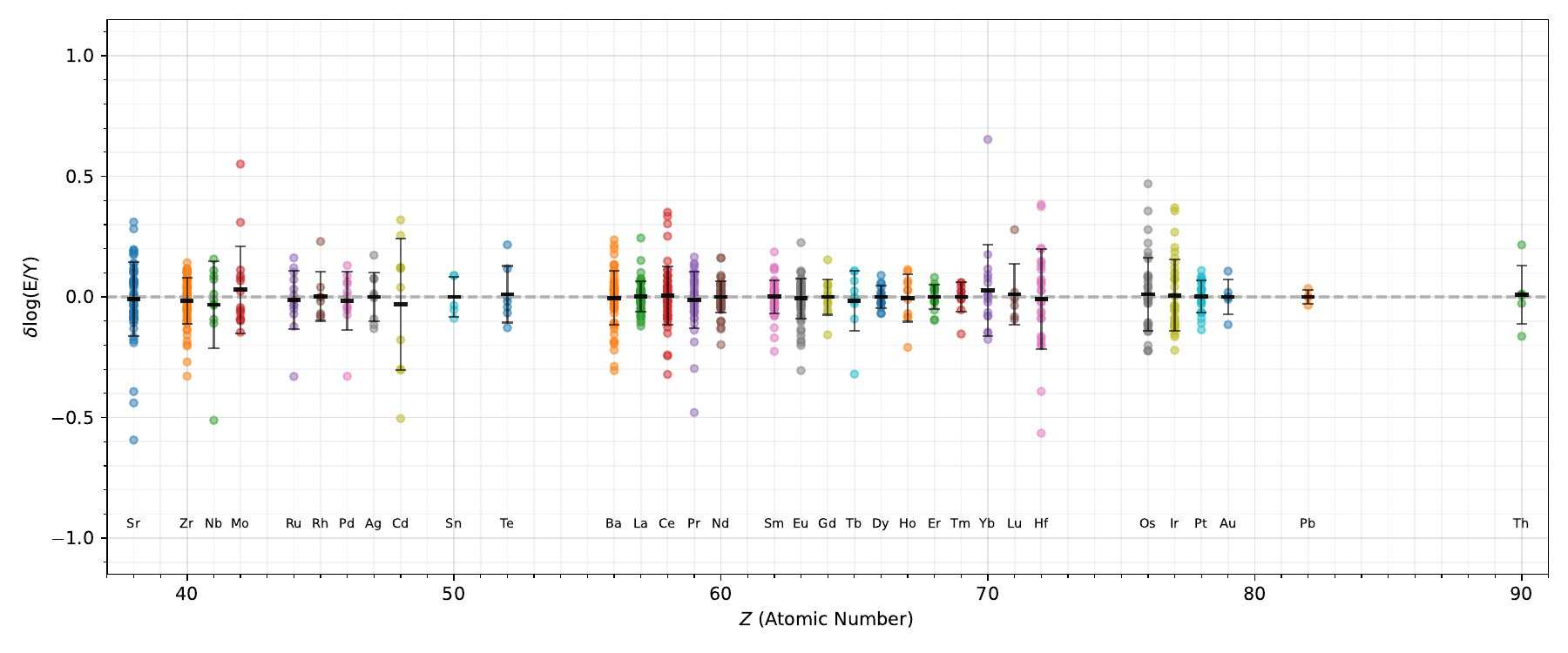}
\caption{Residuals for the three-pattern fits to the combined sample. For each element, points show $\delta\log({\rm E/Y})$, defined as the predicted value minus the measured value. The thick horizontal bar marks the sample mean, and the error bar shows the $1\sigma$ scatter about that mean.}
\label{fig:three_pattern_residuals}
\end{figure*}

\begin{figure*}[htbp]
\centering
\includegraphics[width=\textwidth]{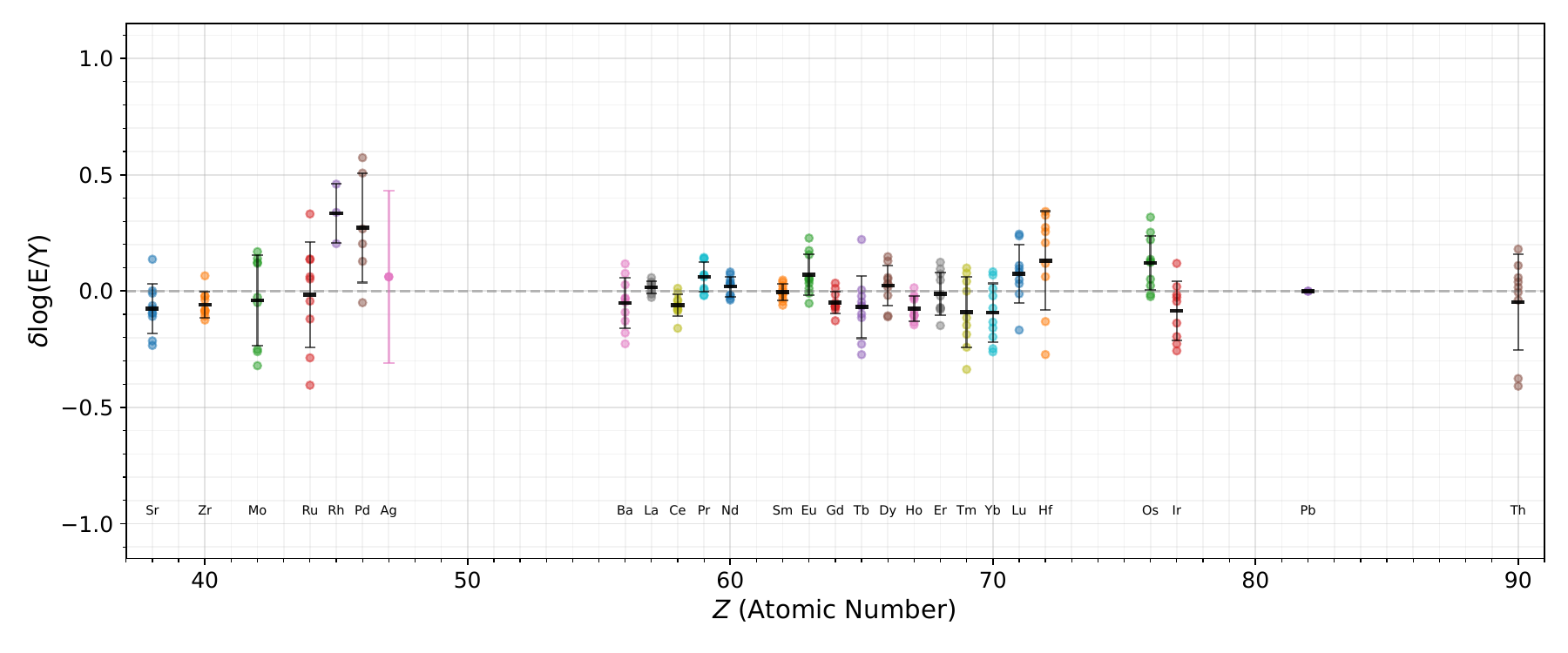}
\caption{Residuals for the three-pattern fits to the 10 stars in the RPA sample \citep{2025A&A...704A.282R}. Points show $\delta\log({\rm E/Y})$, defined as the predicted value minus the measured value. For each element except Ag, the thick horizontal bar marks the sample mean and the error bar shows the $1\sigma$ scatter. Only one Ag measurement is available, and its pink error bar gives the measurement uncertainty. For Pb, there are two almost coincident points close to zero with invisible scatter.}
\label{fig:three_pattern_rpa_residuals}
\end{figure*}

The optimized Patterns $1'$, $2'$, and $3'$ on the Y-reference scale are shown in Fig.~\ref{fig:three_patterns} and compared with Patterns 1 and 2 in Fig.~\ref{fig:model_pattern_comparison}. For clarity, elements with $\log({\rm E/Y})_{i'}<-2.5$ ($i=1$, 2, and 3) are omitted in Fig.~\ref{fig:three_patterns}. The residuals for the three-pattern fits to the combined sample and the 10-star RPA sample \citep{2025A&A...704A.282R} are shown in Figs.~\ref{fig:three_pattern_residuals} and \ref{fig:three_pattern_rpa_residuals}, respectively.

\clearpage
\section{Cross Validation}
\label{append:validation}
\restartappendixnumbering
\renewcommand{\theHfigure}{H\arabic{figure}}
\renewcommand{\theHtable}{H\arabic{table}}
Here we use 10-fold cross validation to compare the predictive performance and stability of the models involving two and three patterns. Because the models depend on both universal production patterns for all stars and star-specific mixing parameters, we perform measurement-level cross validation by organizing the data into 10 folds in terms of individual star-element entries. Specifically, an entry consists of star $j$, element E, the observed $\log({\rm E/Y})_{*,j}^{\rm obs}$, and the measurement uncertainty $\sigma_{\log({\rm E/Y})_{*,j}}$. Nine folds are selected for training, and they always contain all the stars and elements so that all the mixing parameters can be inferred along with the whole patterns. Then these results are used to make predictions for the single remaining test fold. This procedure is repeated by choosing each of the 10 folds for test in turn. 

Models of two and three patterns are evaluated on the same training and test folds. To ensure that training folds effectively constrain parameters of the three-pattern model, we require that each star have at least three elements measured (in addition to Y) and that each element be measured in at least six stars. As a result, we exclude the star CES2103-6505 from the CERES sample and the elements In, Sn, Sb, Ta, W, Re, Pb, Bi, and U from the literature sample (Table~\ref{tab:lit}). With 21 measurements removed by these cuts, we obtain a new combined sample of 844 measurements for 67 stars and 31 elements. In assigning measurements from the CERES and literature samples to folds, we preserve the different elemental coverage of the two samples. In the end, every measurement appears in a test fold exactly once.

\begin{figure*}[htbp]
\centering
\includegraphics[width=0.48\textwidth]{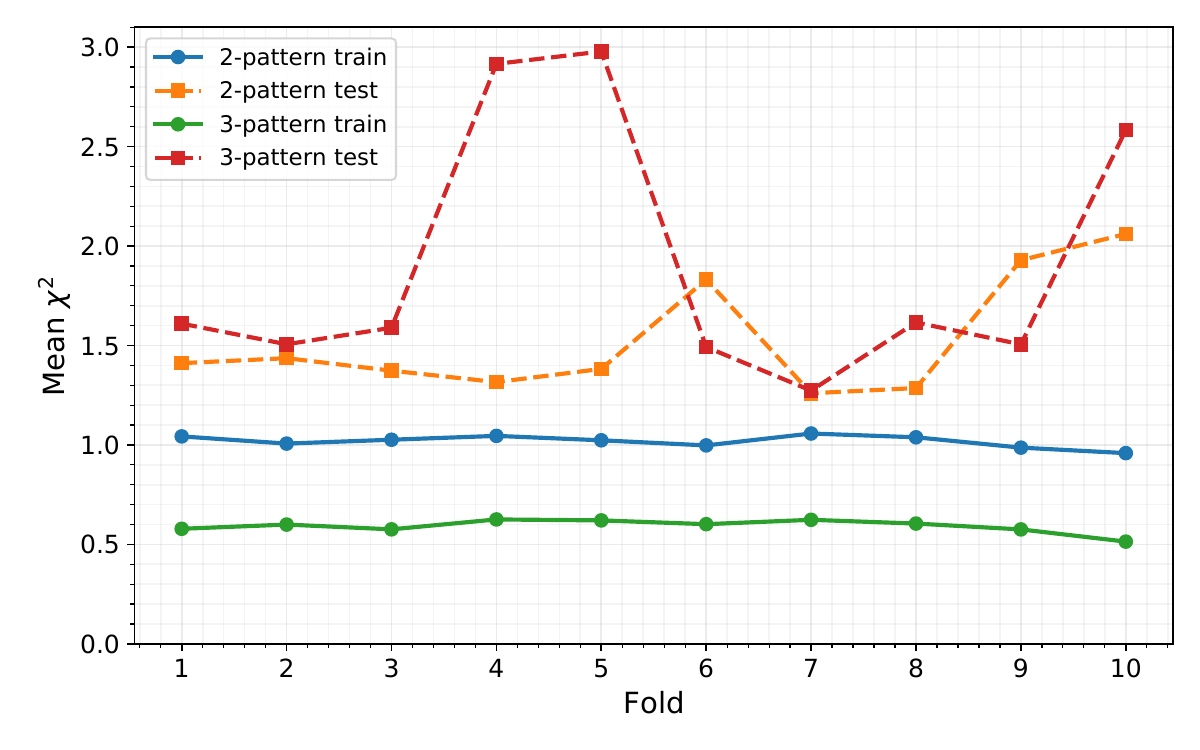}
\caption{Comparison of models involving two and three patterns based on 10-fold cross validation. For each test fold, the mean square of the standardized residuals $\langle\chi^2\rangle_{\rm test}$ is shown along with the corresponding $\langle\chi^2\rangle_{\rm train}$ for the training folds.}
\label{fig:cv_compare}
\end{figure*}

\begin{figure*}[htbp]
\centering
\includegraphics[width=\textwidth]{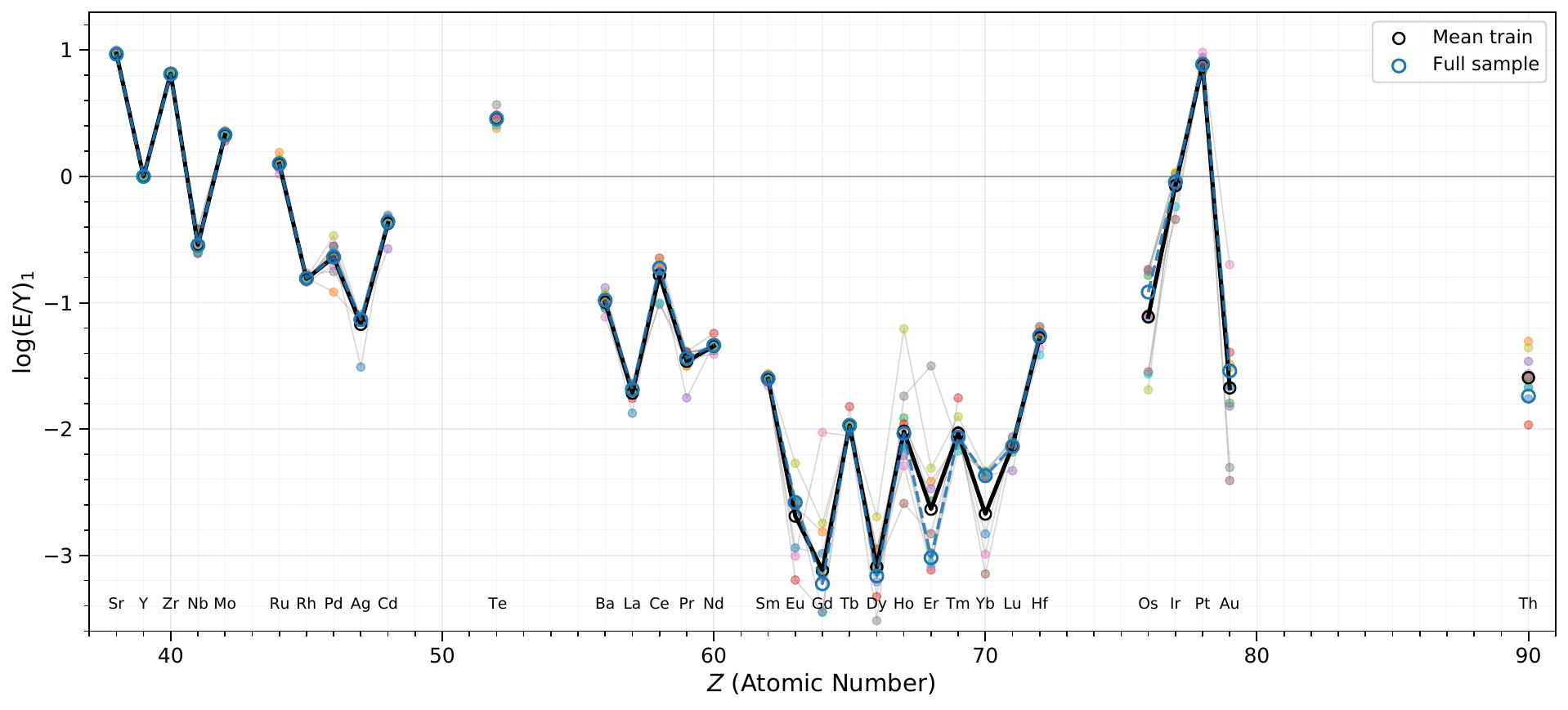}\\[1em]
\includegraphics[width=\textwidth]{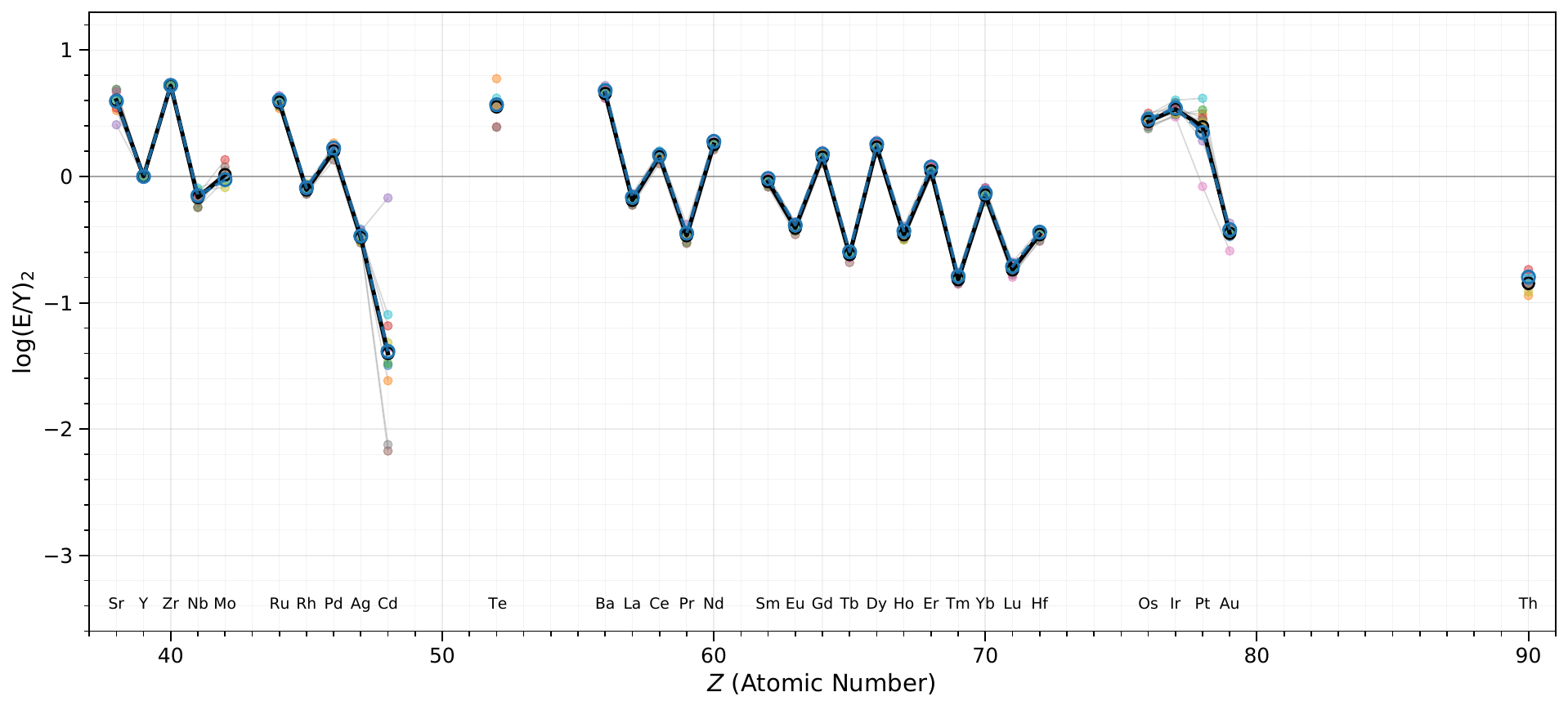}
\caption{Patterns 1 and 2 inferred during 10-fold cross validation (colored points) compared with those inferred from the full new combined sample (blue circles connected by dashed lines). The black circles (connected by solid lines) correspond to the averages of the colored points.}
\label{fig:cv_2patterns}
\end{figure*}

For each test fold, we train models of two and three patterns using the procedures described in Appendices~\ref{append:optimization} and \ref{append:3pattern}, respectively. For element E in star $j$, we define the standardized residual
\begin{equation}
\chi_{*,j}({\rm E})
\equiv
\frac{\log({\rm E/Y})_{*,j}-\log({\rm E/Y})_{*,j}^{\rm obs}}{\sigma_{\log({\rm E/Y})_{*,j}}},
\label{eq:cv_chi}
\end{equation}
where $\log({\rm E/Y})_{*,j}$ is obtained from the model. We calculate $\langle\chi^2\rangle_{\rm test}$ and $\langle\chi^2\rangle_{\rm train}$, which are the mean squares of the standardized residuals for all the entries in each test fold and the corresponding training folds, respectively. In Fig.~\ref{fig:cv_compare}, we compare these quantities for models involving two and three patterns. To demonstrate the stability of the two-pattern model, we show in Fig.~\ref{fig:cv_2patterns} Patterns 1 and 2 inferred during 10-fold cross validation. In contrast, the three-pattern model varies so widely across different training folds that Patterns $1'$, $2'$, and $3'$ cannot be assigned in a straightforward manner, and therefore, are not shown.

\clearpage
\section{Astrophysical Conditions for Templates Matching Patterns 1 and 2}
\label{append:cond}
\restartappendixnumbering
\renewcommand{\theHfigure}{I\arabic{figure}}
\renewcommand{\theHtable}{I\arabic{table}}

\begin{figure*}[htbp]
\centering
\includegraphics[width=0.48\textwidth]{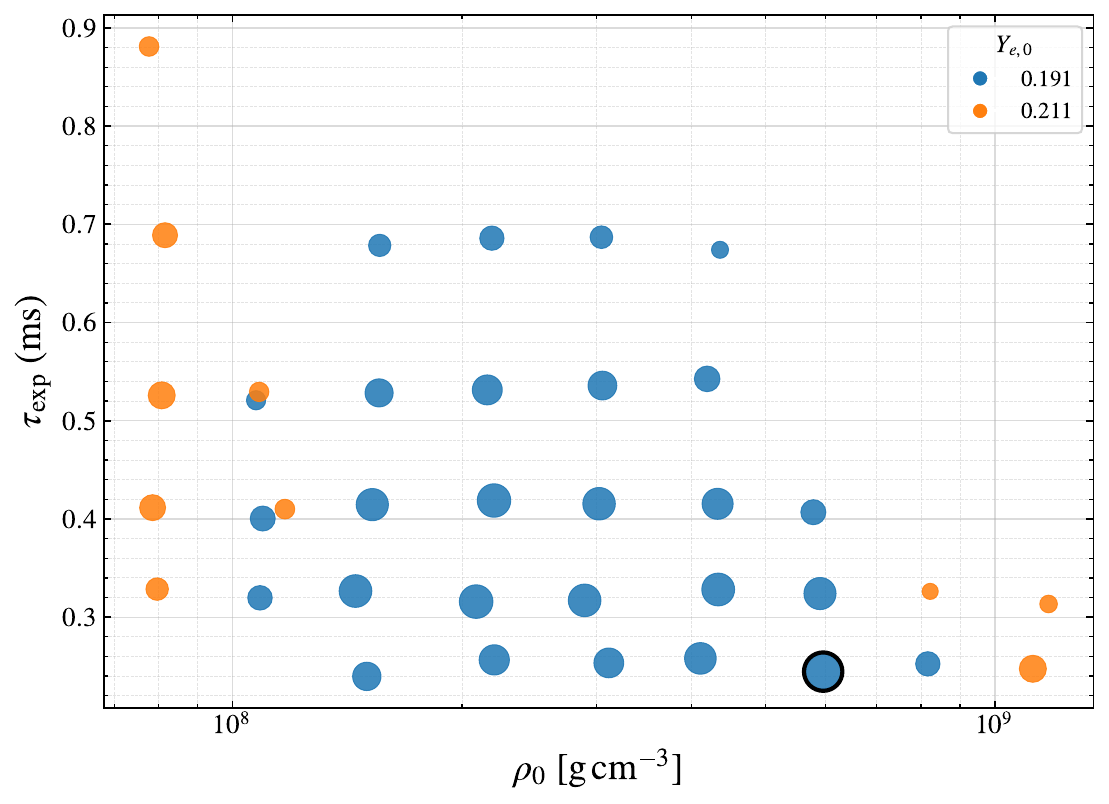}
\hfill
\includegraphics[width=0.48\textwidth]{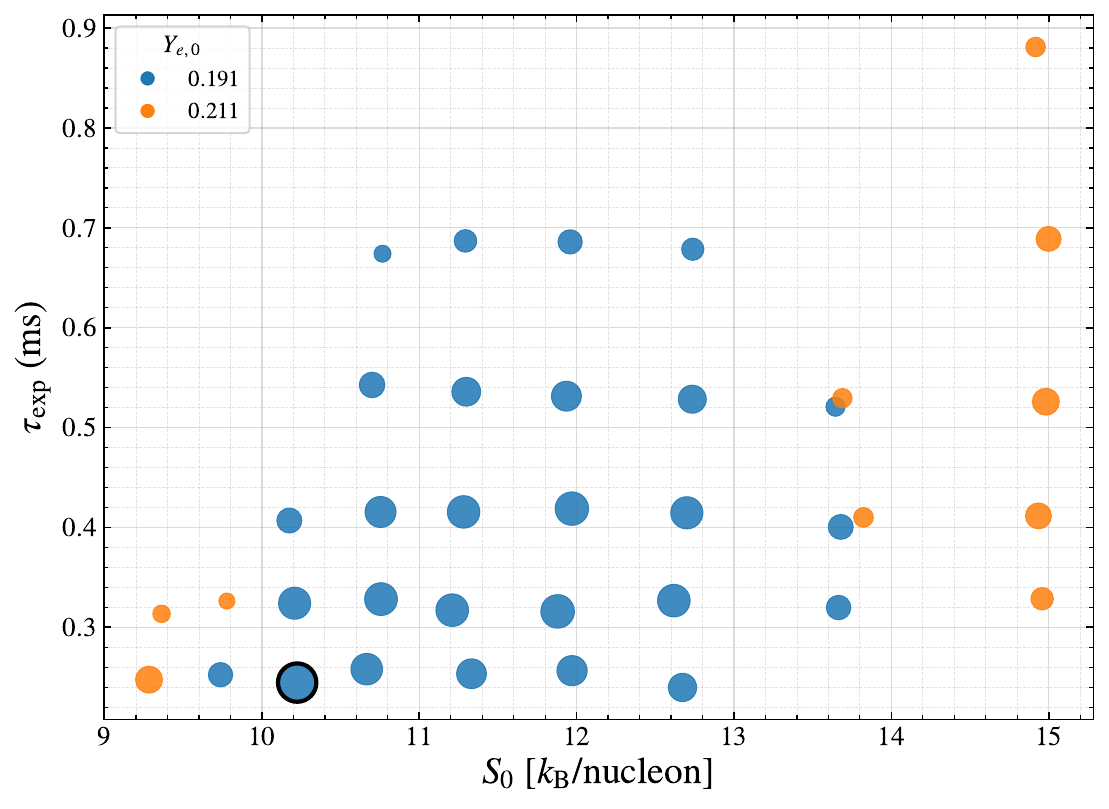}
\caption{Parameters of the templates matching Pattern 2, shown in terms of $Y_{e,0}$, $\rho_0$, and $\tau_{\rm exp}$ (left) and $Y_{e,0}$, $S_0$, and $\tau_{\rm exp}$ (right). For each of the 37 templates, the RMSD from Pattern 2 in $\log({\rm E/Y})_2$ is 0.39--0.45 dex, and the total mass fraction of Sr and heavier elements exceeds 0.1. Symbol size increases as the RMSD decreases, and a black circle encloses the best match shown in Fig.~\ref{fig:cond-p2}. Small random offsets are applied to reduce symbol overlap.}
\label{fig:para_pattern2}
\end{figure*}

\begin{figure*}[htbp]
\centering
\includegraphics[width=\textwidth]{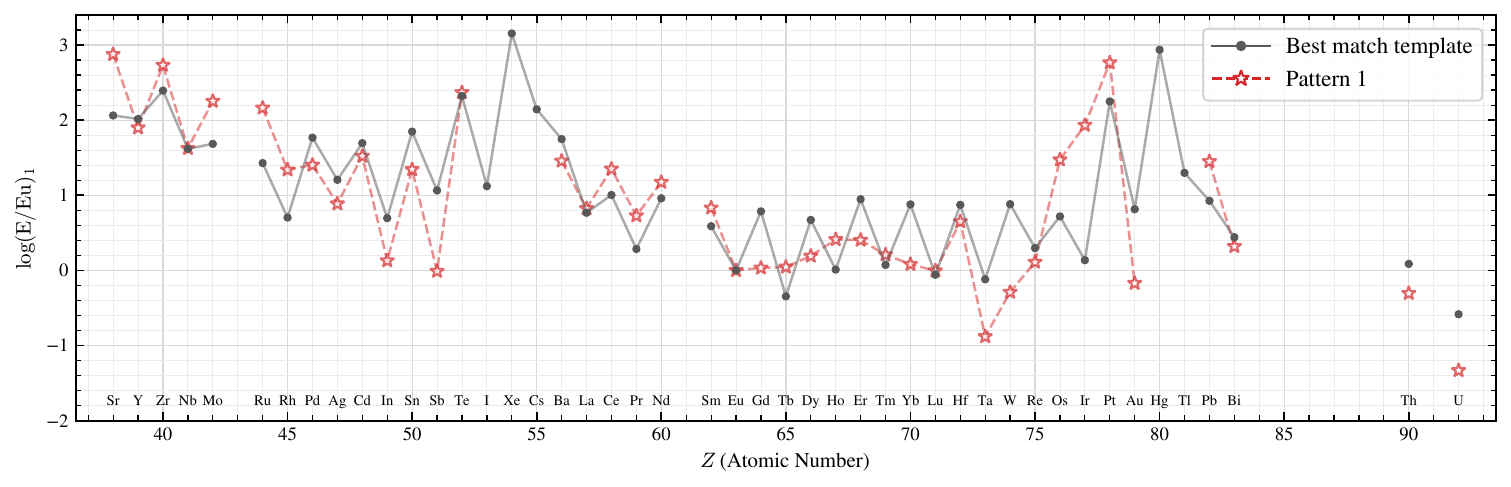}\\[1em]
\includegraphics[width=\textwidth]{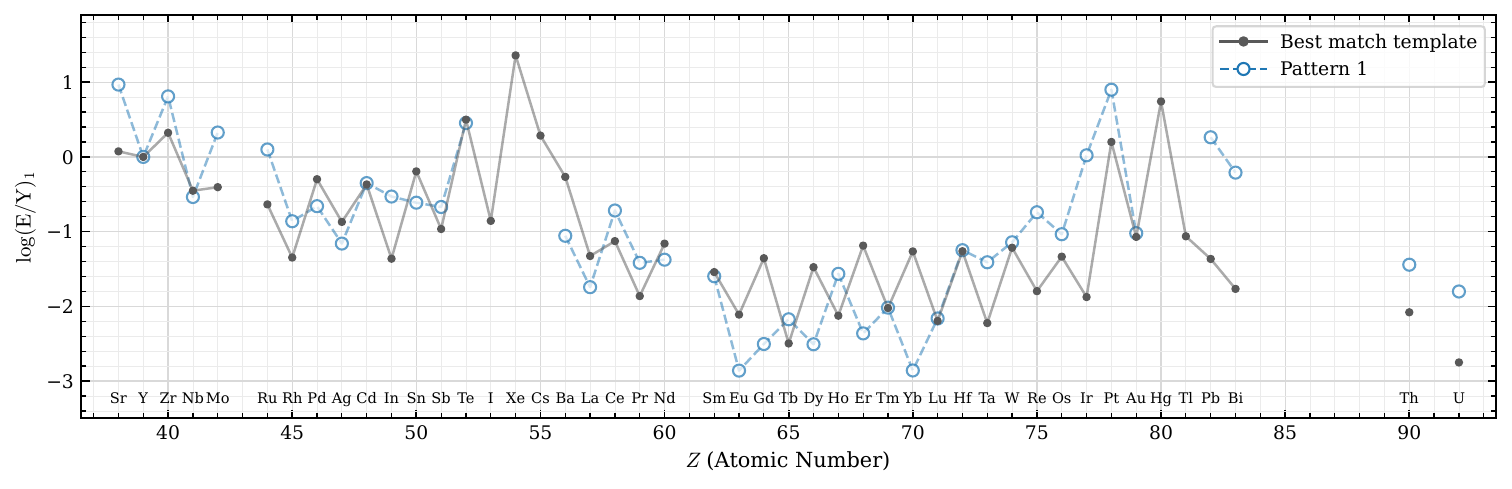}
\caption{Template (black filled circles) that best matches Pattern 1 on the Eu-reference scale (red stars) and is close to the best match on the Y-reference scale (light blue circles). The parameters are $Y_{e,0}=0.46$, $\rho_0=1.95\times10^6$~g~cm$^{-3}$, and $\tau_{\rm exp}=0.25$~ms.}
\label{fig:para_1_Eu_Y}
\end{figure*}

\begin{figure*}[htbp]
\centering
\includegraphics[width=0.48\textwidth]{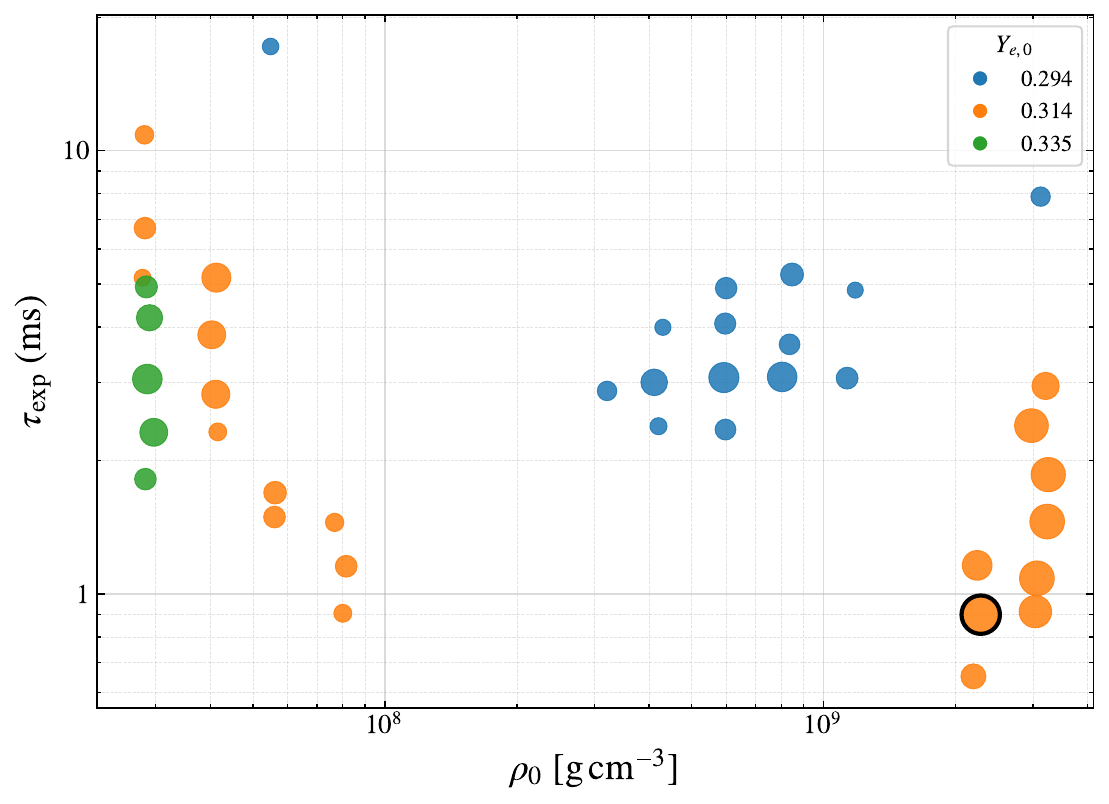}
\hfill
\includegraphics[width=0.48\textwidth]{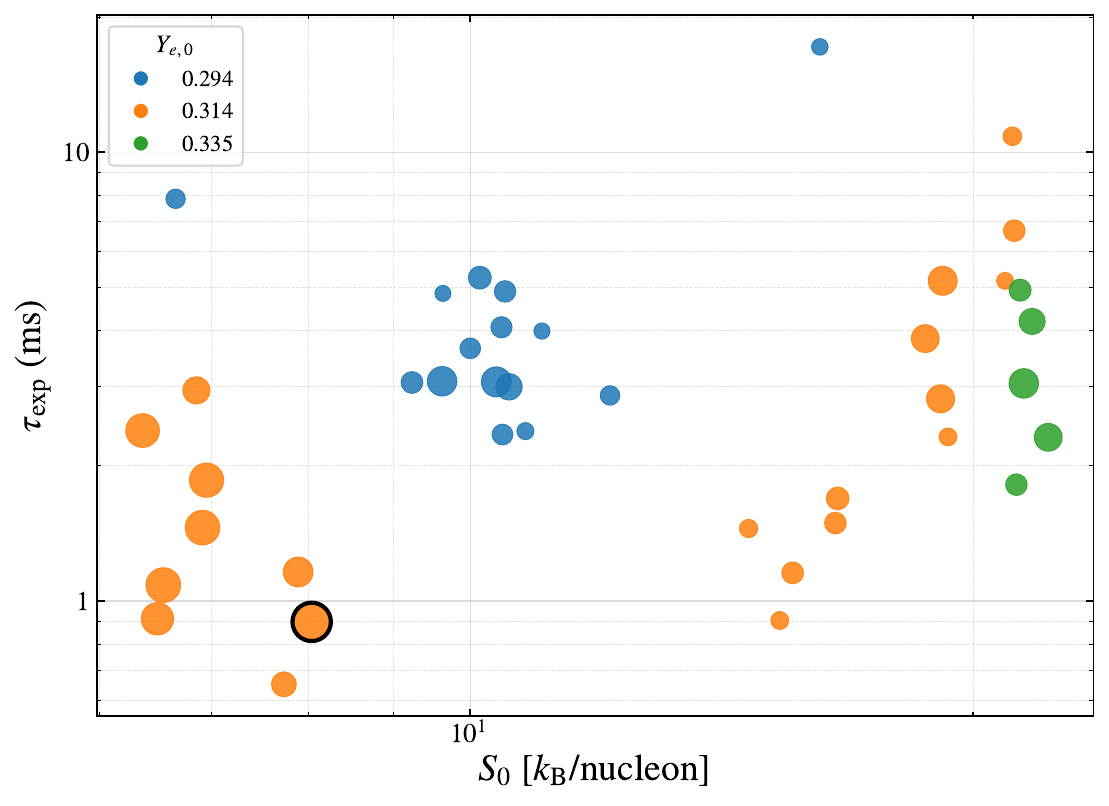}
\caption{Same as Fig.~\ref{fig:para_pattern2}, but for templates matching Pattern 1 from Sr to Cd. For each of the 41 templates, the RMSD from the matched pattern in $\log({\rm E/Y})_1$ is 0.31--0.40 dex, and the total mass fraction of Sr to Cd exceeds 0.1. The best match is shown in Fig.~\ref{fig:para_1}.}
\label{fig:para_pattern1}
\end{figure*}

We show 37 sets of conditions for templates matching Pattern 2 in Fig.~\ref{fig:para_pattern2}, compare a single template with Pattern 1 in Fig.~\ref{fig:para_1_Eu_Y}, and show 41 sets of conditions for templates matching Pattern 1 from Sr to Cd in Fig.~\ref{fig:para_pattern1}. No single template provides a quantitative match to Pattern 1 over the full elemental range.

\clearpage
\bibliography{ref}

@ARTICLE{2026arXiv260711323K,
       author = {{Kuske}, Jan and {Arcones}, Almudena and {Svensson}, Isak},
        title = "{Bayesian Inference of Stellar r-Process Abundances}",
      journal = {arXiv e-prints},
         year = 2026,
        month = jul,
          eid = {arXiv:2607.11323},
        pages = {arXiv:2607.11323},
          doi = {10.48550/arXiv.2607.11323},
archivePrefix = {arXiv},
       eprint = {2607.11323},
 primaryClass = {astro-ph.SR},
       adsurl = {https://ui.adsabs.harvard.edu/abs/2026arXiv260711323K}
}

@ARTICLE{2018ApJ...858...92H,
       author = {{Hansen}, Terese T. and {Holmbeck}, Erika M. and {Beers}, Timothy C. and {Placco}, Vinicius M. and {Roederer}, Ian U. and {Frebel}, Anna and {Sakari}, Charli M. and {Simon}, Joshua D. and {Thompson}, Ian B.},
        title = "{The R-process Alliance: First Release from the Southern Search for R-process-enhanced Stars in the Galactic Halo}",
      journal = {\apj},
         year = 2018,
        month = may,
       volume = {858},
       number = {2},
          eid = {92},
        pages = {92},
          doi = {10.3847/1538-4357/aabacc},
archivePrefix = {arXiv},
       eprint = {1804.03114},
 primaryClass = {astro-ph.SR},
       adsurl = {https://ui.adsabs.harvard.edu/abs/2018ApJ...858...92H}
}

@ARTICLE{2018ApJ...868..110S,
       author = {{Sakari}, Charli M. and {Placco}, Vinicius M. and {Farrell}, Elizabeth M. and {Roederer}, Ian U. and {Wallerstein}, George and {Beers}, Timothy C. and {Ezzeddine}, Rana and {Frebel}, Anna and {Hansen}, Terese and {Holmbeck}, Erika M. and {Sneden}, Christopher and {Cowan}, John J. and {Venn}, Kim A. and {Davis}, Christopher Evan and {Matijevi{\v{c}}}, Gal and {Wyse}, Rosemary F.~G. and {Bland-Hawthorn}, Joss and {Chiappini}, Cristina and {Freeman}, Kenneth C. and {Gibson}, Brad K. and {Grebel}, Eva K. and {Helmi}, Amina and {Kordopatis}, Georges and {Kunder}, Andrea and {Navarro}, Julio and {Reid}, Warren and {Seabroke}, George and {Steinmetz}, Matthias and {Watson}, Fred},
        title = "{The R-Process Alliance: First Release from the Northern Search for r-process-enhanced Metal-poor Stars in the Galactic Halo}",
      journal = {\apj},
         year = 2018,
        month = dec,
       volume = {868},
       number = {2},
          eid = {110},
        pages = {110},
          doi = {10.3847/1538-4357/aae9df},
archivePrefix = {arXiv},
       eprint = {1809.09156},
 primaryClass = {astro-ph.SR},
       adsurl = {https://ui.adsabs.harvard.edu/abs/2018ApJ...868..110S}
}

@ARTICLE{2020ApJ...898..150E,
       author = {{Ezzeddine}, Rana and {Rasmussen}, Kaitlin and {Frebel}, Anna and {Chiti}, Anirudh and {Hinojisa}, Karina and {Placco}, Vinicius M. and {Ji}, Alexander P. and {Beers}, Timothy C. and {Hansen}, Terese T. and {Roederer}, Ian U. and {Sakari}, Charli M. and {Melendez}, Jorge},
        title = "{The R-Process Alliance: First Magellan/MIKE Release from the Southern Search for R-process-enhanced Stars}",
      journal = {\apj},
         year = 2020,
        month = aug,
       volume = {898},
       number = {2},
          eid = {150},
        pages = {150},
          doi = {10.3847/1538-4357/ab9d1a},
archivePrefix = {arXiv},
       eprint = {2006.07731},
 primaryClass = {astro-ph.SR},
       adsurl = {https://ui.adsabs.harvard.edu/abs/2020ApJ...898..150E}
}

@ARTICLE{2013MNRAS.435..502F,
       author = {{Fern{\'a}ndez}, Rodrigo and {Metzger}, Brian D.},
        title = "{Delayed outflows from black hole accretion tori following neutron star binary coalescence}",
      journal = {\mnras},
         year = 2013,
        month = oct,
       volume = {435},
       number = {1},
        pages = {502-517},
          doi = {10.1093/mnras/stt1312},
archivePrefix = {arXiv},
       eprint = {1304.6720},
 primaryClass = {astro-ph.HE},
       adsurl = {https://ui.adsabs.harvard.edu/abs/2013MNRAS.435..502F}
}

@ARTICLE{2013MNRAS.430.2585R,
       author = {{Rosswog}, S. and {Piran}, T. and {Nakar}, E.},
        title = "{The multimessenger picture of compact object encounters: binary mergers versus dynamical collisions}",
      journal = {\mnras},
         year = 2013,
        month = apr,
       volume = {430},
       number = {4},
        pages = {2585-2604},
          doi = {10.1093/mnras/sts708},
archivePrefix = {arXiv},
       eprint = {1204.6240},
 primaryClass = {astro-ph.HE},
       adsurl = {https://ui.adsabs.harvard.edu/abs/2013MNRAS.430.2585R}
}

@ARTICLE{2013MNRAS.430.2121P,
       author = {{Piran}, Tsvi and {Nakar}, Ehud and {Rosswog}, Stephan},
        title = "{The electromagnetic signals of compact binary mergers}",
      journal = {\mnras},
         year = 2013,
        month = apr,
       volume = {430},
       number = {3},
        pages = {2121-2136},
          doi = {10.1093/mnras/stt037},
archivePrefix = {arXiv},
       eprint = {1204.6242},
 primaryClass = {astro-ph.HE},
       adsurl = {https://ui.adsabs.harvard.edu/abs/2013MNRAS.430.2121P}
}

@ARTICLE{2012MNRAS.426.1940K,
       author = {{Korobkin}, O. and {Rosswog}, S. and {Arcones}, A. and {Winteler}, C.},
        title = "{On the astrophysical robustness of the neutron star merger r-process}",
      journal = {\mnras},
         year = 2012,
        month = nov,
       volume = {426},
       number = {3},
        pages = {1940-1949},
          doi = {10.1111/j.1365-2966.2012.21859.x},
archivePrefix = {arXiv},
       eprint = {1206.2379},
 primaryClass = {astro-ph.SR},
       adsurl = {https://ui.adsabs.harvard.edu/abs/2012MNRAS.426.1940K}
}

@ARTICLE{2016MNRAS.463.2323W,
       author = {{Wu}, Meng-Ru and {Fern{\'a}ndez}, Rodrigo and {Mart{\'\i}nez-Pinedo}, Gabriel and {Metzger}, Brian D.},
        title = "{Production of the entire range of r-process nuclides by black hole accretion disc outflows from neutron star mergers}",
      journal = {\mnras},
         year = 2016,
        month = dec,
       volume = {463},
       number = {3},
        pages = {2323-2334},
          doi = {10.1093/mnras/stw2156},
archivePrefix = {arXiv},
       eprint = {1607.05290},
 primaryClass = {astro-ph.HE},
       adsurl = {https://ui.adsabs.harvard.edu/abs/2016MNRAS.463.2323W}
}

@ARTICLE{2014MNRAS.443.3134P,
       author = {{Perego}, A. and {Rosswog}, S. and {Cabez{\'o}n}, R.~M. and {Korobkin}, O. and {K{\"a}ppeli}, R. and {Arcones}, A. and {Liebend{\"o}rfer}, M.},
        title = "{Neutrino-driven winds from neutron star merger remnants}",
      journal = {\mnras},
         year = 2014,
        month = oct,
       volume = {443},
       number = {4},
        pages = {3134-3156},
          doi = {10.1093/mnras/stu1352},
archivePrefix = {arXiv},
       eprint = {1405.6730},
 primaryClass = {astro-ph.HE},
       adsurl = {https://ui.adsabs.harvard.edu/abs/2014MNRAS.443.3134P}
}

@ARTICLE{2015ApJ...813....2M,
       author = {{Martin}, D. and {Perego}, A. and {Arcones}, A. and {Thielemann}, F.-K. and {Korobkin}, O. and {Rosswog}, S.},
        title = "{Neutrino-driven Winds in the Aftermath of a Neutron Star Merger: Nucleosynthesis and Electromagnetic Transients}",
      journal = {\apj},
         year = 2015,
        month = nov,
       volume = {813},
       number = {1},
          eid = {2},
        pages = {2},
          doi = {10.1088/0004-637X/813/1/2},
archivePrefix = {arXiv},
       eprint = {1506.05048},
 primaryClass = {astro-ph.SR},
       adsurl = {https://ui.adsabs.harvard.edu/abs/2015ApJ...813....2M}
}

@ARTICLE{2024MNRAS.533.2096R,
       author = {{Ricigliano}, Giacomo and {Jacobi}, Maximilian and {Arcones}, Almudena},
        title = "{Impact of nuclear matter properties on the nucleosynthesis and the kilonova from binary neutron star merger ejecta}",
      journal = {\mnras},
         year = 2024,
        month = sep,
       volume = {533},
       number = {2},
        pages = {2096-2112},
          doi = {10.1093/mnras/stae1979},
archivePrefix = {arXiv},
       eprint = {2406.03649},
 primaryClass = {astro-ph.HE},
       adsurl = {https://ui.adsabs.harvard.edu/abs/2024MNRAS.533.2096R}
}

@ARTICLE{2024MNRAS.527.8812J,
       author = {{Jacobi}, M. and {Guercilena}, F.~M. and {Huth}, S. and {Ricigliano}, G. and {Arcones}, A. and {Schwenk}, A.},
        title = "{Effects of nuclear matter properties in neutron star mergers}",
      journal = {\mnras},
         year = 2024,
        month = jan,
       volume = {527},
       number = {3},
        pages = {8812-8828},
          doi = {10.1093/mnras/stad3738},
archivePrefix = {arXiv},
       eprint = {2309.09233},
 primaryClass = {astro-ph.HE},
       adsurl = {https://ui.adsabs.harvard.edu/abs/2024MNRAS.527.8812J}
}

@ARTICLE{2017PhRvD..96l4005B,
       author = {{Bovard}, Luke and {Martin}, Dirk and {Guercilena}, Federico and {Arcones}, Almudena and {Rezzolla}, Luciano and {Korobkin}, Oleg},
        title = "{r -process nucleosynthesis from matter ejected in binary neutron star mergers}",
      journal = {\prd},
         year = 2017,
        month = dec,
       volume = {96},
       number = {12},
          eid = {124005},
        pages = {124005},
          doi = {10.1103/PhysRevD.96.124005},
archivePrefix = {arXiv},
       eprint = {1709.09630},
 primaryClass = {gr-qc},
       adsurl = {https://ui.adsabs.harvard.edu/abs/2017PhRvD..96l4005B}
}

@ARTICLE{2012ApJ...750L..22W,
       author = {{Winteler}, C. and {K{\"a}ppeli}, R. and {Perego}, A. and {Arcones}, A. and {Vasset}, N. and {Nishimura}, N. and {Liebend{\"o}rfer}, M. and {Thielemann}, F.-K.},
        title = "{Magnetorotationally Driven Supernovae as the Origin of Early Galaxy r-process Elements?}",
      journal = {\apjl},
         year = 2012,
        month = may,
       volume = {750},
       number = {1},
          eid = {L22},
        pages = {L22},
          doi = {10.1088/2041-8205/750/1/L22},
archivePrefix = {arXiv},
       eprint = {1203.0616},
 primaryClass = {astro-ph.SR},
       adsurl = {https://ui.adsabs.harvard.edu/abs/2012ApJ...750L..22W}
}

@ARTICLE{2021MNRAS.501.5733R,
       author = {{Reichert}, M. and {Obergaulinger}, M. and {Eichler}, M. and {Aloy}, M. {\'A}. and {Arcones}, A.},
        title = "{Nucleosynthesis in magneto-rotational supernovae}",
      journal = {\mnras},
         year = 2021,
        month = mar,
       volume = {501},
       number = {4},
        pages = {5733-5745},
          doi = {10.1093/mnras/stab029},
archivePrefix = {arXiv},
       eprint = {2010.02227},
 primaryClass = {astro-ph.HE},
       adsurl = {https://ui.adsabs.harvard.edu/abs/2021MNRAS.501.5733R}
}

@ARTICLE{2017MNRAS.469L..43O,
       author = {{Obergaulinger}, M. and {Aloy}, M. {\'A}.},
        title = "{Protomagnetar and black hole formation in high-mass stars}",
      journal = {\mnras},
         year = 2017,
        month = jul,
       volume = {469},
       number = {1},
        pages = {L43-L47},
          doi = {10.1093/mnrasl/slx046},
archivePrefix = {arXiv},
       eprint = {1703.09893},
 primaryClass = {astro-ph.SR},
       adsurl = {https://ui.adsabs.harvard.edu/abs/2017MNRAS.469L..43O}
}

@ARTICLE{1996ApJ...471..331Q,
       author = {{Qian}, Y.-Z. and {Woosley}, S.~E.},
        title = "{Nucleosynthesis in Neutrino-driven Winds. I. The Physical Conditions}",
      journal = {\apj},
         year = 1996,
        month = nov,
       volume = {471},
        pages = {331},
          doi = {10.1086/177973},
archivePrefix = {arXiv},
       eprint = {astro-ph/9611094},
 primaryClass = {astro-ph},
       adsurl = {https://ui.adsabs.harvard.edu/abs/1996ApJ...471..331Q}
}

@ARTICLE{2025A&A...704A.282R,
       author = {{Racca}, M. and {Hansen}, T.~T. and {Roederer}, I.~U. and {Placco}, V.~M. and {Frebel}, A. and {Beers}, T.~C. and {Ezzeddine}, R. and {Holmbeck}, E.~M. and {Sakari}, C.~M. and {Monty}, S. and {Harket}, {\O}. and {Simon}, J.~D. and {Sneden}, C. and {Thompson}, I.~B.},
        title = "{The R-Process Alliance: Exploring the cosmic scatter among ten r-process sites with stellar abundances}",
      journal = {\aap},
         year = 2025,
        month = dec,
       volume = {704},
          eid = {A282},
        pages = {A282},
          doi = {10.1051/0004-6361/202556947},
archivePrefix = {arXiv},
       eprint = {2510.25500},
 primaryClass = {astro-ph.SR},
       adsurl = {https://ui.adsabs.harvard.edu/abs/2025A&A...704A.282R}
}

@ARTICLE{2005ARA&A..43..531B,
       author = {{Beers}, Timothy C. and {Christlieb}, Norbert},
        title = "{The Discovery and Analysis of Very Metal-Poor Stars in the Galaxy}",
      journal = {\araa},
         year = 2005,
        month = sep,
       volume = {43},
       number = {1},
        pages = {531-580},
          doi = {10.1146/annurev.astro.42.053102.134057},
       adsurl = {https://ui.adsabs.harvard.edu/abs/2005ARA&A..43..531B}
}

@ARTICLE{2018ARNPS..68..237F,
       author = {{Frebel}, Anna},
        title = "{From Nuclei to the Cosmos: Tracing Heavy-Element Production with the Oldest Stars}",
      journal = {Annual Review of Nuclear and Particle Science},
         year = 2018,
        month = oct,
       volume = {68},
       number = {1},
        pages = {237-269},
          doi = {10.1146/annurev-nucl-101917-021141},
archivePrefix = {arXiv},
       eprint = {1806.08955},
 primaryClass = {astro-ph.SR},
       adsurl = {https://ui.adsabs.harvard.edu/abs/2018ARNPS..68..237F}
}

@ARTICLE{2013ApJ...774...17S,
       author = {{Steiner}, A.~W. and {Hempel}, M. and {Fischer}, T.},
        title = "{Core-collapse Supernova Equations of State Based on Neutron Star Observations}",
      journal = {\apj},
         year = 2013,
        month = sep,
       volume = {774},
       number = {1},
          eid = {17},
        pages = {17},
          doi = {10.1088/0004-637X/774/1/17},
archivePrefix = {arXiv},
       eprint = {1207.2184},
 primaryClass = {astro-ph.SR},
       adsurl = {https://ui.adsabs.harvard.edu/abs/2013ApJ...774...17S}
}

@ARTICLE{Vassh2019,
       author = {{Vassh}, N. and {Vogt}, R. and {Surman}, R. and {Randrup}, J. and {Sprouse}, T.~M. and {Mumpower}, M.~R. and {Jaffke}, P. and {Shaw}, D. and {Holmbeck}, E.~M. and {Zhu}, Y. and {McLaughlin}, G.~C.},
        title = "{Using excitation-energy dependent fission yields to identify key fissioning nuclei in r-process nucleosynthesis}",
      journal = {Journal of Physics G Nuclear Physics},
         year = 2019,
        month = jun,
       volume = {46},
       number = {6},
        pages = {065202},
          doi = {10.1088/1361-6471/ab0bea},
archivePrefix = {arXiv},
       eprint = {1810.08133},
 primaryClass = {nucl-th},
       adsurl = {https://ui.adsabs.harvard.edu/abs/2019JPhG...46f5202V}
}

@ARTICLE{2010ApJS..189..240C,
       author = {{Cyburt}, Richard H. and {Amthor}, A. Matthew and {Ferguson}, Ryan and {Meisel}, Zach and {Smith}, Karl and {Warren}, Scott and {Heger}, Alexander and {Hoffman}, R.~D. and {Rauscher}, Thomas and {Sakharuk}, Alexander and {Schatz}, Hendrik and {Thielemann}, F.~K. and {Wiescher}, Michael},
        title = "{The JINA REACLIB Database: Its Recent Updates and Impact on Type-I X-ray Bursts}",
      journal = {\apjs},
         year = 2010,
        month = jul,
       volume = {189},
       number = {1},
        pages = {240-252},
          doi = {10.1088/0067-0049/189/1/240},
       adsurl = {https://ui.adsabs.harvard.edu/abs/2010ApJS..189..240C}
}

@ARTICLE{2016PhRvC..94f4317M,
       author = {{Mumpower}, M.~R. and {Kawano}, T. and {M{\"o}ller}, P.},
        title = "{Neutron-{\ensuremath{\gamma}} competition for {\ensuremath{\beta}} -delayed neutron emission}",
      journal = {\prc},
         year = 2016,
        month = dec,
       volume = {94},
       number = {6},
          eid = {064317},
        pages = {064317},
          doi = {10.1103/PhysRevC.94.064317},
archivePrefix = {arXiv},
       eprint = {1608.01956},
 primaryClass = {nucl-th},
       adsurl = {https://ui.adsabs.harvard.edu/abs/2016PhRvC..94f4317M}
}

@ARTICLE{2018ApJ...869...14M,
       author = {{Mumpower}, M.~R. and {Kawano}, T. and {Sprouse}, T.~M. and {Vassh}, N. and {Holmbeck}, E.~M. and {Surman}, R. and {M{\"o}ller}, P.},
        title = "{{\ensuremath{\beta}}-delayed Fission in r-process Nucleosynthesis}",
      journal = {\apj},
         year = 2018,
        month = dec,
       volume = {869},
       number = {1},
          eid = {14},
        pages = {14},
          doi = {10.3847/1538-4357/aaeaca},
archivePrefix = {arXiv},
       eprint = {1802.04398},
 primaryClass = {nucl-th},
       adsurl = {https://ui.adsabs.harvard.edu/abs/2018ApJ...869...14M}
}

@ARTICLE{2021EPJA...57...16K,
       author = {{Kawano}, Toshihiko},
        title = "{Unified description of the coupled-channels and statistical Hauser-Feshbach nuclear reaction theories for low energy neutron incident reactions}",
      journal = {European Physical Journal A},
         year = 2021,
        month = jan,
       volume = {57},
       number = {1},
          eid = {16},
        pages = {16},
          doi = {10.1140/epja/s10050-020-00311-9},
archivePrefix = {arXiv},
       eprint = {2009.12736},
 primaryClass = {nucl-th},
       adsurl = {https://ui.adsabs.harvard.edu/abs/2021EPJA...57...16K}
}

@ARTICLE{2019arXiv190105641K,
       author = {{Kawano}, Toshihiko},
        title = "{Unified Coupled-Channels and Hauser-Feshbach Model Calculation for Nuclear Data Evaluation}",
      journal = {arXiv e-prints},
         year = 2019,
        month = jan,
          eid = {arXiv:1901.05641},
        pages = {arXiv:1901.05641},
          doi = {10.48550/arXiv.1901.05641},
archivePrefix = {arXiv},
       eprint = {1901.05641},
 primaryClass = {nucl-th},
       adsurl = {https://ui.adsabs.harvard.edu/abs/2019arXiv190105641K}
}

@ARTICLE{2016ADNDT.109....1M,
       author = {{M{\"o}ller}, P. and {Sierk}, A.~J. and {Ichikawa}, T. and {Sagawa}, H.},
        title = "{Nuclear ground-state masses and deformations: FRDM(2012)}",
      journal = {Atomic Data and Nuclear Data Tables},
         year = 2016,
        month = may,
       volume = {109},
        pages = {1-204},
          doi = {10.1016/j.adt.2015.10.002},
archivePrefix = {arXiv},
       eprint = {1508.06294},
 primaryClass = {nucl-th},
       adsurl = {https://ui.adsabs.harvard.edu/abs/2016ADNDT.109....1M}
}

@ARTICLE{2012PhRvL.108e2501M,
       author = {{M{\"o}ller}, Peter and {Myers}, William D. and {Sagawa}, Hiroyuki and {Yoshida}, Satoshi},
        title = "{New Finite-Range Droplet Mass Model and Equation-of-State Parameters}",
      journal = {\prl},
         year = 2012,
        month = feb,
       volume = {108},
       number = {5},
          eid = {052501},
        pages = {052501},
          doi = {10.1103/PhysRevLett.108.052501},
       adsurl = {https://ui.adsabs.harvard.edu/abs/2012PhRvL.108e2501M}
}

@ARTICLE{2021PhRvC.104a5803S,
       author = {{Sprouse}, T.~M. and {Mumpower}, M.~R. and {Surman}, R.},
        title = "{Following nuclei through nucleosynthesis: A novel tracing technique}",
      journal = {\prc},
         year = 2021,
        month = jul,
       volume = {104},
       number = {1},
          eid = {015803},
        pages = {015803},
          doi = {10.1103/PhysRevC.104.015803},
archivePrefix = {arXiv},
       eprint = {2008.06075},
 primaryClass = {nucl-th},
       adsurl = {https://ui.adsabs.harvard.edu/abs/2021PhRvC.104a5803S}
}

@ARTICLE{2009ARA&A..47..481A,
       author = {{Asplund}, Martin and {Grevesse}, Nicolas and {Sauval}, A. Jacques and {Scott}, Pat},
        title = "{The Chemical Composition of the Sun}",
      journal = {\araa},
         year = 2009,
        month = sep,
       volume = {47},
       number = {1},
        pages = {481-522},
          doi = {10.1146/annurev.astro.46.060407.145222},
archivePrefix = {arXiv},
       eprint = {0909.0948},
 primaryClass = {astro-ph.SR},
       adsurl = {https://ui.adsabs.harvard.edu/abs/2009ARA&A..47..481A}
}

@ARTICLE{2021Natur.595..223Y,
       author = {{Yong}, D. and {Kobayashi}, C. and {Da Costa}, G.~S. and {Bessell}, M.~S. and {Chiti}, A. and {Frebel}, A. and {Lind}, K. and {Mackey}, A.~D. and {Nordlander}, T. and {Asplund}, M. and {Casey}, A.~R. and {Marino}, A.~F. and {Murphy}, S.~J. and {Schmidt}, B.~P.},
        title = "{r-Process elements from magnetorotational hypernovae}",
      journal = {\nat},
         year = 2021,
        month = jul,
       volume = {595},
       number = {7866},
        pages = {223-226},
          doi = {10.1038/s41586-021-03611-2},
archivePrefix = {arXiv},
       eprint = {2107.03010},
 primaryClass = {astro-ph.SR},
       adsurl = {https://ui.adsabs.harvard.edu/abs/2021Natur.595..223Y}
}

@ARTICLE{2003ApJ...591..936S,
       author = {{Sneden}, Christopher and {Cowan}, John J. and {Lawler}, James E. and {Ivans}, Inese I. and {Burles}, Scott and {Beers}, Timothy C. and {Primas}, Francesca and {Hill}, Vanessa and {Truran}, James W. and {Fuller}, George M. and {Pfeiffer}, Bernd and {Kratz}, Karl-Ludwig},
        title = "{The Extremely Metal-poor, Neutron Capture-rich Star CS 22892-052: A Comprehensive Abundance Analysis}",
      journal = {\apj},
         year = 2003,
        month = jul,
       volume = {591},
       number = {2},
        pages = {936-953},
          doi = {10.1086/375491},
archivePrefix = {arXiv},
       eprint = {astro-ph/0303542},
 primaryClass = {astro-ph},
       adsurl = {https://ui.adsabs.harvard.edu/abs/2003ApJ...591..936S}
}

@ARTICLE{2012ApJS..203...27R,
       author = {{Roederer}, Ian U. and {Lawler}, James E. and {Sobeck}, Jennifer S. and {Beers}, Timothy C. and {Cowan}, John J. and {Frebel}, Anna and {Ivans}, Inese I. and {Schatz}, Hendrik and {Sneden}, Christopher and {Thompson}, Ian B.},
        title = "{New Hubble Space Telescope Observations of Heavy Elements in Four Metal-Poor Stars}",
      journal = {\apjs},
         year = 2012,
        month = dec,
       volume = {203},
       number = {2},
          eid = {27},
        pages = {27},
          doi = {10.1088/0067-0049/203/2/27},
archivePrefix = {arXiv},
       eprint = {1210.6387},
 primaryClass = {astro-ph.SR},
       adsurl = {https://ui.adsabs.harvard.edu/abs/2012ApJS..203...27R}
}

@ARTICLE{2013A&A...550A.122S,
       author = {{Siqueira Mello}, C. and {Spite}, M. and {Barbuy}, B. and {Spite}, F. and {Caffau}, E. and {Hill}, V. and {Wanajo}, S. and {Primas}, F. and {Plez}, B. and {Cayrel}, R. and {Andersen}, J. and {Nordstr{\"o}m}, B. and {Sneden}, C. and {Beers}, T.~C. and {Bonifacio}, P. and {Fran{\c{c}}ois}, P. and {Molaro}, P.},
        title = "{First stars. XVI. HST/STIS abundances of heavy elements in the uranium-rich metal-poor star CS 31082-001}",
      journal = {\aap},
         year = 2013,
        month = feb,
       volume = {550},
          eid = {A122},
        pages = {A122},
          doi = {10.1051/0004-6361/201219949},
archivePrefix = {arXiv},
       eprint = {1212.0211},
 primaryClass = {astro-ph.SR},
       adsurl = {https://ui.adsabs.harvard.edu/abs/2013A&A...550A.122S}
}

@ARTICLE{2020ApJ...898...40C,
       author = {{Cain}, Madelyn and {Frebel}, Anna and {Ji}, Alexander P. and {Placco}, Vinicius M. and {Ezzeddine}, Rana and {Roederer}, Ian U. and {Hattori}, Kohei and {Beers}, Timothy C. and {Mel{\'e}ndez}, Jorge and {Hansen}, Terese T. and {Sakari}, Charli M.},
        title = "{The R-Process Alliance: A Very Metal-poor, Extremely r-process-enhanced Star with [Eu/Fe] = + 2.2, and the Class of r-III Stars}",
      journal = {\apj},
         year = 2020,
        month = jul,
       volume = {898},
       number = {1},
          eid = {40},
        pages = {40},
          doi = {10.3847/1538-4357/ab97ba},
archivePrefix = {arXiv},
       eprint = {2006.08080},
 primaryClass = {astro-ph.SR},
       adsurl = {https://ui.adsabs.harvard.edu/abs/2020ApJ...898...40C}
}

@ARTICLE{2022ApJS..260...27R,
       author = {{Roederer}, Ian U. and {Lawler}, James E. and {Den Hartog}, Elizabeth A. and {Placco}, Vinicius M. and {Surman}, Rebecca and {Beers}, Timothy C. and {Ezzeddine}, Rana and {Frebel}, Anna and {Hansen}, Terese T. and {Hattori}, Kohei and {Holmbeck}, Erika M. and {Sakari}, Charli M.},
        title = "{The R-process Alliance: A Nearly Complete R-process Abundance Template Derived from Ultraviolet Spectroscopy of the R-process-enhanced Metal-poor Star HD 222925}",
      journal = {\apjs},
         year = 2022,
        month = jun,
       volume = {260},
       number = {2},
          eid = {27},
        pages = {27},
          doi = {10.3847/1538-4365/ac5cbc},
archivePrefix = {arXiv},
       eprint = {2205.03426},
 primaryClass = {astro-ph.SR},
       adsurl = {https://ui.adsabs.harvard.edu/abs/2022ApJS..260...27R}
}

@ARTICLE{2024MNRAS.529.1917S,
       author = {{Shah}, Shivani P. and {Ezzeddine}, Rana and {Roederer}, Ian U. and {Hansen}, Terese T. and {Placco}, Vinicius M. and {Beers}, Timothy C. and {Frebel}, Anna and {Ji}, Alexander P. and {Holmbeck}, Erika M. and {Marshall}, Jennifer and {Sakari}, Charli M.},
        title = "{The R-Process Alliance: detailed chemical composition of an r-process enhanced star with UV and optical spectroscopy}",
      journal = {\mnras},
         year = 2024,
        month = apr,
       volume = {529},
       number = {3},
        pages = {1917-1940},
          doi = {10.1093/mnras/stae255},
archivePrefix = {arXiv},
       eprint = {2401.12311},
 primaryClass = {astro-ph.SR},
       adsurl = {https://ui.adsabs.harvard.edu/abs/2024MNRAS.529.1917S}
}

@ARTICLE{2024A&A...688A.123X,
       author = {{Xylakis-Dornbusch}, T. and {Hansen}, T.~T. and {Beers}, T.~C. and {Christlieb}, N. and {Ezzeddine}, R. and {Frebel}, A. and {Holmbeck}, E. and {Placco}, V.~M. and {Roederer}, I.~U. and {Sakari}, C.~M. and {Sneden}, C.},
        title = "{The R-Process Alliance: Analysis of limited-r stars}",
      journal = {\aap},
         year = 2024,
        month = aug,
       volume = {688},
          eid = {A123},
        pages = {A123},
          doi = {10.1051/0004-6361/202449376},
archivePrefix = {arXiv},
       eprint = {2404.03379},
 primaryClass = {astro-ph.SR},
       adsurl = {https://ui.adsabs.harvard.edu/abs/2024A&A...688A.123X}
}

@ARTICLE{2024ApJ...971..158R,
       author = {{Roederer}, Ian U. and {Beers}, Timothy C. and {Hattori}, Kohei and {Placco}, Vinicius M. and {Hansen}, Terese T. and {Ezzeddine}, Rana and {Frebel}, Anna and {Holmbeck}, Erika M. and {Sakari}, Charli M.},
        title = "{The R-Process Alliance: 2MASS J22132050{\textendash}5137385, the Star with the Highest-known r-process Enhancement at [Eu/Fe] = +2.45}",
      journal = {\apj},
         year = 2024,
        month = aug,
       volume = {971},
       number = {2},
          eid = {158},
        pages = {158},
          doi = {10.3847/1538-4357/ad57bf},
archivePrefix = {arXiv},
       eprint = {2406.02691},
 primaryClass = {astro-ph.SR},
       adsurl = {https://ui.adsabs.harvard.edu/abs/2024ApJ...971..158R}
}

@ARTICLE{2025A&A...697A.127H,
       author = {{Hansen}, Terese T. and {Roederer}, Ian U. and {Shah}, Shivani P. and {Ezzeddine}, Rana and {Beers}, Timothy C. and {Frebel}, Anna and {Holmbeck}, Erika M. and {Placco}, Vinicius M. and {Sakari}, Charli M. and {Ji}, Alexander and {Marshall}, Jennifer L. and {Mardini}, Mohammad K. and {Chiti}, Anirudh},
        title = "{The R-Process Alliance: Hunting for gold in the near-UV spectrum of 2MASS J05383296{\textendash}5904280}",
      journal = {\aap},
         year = 2025,
        month = may,
       volume = {697},
          eid = {A127},
        pages = {A127},
          doi = {10.1051/0004-6361/202554123},
archivePrefix = {arXiv},
       eprint = {2503.13426},
 primaryClass = {astro-ph.SR},
       adsurl = {https://ui.adsabs.harvard.edu/abs/2025A&A...697A.127H}
}

@ARTICLE{2025A&A...693A.294A,
       author = {{Alencastro Puls}, Arthur and {Kuske}, Jan and {Hansen}, Camilla Juul and {Lombardo}, Linda and {Visentin}, Giorgio and {Arcones}, Almudena and {Fernandes de Melo}, Raphaela and {Reichert}, Moritz and {Bonifacio}, Piercarlo and {Caffau}, Elisabetta and {Fritzsche}, Stephan},
        title = "{Chemical Evolution of R-process Elements in Stars (CERES): IV. An observational run-up of the third r-process peak with Hf, Os, Ir, and Pt}",
      journal = {\aap},
         year = 2025,
        month = jan,
       volume = {693},
          eid = {A294},
        pages = {A294},
          doi = {10.1051/0004-6361/202452537},
archivePrefix = {arXiv},
       eprint = {2412.00195},
 primaryClass = {astro-ph.SR},
       adsurl = {https://ui.adsabs.harvard.edu/abs/2025A&A...693A.294A}
}

@ARTICLE{2025A&A...693A.293L,
       author = {{Lombardo}, L. and {Hansen}, C.~J. and {Rizzuti}, F. and {Cescutti}, G. and {Mashonkina}, L.~I. and {Fran{\c{c}}ois}, P. and {Bonifacio}, P. and {Caffau}, E. and {Alencastro Puls}, A. and {Fernandes de Melo}, R. and {Gallagher}, A.~J. and {Sk{\'u}lad{\'o}ttir}, {\'A}. and {Koch-Hansen}, A.~J. and {Sbordone}, L.},
        title = "{Chemical Evolution of R-process Elements in Stars (CERES): III. Chemical abundances of neutron capture elements from Ba to Eu}",
      journal = {\aap},
         year = 2025,
        month = jan,
       volume = {693},
          eid = {A293},
        pages = {A293},
          doi = {10.1051/0004-6361/202452283},
archivePrefix = {arXiv},
       eprint = {2412.09141},
 primaryClass = {astro-ph.GA},
       adsurl = {https://ui.adsabs.harvard.edu/abs/2025A&A...693A.293L}
}

@ARTICLE{2022A&A...665A..10L,
       author = {{Lombardo}, Linda and {Bonifacio}, Piercarlo and {Fran{\c{c}}ois}, Patrick and {Hansen}, Camilla J. and {Caffau}, Elisabetta and {Hanke}, Michael and {Sk{\'u}lad{\'o}ttir}, {\'A}sa and {Arcones}, Almudena and {Eichler}, Marius and {Reichert}, Moritz and {Psaltis}, Athanasios and {Koch Hansen}, Andreas J. and {Sbordone}, Luca},
        title = "{Chemical Evolution of R-process Elements in Stars (CERES). I. Stellar parameters and chemical abundances from Na to Zr}",
      journal = {\aap},
         year = 2022,
        month = sep,
       volume = {665},
          eid = {A10},
        pages = {A10},
          doi = {10.1051/0004-6361/202243932},
archivePrefix = {arXiv},
       eprint = {2206.13836},
 primaryClass = {astro-ph.GA},
       adsurl = {https://ui.adsabs.harvard.edu/abs/2022A&A...665A..10L}
}

@ARTICLE{2024PhRvL.133x1201W,
       author = {{Wanajo}, Shinya and {Fujibayashi}, Sho and {Hayashi}, Kota and {Kiuchi}, Kenta and {Sekiguchi}, Yuichiro and {Shibata}, Masaru},
        title = "{Actinide-Boosting r Process in Black-Hole{\textendash}Neutron-Star Merger Ejecta}",
      journal = {\prl},
         year = 2024,
        month = dec,
       volume = {133},
       number = {24},
          eid = {241201},
        pages = {241201},
          doi = {10.1103/PhysRevLett.133.241201},
archivePrefix = {arXiv},
       eprint = {2212.04507},
 primaryClass = {astro-ph.HE},
       adsurl = {https://ui.adsabs.harvard.edu/abs/2024PhRvL.133x1201W}
}

@ARTICLE{2023MNRAS.518.1557R,
       author = {{Reichert}, M. and {Obergaulinger}, M. and {Aloy}, M. {\'A}. and {Gabler}, M. and {Arcones}, A. and {Thielemann}, F.~K.},
        title = "{Magnetorotational supernovae: a nucleosynthetic analysis of sophisticated 3D models}",
      journal = {\mnras},
         year = 2023,
        month = jan,
       volume = {518},
       number = {1},
        pages = {1557-1583},
          doi = {10.1093/mnras/stac3185},
archivePrefix = {arXiv},
       eprint = {2206.11914},
 primaryClass = {astro-ph.HE},
       adsurl = {https://ui.adsabs.harvard.edu/abs/2023MNRAS.518.1557R}
}

@ARTICLE{2007PhR...442..237Q,
       author = {{Qian}, Y. -Z. and {Wasserburg}, G.~J.},
        title = "{Where, oh where has the r-process gone?}",
      journal = {\physrep},
         year = 2007,
        month = apr,
       volume = {442},
       number = {1-6},
        pages = {237-268},
          doi = {10.1016/j.physrep.2007.02.006},
archivePrefix = {arXiv},
       eprint = {0708.1767},
 primaryClass = {astro-ph},
       adsurl = {https://ui.adsabs.harvard.edu/abs/2007PhR...442..237Q}
}

@ARTICLE{2021RvMP...93a5002C,
       author = {{Cowan}, John J. and {Sneden}, Christopher and {Lawler}, James E. and {Aprahamian}, Ani and {Wiescher}, Michael and {Langanke}, Karlheinz and {Mart{\'\i}nez-Pinedo}, Gabriel and {Thielemann}, Friedrich-Karl},
        title = "{Origin of the heaviest elements: The rapid neutron-capture process}",
      journal = {Reviews of Modern Physics},
         year = 2021,
        month = jan,
       volume = {93},
       number = {1},
          eid = {015002},
        pages = {015002},
          doi = {10.1103/RevModPhys.93.015002},
archivePrefix = {arXiv},
       eprint = {1901.01410},
 primaryClass = {astro-ph.HE},
       adsurl = {https://ui.adsabs.harvard.edu/abs/2021RvMP...93a5002C}
}

@ARTICLE{2000ApJ...533L.139S,
       author = {{Sneden}, Christopher and {Cowan}, John J. and {Ivans}, Inese I. and {Fuller}, George M. and {Burles}, Scott and {Beers}, Timothy C. and {Lawler}, James E.},
        title = "{Evidence of Multiple R-Process Sites in the Early Galaxy: New Observations of CS 22892-052}",
      journal = {\apjl},
         year = 2000,
        month = apr,
       volume = {533},
       number = {2},
        pages = {L139-L142},
          doi = {10.1086/312631},
archivePrefix = {arXiv},
       eprint = {astro-ph/0003086},
 primaryClass = {astro-ph},
       adsurl = {https://ui.adsabs.harvard.edu/abs/2000ApJ...533L.139S}
}

@ARTICLE{1996ApJ...466L.109W,
       author = {{Wasserburg}, G.~J. and {Busso}, M. and {Gallino}, R.},
        title = "{Abundances of Actinides and Short-lived Nonactinides in the Interstellar Medium: Diverse Supernova Sources for the r-Processes}",
      journal = {\apjl},
         year = 1996,
        month = aug,
       volume = {466},
        pages = {L109},
          doi = {10.1086/310177},
       adsurl = {https://ui.adsabs.harvard.edu/abs/1996ApJ...466L.109W}
}

@ARTICLE{2006ApJ...643.1180H,
       author = {{Honda}, S. and {Aoki}, W. and {Ishimaru}, Y. and {Wanajo}, S. and {Ryan}, S.~G.},
        title = "{Neutron-Capture Elements in the Very Metal Poor Star HD 122563}",
      journal = {\apj},
         year = 2006,
        month = jun,
       volume = {643},
       number = {2},
        pages = {1180-1189},
          doi = {10.1086/503195},
archivePrefix = {arXiv},
       eprint = {astro-ph/0602107},
 primaryClass = {astro-ph},
       adsurl = {https://ui.adsabs.harvard.edu/abs/2006ApJ...643.1180H}
}

@ARTICLE{2007ApJ...666.1189H,
       author = {{Honda}, Satoshi and {Aoki}, Wako and {Ishimaru}, Yuhri and {Wanajo}, Shinya},
        title = "{Neutron-Capture Elements in the Very Metal-poor Star HD 88609: Another Star with Excesses of Light Neutron-Capture Elements}",
      journal = {\apj},
         year = 2007,
        month = sep,
       volume = {666},
       number = {2},
        pages = {1189-1197},
          doi = {10.1086/520034},
archivePrefix = {arXiv},
       eprint = {0705.3975},
 primaryClass = {astro-ph},
       adsurl = {https://ui.adsabs.harvard.edu/abs/2007ApJ...666.1189H}
}

@ARTICLE{qian2001,
       author = {{Qian}, Y. -Z. and {Wasserburg}, G.~J.},
        title = "{A Model for Abundances in Metal-poor Stars}",
      journal = {\apj},
         year = 2001,
        month = oct,
       volume = {559},
       number = {2},
        pages = {925-941},
          doi = {10.1086/322367},
archivePrefix = {arXiv},
       eprint = {astro-ph/0106085},
 primaryClass = {astro-ph},
       adsurl = {https://ui.adsabs.harvard.edu/abs/2001ApJ...559..925Q}
}

@ARTICLE{qian2008,
       author = {{Qian}, Y. -Z. and {Wasserburg}, G.~J.},
        title = "{Abundances of Sr, Y, and Zr in Metal-Poor Stars and Implications for Chemical Evolution in the Early Galaxy}",
      journal = {\apj},
         year = 2008,
        month = nov,
       volume = {687},
       number = {1},
        pages = {272-286},
          doi = {10.1086/591545},
archivePrefix = {arXiv},
       eprint = {0807.0809},
 primaryClass = {astro-ph},
       adsurl = {https://ui.adsabs.harvard.edu/abs/2008ApJ...687..272Q}
}

@ARTICLE{2014ApJ...797..123H,
       author = {{Hansen}, C.~J. and {Montes}, F. and {Arcones}, A.},
        title = "{How Many Nucleosynthesis Processes Exist at Low Metallicity?}",
      journal = {\apj},
         year = 2014,
        month = dec,
       volume = {797},
       number = {2},
          eid = {123},
        pages = {123},
          doi = {10.1088/0004-637X/797/2/123},
archivePrefix = {arXiv},
       eprint = {1408.4135},
 primaryClass = {astro-ph.SR},
       adsurl = {https://ui.adsabs.harvard.edu/abs/2014ApJ...797..123H}
}

@ARTICLE{2022ApJ...936...84R,
       author = {{Roederer}, Ian U. and {Cowan}, John J. and {Pignatari}, Marco and {Beers}, Timothy C. and {Den Hartog}, Elizabeth A. and {Ezzeddine}, Rana and {Frebel}, Anna and {Hansen}, Terese T. and {Holmbeck}, Erika M. and {Mumpower}, Matthew R. and {Placco}, Vinicius M. and {Sakari}, Charli M. and {Surman}, Rebecca and {Vassh}, Nicole},
        title = "{The R-Process Alliance: Abundance Universality among Some Elements at and between the First and Second R-Process Peaks}",
      journal = {\apj},
         year = 2022,
        month = sep,
       volume = {936},
       number = {1},
          eid = {84},
        pages = {84},
          doi = {10.3847/1538-4357/ac85bc},
archivePrefix = {arXiv},
       eprint = {2210.15105},
 primaryClass = {astro-ph.SR},
       adsurl = {https://ui.adsabs.harvard.edu/abs/2022ApJ...936...84R}
}

@ARTICLE{2023Sci...382.1177R,
       author = {{Roederer}, Ian U. and {Vassh}, Nicole and {Holmbeck}, Erika M. and {Mumpower}, Matthew R. and {Surman}, Rebecca and {Cowan}, John J. and {Beers}, Timothy C. and {Ezzeddine}, Rana and {Frebel}, Anna and {Hansen}, Terese T. and {Placco}, Vinicius M. and {Sakari}, Charli M.},
        title = "{Element abundance patterns in stars indicate fission of nuclei heavier than uranium}",
      journal = {Science},
         year = 2023,
        month = dec,
       volume = {382},
       number = {6675},
        pages = {1177-1180},
          doi = {10.1126/science.adf1341},
archivePrefix = {arXiv},
       eprint = {2312.06844},
 primaryClass = {astro-ph.SR},
       adsurl = {https://ui.adsabs.harvard.edu/abs/2023Sci...382.1177R}
}

@ARTICLE{2022ApJ...935...27P,
       author = {{Psaltis}, A. and {Arcones}, A. and {Montes}, F. and {Mohr}, P. and {Hansen}, C.~J. and {Jacobi}, M. and {Schatz}, H.},
        title = "{Constraining Nucleosynthesis in Neutrino-driven Winds: Observations, Simulations, and Nuclear Physics}",
      journal = {\apj},
         year = 2022,
        month = aug,
       volume = {935},
       number = {1},
          eid = {27},
        pages = {27},
          doi = {10.3847/1538-4357/ac7da7},
archivePrefix = {arXiv},
       eprint = {2204.07136},
 primaryClass = {astro-ph.HE},
       adsurl = {https://ui.adsabs.harvard.edu/abs/2022ApJ...935...27P}
}

@ARTICLE{2024ApJ...966...11P,
       author = {{Psaltis}, A. and {Jacobi}, M. and {Montes}, F. and {Arcones}, A. and {Hansen}, C.~J. and {Schatz}, H.},
        title = "{Neutrino-driven Outflows and the Elemental Abundance Patterns of Very Metal-poor Stars}",
      journal = {\apj},
         year = 2024,
        month = may,
       volume = {966},
       number = {1},
          eid = {11},
        pages = {11},
          doi = {10.3847/1538-4357/ad2dfb},
archivePrefix = {arXiv},
       eprint = {2312.12306},
 primaryClass = {astro-ph.HE},
       adsurl = {https://ui.adsabs.harvard.edu/abs/2024ApJ...966...11P}
}

@ARTICLE{2022A&A...663A..70F,
       author = {{Farouqi}, K. and {Thielemann}, F. -K. and {Rosswog}, S. and {Kratz}, K. -L.},
        title = "{Correlations of r-process elements in very metal-poor stars as clues to their nucleosynthesis sites}",
      journal = {\aap},
         year = 2022,
        month = jul,
       volume = {663},
          eid = {A70},
        pages = {A70},
          doi = {10.1051/0004-6361/202141038},
archivePrefix = {arXiv},
       eprint = {2107.03486},
 primaryClass = {astro-ph.SR},
       adsurl = {https://ui.adsabs.harvard.edu/abs/2022A&A...663A..70F}
}

@ARTICLE{2008PASJ...60.1159S,
       author = {{Suda}, Takuma and {Katsuta}, Yutaka and {Yamada}, Shimako and {Suwa}, Tamon and {Ishizuka}, Chikako and {Komiya}, Yutaka and {Sorai}, Kazuo and {Aikawa}, Masayuki and {Fujimoto}, Masayuki Y.},
        title = "{Stellar Abundances for the Galactic Archeology (SAGA) Database --- Compilation of the Characteristics of Known Extremely Metal-Poor Stars}",
      journal = {\pasj},
         year = 2008,
        month = oct,
       volume = {60},
        pages = {1159},
          doi = {10.1093/pasj/60.5.1159},
archivePrefix = {arXiv},
       eprint = {0806.3697},
 primaryClass = {astro-ph},
       adsurl = {https://ui.adsabs.harvard.edu/abs/2008PASJ...60.1159S}
}

@ARTICLE{2018ApJS..238...36A,
       author = {{Abohalima}, Abdu and {Frebel}, Anna},
        title = "{JINAbase{\textemdash}A Database for Chemical Abundances of Metal-poor Stars}",
      journal = {\apjs},
         year = 2018,
        month = oct,
       volume = {238},
       number = {2},
          eid = {36},
        pages = {36},
          doi = {10.3847/1538-4365/aadfe9},
archivePrefix = {arXiv},
       eprint = {1711.04410},
 primaryClass = {astro-ph.SR},
       adsurl = {https://ui.adsabs.harvard.edu/abs/2018ApJS..238...36A}
}

@ARTICLE{2025EPJA...61..207F,
       author = {{Farouqi}, Khalil and {Frebel}, Anna and {Thielemann}, Friedrich-Karl},
        title = "{Deciphering the origins of the elements through galactic archeology}",
      journal = {European Physical Journal A},
         year = 2025,
        month = sep,
       volume = {61},
       number = {9},
          eid = {207},
        pages = {207},
          doi = {10.1140/epja/s10050-025-01668-5},
archivePrefix = {arXiv},
       eprint = {2503.18233},
 primaryClass = {astro-ph.SR},
       adsurl = {https://ui.adsabs.harvard.edu/abs/2025EPJA...61..207F}
}

@ARTICLE{2023arXiv230909385G,
       author = {{Gross}, Axel and {Xiong}, Zewei and {Qian}, Yong-Zhong},
        title = "{A Data-Driven Model for Abundances in Metal-poor Stars and Implications for Nucleosynthetic Sources}",
      journal = {arXiv e-prints},
         year = 2023,
        month = sep,
          eid = {arXiv:2309.09385},
        pages = {arXiv:2309.09385},
          doi = {10.48550/arXiv.2309.09385},
archivePrefix = {arXiv},
       eprint = {2309.09385},
 primaryClass = {astro-ph.SR},
       adsurl = {https://ui.adsabs.harvard.edu/abs/2023arXiv230909385G}
}

@ARTICLE{2025ApJ...990...37K,
       author = {{Kuske}, Jan and {Arcones}, Almudena and {Reichert}, Moritz},
        title = "{Complete Survey of r-process Conditions: The (Un)robustness of the r-process(es)}",
      journal = {\apj},
         year = 2025,
        month = sep,
       volume = {990},
       number = {1},
          eid = {37},
        pages = {37},
          doi = {10.3847/1538-4357/adf0f7},
archivePrefix = {arXiv},
       eprint = {2506.00092},
 primaryClass = {astro-ph.HE},
       adsurl = {https://ui.adsabs.harvard.edu/abs/2025ApJ...990...37K}
}

@ARTICLE{holmbeck2020,
       author = {{Holmbeck}, Erika M. and {Hansen}, Terese T. and {Beers}, Timothy C. and {Placco}, Vinicius M. and {Whitten}, Devin D. and {Rasmussen}, Kaitlin C. and {Roederer}, Ian U. and {Ezzeddine}, Rana and {Sakari}, Charli M. and {Frebel}, Anna and {Drout}, Maria R. and {Simon}, Joshua D. and {Thompson}, Ian B. and {Bland-Hawthorn}, Joss and {Gibson}, Brad K. and {Grebel}, Eva K. and {Kordopatis}, Georges and {Kunder}, Andrea and {Mel{\'e}ndez}, Jorge and {Navarro}, Julio F. and {Reid}, Warren A. and {Seabroke}, George and {Steinmetz}, Matthias and {Watson}, Fred and {Wyse}, Rosemary F. {\.{G}}.},
        title = "{The R-Process Alliance: Fourth Data Release from the Search for R-process-enhanced Stars in the Galactic Halo}",
      journal = {\apjs},
         year = 2020,
        month = aug,
       volume = {249},
       number = {2},
          eid = {30},
        pages = {30},
          doi = {10.3847/1538-4365/ab9c19},
archivePrefix = {arXiv},
       eprint = {2007.00749},
 primaryClass = {astro-ph.SR},
       adsurl = {https://ui.adsabs.harvard.edu/abs/2020ApJS..249...30H}
}

@ARTICLE{arcones2023,
       author = {{Arcones}, Almudena and {Thielemann}, Friedrich-Karl},
        title = "{Origin of the elements}",
      journal = {\aapr},
         year = 2023,
        month = dec,
       volume = {31},
       number = {1},
          eid = {1},
        pages = {1},
          doi = {10.1007/s00159-022-00146-x},
       adsurl = {https://ui.adsabs.harvard.edu/abs/2023A&ARv..31....1A}
}

@ARTICLE{fischer2020,
       author = {{Fischer}, Tobias and {Wu}, Meng-Ru and {Wehmeyer}, Benjamin and {Bastian}, Niels-Uwe F. and {Mart{\'\i}nez-Pinedo}, Gabriel and {Thielemann}, Friedrich-Karl},
        title = "{Core-collapse Supernova Explosions Driven by the Hadron-quark Phase Transition as a Rare r-process Site}",
      journal = {\apj},
         year = 2020,
        month = may,
       volume = {894},
       number = {1},
          eid = {9},
        pages = {9},
          doi = {10.3847/1538-4357/ab86b0},
archivePrefix = {arXiv},
       eprint = {2003.00972},
 primaryClass = {astro-ph.HE},
       adsurl = {https://ui.adsabs.harvard.edu/abs/2020ApJ...894....9F}
}

@ARTICLE{nishimura2015,
       author = {{Nishimura}, Nobuya and {Takiwaki}, Tomoya and {Thielemann}, Friedrich-Karl},
        title = "{The r-process Nucleosynthesis in the Various Jet-like Explosions of Magnetorotational Core-collapse Supernovae}",
      journal = {\apj},
         year = 2015,
        month = sep,
       volume = {810},
       number = {2},
          eid = {109},
        pages = {109},
          doi = {10.1088/0004-637X/810/2/109},
archivePrefix = {arXiv},
       eprint = {1501.06567},
 primaryClass = {astro-ph.SR},
       adsurl = {https://ui.adsabs.harvard.edu/abs/2015ApJ...810..109N}
}

@ARTICLE{siegel2019,
       author = {{Siegel}, Daniel M. and {Barnes}, Jennifer and {Metzger}, Brian D.},
        title = "{Collapsars as a major source of r-process elements}",
      journal = {\nat},
         year = 2019,
        month = may,
       volume = {569},
       number = {7755},
        pages = {241-244},
          doi = {10.1038/s41586-019-1136-0},
archivePrefix = {arXiv},
       eprint = {1810.00098},
 primaryClass = {astro-ph.HE},
       adsurl = {https://ui.adsabs.harvard.edu/abs/2019Natur.569..241S}
}

@ARTICLE{just2023,
       author = {{Just}, O. and {Vijayan}, V. and {Xiong}, Z. and {Goriely}, S. and {Soultanis}, T. and {Bauswein}, A. and {Guilet}, J. and {Janka}, H. -Th. and {Mart{\'\i}nez-Pinedo}, G.},
        title = "{End-to-end Kilonova Models of Neutron Star Mergers with Delayed Black Hole Formation}",
      journal = {\apjl},
         year = 2023,
        month = jul,
       volume = {951},
       number = {1},
          eid = {L12},
        pages = {L12},
          doi = {10.3847/2041-8213/acdad2},
archivePrefix = {arXiv},
       eprint = {2302.10928},
 primaryClass = {astro-ph.HE},
       adsurl = {https://ui.adsabs.harvard.edu/abs/2023ApJ...951L..12J}
}

@ARTICLE{goriely1999,
       author = {{Goriely}, S.},
        title = "{Uncertainties in the solar system r-abundance distribution}",
      journal = {\aap},
         year = 1999,
        month = feb,
       volume = {342},
        pages = {881-891},
       adsurl = {https://ui.adsabs.harvard.edu/abs/1999A&A...342..881G}
}

@ARTICLE{lippuner2015,
       author = {{Lippuner}, Jonas and {Roberts}, Luke F.},
        title = "{r-process Lanthanide Production and Heating Rates in Kilonovae}",
      journal = {\apj},
         year = 2015,
        month = dec,
       volume = {815},
       number = {2},
          eid = {82},
        pages = {82},
          doi = {10.1088/0004-637X/815/2/82},
archivePrefix = {arXiv},
       eprint = {1508.03133},
 primaryClass = {astro-ph.HE},
       adsurl = {https://ui.adsabs.harvard.edu/abs/2015ApJ...815...82L}
}

@ARTICLE{wanajo2018,
       author = {{Wanajo}, Shinya and {M{\"u}ller}, Bernhard and {Janka}, Hans-Thomas and {Heger}, Alexander},
        title = "{Nucleosynthesis in the Innermost Ejecta of Neutrino-driven Supernova Explosions in Two Dimensions}",
      journal = {\apj},
         year = 2018,
        month = jan,
       volume = {852},
       number = {1},
          eid = {40},
        pages = {40},
          doi = {10.3847/1538-4357/aa9d97},
archivePrefix = {arXiv},
       eprint = {1701.06786},
 primaryClass = {astro-ph.SR},
       adsurl = {https://ui.adsabs.harvard.edu/abs/2018ApJ...852...40W}
}

@ARTICLE{wang2023,
       author = {{Wang}, Tianshu and {Burrows}, Adam},
        title = "{Neutrino-driven Winds in Three-dimensional Core-collapse Supernova Simulations}",
      journal = {\apj},
         year = 2023,
        month = sep,
       volume = {954},
       number = {2},
          eid = {114},
        pages = {114},
          doi = {10.3847/1538-4357/ace7b2},
archivePrefix = {arXiv},
       eprint = {2306.13712},
 primaryClass = {astro-ph.SR},
       adsurl = {https://ui.adsabs.harvard.edu/abs/2023ApJ...954..114W}
}

@ARTICLE{kiuchi2023,
       author = {{Kiuchi}, Kenta and {Fujibayashi}, Sho and {Hayashi}, Kota and {Kyutoku}, Koutarou and {Sekiguchi}, Yuichiro and {Shibata}, Masaru},
        title = "{Self-Consistent Picture of the Mass Ejection from a One Second Long Binary Neutron Star Merger Leaving a Short-Lived Remnant in a General-Relativistic Neutrino-Radiation Magnetohydrodynamic Simulation}",
      journal = {\prl},
         year = 2023,
        month = jul,
       volume = {131},
       number = {1},
          eid = {011401},
        pages = {011401},
          doi = {10.1103/PhysRevLett.131.011401},
archivePrefix = {arXiv},
       eprint = {2211.07637},
 primaryClass = {astro-ph.HE},
       adsurl = {https://ui.adsabs.harvard.edu/abs/2023PhRvL.131a1401K}
}

@ARTICLE{curtis2023,
       author = {{Curtis}, Sanjana and {M{\"o}sta}, Philipp and {Wu}, Zhenyu and {Radice}, David and {Roberts}, Luke and {Ricigliano}, Giacomo and {Perego}, Albino},
        title = "{r-process nucleosynthesis and kilonovae from hypermassive neutron star post-merger remnants}",
      journal = {\mnras},
         year = 2023,
        month = feb,
       volume = {518},
       number = {4},
        pages = {5313-5322},
          doi = {10.1093/mnras/stac3128},
archivePrefix = {arXiv},
       eprint = {2112.00772},
 primaryClass = {astro-ph.HE},
       adsurl = {https://ui.adsabs.harvard.edu/abs/2023MNRAS.518.5313C}
}

@ARTICLE{hoffman1997,
       author = {{Hoffman}, R.~D. and {Woosley}, S.~E. and {Qian}, Y. -Z.},
        title = "{Nucleosynthesis in Neutrino-driven Winds. II. Implications for Heavy Element Synthesis}",
      journal = {\apj},
         year = 1997,
        month = jun,
       volume = {482},
       number = {2},
        pages = {951-962},
          doi = {10.1086/304181},
archivePrefix = {arXiv},
       eprint = {astro-ph/9611097},
 primaryClass = {astro-ph},
       adsurl = {https://ui.adsabs.harvard.edu/abs/1997ApJ...482..951H}
}

@ARTICLE{woosley1992,
       author = {{Woosley}, S.~E. and {Hoffman}, Robert D.},
        title = "{The alpha -Process and the r-Process}",
      journal = {\apj},
         year = 1992,
        month = aug,
       volume = {395},
        pages = {202},
          doi = {10.1086/171644},
       adsurl = {https://ui.adsabs.harvard.edu/abs/1992ApJ...395..202W}
}

@ARTICLE{thielemann2017,
       author = {{Thielemann}, F. -K. and {Eichler}, M. and {Panov}, I.~V. and {Wehmeyer}, B.},
        title = "{Neutron Star Mergers and Nucleosynthesis of Heavy Elements}",
      journal = {ARNPS},
         year = 2017,
        month = oct,
       volume = {67},
        pages = {253-274},
          doi = {10.1146/annurev-nucl-101916-123246},
archivePrefix = {arXiv},
       eprint = {1710.02142},
 primaryClass = {astro-ph.HE},
       adsurl = {https://ui.adsabs.harvard.edu/abs/2017ARNPS..67..253T}
}

@ARTICLE{kasen2017,
       author = {{Kasen}, Daniel and {Metzger}, Brian and {Barnes}, Jennifer and {Quataert}, Eliot and {Ramirez-Ruiz}, Enrico},
        title = "{Origin of the heavy elements in binary neutron-star mergers from a gravitational-wave event}",
      journal = {\nat},
         year = 2017,
        month = nov,
       volume = {551},
       number = {7678},
        pages = {80-84},
          doi = {10.1038/nature24453},
archivePrefix = {arXiv},
       eprint = {1710.05463},
 primaryClass = {astro-ph.HE},
       adsurl = {https://ui.adsabs.harvard.edu/abs/2017Natur.551...80K}
}

@ARTICLE{abbott2017,
       author = {{Abbott}, B.~P. and {Abbott}, R. and {Abbott}, T.~D. and {Acernese}, F. and {Ackley}, K. and {Adams}, C. and {Adams}, T. and {Addesso}, P. and {Adhikari}, R.~X. and {Adya}, V.~B. and {Affeldt}, C. and {Afrough}, M. and {Agarwal}, B. and {Agathos}, M. and {Agatsuma}, K. and {Aggarwal}, N. and {Aguiar}, O.~D. and {Aiello}, L. and {Ain}, A. and {Ajith}, P. and {Allen}, B. and {Allen}, G. and {Allocca}, A. and {Altin}, P.~A. and {Amato}, A. and {Ananyeva}, A. and {Anderson}, S.~B. and {Anderson}, W.~G. and {Angelova}, S.~V. and {Antier}, S. and {Appert}, S. and {Arai}, K. and {Araya}, M.~C. and {Areeda}, J.~S. and {Arnaud}, N. and {Arun}, K.~G. and {Ascenzi}, S. and {Ashton}, G. and {Ast}, M. and {Aston}, S.~M. and {Astone}, P. and {Atallah}, D.~V. and {Aufmuth}, P. and {Aulbert}, C. and {AultONeal}, K. and {Austin}, C. and {Avila-Alvarez}, A. and {Babak}, S. and {Bacon}, P. and {Bader}, M.~K.~M. and {Bae}, S. and {Baker}, P.~T. and {Baldaccini}, F. and {Ballardin}, G. and {Ballmer}, S.~W. and {Banagiri}, S. and {Barayoga}, J.~C. and {Barclay}, S.~E. and {Barish}, B.~C. and {Barker}, D. and {Barkett}, K. and {Barone}, F. and {Barr}, B. and {Barsotti}, L. and {Barsuglia}, M. and {Barta}, D. and {Barthelmy}, S.~D. and {Bartlett}, J. and {Bartos}, I. and {Bassiri}, R. and {Basti}, A. and {Batch}, J.~C. and {Bawaj}, M. and {Bayley}, J.~C. and {Bazzan}, M. and {B{\'e}csy}, B. and {Beer}, C. and {Bejger}, M. and {Belahcene}, I. and {Bell}, A.~S. and {Berger}, B.~K. and {Bergmann}, G. and {Bero}, J.~J. and {Berry}, C.~P.~L. and {Bersanetti}, D. and {Bertolini}, A. and {Betzwieser}, J. and {Bhagwat}, S. and {Bhandare}, R. and {Bilenko}, I.~A. and {Billingsley}, G. and {Billman}, C.~R. and {Birch}, J. and {Birney}, R. and {Birnholtz}, O. and {Biscans}, S. and {Biscoveanu}, S. and {Bisht}, A. and {Bitossi}, M. and {Biwer}, C. and {Bizouard}, M.~A. and {Blackburn}, J.~K. and {Blackman}, J. and {Blair}, C.~D. and {Blair}, D.~G. and {Blair}, R.~M. and {Bloemen}, S. and {Bock}, O. and {Bode}, N. and {Boer}, M. and {Bogaert}, G. and {Bohe}, A. and {Bondu}, F. and {Bonilla}, E. and {Bonnand}, R. and {Boom}, B.~A. and {Bork}, R. and {Boschi}, V. and {Bose}, S. and {Bossie}, K. and {Bouffanais}, Y. and {Bozzi}, A. and {Bradaschia}, C. and {Brady}, P.~R. and {Branchesi}, M. and {Brau}, J.~E. and {Briant}, T. and {Brillet}, A. and {Brinkmann}, M. and {Brisson}, V. and {Brockill}, P. and {Broida}, J.~E. and {Brooks}, A.~F. and {Brown}, D.~A. and {Brown}, D.~D. and {Brunett}, S. and {Buchanan}, C.~C. and {Buikema}, A. and {Bulik}, T. and {Bulten}, H.~J. and {Buonanno}, A. and {Buskulic}, D. and {Buy}, C. and {Byer}, R.~L. and {Cabero}, M. and {Cadonati}, L. and {Cagnoli}, G. and {Cahillane}, C. and {Calder{\'o}n Bustillo}, J. and {Callister}, T.~A. and {Calloni}, E. and {Camp}, J.~B. and {Canepa}, M. and {Canizares}, P. and {Cannon}, K.~C. and {Cao}, H. and {Cao}, J. and {Capano}, C.~D. and {Capocasa}, E. and {Carbognani}, F. and {Caride}, S. and {Carney}, M.~F. and {Casanueva Diaz}, J. and {Casentini}, C. and {Caudill}, S. and {Cavagli{\`a}}, M. and {Cavalier}, F. and {Cavalieri}, R. and {Cella}, G. and {Cepeda}, C.~B. and {Cerd{\'a}-Dur{\'a}n}, P. and {Cerretani}, G. and {Cesarini}, E. and {Chamberlin}, S.~J. and {Chan}, M. and {Chao}, S. and {Charlton}, P. and {Chase}, E. and {Chassande-Mottin}, E. and {Chatterjee}, D. and {Chatziioannou}, K. and {Cheeseboro}, B.~D. and {Chen}, H.~Y. and {Chen}, X. and {Chen}, Y. and {Cheng}, H. -P. and {Chia}, H. and {Chincarini}, A. and {Chiummo}, A. and {Chmiel}, T. and {Cho}, H.~S. and {Cho}, M. and {Chow}, J.~H. and {Christensen}, N. and {Chu}, Q. and {Chua}, A.~J.~K. and {Chua}, S. and {Chung}, A.~K.~W. and {Chung}, S. and {Ciani}, G. and {Ciolfi}, R. and {Cirelli}, C.~E. and {Cirone}, A. and {Clara}, F. and {Clark}, J.~A. and {Clearwater}, P. and {Cleva}, F. and {Cocchieri}, C. and {Coccia}, E. and {Cohadon}, P. -F. and {Cohen}, D. and {Colla}, A. and {Collette}, C.~G. and {Cominsky}, L.~R. and {Constancio}, M., Jr. and {Conti}, L. and {Cooper}, S.~J. and {Corban}, P. and {Corbitt}, T.~R. and {Cordero-Carri{\'o}n}, I. and {Corley}, K.~R. and {Cornish}, N. and {Corsi}, A. and {Cortese}, S. and {Costa}, C.~A. and {Coughlin}, M.~W. and {Coughlin}, S.~B. and {Coulon}, J. -P. and {Countryman}, S.~T. and {Couvares}, P. and {Covas}, P.~B. and {Cowan}, E.~E. and {Coward}, D.~M. and {Cowart}, M.~J. and {Coyne}, D.~C. and {Coyne}, R. and {Creighton}, J.~D.~E. and {Creighton}, T.~D. and {Cripe}, J. and {Crowder}, S.~G. and {Cullen}, T.~J. and {Cumming}, A. and {Cunningham}, L. and {Cuoco}, E. and {Dal Canton}, T. and {D{\'a}lya}, G. and {Danilishin}, S.~L. and {D'Antonio}, S. and {Danzmann}, K. and {Dasgupta}, A. and {Da Silva Costa}, C.~F. and {Dattilo}, V. and {Dave}, I. and {Davier}, M. and {Davis}, D. and {Daw}, E.~J. and {Day}, B. and {De}, S. and {DeBra}, D. and {Degallaix}, J. and {De Laurentis}, M. and {Del{\'e}glise}, S. and {Del Pozzo}, W. and {Demos}, N. and {Denker}, T. and {Dent}, T. and {De Pietri}, R. and {Dergachev}, V. and {De Rosa}, R. and {DeRosa}, R.~T. and {De Rossi}, C. and {DeSalvo}, R. and {de Varona}, O. and {Devenson}, J. and {Dhurandhar}, S. and {D{\'\i}az}, M.~C. and {Di Fiore}, L. and {Di Giovanni}, M. and {Di Girolamo}, T. and {Di Lieto}, A. and {Di Pace}, S. and {Di Palma}, I. and {Di Renzo}, F. and {Doctor}, Z. and {Dolique}, V. and {Donovan}, F. and {Dooley}, K.~L. and {Doravari}, S. and {Dorrington}, I. and {Douglas}, R. and {Dovale {\'A}lvarez}, M. and {Downes}, T.~P. and {Drago}, M. and {Dreissigacker}, C. and {Driggers}, J.~C. and {Du}, Z. and {Ducrot}, M. and {Dupej}, P. and {Dwyer}, S.~E. and {Edo}, T.~B. and {Edwards}, M.~C. and {Effler}, A. and {Ehrens}, P. and {Eichholz}, J. and {Eikenberry}, S.~S. and {Eisenstein}, R.~A. and {Essick}, R.~C. and {Estevez}, D. and {Etienne}, Z.~B. and {Etzel}, T. and {Evans}, M. and {Evans}, T.~M. and {Factourovich}, M. and {Fafone}, V. and {Fair}, H. and {Fairhurst}, S. and {Fan}, X. and {Farinon}, S. and {Farr}, B. and {Farr}, W.~M. and {Fauchon-Jones}, E.~J. and {Favata}, M. and {Fays}, M. and {Fee}, C. and {Fehrmann}, H. and {Feicht}, J. and {Fejer}, M.~M. and {Fernandez-Galiana}, A. and {Ferrante}, I. and {Ferreira}, E.~C. and {Ferrini}, F. and {Fidecaro}, F. and {Finstad}, D. and {Fiori}, I. and {Fiorucci}, D. and {Fishbach}, M. and {Fisher}, R.~P. and {Fitz-Axen}, M. and {Flaminio}, R. and {Fletcher}, M. and {Fong}, H. and {Font}, J.~A. and {Forsyth}, P.~W.~F. and {Forsyth}, S.~S. and {Fournier}, J. -D. and {Frasca}, S. and {Frasconi}, F. and {Frei}, Z. and {Freise}, A. and {Frey}, R. and {Frey}, V. and {Fries}, E.~M. and {Fritschel}, P. and {Frolov}, V.~V. and {Fulda}, P. and {Fyffe}, M. and {Gabbard}, H. and {Gadre}, B.~U. and {Gaebel}, S.~M. and {Gair}, J.~R. and {Gammaitoni}, L. and {Ganija}, M.~R. and {Gaonkar}, S.~G. and {Garcia-Quiros}, C. and {Garufi}, F. and {Gateley}, B. and {Gaudio}, S. and {Gaur}, G. and {Gayathri}, V. and {Gehrels}, N. and {Gemme}, G. and {Genin}, E. and {Gennai}, A. and {George}, D. and {George}, J. and {Gergely}, L. and {Germain}, V. and {Ghonge}, S. and {Ghosh}, Abhirup and {Ghosh}, Archisman and {Ghosh}, S. and {Giaime}, J.~A. and {Giardina}, K.~D. and {Giazotto}, A. and {Gill}, K. and {Glover}, L. and {Goetz}, E. and {Goetz}, R. and {Gomes}, S. and {Goncharov}, B. and {Gonz{\'a}lez}, G. and {Gonzalez Castro}, J.~M. and {Gopakumar}, A. and {Gorodetsky}, M.~L. and {Gossan}, S.~E. and {Gosselin}, M. and {Gouaty}, R. and {Grado}, A. and {Graef}, C. and {Granata}, M. and {Grant}, A. and {Gras}, S. and {Gray}, C. and {Greco}, G. and {Green}, A.~C. and {Gretarsson}, E.~M. and {Griswold}, B. and {Groot}, P. and {Grote}, H. and {Grunewald}, S. and {Gruning}, P. and {Guidi}, G.~M. and {Guo}, X. and {Gupta}, A. and {Gupta}, M.~K. and {Gushwa}, K.~E. and {Gustafson}, E.~K. and {Gustafson}, R. and {Halim}, O. and {Hall}, B.~R. and {Hall}, E.~D. and {Hamilton}, E.~Z. and {Hammond}, G. and {Haney}, M. and {Hanke}, M.~M. and {Hanks}, J. and {Hanna}, C. and {Hannam}, M.~D. and {Hannuksela}, O.~A. and {Hanson}, J. and {Hardwick}, T. and {Harms}, J. and {Harry}, G.~M. and {Harry}, I.~W. and {Hart}, M.~J. and {Haster}, C. -J. and {Haughian}, K. and {Healy}, J. and {Heidmann}, A. and {Heintze}, M.~C. and {Heitmann}, H. and {Hello}, P. and {Hemming}, G. and {Hendry}, M. and {Heng}, I.~S. and {Hennig}, J. and {Heptonstall}, A.~W. and {Heurs}, M. and {Hild}, S. and {Hinderer}, T. and {Hoak}, D. and {Hofman}, D. and {Holt}, K. and {Holz}, D.~E. and {Hopkins}, P. and {Horst}, C. and {Hough}, J. and {Houston}, E.~A. and {Howell}, E.~J. and {Hreibi}, A. and {Hu}, Y.~M. and {Huerta}, E.~A. and {Huet}, D. and {Hughey}, B. and {Husa}, S. and {Huttner}, S.~H. and {Huynh-Dinh}, T. and {Indik}, N. and {Inta}, R. and {Intini}, G. and {Isa}, H.~N. and {Isac}, J. -M. and {Isi}, M. and {Iyer}, B.~R. and {Izumi}, K. and {Jacqmin}, T. and {Jani}, K. and {Jaranowski}, P. and {Jawahar}, S. and {Jim{\'e}nez-Forteza}, F. and {Johnson}, W.~W. and {Jones}, D.~I. and {Jones}, R. and {Jonker}, R.~J.~G. and {Ju}, L. and {Junker}, J. and {Kalaghatgi}, C.~V. and {Kalogera}, V. and {Kamai}, B. and {Kandhasamy}, S. and {Kang}, G. and {Kanner}, J.~B. and {Kapadia}, S.~J. and {Karki}, S. and {Karvinen}, K.~S. and {Kasprzack}, M. and {Katolik}, M. and {Katsavounidis}, E. and {Katzman}, W. and {Kaufer}, S. and {Kawabe}, K. and {K{\'e}f{\'e}lian}, F. and {Keitel}, D. and {Kemball}, A.~J. and {Kennedy}, R. and {Kent}, C. and {Key}, J.~S. and {Khalili}, F.~Y. and {Khan}, I. and {Khan}, S. and {Khan}, Z. and {Khazanov}, E.~A. and {Kijbunchoo}, N. and {Kim}, Chunglee and {Kim}, J.~C. and {Kim}, K. and {Kim}, W. and {Kim}, W.~S. and {Kim}, Y. -M. and {Kimbrell}, S.~J. and {King}, E.~J. and {King}, P.~J. and {Kinley-Hanlon}, M. and {Kirchhoff}, R. and {Kissel}, J.~S. and {Kleybolte}, L. and {Klimenko}, S. and {Knowles}, T.~D. and {Koch}, P. and {Koehlenbeck}, S.~M. and {Koley}, S. and {Kondrashov}, V. and {Kontos}, A. and {Korobko}, M. and {Korth}, W.~Z. and {Kowalska}, I. and {Kozak}, D.~B. and {Kr{\"a}mer}, C. and {Kringel}, V. and {Krishnan}, B. and {Kr{\'o}lak}, A. and {Kuehn}, G. and {Kumar}, P. and {Kumar}, R. and {Kumar}, S. and {Kuo}, L. and {Kutynia}, A. and {Kwang}, S. and {Lackey}, B.~D. and {Lai}, K.~H. and {Landry}, M. and {Lang}, R.~N. and {Lange}, J. and {Lantz}, B. and {Lanza}, R.~K. and {Larson}, S.~L. and {Lartaux-Vollard}, A. and {Lasky}, P.~D. and {Laxen}, M. and {Lazzarini}, A. and {Lazzaro}, C. and {Leaci}, P. and {Leavey}, S. and {Lee}, C.~H. and {Lee}, H.~K. and {Lee}, H.~M. and {Lee}, H.~W. and {Lee}, K. and {Lehmann}, J. and {Lenon}, A. and {Leonardi}, M. and {Leroy}, N. and {Letendre}, N. and {Levin}, Y. and {Li}, T.~G.~F. and {Linker}, S.~D. and {Littenberg}, T.~B. and {Liu}, J. and {Lo}, R.~K.~L. and {Lockerbie}, N.~A. and {London}, L.~T. and {Lord}, J.~E. and {Lorenzini}, M. and {Loriette}, V. and {Lormand}, M. and {Losurdo}, G. and {Lough}, J.~D. and {Lousto}, C.~O. and {Lovelace}, G. and {L{\"u}ck}, H. and {Lumaca}, D. and {Lundgren}, A.~P. and {Lynch}, R. and {Ma}, Y. and {Macas}, R. and {Macfoy}, S. and {Machenschalk}, B. and {MacInnis}, M. and {Macleod}, D.~M. and {Maga{\~n}a Hernandez}, I. and {Maga{\~n}a-Sandoval}, F. and {Maga{\~n}a Zertuche}, L. and {Magee}, R.~M. and {Majorana}, E. and {Maksimovic}, I. and {Man}, N. and {Mandic}, V. and {Mangano}, V. and {Mansell}, G.~L. and {Manske}, M. and {Mantovani}, M. and {Marchesoni}, F. and {Marion}, F. and {M{\'a}rka}, S. and {M{\'a}rka}, Z. and {Markakis}, C. and {Markosyan}, A.~S. and {Markowitz}, A. and {Maros}, E. and {Marquina}, A. and {Marsh}, P. and {Martelli}, F. and {Martellini}, L. and {Martin}, I.~W. and {Martin}, R.~M. and {Martynov}, D.~V. and {Mason}, K. and {Massera}, E. and {Masserot}, A. and {Massinger}, T.~J. and {Masso-Reid}, M. and {Mastrogiovanni}, S. and {Matas}, A. and {Matichard}, F. and {Matone}, L. and {Mavalvala}, N. and {Mazumder}, N. and {McCarthy}, R. and {McClelland}, D.~E. and {McCormick}, S. and {McCuller}, L. and {McGuire}, S.~C. and {McIntyre}, G. and {McIver}, J. and {McManus}, D.~J. and {McNeill}, L. and {McRae}, T. and {McWilliams}, S.~T. and {Meacher}, D. and {Meadors}, G.~D. and {Mehmet}, M. and {Meidam}, J. and {Mejuto-Villa}, E. and {Melatos}, A. and {Mendell}, G. and {Mercer}, R.~A. and {Merilh}, E.~L. and {Merzougui}, M. and {Meshkov}, S. and {Messenger}, C. and {Messick}, C. and {Metzdorff}, R. and {Meyers}, P.~M. and {Miao}, H. and {Michel}, C. and {Middleton}, H. and {Mikhailov}, E.~E. and {Milano}, L. and {Miller}, A.~L. and {Miller}, B.~B. and {Miller}, J. and {Millhouse}, M. and {Milovich-Goff}, M.~C. and {Minazzoli}, O. and {Minenkov}, Y. and {Ming}, J. and {Mishra}, C. and {Mitra}, S. and {Mitrofanov}, V.~P. and {Mitselmakher}, G. and {Mittleman}, R. and {Moffa}, D. and {Moggi}, A. and {Mogushi}, K. and {Mohan}, M. and {Mohapatra}, S.~R.~P. and {Montani}, M. and {Moore}, C.~J. and {Moraru}, D. and {Moreno}, G. and {Morriss}, S.~R. and {Mours}, B. and {Mow-Lowry}, C.~M. and {Mueller}, G. and {Muir}, A.~W. and {Mukherjee}, Arunava and {Mukherjee}, D. and {Mukherjee}, S. and {Mukund}, N. and {Mullavey}, A. and {Munch}, J. and {Mu{\~n}iz}, E.~A. and {Muratore}, M. and {Murray}, P.~G. and {Napier}, K. and {Nardecchia}, I. and {Naticchioni}, L. and {Nayak}, R.~K. and {Neilson}, J. and {Nelemans}, G. and {Nelson}, T.~J.~N. and {Nery}, M. and {Neunzert}, A. and {Nevin}, L. and {Newport}, J.~M. and {Newton}, G. and {Ng}, K.~K.~Y. and {Nguyen}, P. and {Nguyen}, T.~T. and {Nichols}, D. and {Nielsen}, A.~B. and {Nissanke}, S. and {Nitz}, A. and {Noack}, A. and {Nocera}, F. and {Nolting}, D. and {North}, C. and {Nuttall}, L.~K. and {Oberling}, J. and {O'Dea}, G.~D. and {Ogin}, G.~H. and {Oh}, J.~J. and {Oh}, S.~H. and {Ohme}, F. and {Okada}, M.~A. and {Oliver}, M. and {Oppermann}, P. and {Oram}, Richard J. and {O'Reilly}, B. and {Ormiston}, R. and {Ortega}, L.~F. and {O'Shaughnessy}, R. and {Ossokine}, S. and {Ottaway}, D.~J. and {Overmier}, H. and {Owen}, B.~J. and {Pace}, A.~E. and {Page}, J. and {Page}, M.~A. and {Pai}, A. and {Pai}, S.~A. and {Palamos}, J.~R. and {Palashov}, O. and {Palomba}, C. and {Pal-Singh}, A. and {Pan}, Howard and {Pan}, Huang-Wei and {Pang}, B. and {Pang}, P.~T.~H. and {Pankow}, C. and {Pannarale}, F. and {Pant}, B.~C. and {Paoletti}, F. and {Paoli}, A. and {Papa}, M.~A. and {Parida}, A. and {Parker}, W. and {Pascucci}, D. and {Pasqualetti}, A. and {Passaquieti}, R. and {Passuello}, D. and {Patil}, M. and {Patricelli}, B. and {Pearlstone}, B.~L. and {Pedraza}, M. and {Pedurand}, R. and {Pekowsky}, L. and {Pele}, A. and {Penn}, S. and {Perez}, C.~J. and {Perreca}, A. and {Perri}, L.~M. and {Pfeiffer}, H.~P. and {Phelps}, M. and {Piccinni}, O.~J. and {Pichot}, M. and {Piergiovanni}, F. and {Pierro}, V. and {Pillant}, G. and {Pinard}, L. and {Pinto}, I.~M. and {Pirello}, M. and {Pitkin}, M. and {Poe}, M. and {Poggiani}, R. and {Popolizio}, P. and {Porter}, E.~K. and {Post}, A. and {Powell}, J. and {Prasad}, J. and {Pratt}, J.~W.~W. and {Pratten}, G. and {Predoi}, V. and {Prestegard}, T. and {Price}, L.~R. and {Prijatelj}, M. and {Principe}, M. and {Privitera}, S. and {Prodi}, G.~A. and {Prokhorov}, L.~G. and {Puncken}, O. and {Punturo}, M. and {Puppo}, P. and {P{\"u}rrer}, M. and {Qi}, H. and {Quetschke}, V. and {Quintero}, E.~A. and {Quitzow-James}, R. and {Raab}, F.~J. and {Rabeling}, D.~S. and {Radkins}, H. and {Raffai}, P. and {Raja}, S. and {Rajan}, C. and {Rajbhandari}, B. and {Rakhmanov}, M. and {Ramirez}, K.~E. and {Ramos-Buades}, A. and {Rapagnani}, P. and {Raymond}, V. and {Razzano}, M. and {Read}, J. and {Regimbau}, T. and {Rei}, L. and {Reid}, S. and {Reitze}, D.~H. and {Ren}, W. and {Reyes}, S.~D. and {Ricci}, F. and {Ricker}, P.~M. and {Rieger}, S. and {Riles}, K. and {Rizzo}, M. and {Robertson}, N.~A. and {Robie}, R. and {Robinet}, F. and {Rocchi}, A. and {Rolland}, L. and {Rollins}, J.~G. and {Roma}, V.~J. and {Romano}, R. and {Romel}, C.~L. and {Romie}, J.~H. and {Rosi{\'n}ska}, D. and {Ross}, M.~P. and {Rowan}, S. and {R{\"u}diger}, A. and {Ruggi}, P. and {Rutins}, G. and {Ryan}, K. and {Sachdev}, S. and {Sadecki}, T. and {Sadeghian}, L. and {Sakellariadou}, M. and {Salconi}, L. and {Saleem}, M. and {Salemi}, F. and {Samajdar}, A. and {Sammut}, L. and {Sampson}, L.~M. and {Sanchez}, E.~J. and {Sanchez}, L.~E. and {Sanchis-Gual}, N. and {Sandberg}, V. and {Sanders}, J.~R. and {Sassolas}, B. and {Sathyaprakash}, B.~S. and {Saulson}, P.~R. and {Sauter}, O. and {Savage}, R.~L. and {Sawadsky}, A. and {Schale}, P. and {Scheel}, M. and {Scheuer}, J. and {Schmidt}, J. and {Schmidt}, P. and {Schnabel}, R. and {Schofield}, R.~M.~S. and {Sch{\"o}nbeck}, A. and {Schreiber}, E. and {Schuette}, D. and {Schulte}, B.~W. and {Schutz}, B.~F. and {Schwalbe}, S.~G. and {Scott}, J. and {Scott}, S.~M. and {Seidel}, E. and {Sellers}, D. and {Sengupta}, A.~S. and {Sentenac}, D. and {Sequino}, V. and {Sergeev}, A. and {Shaddock}, D.~A. and {Shaffer}, T.~J. and {Shah}, A.~A. and {Shahriar}, M.~S. and {Shaner}, M.~B. and {Shao}, L. and {Shapiro}, B. and {Shawhan}, P. and {Sheperd}, A. and {Shoemaker}, D.~H. and {Shoemaker}, D.~M. and {Siellez}, K. and {Siemens}, X. and {Sieniawska}, M. and {Sigg}, D. and {Silva}, A.~D. and {Singer}, L.~P. and {Singh}, A. and {Singhal}, A. and {Sintes}, A.~M. and {Slagmolen}, B.~J.~J. and {Smith}, B. and {Smith}, J.~R. and {Smith}, R.~J.~E. and {Somala}, S. and {Son}, E.~J. and {Sonnenberg}, J.~A. and {Sorazu}, B. and {Sorrentino}, F. and {Souradeep}, T. and {Spencer}, A.~P. and {Srivastava}, A.~K. and {Staats}, K. and {Staley}, A. and {Steinke}, M. and {Steinlechner}, J. and {Steinlechner}, S. and {Steinmeyer}, D. and {Stevenson}, S.~P. and {Stone}, R. and {Stops}, D.~J. and {Strain}, K.~A. and {Stratta}, G. and {Strigin}, S.~E. and {Strunk}, A. and {Sturani}, R. and {Stuver}, A.~L. and {Summerscales}, T.~Z. and {Sun}, L. and {Sunil}, S. and {Suresh}, J. and {Sutton}, P.~J. and {Swinkels}, B.~L. and {Szczepa{\'n}czyk}, M.~J. and {Tacca}, M. and {Tait}, S.~C. and {Talbot}, C. and {Talukder}, D. and {Tanner}, D.~B. and {T{\'a}pai}, M. and {Taracchini}, A. and {Tasson}, J.~D. and {Taylor}, J.~A. and {Taylor}, R. and {Tewari}, S.~V. and {Theeg}, T. and {Thies}, F. and {Thomas}, E.~G. and {Thomas}, M. and {Thomas}, P. and {Thorne}, K.~A. and {Thorne}, K.~S. and {Thrane}, E. and {Tiwari}, S. and {Tiwari}, V. and {Tokmakov}, K.~V. and {Toland}, K. and {Tonelli}, M. and {Tornasi}, Z. and {Torres-Forn{\'e}}, A. and {Torrie}, C.~I. and {T{\"o}yr{\"a}}, D. and {Travasso}, F. and {Traylor}, G. and {Trinastic}, J. and {Tringali}, M.~C. and {Trozzo}, L. and {Tsang}, K.~W. and {Tse}, M. and {Tso}, R. and {Tsukada}, L. and {Tsuna}, D. and {Tuyenbayev}, D. and {Ueno}, K. and {Ugolini}, D. and {Unnikrishnan}, C.~S. and {Urban}, A.~L. and {Usman}, S.~A. and {Vahlbruch}, H. and {Vajente}, G. and {Valdes}, G. and {van Bakel}, N. and {van Beuzekom}, M. and {van den Brand}, J.~F.~J. and {Van Den Broeck}, C. and {Vander-Hyde}, D.~C. and {van der Schaaf}, L. and {van Heijningen}, J.~V. and {van Veggel}, A.~A. and {Vardaro}, M. and {Varma}, V. and {Vass}, S. and {Vas{\'u}th}, M. and {Vecchio}, A. and {Vedovato}, G. and {Veitch}, J. and {Veitch}, P.~J. and {Venkateswara}, K. and {Venugopalan}, G. and {Verkindt}, D. and {Vetrano}, F. and {Vicer{\'e}}, A. and {Viets}, A.~D. and {Vinciguerra}, S. and {Vine}, D.~J. and {Vinet}, J. -Y. and {Vitale}, S. and {Vo}, T. and {Vocca}, H. and {Vorvick}, C. and {Vyatchanin}, S.~P. and {Wade}, A.~R. and {Wade}, L.~E. and {Wade}, M. and {Walet}, R. and {Walker}, M. and {Wallace}, L. and {Walsh}, S. and {Wang}, G. and {Wang}, H. and {Wang}, J.~Z. and {Wang}, W.~H. and {Wang}, Y.~F. and {Ward}, R.~L. and {Warner}, J. and {Was}, M. and {Watchi}, J. and {Weaver}, B. and {Wei}, L. -W. and {Weinert}, M. and {Weinstein}, A.~J. and {Weiss}, R. and {Wen}, L. and {Wessel}, E.~K. and {Wessels}, P. and {Westerweck}, J. and {Westphal}, T. and {Wette}, K. and {Whelan}, J.~T. and {Whitcomb}, S.~E. and {Whiting}, B.~F. and {Whittle}, C. and {Wilken}, D. and {Williams}, D. and {Williams}, R.~D. and {Williamson}, A.~R. and {Willis}, J.~L. and {Willke}, B. and {Wimmer}, M.~H. and {Winkler}, W. and {Wipf}, C.~C. and {Wittel}, H. and {Woan}, G. and {Woehler}, J. and {Wofford}, J. and {Wong}, K.~W.~K. and {Worden}, J. and {Wright}, J.~L. and {Wu}, D.~S. and {Wysocki}, D.~M. and {Xiao}, S. and {Yamamoto}, H. and {Yancey}, C.~C. and {Yang}, L. and {Yap}, M.~J. and {Yazback}, M. and {Yu}, Hang and {Yu}, Haocun and {Yvert}, M. and {Zadro{\.z}ny}, A. and {Zanolin}, M. and {Zelenova}, T. and {Zendri}, J. -P. and {Zevin}, M. and {Zhang}, L. and {Zhang}, M. and {Zhang}, T. and {Zhang}, Y. -H. and {Zhao}, C. and {Zhou}, M. and {Zhou}, Z. and {Zhu}, S.~J. and {Zhu}, X.~J. and {Zimmerman}, A.~B. and {Zucker}, M.~E. and {Zweizig}, J. and {LIGO Scientific Collaboration} and {Virgo Collaboration} and {Wilson-Hodge}, C.~A. and {Bissaldi}, E. and {Blackburn}, L. and {Briggs}, M.~S. and {Burns}, E. and {Cleveland}, W.~H. and {Connaughton}, V. and {Gibby}, M.~H. and {Giles}, M.~M. and {Goldstein}, A. and {Hamburg}, R. and {Jenke}, P. and {Hui}, C.~M. and {Kippen}, R.~M. and {Kocevski}, D. and {McBreen}, S. and {Meegan}, C.~A. and {Paciesas}, W.~S. and {Poolakkil}, S. and {Preece}, R.~D. and {Racusin}, J. and {Roberts}, O.~J. and {Stanbro}, M. and {Veres}, P. and {von Kienlin}, A. and {GBM}, Fermi and {Savchenko}, V. and {Ferrigno}, C. and {Kuulkers}, E. and {Bazzano}, A. and {Bozzo}, E. and {Brandt}, S. and {Chenevez}, J. and {Courvoisier}, T.~J. -L. and {Diehl}, R. and {Domingo}, A. and {Hanlon}, L. and {Jourdain}, E. and {Laurent}, P. and {Lebrun}, F. and {Lutovinov}, A. and {Martin-Carrillo}, A. and {Mereghetti}, S. and {Natalucci}, L. and {Rodi}, J. and {Roques}, J. -P. and {Sunyaev}, R. and {Ubertini}, P. and {INTEGRAL} and {Aartsen}, M.~G. and {Ackermann}, M. and {Adams}, J. and {Aguilar}, J.~A. and {Ahlers}, M. and {Ahrens}, M. and {Samarai}, I. Al and {Altmann}, D. and {Andeen}, K. and {Anderson}, T. and {Ansseau}, I. and {Anton}, G. and {Arg{\"u}elles}, C. and {Auffenberg}, J. and {Axani}, S. and {Bagherpour}, H. and {Bai}, X. and {Barron}, J.~P. and {Barwick}, S.~W. and {Baum}, V. and {Bay}, R. and {Beatty}, J.~J. and {Becker Tjus}, J. and {Bernardini}, E. and {Besson}, D.~Z. and {Binder}, G. and {Bindig}, D. and {Blaufuss}, E. and {Blot}, S. and {Bohm}, C. and {B{\"o}rner}, M. and {Bos}, F. and {Bose}, D. and {B{\"o}ser}, S. and {Botner}, O. and {Bourbeau}, E. and {Bourbeau}, J. and {Bradascio}, F. and {Braun}, J. and {Brayeur}, L. and {Brenzke}, M. and {Bretz}, H. -P. and {Bron}, S. and {Brostean-Kaiser}, J. and {Burgman}, A. and {Carver}, T. and {Casey}, J. and {Casier}, M. and {Cheung}, E. and {Chirkin}, D. and {Christov}, A. and {Clark}, K. and {Classen}, L. and {Coenders}, S. and {Collin}, G.~H. and {Conrad}, J.~M. and {Cowen}, D.~F. and {Cross}, R. and {Day}, M. and {de Andr{\'e}}, J.~P.~A.~M. and {De Clercq}, C. and {DeLaunay}, J.~J. and {Dembinski}, H. and {De Ridder}, S. and {Desiati}, P. and {de Vries}, K.~D. and {de Wasseige}, G. and {de With}, M. and {DeYoung}, T. and {D{\'\i}az-V{\'e}lez}, J.~C. and {di Lorenzo}, V. and {Dujmovic}, H. and {Dumm}, J.~P. and {Dunkman}, M. and {Dvorak}, E. and {Eberhardt}, B. and {Ehrhardt}, T. and {Eichmann}, B. and {Eller}, P. and {Evenson}, P.~A. and {Fahey}, S. and {Fazely}, A.~R. and {Felde}, J. and {Filimonov}, K. and {Finley}, C. and {Flis}, S. and {Franckowiak}, A. and {Friedman}, E. and {Fuchs}, T. and {Gaisser}, T.~K. and {Gallagher}, J. and {Gerhardt}, L. and {Ghorbani}, K. and {Giang}, W. and {Glauch}, T. and {Gl{\"u}senkamp}, T. and {Goldschmidt}, A. and {Gonzalez}, J.~G. and {Grant}, D. and {Griffith}, Z. and {Haack}, C. and {Hallgren}, A. and {Halzen}, F. and {Hanson}, K. and {Hebecker}, D. and {Heereman}, D. and {Helbing}, K. and {Hellauer}, R. and {Hickford}, S. and {Hignight}, J. and {Hill}, G.~C. and {Hoffman}, K.~D. and {Hoffmann}, R. and {Hokanson-Fasig}, B. and {Hoshina}, K. and {Huang}, F. and {Huber}, M. and {Hultqvist}, K. and {H{\"u}nnefeld}, M. and {In}, S. and {Ishihara}, A. and {Jacobi}, E. and {Japaridze}, G.~S. and {Jeong}, M. and {Jero}, K. and {Jones}, B.~J.~P. and {Kalaczynski}, P. and {Kang}, W. and {Kappes}, A. and {Karg}, T. and {Karle}, A. and {Kauer}, M. and {Keivani}, A. and {Kelley}, J.~L. and {Kheirandish}, A. and {Kim}, J. and {Kim}, M. and {Kintscher}, T. and {Kiryluk}, J. and {Kittler}, T. and {Klein}, S.~R. and {Kohnen}, G. and {Koirala}, R. and {Kolanoski}, H. and {K{\"o}pke}, L. and {Kopper}, C. and {Kopper}, S. and {Koschinsky}, J.~P. and {Koskinen}, D.~J. and {Kowalski}, M. and {Krings}, K. and {Kroll}, M. and {Kr{\"u}ckl}, G. and {Kunnen}, J. and {Kunwar}, S. and {Kurahashi}, N. and {Kuwabara}, T. and {Kyriacou}, A. and {Labare}, M. and {Lanfranchi}, J.~L. and {Larson}, M.~J. and {Lauber}, F. and {Lesiak-Bzdak}, M. and {Leuermann}, M. and {Liu}, Q.~R. and {Lu}, L. and {L{\"u}nemann}, J. and {Luszczak}, W. and {Madsen}, J. and {Maggi}, G. and {Mahn}, K.~B.~M. and {Mancina}, S. and {Maruyama}, R. and {Mase}, K. and {Maunu}, R. and {McNally}, F. and {Meagher}, K. and {Medici}, M. and {Meier}, M. and {Menne}, T. and {Merino}, G. and {Meures}, T. and {Miarecki}, S. and {Micallef}, J. and {Moment{\'e}}, G. and {Montaruli}, T. and {Moore}, R.~W. and {Moulai}, M. and {Nahnhauer}, R. and {Nakarmi}, P. and {Naumann}, U. and {Neer}, G. and {Niederhausen}, H. and {Nowicki}, S.~C. and {Nygren}, D.~R. and {Obertacke Pollmann}, A. and {Olivas}, A. and {O'Murchadha}, A. and {Palczewski}, T. and {Pandya}, H. and {Pankova}, D.~V. and {Peiffer}, P. and {Pepper}, J.~A. and {P{\'e}rez de los Heros}, C. and {Pieloth}, D. and {Pinat}, E. and {Price}, P.~B. and {Przybylski}, G.~T. and {Raab}, C. and {R{\"a}del}, L. and {Rameez}, M. and {Rawlins}, K. and {Rea}, I.~C. and {Reimann}, R. and {Relethford}, B. and {Relich}, M. and {Resconi}, E. and {Rhode}, W. and {Richman}, M. and {Robertson}, S. and {Rongen}, M. and {Rott}, C. and {Ruhe}, T. and {Ryckbosch}, D. and {Rysewyk}, D. and {S{\"a}lzer}, T. and {Sanchez Herrera}, S.~E. and {Sandrock}, A. and {Sandroos}, J. and {Santander}, M. and {Sarkar}, S. and {Sarkar}, S. and {Satalecka}, K. and {Schlunder}, P. and {Schmidt}, T. and {Schneider}, A. and {Schoenen}, S. and {Sch{\"o}neberg}, S. and {Schumacher}, L. and {Seckel}, D. and {Seunarine}, S. and {Soedingrekso}, J. and {Soldin}, D. and {Song}, M. and {Spiczak}, G.~M. and {Spiering}, C. and {Stachurska}, J. and {Stamatikos}, M. and {Stanev}, T. and {Stasik}, A. and {Stettner}, J. and {Steuer}, A. and {Stezelberger}, T. and {Stokstad}, R.~G. and {St{\"o}ssl}, A. and {Strotjohann}, N.~L. and {Stuttard}, T. and {Sullivan}, G.~W. and {Sutherland}, M. and {Taboada}, I. and {Tatar}, J. and {Tenholt}, F. and {Ter-Antonyan}, S. and {Terliuk}, A. and {Te{\v{s}}i{\'c}}, G. and {Tilav}, S. and {Toale}, P.~A. and {Tobin}, M.~N. and {Toscano}, S. and {Tosi}, D. and {Tselengidou}, M. and {Tung}, C.~F. and {Turcati}, A. and {Turley}, C.~F. and {Ty}, B. and {Unger}, E. and {Usner}, M. and {Vandenbroucke}, J. and {Van Driessche}, W. and {van Eijndhoven}, N. and {Vanheule}, S. and {van Santen}, J. and {Vehring}, M. and {Vogel}, E. and {Vraeghe}, M. and {Walck}, C. and {Wallace}, A. and {Wallraff}, M. and {Wandler}, F.~D. and {Wandkowsky}, N. and {Waza}, A. and {Weaver}, C. and {Weiss}, M.~J. and {Wendt}, C. and {Werthebach}, J. and {Whelan}, B.~J. and {Wiebe}, K. and {Wiebusch}, C.~H. and {Wille}, L. and {Williams}, D.~R. and {Wills}, L. and {Wolf}, M. and {Wood}, T.~R. and {Woolsey}, E. and {Woschnagg}, K. and {Xu}, D.~L. and {Xu}, X.~W. and {Xu}, Y. and {Yanez}, J.~P. and {Yodh}, G. and {Yoshida}, S. and {Yuan}, T. and {Zoll}, M. and {IceCube Collaboration} and {Balasubramanian}, A. and {Mate}, S. and {Bhalerao}, V. and {Bhattacharya}, D. and {Vibhute}, A. and {Dewangan}, G.~C. and {Rao}, A.~R. and {Vadawale}, S.~V. and {AstroSat Cadmium Zinc Telluride Imager Team} and {Svinkin}, D.~S. and {Hurley}, K. and {Aptekar}, R.~L. and {Frederiks}, D.~D. and {Golenetskii}, S.~V. and {Kozlova}, A.~V. and {Lysenko}, A.~L. and {Oleynik}, Ph. P. and {Tsvetkova}, A.~E. and {Ulanov}, M.~V. and {Cline}, T. and {IPN Collaboration} and {Li}, T.~P. and {Xiong}, S.~L. and {Zhang}, S.~N. and {Lu}, F.~J. and {Song}, L.~M. and {Cao}, X.~L. and {Chang}, Z. and {Chen}, G. and {Chen}, L. and {Chen}, T.~X. and {Chen}, Y. and {Chen}, Y.~B. and {Chen}, Y.~P. and {Cui}, W. and {Cui}, W.~W. and {Deng}, J.~K. and {Dong}, Y.~W. and {Du}, Y.~Y. and {Fu}, M.~X. and {Gao}, G.~H. and {Gao}, H. and {Gao}, M. and {Ge}, M.~Y. and {Gu}, Y.~D. and {Guan}, J. and {Guo}, C.~C. and {Han}, D.~W. and {Hu}, W. and {Huang}, Y. and {Huo}, J. and {Jia}, S.~M. and {Jiang}, L.~H. and {Jiang}, W.~C. and {Jin}, J. and {Jin}, Y.~J. and {Li}, B. and {Li}, C.~K. and {Li}, G. and {Li}, M.~S. and {Li}, W. and {Li}, X. and {Li}, X.~B. and {Li}, X.~F. and {Li}, Y.~G. and {Li}, Z.~J. and {Li}, Z.~W. and {Liang}, X.~H. and {Liao}, J.~Y. and {Liu}, C.~Z. and {Liu}, G.~Q. and {Liu}, H.~W. and {Liu}, S.~Z. and {Liu}, X.~J. and {Liu}, Y. and {Liu}, Y.~N. and {Lu}, B. and {Lu}, X.~F. and {Luo}, T. and {Ma}, X. and {Meng}, B. and {Nang}, Y. and {Nie}, J.~Y. and {Ou}, G. and {Qu}, J.~L. and {Sai}, N. and {Sun}, L. and {Tan}, Y. and {Tao}, L. and {Tao}, W.~H. and {Tuo}, Y.~L. and {Wang}, G.~F. and {Wang}, H.~Y. and {Wang}, J. and {Wang}, W.~S. and {Wang}, Y.~S. and {Wen}, X.~Y. and {Wu}, B.~B. and {Wu}, M. and {Xiao}, G.~C. and {Xu}, H. and {Xu}, Y.~P. and {Yan}, L.~L. and {Yang}, J.~W. and {Yang}, S. and {Yang}, Y.~J. and {Zhang}, A.~M. and {Zhang}, C.~L. and {Zhang}, C.~M. and {Zhang}, F. and {Zhang}, H.~M. and {Zhang}, J. and {Zhang}, Q. and {Zhang}, S. and {Zhang}, T. and {Zhang}, W. and {Zhang}, W.~C. and {Zhang}, W.~Z. and {Zhang}, Y. and {Zhang}, Y. and {Zhang}, Y.~F. and {Zhang}, Y.~J. and {Zhang}, Z. and {Zhang}, Z.~L. and {Zhao}, H.~S. and {Zhao}, J.~L. and {Zhao}, X.~F. and {Zheng}, S.~J. and {Zhu}, Y. and {Zhu}, Y.~X. and {Zou}, C.~L. and {Insight-HXMT Collaboration} and {Albert}, A. and {Andr{\'e}}, M. and {Anghinolfi}, M. and {Ardid}, M. and {Aubert}, J. -J. and {Aublin}, J. and {Avgitas}, T. and {Baret}, B. and {Barrios-Mart{\'\i}}, J. and {Basa}, S. and {Belhorma}, B. and {Bertin}, V. and {Biagi}, S. and {Bormuth}, R. and {Bourret}, S. and {Bouwhuis}, M.~C. and {Br{\^a}nza{\c{s}}}, H. and {Bruijn}, R. and {Brunner}, J. and {Busto}, J. and {Capone}, A. and {Caramete}, L. and {Carr}, J. and {Celli}, S. and {Cherkaoui El Moursli}, R. and {Chiarusi}, T. and {Circella}, M. and {Coelho}, J.~A.~B. and {Coleiro}, A. and {Coniglione}, R. and {Costantini}, H. and {Coyle}, P. and {Creusot}, A. and {D{\'\i}az}, A.~F. and {Deschamps}, A. and {De Bonis}, G. and {Distefano}, C. and {Di Palma}, I. and {Domi}, A. and {Donzaud}, C. and {Dornic}, D. and {Drouhin}, D. and {Eberl}, T. and {El Bojaddaini}, I. and {El Khayati}, N. and {Els{\"a}sser}, D. and {Enzenh{\"o}fer}, A. and {Ettahiri}, A. and {Fassi}, F. and {Felis}, I. and {Fusco}, L.~A. and {Gay}, P. and {Giordano}, V. and {Glotin}, H. and {Gr{\'e}goire}, T. and {Ruiz}, R. Gracia and {Graf}, K. and {Hallmann}, S. and {van Haren}, H. and {Heijboer}, A.~J. and {Hello}, Y. and {Hern{\'a}ndez-Rey}, J.~J. and {H{\"o}ssl}, J. and {Hofest{\"a}dt}, J. and {Hugon}, C. and {Illuminati}, G. and {James}, C.~W. and {de Jong}, M. and {Jongen}, M. and {Kadler}, M. and {Kalekin}, O. and {Katz}, U. and {Kiessling}, D. and {Kouchner}, A. and {Kreter}, M. and {Kreykenbohm}, I. and {Kulikovskiy}, V. and {Lachaud}, C. and {Lahmann}, R. and {Lef{\`e}vre}, D. and {Leonora}, E. and {Lotze}, M. and {Loucatos}, S. and {Marcelin}, M. and {Margiotta}, A. and {Marinelli}, A. and {Mart{\'\i}nez-Mora}, J.~A. and {Mele}, R. and {Melis}, K. and {Michael}, T. and {Migliozzi}, P. and {Moussa}, A. and {Navas}, S. and {Nezri}, E. and {Organokov}, M. and {P{\u{a}}v{\u{a}}la{\c{s}}}, G.~E. and {Pellegrino}, C. and {Perrina}, C. and {Piattelli}, P. and {Popa}, V. and {Pradier}, T. and {Quinn}, L. and {Racca}, C. and {Riccobene}, G. and {S{\'a}nchez-Losa}, A. and {Salda{\~n}a}, M. and {Salvadori}, I. and {Samtleben}, D.~F.~E. and {Sanguineti}, M. and {Sapienza}, P. and {Sieger}, C. and {Spurio}, M. and {Stolarczyk}, Th. and {Taiuti}, M. and {Tayalati}, Y. and {Trovato}, A. and {Turpin}, D. and {T{\"o}nnis}, C. and {Vallage}, B. and {Van Elewyck}, V. and {Versari}, F. and {Vivolo}, D. and {Vizzoca}, A. and {Wilms}, J. and {Zornoza}, J.~D. and {Z{\'u}{\~n}iga}, J. and {ANTARES Collaboration} and {Beardmore}, A.~P. and {Breeveld}, A.~A. and {Burrows}, D.~N. and {Cenko}, S.~B. and {Cusumano}, G. and {D'A{\`\i}}, A. and {de Pasquale}, M. and {Emery}, S.~W.~K. and {Evans}, P.~A. and {Giommi}, P. and {Gronwall}, C. and {Kennea}, J.~A. and {Krimm}, H.~A. and {Kuin}, N.~P.~M. and {Lien}, A. and {Marshall}, F.~E. and {Melandri}, A. and {Nousek}, J.~A. and {Oates}, S.~R. and {Osborne}, J.~P. and {Pagani}, C. and {Page}, K.~L. and {Palmer}, D.~M. and {Perri}, M. and {Siegel}, M.~H. and {Sbarufatti}, B. and {Tagliaferri}, G. and {Tohuvavohu}, A. and {Swift Collaboration} and {Tavani}, M. and {Verrecchia}, F. and {Bulgarelli}, A. and {Evangelista}, Y. and {Pacciani}, L. and {Feroci}, M. and {Pittori}, C. and {Giuliani}, A. and {Del Monte}, E. and {Donnarumma}, I. and {Argan}, A. and {Trois}, A. and {Ursi}, A. and {Cardillo}, M. and {Piano}, G. and {Longo}, F. and {Lucarelli}, F. and {Munar-Adrover}, P. and {Fuschino}, F. and {Labanti}, C. and {Marisaldi}, M. and {Minervini}, G. and {Fioretti}, V. and {Parmiggiani}, N. and {Gianotti}, F. and {Trifoglio}, M. and {Di Persio}, G. and {Antonelli}, L.~A. and {Barbiellini}, G. and {Caraveo}, P. and {Cattaneo}, P.~W. and {Costa}, E. and {Colafrancesco}, S. and {D'Amico}, F. and {Ferrari}, A. and {Morselli}, A. and {Paoletti}, F. and {Picozza}, P. and {Pilia}, M. and {Rappoldi}, A. and {Soffitta}, P. and {Vercellone}, S. and {AGILE Team} and {Foley}, R.~J. and {Coulter}, D.~A. and {Kilpatrick}, C.~D. and {Drout}, M.~R. and {Piro}, A.~L. and {Shappee}, B.~J. and {Siebert}, M.~R. and {Simon}, J.~D. and {Ulloa}, N. and {Kasen}, D. and {Madore}, B.~F. and {Murguia-Berthier}, A. and {Pan}, Y. -C. and {Prochaska}, J.~X. and {Ramirez-Ruiz}, E. and {Rest}, A. and {Rojas-Bravo}, C. and {1M2H Team} and {Berger}, E. and {Soares-Santos}, M. and {Annis}, J. and {Alexander}, K.~D. and {Allam}, S. and {Balbinot}, E. and {Blanchard}, P. and {Brout}, D. and {Butler}, R.~E. and {Chornock}, R. and {Cook}, E.~R. and {Cowperthwaite}, P. and {Diehl}, H.~T. and {Drlica-Wagner}, A. and {Drout}, M.~R. and {Durret}, F. and {Eftekhari}, T. and {Finley}, D.~A. and {Fong}, W. and {Frieman}, J.~A. and {Fryer}, C.~L. and {Garc{\'\i}a-Bellido}, J. and {Gruendl}, R.~A. and {Hartley}, W. and {Herner}, K. and {Kessler}, R. and {Lin}, H. and {Lopes}, P.~A.~A. and {Louren{\c{c}}o}, A.~C.~C. and {Margutti}, R. and {Marshall}, J.~L. and {Matheson}, T. and {Medina}, G.~E. and {Metzger}, B.~D. and {Mu{\~n}oz}, R.~R. and {Muir}, J. and {Nicholl}, M. and {Nugent}, P. and {Palmese}, A. and {Paz-Chinch{\'o}n}, F. and {Quataert}, E. and {Sako}, M. and {Sauseda}, M. and {Schlegel}, D.~J. and {Scolnic}, D. and {Secco}, L.~F. and {Smith}, N. and {Sobreira}, F. and {Villar}, V.~A. and {Vivas}, A.~K. and {Wester}, W. and {Williams}, P.~K.~G. and {Yanny}, B. and {Zenteno}, A. and {Zhang}, Y. and {Abbott}, T.~M.~C. and {Banerji}, M. and {Bechtol}, K. and {Benoit-L{\'e}vy}, A. and {Bertin}, E. and {Brooks}, D. and {Buckley-Geer}, E. and {Burke}, D.~L. and {Capozzi}, D. and {Carnero Rosell}, A. and {Carrasco Kind}, M. and {Castander}, F.~J. and {Crocce}, M. and {Cunha}, C.~E. and {D'Andrea}, C.~B. and {da Costa}, L.~N. and {Davis}, C. and {DePoy}, D.~L. and {Desai}, S. and {Dietrich}, J.~P. and {Eifler}, T.~F. and {Fernandez}, E. and {Flaugher}, B. and {Fosalba}, P. and {Gaztanaga}, E. and {Gerdes}, D.~W. and {Giannantonio}, T. and {Goldstein}, D.~A. and {Gruen}, D. and {Gschwend}, J. and {Gutierrez}, G. and {Honscheid}, K. and {James}, D.~J. and {Jeltema}, T. and {Johnson}, M.~W.~G. and {Johnson}, M.~D. and {Kent}, S. and {Krause}, E. and {Kron}, R. and {Kuehn}, K. and {Lahav}, O. and {Lima}, M. and {Maia}, M.~A.~G. and {March}, M. and {Martini}, P. and {McMahon}, R.~G. and {Menanteau}, F. and {Miller}, C.~J. and {Miquel}, R. and {Mohr}, J.~J. and {Nichol}, R.~C. and {Ogando}, R.~L.~C. and {Plazas}, A.~A. and {Romer}, A.~K. and {Roodman}, A. and {Rykoff}, E.~S. and {Sanchez}, E. and {Scarpine}, V. and {Schindler}, R. and {Schubnell}, M. and {Sevilla-Noarbe}, I. and {Sheldon}, E. and {Smith}, M. and {Smith}, R.~C. and {Stebbins}, A. and {Suchyta}, E. and {Swanson}, M.~E.~C. and {Tarle}, G. and {Thomas}, R.~C. and {Troxel}, M.~A. and {Tucker}, D.~L. and {Vikram}, V. and {Walker}, A.~R. and {Wechsler}, R.~H. and {Weller}, J. and {Carlin}, J.~L. and {Gill}, M.~S.~S. and {Li}, T.~S. and {Marriner}, J. and {Neilsen}, E. and {Dark Energy Camera GW-EM Collaboration} and {DES Collaboration} and {Haislip}, J.~B. and {Kouprianov}, V.~V. and {Reichart}, D.~E. and {Sand}, D.~J. and {Tartaglia}, L. and {Valenti}, S. and {Yang}, S. and {DLT40 Collaboration} and {Benetti}, S. and {Brocato}, E. and {Campana}, S. and {Cappellaro}, E. and {Covino}, S. and {D'Avanzo}, P. and {D'Elia}, V. and {Getman}, F. and {Ghirlanda}, G. and {Ghisellini}, G. and {Limatola}, L. and {Nicastro}, L. and {Palazzi}, E. and {Pian}, E. and {Piranomonte}, S. and {Possenti}, A. and {Rossi}, A. and {Salafia}, O.~S. and {Tomasella}, L. and {Amati}, L. and {Antonelli}, L.~A. and {Bernardini}, M.~G. and {Bufano}, F. and {Capaccioli}, M. and {Casella}, P. and {Dadina}, M. and {De Cesare}, G. and {Di Paola}, A. and {Giuffrida}, G. and {Giunta}, A. and {Israel}, G.~L. and {Lisi}, M. and {Maiorano}, E. and {Mapelli}, M. and {Masetti}, N. and {Pescalli}, A. and {Pulone}, L. and {Salvaterra}, R. and {Schipani}, P. and {Spera}, M. and {Stamerra}, A. and {Stella}, L. and {Testa}, V. and {Turatto}, M. and {Vergani}, D. and {Aresu}, G. and {Bachetti}, M. and {Buffa}, F. and {Burgay}, M. and {Buttu}, M. and {Caria}, T. and {Carretti}, E. and {Casasola}, V. and {Castangia}, P. and {Carboni}, G. and {Casu}, S. and {Concu}, R. and {Corongiu}, A. and {Deiana}, G.~L. and {Egron}, E. and {Fara}, A. and {Gaudiomonte}, F. and {Gusai}, V. and {Ladu}, A. and {Loru}, S. and {Leurini}, S. and {Marongiu}, L. and {Melis}, A. and {Melis}, G. and {Migoni}, Carlo and {Milia}, Sabrina and {Navarrini}, Alessandro and {Orlati}, A. and {Ortu}, P. and {Palmas}, S. and {Pellizzoni}, A. and {Perrodin}, D. and {Pisanu}, T. and {Poppi}, S. and {Righini}, S. and {Saba}, A. and {Serra}, G. and {Serrau}, M. and {Stagni}, M. and {Surcis}, G. and {Vacca}, V. and {Vargiu}, G.~P. and {Hunt}, L.~K. and {Jin}, Z.~P. and {Klose}, S. and {Kouveliotou}, C. and {Mazzali}, P.~A. and {M{\o}ller}, P. and {Nava}, L. and {Piran}, T. and {Selsing}, J. and {Vergani}, S.~D. and {Wiersema}, K. and {Toma}, K. and {Higgins}, A.~B. and {Mundell}, C.~G. and {di Serego Alighieri}, S. and {G{\'o}tz}, D. and {Gao}, W. and {Gomboc}, A. and {Kaper}, L. and {Kobayashi}, S. and {Kopac}, D. and {Mao}, J. and {Starling}, R.~L.~C. and {Steele}, I. and {van der Horst}, A.~J. and {GRAWITA: GRAvitational Wave Inaf TeAm} and {Acero}, F. and {Atwood}, W.~B. and {Baldini}, L. and {Barbiellini}, G. and {Bastieri}, D. and {Berenji}, B. and {Bellazzini}, R. and {Bissaldi}, E. and {Blandford}, R.~D. and {Bloom}, E.~D. and {Bonino}, R. and {Bottacini}, E. and {Bregeon}, J. and {Buehler}, R. and {Buson}, S. and {Cameron}, R.~A. and {Caputo}, R. and {Caraveo}, P.~A. and {Cavazzuti}, E. and {Chekhtman}, A. and {Cheung}, C.~C. and {Chiang}, J. and {Ciprini}, S. and {Cohen-Tanugi}, J. and {Cominsky}, L.~R. and {Costantin}, D. and {Cuoco}, A. and {D'Ammando}, F. and {de Palma}, F. and {Digel}, S.~W. and {Di Lalla}, N. and {Di Mauro}, M. and {Di Venere}, L. and {Dubois}, R. and {Fegan}, S.~J. and {Focke}, W.~B. and {Franckowiak}, A. and {Fukazawa}, Y. and {Funk}, S. and {Fusco}, P. and {Gargano}, F. and {Gasparrini}, D. and {Giglietto}, N. and {Giordano}, F. and {Giroletti}, M. and {Glanzman}, T. and {Green}, D. and {Grondin}, M. -H. and {Guillemot}, L. and {Guiriec}, S. and {Harding}, A.~K. and {Horan}, D. and {J{\'o}hannesson}, G. and {Kamae}, T. and {Kensei}, S. and {Kuss}, M. and {La Mura}, G. and {Latronico}, L. and {Lemoine-Goumard}, M. and {Longo}, F. and {Loparco}, F. and {Lovellette}, M.~N. and {Lubrano}, P. and {Magill}, J.~D. and {Maldera}, S. and {Manfreda}, A. and {Mazziotta}, M.~N. and {McEnery}, J.~E. and {Meyer}, M. and {Michelson}, P.~F. and {Mirabal}, N. and {Monzani}, M.~E. and {Moretti}, E. and {Morselli}, A. and {Moskalenko}, I.~V. and {Negro}, M. and {Nuss}, E. and {Ojha}, R. and {Omodei}, N. and {Orienti}, M. and {Orlando}, E. and {Palatiello}, M. and {Paliya}, V.~S. and {Paneque}, D. and {Pesce-Rollins}, M. and {Piron}, F. and {Porter}, T.~A. and {Principe}, G. and {Rain{\`o}}, S. and {Rando}, R. and {Razzano}, M. and {Razzaque}, S. and {Reimer}, A. and {Reimer}, O. and {Reposeur}, T. and {Rochester}, L.~S. and {Saz Parkinson}, P.~M. and {Sgr{\`o}}, C. and {Siskind}, E.~J. and {Spada}, F. and {Spandre}, G. and {Suson}, D.~J. and {Takahashi}, M. and {Tanaka}, Y. and {Thayer}, J.~G. and {Thayer}, J.~B. and {Thompson}, D.~J. and {Tibaldo}, L. and {Torres}, D.~F. and {Torresi}, E. and {Troja}, E. and {Venters}, T.~M. and {Vianello}, G. and {Zaharijas}, G. and {Fermi Large Area Telescope Collaboration} and {Allison}, J.~R. and {Bannister}, K.~W. and {Dobie}, D. and {Kaplan}, D.~L. and {Lenc}, E. and {Lynch}, C. and {Murphy}, T. and {Sadler}, E.~M. and {Australia Telescope Compact Array}, ATCA: and {Hotan}, A. and {James}, C.~W. and {Oslowski}, S. and {Raja}, W. and {Shannon}, R.~M. and {Whiting}, M. and {Australian SKA Pathfinder}, ASKAP: and {Arcavi}, I. and {Howell}, D.~A. and {McCully}, C. and {Hosseinzadeh}, G. and {Hiramatsu}, D. and {Poznanski}, D. and {Barnes}, J. and {Zaltzman}, M. and {Vasylyev}, S. and {Maoz}, D. and {Las Cumbres Observatory Group} and {Cooke}, J. and {Bailes}, M. and {Wolf}, C. and {Deller}, A.~T. and {Lidman}, C. and {Wang}, L. and {Gendre}, B. and {Andreoni}, I. and {Ackley}, K. and {Pritchard}, T.~A. and {Bessell}, M.~S. and {Chang}, S. -W. and {M{\"o}ller}, A. and {Onken}, C.~A. and {Scalzo}, R.~A. and {Ridden-Harper}, R. and {Sharp}, R.~G. and {Tucker}, B.~E. and {Farrell}, T.~J. and {Elmer}, E. and {Johnston}, S. and {Venkatraman Krishnan}, V. and {Keane}, E.~F. and {Green}, J.~A. and {Jameson}, A. and {Hu}, L. and {Ma}, B. and {Sun}, T. and {Wu}, X. and {Wang}, X. and {Shang}, Z. and {Hu}, Y. and {Ashley}, M.~C.~B. and {Yuan}, X. and {Li}, X. and {Tao}, C. and {Zhu}, Z. and {Zhang}, H. and {Suntzeff}, N.~B. and {Zhou}, J. and {Yang}, J. and {Orange}, B. and {Morris}, D. and {Cucchiara}, A. and {Giblin}, T. and {Klotz}, A. and {Staff}, J. and {Thierry}, P. and {Schmidt}, B.~P. and {OzGrav} and {(Deeper}, DWF and {Wider} and {program}, Faster and {AST3} and {CAASTRO Collaborations} and {Tanvir}, N.~R. and {Levan}, A.~J. and {Cano}, Z. and {de Ugarte-Postigo}, A. and {Gonz{\'a}lez-Fern{\'a}ndez}, C. and {Greiner}, J. and {Hjorth}, J. and {Irwin}, M. and {Kr{\"u}hler}, T. and {Mandel}, I. and {Milvang-Jensen}, B. and {O'Brien}, P. and {Rol}, E. and {Rosetti}, S. and {Rosswog}, S. and {Rowlinson}, A. and {Steeghs}, D.~T.~H. and {Th{\"o}ne}, C.~C. and {Ulaczyk}, K. and {Watson}, D. and {Bruun}, S.~H. and {Cutter}, R. and {Figuera Jaimes}, R. and {Fujii}, Y.~I. and {Fruchter}, A.~S. and {Gompertz}, B. and {Jakobsson}, P. and {Hodosan}, G. and {J{\`e}rgensen}, U.~G. and {Kangas}, T. and {Kann}, D.~A. and {Rabus}, M. and {Schr{\o}der}, S.~L. and {Stanway}, E.~R. and {Wijers}, R.~A.~M.~J. and {VINROUGE Collaboration} and {Lipunov}, V.~M. and {Gorbovskoy}, E.~S. and {Kornilov}, V.~G. and {Tyurina}, N.~V. and {Balanutsa}, P.~V. and {Kuznetsov}, A.~S. and {Vlasenko}, D.~M. and {Podesta}, R.~C. and {Lopez}, C. and {Podesta}, F. and {Levato}, H.~O. and {Saffe}, C. and {Mallamaci}, C.~C. and {Budnev}, N.~M. and {Gress}, O.~A. and {Kuvshinov}, D.~A. and {Gorbunov}, I.~A. and {Vladimirov}, V.~V. and {Zimnukhov}, D.~S. and {Gabovich}, A.~V. and {Yurkov}, V.~V. and {Sergienko}, Yu. P. and {Rebolo}, R. and {Serra-Ricart}, M. and {Tlatov}, A.~G. and {Ishmuhametova}, Yu. V. and {MASTER Collaboration} and {Abe}, F. and {Aoki}, K. and {Aoki}, W. and {Asakura}, Y. and {Baar}, S. and {Barway}, S. and {Bond}, I.~A. and {Doi}, M. and {Finet}, F. and {Fujiyoshi}, T. and {Furusawa}, H. and {Honda}, S. and {Itoh}, R. and {Kanda}, N. and {Kawabata}, K.~S. and {Kawabata}, M. and {Kim}, J.~H. and {Koshida}, S. and {Kuroda}, D. and {Lee}, C. -H. and {Liu}, W. and {Matsubayashi}, K. and {Miyazaki}, S. and {Morihana}, K. and {Morokuma}, T. and {Motohara}, K. and {Murata}, K.~L. and {Nagai}, H. and {Nagashima}, H. and {Nagayama}, T. and {Nakaoka}, T. and {Nakata}, F. and {Ohsawa}, R. and {Ohshima}, T. and {Ohta}, K. and {Okita}, H. and {Saito}, T. and {Saito}, Y. and {Sako}, S. and {Sekiguchi}, Y. and {Sumi}, T. and {Tajitsu}, A. and {Takahashi}, J. and {Takayama}, M. and {Tamura}, Y. and {Tanaka}, I. and {Tanaka}, M. and {Terai}, T. and {Tominaga}, N. and {Tristram}, P.~J. and {Uemura}, M. and {Utsumi}, Y. and {Yamaguchi}, M.~S. and {Yasuda}, N. and {Yoshida}, M. and {Zenko}, T. and {J-GEM} and {Adams}, S.~M. and {Anupama}, G.~C. and {Bally}, J. and {Barway}, S. and {Bellm}, E. and {Blagorodnova}, N. and {Cannella}, C. and {Chandra}, P. and {Chatterjee}, D. and {Clarke}, T.~E. and {Cobb}, B.~E. and {Cook}, D.~O. and {Copperwheat}, C. and {De}, K. and {Emery}, S.~W.~K. and {Feindt}, U. and {Foster}, K. and {Fox}, O.~D. and {Frail}, D.~A. and {Fremling}, C. and {Frohmaier}, C. and {Garcia}, J.~A. and {Ghosh}, S. and {Giacintucci}, S. and {Goobar}, A. and {Gottlieb}, O. and {Grefenstette}, B.~W. and {Hallinan}, G. and {Harrison}, F. and {Heida}, M. and {Helou}, G. and {Ho}, A.~Y.~Q. and {Horesh}, A. and {Hotokezaka}, K. and {Ip}, W. -H. and {Itoh}, R. and {Jacobs}, Bob and {Jencson}, J.~E. and {Kasen}, D. and {Kasliwal}, M.~M. and {Kassim}, N.~E. and {Kim}, H. and {Kiran}, B.~S. and {Kuin}, N.~P.~M. and {Kulkarni}, S.~R. and {Kupfer}, T. and {Lau}, R.~M. and {Madsen}, K. and {Mazzali}, P.~A. and {Miller}, A.~A. and {Miyasaka}, H. and {Mooley}, K. and {Myers}, S.~T. and {Nakar}, E. and {Ngeow}, C. -C. and {Nugent}, P. and {Ofek}, E.~O. and {Palliyaguru}, N. and {Pavana}, M. and {Perley}, D.~A. and {Peters}, W.~M. and {Pike}, S. and {Piran}, T. and {Qi}, H. and {Quimby}, R.~M. and {Rana}, J. and {Rosswog}, S. and {Rusu}, F. and {Sadler}, E.~M. and {Van Sistine}, A. and {Sollerman}, J. and {Xu}, Y. and {Yan}, L. and {Yatsu}, Y. and {Yu}, P. -C. and {Zhang}, C. and {Zhao}, W. and {GROWTH} and {JAGWAR} and {Caltech-NRAO} and {TTU-NRAO} and {NuSTAR Collaborations} and {Chambers}, K.~C. and {Huber}, M.~E. and {Schultz}, A.~S.~B. and {Bulger}, J. and {Flewelling}, H. and {Magnier}, E.~A. and {Lowe}, T.~B. and {Wainscoat}, R.~J. and {Waters}, C. and {Willman}, M. and {Pan-STARRS} and {Ebisawa}, K. and {Hanyu}, C. and {Harita}, S. and {Hashimoto}, T. and {Hidaka}, K. and {Hori}, T. and {Ishikawa}, M. and {Isobe}, N. and {Iwakiri}, W. and {Kawai}, H. and {Kawai}, N. and {Kawamuro}, T. and {Kawase}, T. and {Kitaoka}, Y. and {Makishima}, K. and {Matsuoka}, M. and {Mihara}, T. and {Morita}, T. and {Morita}, K. and {Nakahira}, S. and {Nakajima}, M. and {Nakamura}, Y. and {Negoro}, H. and {Oda}, S. and {Sakamaki}, A. and {Sasaki}, R. and {Serino}, M. and {Shidatsu}, M. and {Shimomukai}, R. and {Sugawara}, Y. and {Sugita}, S. and {Sugizaki}, M. and {Tachibana}, Y. and {Takao}, Y. and {Tanimoto}, A. and {Tomida}, H. and {Tsuboi}, Y. and {Tsunemi}, H. and {Ueda}, Y. and {Ueno}, S. and {Yamada}, S. and {Yamaoka}, K. and {Yamauchi}, M. and {Yatabe}, F. and {Yoneyama}, T. and {Yoshii}, T. and {MAXI Team} and {Coward}, D.~M. and {Crisp}, H. and {Macpherson}, D. and {Andreoni}, I. and {Laugier}, R. and {Noysena}, K. and {Klotz}, A. and {Gendre}, B. and {Thierry}, P. and {Turpin}, D. and {Consortium}, TZAC and {Im}, M. and {Choi}, C. and {Kim}, J. and {Yoon}, Y. and {Lim}, G. and {Lee}, S. -K. and {Lee}, C. -U. and {Kim}, S. -L. and {Ko}, S. -W. and {Joe}, J. and {Kwon}, M. -K. and {Kim}, P. -J. and {Lim}, S. -K. and {Choi}, J. -S. and {KU Collaboration} and {Fynbo}, J.~P.~U. and {Malesani}, D. and {Xu}, D. and {Optical Telescope}, Nordic and {Smartt}, S.~J. and {Jerkstrand}, A. and {Kankare}, E. and {Sim}, S.~A. and {Fraser}, M. and {Inserra}, C. and {Maguire}, K. and {Leloudas}, G. and {Magee}, M. and {Shingles}, L.~J. and {Smith}, K.~W. and {Young}, D.~R. and {Kotak}, R. and {Gal-Yam}, A. and {Lyman}, J.~D. and {Homan}, D.~S. and {Agliozzo}, C. and {Anderson}, J.~P. and {Angus}, C.~R. and {Ashall}, C. and {Barbarino}, C. and {Bauer}, F.~E. and {Berton}, M. and {Botticella}, M.~T. and {Bulla}, M. and {Cannizzaro}, G. and {Cartier}, R. and {Cikota}, A. and {Clark}, P. and {De Cia}, A. and {Della Valle}, M. and {Dennefeld}, M. and {Dessart}, L. and {Dimitriadis}, G. and {Elias-Rosa}, N. and {Firth}, R.~E. and {Fl{\"o}rs}, A. and {Frohmaier}, C. and {Galbany}, L. and {Gonz{\'a}lez-Gait{\'a}n}, S. and {Gromadzki}, M. and {Guti{\'e}rrez}, C.~P. and {Hamanowicz}, A. and {Harmanen}, J. and {Heintz}, K.~E. and {Hernandez}, M. -S. and {Hodgkin}, S.~T. and {Hook}, I.~M. and {Izzo}, L. and {James}, P.~A. and {Jonker}, P.~G. and {Kerzendorf}, W.~E. and {Kostrzewa-Rutkowska}, Z. and {Kromer}, M. and {Kuncarayakti}, H. and {Lawrence}, A. and {Manulis}, I. and {Mattila}, S. and {McBrien}, O. and {M{\"u}ller}, A. and {Nordin}, J. and {O'Neill}, D. and {Onori}, F. and {Palmerio}, J.~T. and {Pastorello}, A. and {Patat}, F. and {Pignata}, G. and {Podsiadlowski}, P. and {Razza}, A. and {Reynolds}, T. and {Roy}, R. and {Ruiter}, A.~J. and {Rybicki}, K.~A. and {Salmon}, L. and {Pumo}, M.~L. and {Prentice}, S.~J. and {Seitenzahl}, I.~R. and {Smith}, M. and {Sollerman}, J. and {Sullivan}, M. and {Szegedi}, H. and {Taddia}, F. and {Taubenberger}, S. and {Terreran}, G. and {Van Soelen}, B. and {Vos}, J. and {Walton}, N.~A. and {Wright}, D.~E. and {Wyrzykowski}, {\L}. and {Yaron}, O. and {pre=''(''>ePESSTO}, <author and {Chen}, T. -W. and {Kr{\"u}hler}, T. and {Schady}, P. and {Wiseman}, P. and {Greiner}, J. and {Rau}, A. and {Schweyer}, T. and {Klose}, S. and {Nicuesa Guelbenzu}, A. and {GROND} and {Palliyaguru}, N.~T. and {Tech University}, Texas and {Shara}, M.~M. and {Williams}, T. and {Vaisanen}, P. and {Potter}, S.~B. and {Romero Colmenero}, E. and {Crawford}, S. and {Buckley}, D.~A.~H. and {Mao}, J. and {SALT Group} and {D{\'\i}az}, M.~C. and {Macri}, L.~M. and {Garc{\'\i}a Lambas}, D. and {Mendes de Oliveira}, C. and {Nilo Castell{\'o}n}, J.~L. and {Ribeiro}, T. and {S{\'a}nchez}, B. and {Schoenell}, W. and {Abramo}, L.~R. and {Akras}, S. and {Alcaniz}, J.~S. and {Artola}, R. and {Beroiz}, M. and {Bonoli}, S. and {Cabral}, J. and {Camuccio}, R. and {Chavushyan}, V. and {Coelho}, P. and {Colazo}, C. and {Costa-Duarte}, M.~V. and {Cuevas Larenas}, H. and {Dom{\'\i}nguez Romero}, M. and {Dultzin}, D. and {Fern{\'a}ndez}, D. and {Garc{\'\i}a}, J. and {Girardini}, C. and {Gon{\c{c}}alves}, D.~R. and {Gon{\c{c}}alves}, T.~S. and {Gurovich}, S. and {Jim{\'e}nez-Teja}, Y. and {Kanaan}, A. and {Lares}, M. and {Lopes de Oliveira}, R. and {L{\'o}pez-Cruz}, O. and {Melia}, R. and {Molino}, A. and {Padilla}, N. and {Pe{\~n}uela}, T. and {Placco}, V.~M. and {Qui{\~n}ones}, C. and {Ram{\'\i}rez Rivera}, A. and {Renzi}, V. and {Riguccini}, L. and {R{\'\i}os-L{\'o}pez}, E. and {Rodriguez}, H. and {Sampedro}, L. and {Schneiter}, M. and {Sodr{\'e}}, L. and {Starck}, M. and {Torres-Flores}, S. and {Tornatore}, M. and {Zadro{\.z}ny}, A. and {Castillo}, M. and {TOROS: Transient Robotic Observatory of South Collaboration} and {Castro-Tirado}, A.~J. and {Tello}, J.~C. and {Hu}, Y. -D. and {Zhang}, B. -B. and {Cunniffe}, R. and {Castell{\'o}n}, A. and {Hiriart}, D. and {Caballero-Garc{\'\i}a}, M.~D. and {Jel{\'\i}nek}, M. and {Kub{\'a}nek}, P. and {P{\'e}rez del Pulgar}, C. and {Park}, I.~H. and {Jeong}, S. and {Castro Cer{\'o}n}, J.~M. and {Pandey}, S.~B. and {Yock}, P.~C. and {Querel}, R. and {Fan}, Y. and {Wang}, C. and {BOOTES Collaboration} and {Beardsley}, A. and {Brown}, I.~S. and {Crosse}, B. and {Emrich}, D. and {Franzen}, T. and {Gaensler}, B.~M. and {Horsley}, L. and {Johnston-Hollitt}, M. and {Kenney}, D. and {Morales}, M.~F. and {Pallot}, D. and {Sokolowski}, M. and {Steele}, K. and {Tingay}, S.~J. and {Trott}, C.~M. and {Walker}, M. and {Wayth}, R. and {Williams}, A. and {Wu}, C. and {Murchison Widefield Array}, MWA: and {Yoshida}, A. and {Sakamoto}, T. and {Kawakubo}, Y. and {Yamaoka}, K. and {Takahashi}, I. and {Asaoka}, Y. and {Ozawa}, S. and {Torii}, S. and {Shimizu}, Y. and {Tamura}, T. and {Ishizaki}, W. and {Cherry}, M.~L. and {Ricciarini}, S. and {Penacchioni}, A.~V. and {Marrocchesi}, P.~S. and {CALET Collaboration} and {Pozanenko}, A.~S. and {Volnova}, A.~A. and {Mazaeva}, E.~D. and {Minaev}, P. Yu. and {Krugov}, M.~A. and {Kusakin}, A.~V. and {Reva}, I.~V. and {Moskvitin}, A.~S. and {Rumyantsev}, V.~V. and {Inasaridze}, R. and {Klunko}, E.~V. and {Tungalag}, N. and {Schmalz}, S.~E. and {Burhonov}, O. and {IKI-GW Follow-up Collaboration} and {Abdalla}, H. and {Abramowski}, A. and {Aharonian}, F. and {Ait Benkhali}, F. and {Ang{\"u}ner}, E.~O. and {Arakawa}, M. and {Arrieta}, M. and {Aubert}, P. and {Backes}, M. and {Balzer}, A. and {Barnard}, M. and {Becherini}, Y. and {Becker Tjus}, J. and {Berge}, D. and {Bernhard}, S. and {Bernl{\"o}hr}, K. and {Blackwell}, R. and {B{\"o}ttcher}, M. and {Boisson}, C. and {Bolmont}, J. and {Bonnefoy}, S. and {Bordas}, P. and {Bregeon}, J. and {Brun}, F. and {Brun}, P. and {Bryan}, M. and {B{\"u}chele}, M. and {Bulik}, T. and {Capasso}, M. and {Caroff}, S. and {Carosi}, A. and {Casanova}, S. and {Cerruti}, M. and {Chakraborty}, N. and {Chaves}, R.~C.~G. and {Chen}, A. and {Chevalier}, J. and {Colafrancesco}, S. and {Condon}, B. and {Conrad}, J. and {Davids}, I.~D. and {Decock}, J. and {Deil}, C. and {Devin}, J. and {deWilt}, P. and {Dirson}, L. and {Djannati-Ata{\"\i}}, A. and {Donath}, A. and {O'C. Drury}, L. and {Dutson}, K. and {Dyks}, J. and {Edwards}, T. and {Egberts}, K. and {Emery}, G. and {Ernenwein}, J. -P. and {Eschbach}, S. and {Farnier}, C. and {Fegan}, S. and {Fernandes}, M.~V. and {Fiasson}, A. and {Fontaine}, G. and {Funk}, S. and {F{\"u}ssling}, M. and {Gabici}, S. and {Gallant}, Y.~A. and {Garrigoux}, T. and {Gat{\'e}}, F. and {Giavitto}, G. and {Giebels}, B. and {Glawion}, D. and {Glicenstein}, J.~F. and {Gottschall}, D. and {Grondin}, M. -H. and {Hahn}, J. and {Haupt}, M. and {Hawkes}, J. and {Heinzelmann}, G. and {Henri}, G. and {Hermann}, G. and {Hinton}, J.~A. and {Hofmann}, W. and {Hoischen}, C. and {Holch}, T.~L. and {Holler}, M. and {Horns}, D. and {Ivascenko}, A. and {Iwasaki}, H. and {Jacholkowska}, A. and {Jamrozy}, M. and {Jankowsky}, D. and {Jankowsky}, F. and {Jingo}, M. and {Jouvin}, L. and {Jung-Richardt}, I. and {Kastendieck}, M.~A. and {Katarzy{\'n}ski}, K. and {Katsuragawa}, M. and {Kerszberg}, D. and {Khangulyan}, D. and {Kh{\'e}lifi}, B. and {King}, J. and {Klepser}, S. and {Klochkov}, D. and {Klu{\'z}niak}, W. and {Komin}, Nu. and {Kosack}, K. and {Krakau}, S. and {Kraus}, M. and {Kr{\"u}ger}, P.~P. and {Laffon}, H. and {Lamanna}, G. and {Lau}, J. and {Lees}, J. -P. and {Lefaucheur}, J. and {Lemi{\`e}re}, A. and {Lemoine-Goumard}, M. and {Lenain}, J. -P. and {Leser}, E. and {Lohse}, T. and {Lorentz}, M. and {Liu}, R. and {Lypova}, I. and {Malyshev}, D. and {Marandon}, V. and {Marcowith}, A. and {Mariaud}, C. and {Marx}, R. and {Maurin}, G. and {Maxted}, N. and {Mayer}, M. and {Meintjes}, P.~J. and {Meyer}, M. and {Mitchell}, A.~M.~W. and {Moderski}, R. and {Mohamed}, M. and {Mohrmann}, L. and {Mor{\r{a}}}, K. and {Moulin}, E. and {Murach}, T. and {Nakashima}, S. and {de Naurois}, M. and {Ndiyavala}, H. and {Niederwanger}, F. and {Niemiec}, J. and {Oakes}, L. and {O'Brien}, P. and {Odaka}, H. and {Ohm}, S. and {Ostrowski}, M. and {Oya}, I. and {Padovani}, M. and {Panter}, M. and {Parsons}, R.~D. and {Pekeur}, N.~W. and {Pelletier}, G. and {Perennes}, C. and {Petrucci}, P. -O. and {Peyaud}, B. and {Piel}, Q. and {Pita}, S. and {Poireau}, V. and {Poon}, H. and {Prokhorov}, D. and {Prokoph}, H. and {P{\"u}hlhofer}, G. and {Punch}, M. and {Quirrenbach}, A. and {Raab}, S. and {Rauth}, R. and {Reimer}, A. and {Reimer}, O. and {Renaud}, M. and {de los Reyes}, R. and {Rieger}, F. and {Rinchiuso}, L. and {Romoli}, C. and {Rowell}, G. and {Rudak}, B. and {Rulten}, C.~B. and {Sahakian}, V. and {Saito}, S. and {Sanchez}, D.~A. and {Santangelo}, A. and {Sasaki}, M. and {Schlickeiser}, R. and {Sch{\"u}ssler}, F. and {Schulz}, A. and {Schwanke}, U. and {Schwemmer}, S. and {Seglar-Arroyo}, M. and {Settimo}, M. and {Seyffert}, A.~S. and {Shafi}, N. and {Shilon}, I. and {Shiningayamwe}, K. and {Simoni}, R. and {Sol}, H. and {Spanier}, F. and {Spir-Jacob}, M. and {Stawarz}, {\L}. and {Steenkamp}, R. and {Stegmann}, C. and {Steppa}, C. and {Sushch}, I. and {Takahashi}, T. and {Tavernet}, J. -P. and {Tavernier}, T. and {Taylor}, A.~M. and {Terrier}, R. and {Tibaldo}, L. and {Tiziani}, D. and {Tluczykont}, M. and {Trichard}, C. and {Tsirou}, M. and {Tsuji}, N. and {Tuffs}, R. and {Uchiyama}, Y. and {van der Walt}, D.~J. and {van Eldik}, C. and {van Rensburg}, C. and {van Soelen}, B. and {Vasileiadis}, G. and {Veh}, J. and {Venter}, C. and {Viana}, A. and {Vincent}, P. and {Vink}, J. and {Voisin}, F. and {V{\"o}lk}, H.~J. and {Vuillaume}, T. and {Wadiasingh}, Z. and {Wagner}, S.~J. and {Wagner}, P. and {Wagner}, R.~M. and {White}, R. and {Wierzcholska}, A. and {Willmann}, P. and {W{\"o}rnlein}, A. and {Wouters}, D. and {Yang}, R. and {Zaborov}, D. and {Zacharias}, M. and {Zanin}, R. and {Zdziarski}, A.~A. and {Zech}, A. and {Zefi}, F. and {Ziegler}, A. and {Zorn}, J. and {{\.Z}ywucka}, N. and {H.~E.~S.~S. Collaboration} and {Fender}, R.~P. and {Broderick}, J.~W. and {Rowlinson}, A. and {Wijers}, R.~A.~M.~J. and {Stewart}, A.~J. and {ter Veen}, S. and {Shulevski}, A. and {LOFAR Collaboration} and {Kavic}, M. and {Simonetti}, J.~H. and {League}, C. and {Tsai}, J. and {Obenberger}, K.~S. and {Nathaniel}, K. and {Taylor}, G.~B. and {Dowell}, J.~D. and {Liebling}, S.~L. and {Estes}, J.~A. and {Lippert}, M. and {Sharma}, I. and {Vincent}, P. and {Farella}, B. and {Wavelength Array}, LWA: Long and {Abeysekara}, A.~U. and {Albert}, A. and {Alfaro}, R. and {Alvarez}, C. and {Arceo}, R. and {Arteaga-Vel{\'a}zquez}, J.~C. and {Avila Rojas}, D. and {Ayala Solares}, H.~A. and {Barber}, A.~S. and {Becerra Gonzalez}, J. and {Becerril}, A. and {Belmont-Moreno}, E. and {BenZvi}, S.~Y. and {Berley}, D. and {Bernal}, A. and {Braun}, J. and {Brisbois}, C. and {Caballero-Mora}, K.~S. and {Capistr{\'a}n}, T. and {Carrami{\~n}ana}, A. and {Casanova}, S. and {Castillo}, M. and {Cotti}, U. and {Cotzomi}, J. and {Couti{\~n}o de Le{\'o}n}, S. and {De Le{\'o}n}, C. and {De la Fuente}, E. and {Diaz Hernandez}, R. and {Dichiara}, S. and {Dingus}, B.~L. and {DuVernois}, M.~A. and {D{\'\i}az-V{\'e}lez}, J.~C. and {Ellsworth}, R.~W. and {Engel}, K. and {Enr{\'\i}quez-Rivera}, O. and {Fiorino}, D.~W. and {Fleischhack}, H. and {Fraija}, N. and {Garc{\'\i}a-Gonz{\'a}lez}, J.~A. and {Garfias}, F. and {Gerhardt}, M. and {Gonz{\~o}lez Mu{\~n}oz}, A. and {Gonz{\'a}lez}, M.~M. and {Goodman}, J.~A. and {Hampel-Arias}, Z. and {Harding}, J.~P. and {Hernandez}, S. and {Hernandez-Almada}, A. and {Hona}, B. and {H{\"u}ntemeyer}, P. and {Iriarte}, A. and {Jardin-Blicq}, A. and {Joshi}, V. and {Kaufmann}, S. and {Kieda}, D. and {Lara}, A. and {Lauer}, R.~J. and {Lennarz}, D. and {Le{\'o}n Vargas}, H. and {Linnemann}, J.~T. and {Longinotti}, A.~L. and {Raya}, G. Luis and {Luna-Garc{\'\i}a}, R. and {L{\'o}pez-Coto}, R. and {Malone}, K. and {Marinelli}, S.~S. and {Martinez}, O. and {Martinez-Castellanos}, I. and {Mart{\'\i}nez-Castro}, J. and {Mart{\'\i}nez-Huerta}, H. and {Matthews}, J.~A. and {Miranda-Romagnoli}, P. and {Moreno}, E. and {Mostaf{\'a}}, M. and {Nellen}, L. and {Newbold}, M. and {Nisa}, M.~U. and {Noriega-Papaqui}, R. and {Pelayo}, R. and {Pretz}, J. and {P{\'e}rez-P{\'e}rez}, E.~G. and {Ren}, Z. and {Rho}, C.~D. and {Rivi{\`e}re}, C. and {Rosa-Gonz{\'a}lez}, D. and {Rosenberg}, M. and {Ruiz-Velasco}, E. and {Salazar}, H. and {Salesa Greus}, F. and {Sandoval}, A. and {Schneider}, M. and {Schoorlemmer}, H. and {Sinnis}, G. and {Smith}, A.~J. and {Springer}, R.~W. and {Surajbali}, P. and {Tibolla}, O. and {Tollefson}, K. and {Torres}, I. and {Ukwatta}, T.~N. and {Weisgarber}, T. and {Westerhoff}, S. and {Wisher}, I.~G. and {Wood}, J. and {Yapici}, T. and {Yodh}, G.~B. and {Younk}, P.~W. and {Zhou}, H. and {{\'A}lvarez}, J.~D. and {HAWC Collaboration} and {Aab}, A. and {Abreu}, P. and {Aglietta}, M. and {Albuquerque}, I.~F.~M. and {Albury}, J.~M. and {Allekotte}, I. and {Almela}, A. and {Alvarez Castillo}, J. and {Alvarez-Mu{\~n}iz}, J. and {Anastasi}, G.~A. and {Anchordoqui}, L. and {Andrada}, B. and {Andringa}, S. and {Aramo}, C. and {Arsene}, N. and {Asorey}, H. and {Assis}, P. and {Avila}, G. and {Badescu}, A.~M. and {Balaceanu}, A. and {Barbato}, F. and {Barreira Luz}, R.~J. and {Becker}, K.~H. and {Bellido}, J.~A. and {Berat}, C. and {Bertaina}, M.~E. and {Bertou}, X. and {Biermann}, P.~L. and {Biteau}, J. and {Blaess}, S.~G. and {Blanco}, A. and {Blazek}, J. and {Bleve}, C. and {Boh{\'a}{\v{c}}ov{\'a}}, M. and {Bonifazi}, C. and {Borodai}, N. and {Botti}, A.~M. and {Brack}, J. and {Brancus}, I. and {Bretz}, T. and {Bridgeman}, A. and {Briechle}, F.~L. and {Buchholz}, P. and {Bueno}, A. and {Buitink}, S. and {Buscemi}, M. and {Caballero-Mora}, K.~S. and {Caccianiga}, L. and {Cancio}, A. and {Canfora}, F. and {Caruso}, R. and {Castellina}, A. and {Catalani}, F. and {Cataldi}, G. and {Cazon}, L. and {Chavez}, A.~G. and {Chinellato}, J.~A. and {Chudoba}, J. and {Clay}, R.~W. and {Cobos Cerutti}, A.~C. and {Colalillo}, R. and {Coleman}, A. and {Collica}, L. and {Coluccia}, M.~R. and {Concei{\c{c}}{\~a}o}, R. and {Consolati}, G. and {Contreras}, F. and {Cooper}, M.~J. and {Coutu}, S. and {Covault}, C.~E. and {Cronin}, J. and {D'Amico}, S. and {Daniel}, B. and {Dasso}, S. and {Daumiller}, K. and {Dawson}, B.~R. and {Day}, J.~A. and {de Almeida}, R.~M. and {de Jong}, S.~J. and {De Mauro}, G. and {de Mello Neto}, J.~R.~T. and {De Mitri}, I. and {de Oliveira}, J. and {de Souza}, V. and {Debatin}, J. and {Deligny}, O. and {D{\'\i}az Castro}, M.~L. and {Diogo}, F. and {Dobrigkeit}, C. and {D'Olivo}, J.~C. and {Dorosti}, Q. and {Dos Anjos}, R.~C. and {Dova}, M.~T. and {Dundovic}, A. and {Ebr}, J. and {Engel}, R. and {Erdmann}, M. and {Erfani}, M. and {Escobar}, C.~O. and {Espadanal}, J. and {Etchegoyen}, A. and {Falcke}, H. and {Farmer}, J. and {Farrar}, G. and {Fauth}, A.~C. and {Fazzini}, N. and {Feldbusch}, F. and {Fenu}, F. and {Fick}, B. and {Figueira}, J.~M. and {Filip{\v{c}}i{\v{c}}}, A. and {Freire}, M.~M. and {Fujii}, T. and {Fuster}, A. and {Ga{\"\i}or}, R. and {Garc{\'\i}a}, B. and {Gat{\'e}}, F. and {Gemmeke}, H. and {Gherghel-Lascu}, A. and {Ghia}, P.~L. and {Giaccari}, U. and {Giammarchi}, M. and {Giller}, M. and {G{\l}as}, D. and {Glaser}, C. and {Golup}, G. and {G{\'o}mez Berisso}, M. and {G{\'o}mez Vitale}, P.~F. and {Gonz{\'a}lez}, N. and {Gorgi}, A. and {Gottowik}, M. and {Grillo}, A.~F. and {Grubb}, T.~D. and {Guarino}, F. and {Guedes}, G.~P. and {Halliday}, R. and {Hampel}, M.~R. and {Hansen}, P. and {Harari}, D. and {Harrison}, T.~A. and {Harvey}, V.~M. and {Haungs}, A. and {Hebbeker}, T. and {Heck}, D. and {Heimann}, P. and {Herve}, A.~E. and {Hill}, G.~C. and {Hojvat}, C. and {Holt}, E. and {Homola}, P. and {H{\"o}randel}, J.~R. and {Horvath}, P. and {Hrabovsk{\'y}}, M. and {Huege}, T. and {Hulsman}, J. and {Insolia}, A. and {Isar}, P.~G. and {Jandt}, I. and {Johnsen}, J.~A. and {Josebachuili}, M. and {Jurysek}, J. and {K{\"a}{\"a}p{\"a}}, A. and {Kampert}, K.~H. and {Keilhauer}, B. and {Kemmerich}, N. and {Kemp}, J. and {Kieckhafer}, R.~M. and {Klages}, H.~O. and {Kleifges}, M. and {Kleinfeller}, J. and {Krause}, R. and {Krohm}, N. and {Kuempel}, D. and {Kukec Mezek}, G. and {Kunka}, N. and {Kuotb Awad}, A. and {Lago}, B.~L. and {LaHurd}, D. and {Lang}, R.~G. and {Lauscher}, M. and {Legumina}, R. and {Leigui de Oliveira}, M.~A. and {Letessier-Selvon}, A. and {Lhenry-Yvon}, I. and {Link}, K. and {Lo Presti}, D. and {Lopes}, L. and {L{\'o}pez}, R. and {L{\'o}pez Casado}, A. and {Lorek}, R. and {Luce}, Q. and {Lucero}, A. and {Malacari}, M. and {Mallamaci}, M. and {Mandat}, D. and {Mantsch}, P. and {Mariazzi}, A.~G. and {Maris}, I.~C. and {Marsella}, G. and {Martello}, D. and {Martinez}, H. and {Mart{\'\i}nez Bravo}, O. and {Mas{\'\i}as Meza}, J.~J. and {Mathes}, H.~J. and {Mathys}, S. and {Matthews}, J. and {Matthiae}, G. and {Mayotte}, E. and {Mazur}, P.~O. and {Medina}, C. and {Medina-Tanco}, G. and {Melo}, D. and {Menshikov}, A. and {Merenda}, K. -D. and {Michal}, S. and {Micheletti}, M.~I. and {Middendorf}, L. and {Miramonti}, L. and {Mitrica}, B. and {Mockler}, D. and {Mollerach}, S. and {Montanet}, F. and {Morello}, C. and {Morlino}, G. and {M{\"u}ller}, A.~L. and {M{\"u}ller}, G. and {Muller}, M.~A. and {M{\"u}ller}, S. and {Mussa}, R. and {Naranjo}, I. and {Nguyen}, P.~H. and {Niculescu-Oglinzanu}, M. and {Niechciol}, M. and {Niemietz}, L. and {Niggemann}, T. and {Nitz}, D. and {Nosek}, D. and {Novotny}, V. and {No{\v{z}}ka}, L. and {N{\'u}{\~n}ez}, L.~A. and {Oikonomou}, F. and {Olinto}, A. and {Palatka}, M. and {Pallotta}, J. and {Papenbreer}, P. and {Parente}, G. and {Parra}, A. and {Paul}, T. and {Pech}, M. and {Pedreira}, F. and {P{\c{e}}kala}, J. and {Pe{\~n}a-Rodriguez}, J. and {Pereira}, L.~A.~S. and {Perlin}, M. and {Perrone}, L. and {Peters}, C. and {Petrera}, S. and {Phuntsok}, J. and {Pierog}, T. and {Pimenta}, M. and {Pirronello}, V. and {Platino}, M. and {Plum}, M. and {Poh}, J. and {Porowski}, C. and {Prado}, R.~R. and {Privitera}, P. and {Prouza}, M. and {Quel}, E.~J. and {Querchfeld}, S. and {Quinn}, S. and {Ramos-Pollan}, R. and {Rautenberg}, J. and {Ravignani}, D. and {Ridky}, J. and {Riehn}, F. and {Risse}, M. and {Ristori}, P. and {Rizi}, V. and {Rodrigues de Carvalho}, W. and {Rodriguez Fernandez}, G. and {Rodriguez Rojo}, J. and {Roncoroni}, M.~J. and {Roth}, M. and {Roulet}, E. and {Rovero}, A.~C. and {Ruehl}, P. and {Saffi}, S.~J. and {Saftoiu}, A. and {Salamida}, F. and {Salazar}, H. and {Saleh}, A. and {Salina}, G. and {S{\'a}nchez}, F. and {Sanchez-Lucas}, P. and {Santos}, E.~M. and {Santos}, E. and {Sarazin}, F. and {Sarmento}, R. and {Sarmiento-Cano}, C. and {Sato}, R. and {Schauer}, M. and {Scherini}, V. and {Schieler}, H. and {Schimp}, M. and {Schmidt}, D. and {Scholten}, O. and {Schov{\'a}nek}, P. and {Schr{\"o}der}, F.~G. and {Schr{\"o}der}, S. and {Schulz}, A. and {Schumacher}, J. and {Sciutto}, S.~J. and {Segreto}, A. and {Shadkam}, A. and {Shellard}, R.~C. and {Sigl}, G. and {Silli}, G. and {{\v{S}}m{\'\i}da}, R. and {Snow}, G.~R. and {Sommers}, P. and {Sonntag}, S. and {Soriano}, J.~F. and {Squartini}, R. and {Stanca}, D. and {Stani{\v{c}}}, S. and {Stasielak}, J. and {Stassi}, P. and {Stolpovskiy}, M. and {Strafella}, F. and {Streich}, A. and {Suarez}, F. and {Suarez-Dur{\'a}n}, M. and {Sudholz}, T. and {Suomij{\"a}rvi}, T. and {Supanitsky}, A.~D. and {{\v{S}}up{\'\i}k}, J. and {Swain}, J. and {Szadkowski}, Z. and {Taboada}, A. and {Taborda}, O.~A. and {Timmermans}, C. and {Todero Peixoto}, C.~J. and {Tomankova}, L. and {Tom{\'e}}, B. and {Torralba Elipe}, G. and {Travnicek}, P. and {Trini}, M. and {Tueros}, M. and {Ulrich}, R. and {Unger}, M. and {Urban}, M. and {Vald{\'e}s Galicia}, J.~F. and {Vali{\~n}o}, I. and {Valore}, L. and {van Aar}, G. and {van Bodegom}, P. and {van den Berg}, A.~M. and {van Vliet}, A. and {Varela}, E. and {Vargas C{\'a}rdenas}, B. and {V{\'a}zquez}, R.~A. and {Veberi{\v{c}}}, D. and {Ventura}, C. and {Vergara Quispe}, I.~D. and {Verzi}, V. and {Vicha}, J. and {Villase{\~n}or}, L. and {Vorobiov}, S. and {Wahlberg}, H. and {Wainberg}, O. and {Walz}, D. and {Watson}, A.~A. and {Weber}, M. and {Weindl}, A. and {Wiede{\'n}ski}, M. and {Wiencke}, L. and {Wilczy{\'n}ski}, H. and {Wirtz}, M. and {Wittkowski}, D. and {Wundheiler}, B. and {Yang}, L. and {Yushkov}, A. and {Zas}, E. and {Zavrtanik}, D. and {Zavrtanik}, M. and {Zepeda}, A. and {Zimmermann}, B. and {Ziolkowski}, M. and {Zong}, Z. and {Zuccarello}, F. and {Pierre Auger Collaboration} and {Kim}, S. and {Schulze}, S. and {Bauer}, F.~E. and {Corral-Santana}, J.~M. and {de Gregorio-Monsalvo}, I. and {Gonz{\'a}lez-L{\'o}pez}, J. and {Hartmann}, D.~H. and {Ishwara-Chandra}, C.~H. and {Mart{\'\i}n}, S. and {Mehner}, A. and {Misra}, K. and {Micha{\l}owski}, M.~J. and {Resmi}, L. and {ALMA Collaboration} and {Paragi}, Z. and {Agudo}, I. and {An}, T. and {Beswick}, R. and {Casadio}, C. and {Frey}, S. and {Jonker}, P. and {Kettenis}, M. and {Marcote}, B. and {Moldon}, J. and {Szomoru}, A. and {van Langevelde}, H.~J. and {Yang}, J. and {Euro VLBI Team} and {Cwiek}, A. and {Cwiok}, M. and {Czyrkowski}, H. and {Dabrowski}, R. and {Kasprowicz}, G. and {Mankiewicz}, L. and {Nawrocki}, K. and {Opiela}, R. and {Piotrowski}, L.~W. and {Wrochna}, G. and {Zaremba}, M. and {{\.Z}arnecki}, A.~F. and {Pi of Sky Collaboration} and {Haggard}, D. and {Nynka}, M. and {Ruan}, J.~J. and {Chandra Team at McGill University} and {Bland}, P.~A. and {Booler}, T. and {Devillepoix}, H.~A.~R. and {de Gois}, J.~S. and {Hancock}, P.~J. and {Howie}, R.~M. and {Paxman}, J. and {Sansom}, E.~K. and {Towner}, M.~C. and {Desert Fireball Network}, DFN: and {Tonry}, J. and {Coughlin}, M. and {Stubbs}, C.~W. and {Denneau}, L. and {Heinze}, A. and {Stalder}, B. and {Weiland}, H. and {ATLAS} and {Eatough}, R.~P. and {Kramer}, M. and {Kraus}, A. and {Time Resolution Universe Survey}, High and {Troja}, E. and {Piro}, L. and {Becerra Gonz{\'a}lez}, J. and {Butler}, N.~R. and {Fox}, O.~D. and {Khandrika}, H.~G. and {Kutyrev}, A. and {Lee}, W.~H. and {Ricci}, R. and {Ryan}, R.~E., Jr. and {S{\'a}nchez-Ram{\'\i}rez}, R. and {Veilleux}, S. and {Watson}, A.~M. and {Wieringa}, M.~H. and {Burgess}, J.~M. and {van Eerten}, H. and {Fontes}, C.~J. and {Fryer}, C.~L. and {Korobkin}, O. and {Wollaeger}, R.~T. and {RIMAS} and {RATIR} and {Camilo}, F. and {Foley}, A.~R. and {Goedhart}, S. and {Makhathini}, S. and {Oozeer}, N. and {Smirnov}, O.~M. and {Fender}, R.~P. and {Woudt}, P.~A. and {South Africa/MeerKAT}, SKA},
        title = "{Multi-messenger Observations of a Binary Neutron Star Merger}",
      journal = {\apjl},
         year = 2017,
        month = oct,
       volume = {848},
       number = {2},
          eid = {L12},
        pages = {L12},
          doi = {10.3847/2041-8213/aa91c9},
archivePrefix = {arXiv},
       eprint = {1710.05833},
 primaryClass = {astro-ph.HE},
       adsurl = {https://ui.adsabs.harvard.edu/abs/2017ApJ...848L..12A}
}

@ARTICLE{harris2020numpy,
  author = {{Harris}, Charles R. and {Millman}, K. Jarrod and
            {van der Walt}, St{\'e}fan J. and {Gommers}, Ralf and
            {Virtanen}, Pauli and {Cournapeau}, David and {Wieser}, Eric and
            {Taylor}, Julian and {Berg}, Sebastian and {Smith}, Nathaniel J. and
            {Kern}, Robert and {Picus}, Matti and {Hoyer}, Stephan and
            {van Kerkwijk}, Marten H. and {Brett}, Matthew and {Haldane}, Allan and
            {Fern{\'a}ndez del R{\'i}o}, Jaime and {Wiebe}, Mark and
            {Peterson}, Pearu and {G{\'e}rard-Marchant}, Pierre and
            {Sheppard}, Kevin and {Reddy}, Tyler and {Weckesser}, Warren and
            {Abbasi}, Hameer and {Gohlke}, Christoph and {Oliphant}, Travis E.},
  title = "{Array programming with {NumPy}}",
  journal = {Nature},
  year = 2020,
  volume = {585},
  pages = {357--362},
  doi = {10.1038/s41586-020-2649-2}
}

@INPROCEEDINGS{mckinney2010pandas,
  author = {{McKinney}, Wes},
  title = "{Data Structures for Statistical Computing in Python}",
  booktitle = {Proceedings of the 9th Python in Science Conference},
  editor = {{van der Walt}, St{\'e}fan and {Millman}, Jarrod},
  year = 2010,
  pages = {51--56},
  doi = {10.25080/Majora-92bf1922-00a}
}

@ARTICLE{hunter2007matplotlib,
  author = {{Hunter}, John D.},
  title = "{Matplotlib: A 2D Graphics Environment}",
  journal = {Computing in Science \& Engineering},
  year = 2007,
  volume = {9},
  number = {3},
  pages = {90--95},
  doi = {10.1109/MCSE.2007.55}
}

@ARTICLE{virtanen2020scipy,
  author = {{Virtanen}, Pauli and {Gommers}, Ralf and {Oliphant}, Travis E. and
            {Haberland}, Matt and {Reddy}, Tyler and {Cournapeau}, David and
            {Burovski}, Evgeni and {Peterson}, Pearu and {Weckesser}, Warren and
            {Bright}, Jonathan and {van der Walt}, St{\'e}fan J. and
            {Brett}, Matthew and {Wilson}, Joshua and {Millman}, K. Jarrod and
            {Mayorov}, Nikolay and {Nelson}, Andrew R. J. and {Jones}, Eric and
            {Kern}, Robert and {Larson}, Eric and {Carey}, C. J. and
            {Polat}, {\.I}lhan and {Feng}, Yu and {Moore}, Eric W. and
            {VanderPlas}, Jake and {Laxalde}, Denis and {Perktold}, Josef and
            {Cimrman}, Robert and {Henriksen}, Ian and {Quintero}, E. A. and
            {Harris}, Charles R. and {Archibald}, Anne M. and
            {Ribeiro}, Ant{\^o}nio H. and {Pedregosa}, Fabian and
            {van Mulbregt}, Paul and {SciPy 1.0 Contributors}},
  title = "{{SciPy} 1.0: Fundamental Algorithms for Scientific Computing in Python}",
  journal = {Nature Methods},
  year = 2020,
  volume = {17},
  number = {3},
  pages = {261--272},
  doi = {10.1038/s41592-019-0686-2}
}

@ARTICLE{abrilpla2023pymc,
  author = {{Abril-Pla}, Oriol and {Andreani}, Virgile and {Carroll}, Colin and
            {Dong}, Larry and {Fonnesbeck}, Christopher J. and {Kochurov}, Maxim and
            {Kumar}, Ravin and {Lao}, Junpeng and {Luhmann}, Christian C. and
            {Martin}, Osvaldo A. and {Osthege}, Michael and {Vieira}, Ricardo and
            {Wiecki}, Thomas and {Zinkov}, Robert},
  title = "{{PyMC}: A Modern and Comprehensive Probabilistic Programming Framework in Python}",
  journal = {PeerJ Computer Science},
  year = 2023,
  volume = {9},
  pages = {e1516},
  doi = {10.7717/peerj-cs.1516}
}

@ARTICLE{kumar2019arviz,
  author = {{Kumar}, Ravin and {Carroll}, Colin and {Hartikainen}, Ari and {Martin}, Osvaldo},
  title = "{ArviZ: A Unified Library for Exploratory Analysis of Bayesian Models in Python}",
  journal = {Journal of Open Source Software},
  year = 2019,
  volume = {4},
  number = {33},
  pages = {1143},
  doi = {10.21105/joss.01143}
}
\end{document}